\documentclass[12pt]{article}
\usepackage{amssymb,amsmath}
\makeatletter

\usepackage{arydshln}

\newcount\comment@nesting

\begingroup
\xdef\comment@begincomment{\string\\begin\string\{comment\string\}}
\xdef\comment@endcomment{\string\\end\string\{comment\string\}}
\endgroup

\begingroup
\catcode`\^^M=\active
\def\@temp{\endgroup\def\comment@processline##1^^M}%
\@temp{%
    \def\comment@curline{#1}%
    \let\@next=\comment@processline
    \ifx\comment@curline\comment@endcomment
        \ifnum\comment@nesting=0
            \def\@next{\end{comment}}%
        \else
            \global\advance\comment@nesting by -1
        \fi
    \else
        \ifx\comment@curline\comment@begincomment
            \global\advance\comment@nesting by 1
        \fi
    \fi
    \@next
}

\makeatother

\usepackage[dvipdfmx]{graphicx}
\usepackage{bm}
\usepackage{color}
\usepackage{here}
\usepackage{enumerate}
\usepackage{here}
\usepackage{tikz}
\usetikzlibrary{arrows.meta,positioning,decorations.pathmorphing,calc,shapes.geometric}
\usepackage[vcentermath]{youngtab}
\usepackage{subfigure}
\usepackage{mathrsfs}
\usepackage{dsfont}
\usepackage{tikz}
\usepackage{tikz-cd}
\usepackage{hyperref}

\DeclareMathOperator*{\Tr}{{\rm Tr}}

\numberwithin{equation}{section}

\allowdisplaybreaks[1]

\definecolor{mygreen}{rgb}{0,0.714,0.286}

\begin{document}

\thispagestyle{empty}

\begin{flushright}
YITP-26-26

\end{flushright}
\vskip1.5cm
\begin{center}
{\Large \bf 
Holographic spectral functions\\
for Sasaki-Einstein 5-manifolds \\
}

\vskip1.5cm
Yu Nakayama\footnote{yu.nakayama@yukawa.kyoto-u.ac.jp} 
and Tadashi Okazaki\footnote{tokazaki@seu.edu.cn}

\bigskip
{\it Center for Gravitational Physics and Quantum Information, Yukawa Institute for Theoretical Physics, Kyoto University,\\
	Kitashirakawa Oiwakecho, Sakyo-ku, Kyoto 606-8502, Japan

School of Physics and Shing-Tung Yau Center, Southeast University,\\
Yifu Architecture Building, No.2 Sipailou, Xuanwu district, \\
Nanjing, Jiangsu, 210096, China

}

\end{center}

\vskip1cm
\begin{abstract}
We investigate holographic spectral functions for general Sasaki-Einstein 5-manifolds dual to four-dimensional superconformal field theories, 
including supersymmetric indices, supersymmetric zeta functions, and supersymmetric determinants.
The analytic structure of the supersymmetric zeta function, particularly its residue and special value, 
allows for the computation of the curvature-squared integral of the Sasaki-Einstein manifold and the subleading holographic anomaly. 
The reach of this spectral framework is not restricted to toric geometries and accommodates non-toric Sasaki-Einstein manifolds. 
For toric Sasaki-Einstein manifolds, we develop a combinatorial method to compute 
the holographic spectral functions and the holographic geometric invariants directly from the toric data. 
\end{abstract}
\newpage
\setcounter{tocdepth}{3}
\tableofcontents

\section{Introduction and conclusions}
The AdS/CFT correspondence \cite{Maldacena:1997re} asserts an equivalence between Type IIB string theory on $AdS_5\times Y$ with appropriately chosen self-dual 5-form flux 
and a four-dimensional superconformal field theory (SCFT), where $Y$ is a Sasaki-Einstein 5-manifold 
\cite{Kehagias:1998gn,Klebanov:1998hh,Acharya:1998db,Morrison:1998cs}. 
In this setup, the SCFT can be realized as a quiver gauge theory 
that captures the low-energy dynamics of a stack of D3-branes probing the Calabi-Yau cone $C(Y)$ over $Y$. 
While global quantities such as central charges provide coarse tests of the correspondence, 
a more stringent probe is the spectrum of protected operators. 

In this paper we refine the spectral analysis by combining the supersymmetric indices \cite{Kinney:2005ej,Romelsberger:2005eg}
with the supersymmetric zeta functions and the supersymmetric determinants recently introduced in \cite{Nakayama:2025hzr}. 
First, while the explicit closed-form expressions for the supersymmetric gravity indices are known only in special cases in the literature \cite{Kinney:2005ej,Nakayama:2005mf,Nakayama:2006ur,Gadde:2010en,Eager:2012hx,Agarwal:2013pba,ArabiArdehali:2014otj}, 
we present an explicit combinatorial formula of the indices for general toric Sasaki-Einstein manifolds in terms of the combinatorial data of the associated toric diagrams. 
We perform explicit computations of the indices for broad classes of the toric Sasaki-Einstein manifolds, 
including the $L^{a,b,c}$ family \cite{Cvetic:2005ft,Cvetic:2005vk}, 
the complex cone over the (Pseudo) del Pezzo geometries, 
the $X^{\mathsf{p},\mathsf{q}}$ \cite{Hanany:2005hq} and $Z^{\mathsf{p},\mathsf{q}}$ \cite{Oota:2006eg} series. 
A refined understanding of the protected spectra in holography can be achieved by introducing tools that complement the indices. 
In \cite{Nakayama:2025hzr} we introduced the supersymmetric zeta functions and supersymmetric determinants as alternative spectral functions in supersymmetric theories. 
They admit systematic and controlled asymptotic analyses of the Cardy-like limit \cite{Cardy:1986ie,DiPietro:2014bca} and of the supersymmetric Casimir energies \cite{Kim:2009wb,Kim:2012ava,Assel:2014paa,Assel:2014tba,Cassani:2014zwa,Lorenzen:2014pna,Assel:2015nca,Bobev:2015kza,Martelli:2015kuk}. 

In \cite{DiPietro:2014bca} Di Pietro and Komargodski showed, using effective field theory methods, 
that the leading behavior of the Cardy-like limit of the 4d superconformal index is universally governed by the difference of central charges. 
In contrast, in \cite{Nakayama:2025hzr} we developed a framework 
that derives the Cardy-like asymptotics directly from the analytic properties of the supersymmetric zeta function and determinant, 
without relying on the existence of an effective field theory. 
In the present work we apply this framework to analyze the Cardy-like limit of the multi-particle gravity index. 
Although perhaps not widely emphasized, the asymptotic formula does not agree with the one proposed by Di Pietro and Komargodski 
because the large $N$ limit and the high-temperature limit do not commute here. 
However, the leading coefficients in the asymptotic expansions of the multi-particle gravity indices are conjectured to 
correspond to geometric invariants of the internal Sasaki-Einstein manifold $Y$ \cite{ArabiArdehali:2015iow}. 
Combining our general formula for the supersymmetric zeta function with the conjectured identification of the leading asymptotic coefficients, 
we are led to propose that $\int_Y \mathrm{Riem}^2$ is determined by the residue of the supersymmetric zeta function at $s=1$ for any Sasaki-Einstein 5-manifold. 
For toric geometries the residue is itself combinatorial, 
and the proposal specializes to a closed formula for $\int_Y \mathrm{Riem}^2$ depending solely on  discrete data in the associated toric diagram. 

Although the invariant $\int_Y \mathrm{Riem}^2$ is in principle computable once an explicit metric is known, performing the integral is typically highly non-trivial in practice. 
As non-trivial evidence for the validity of our formula, we have carried out several independent checks. 
First we verify that our formula yields the curvature-squared invariants for $S^5/\mathbb{Z}_k$. 
A crucial subtlety is that the value of $\int_{S^5/\mathbb{Z}_k} \mathrm{Riem}^2$ is not determined solely by the group order $k$. 
It depends crucially on (i) the supersymmetry class of the embedding of $\mathbb{Z}_k$ $\subset$ $SO(6)$ 
and (ii) the presence or absence of an element of order two in $\mathbb{Z}_k$. 
They control the fixed-point structure and therefore the localized curvature terms. 
Only when the action is free, i.e. for $\mathcal{N}=1$ embedding and for odd $k$, 
is the integral controlled entirely by the bulk $1/k$ contribution without localized curvature contribution. 
As a non-trivial consistency check of our proposal, we confirm that 
our result agrees with the proposal in \cite{Eager:2010dk} that the invariant $\int_Y \mathrm{Riem}^2$ can be extracted from the asymptotic expansion of the Hilbert series. 
For example, for the $Y^{\mathsf{p},\mathsf{q}}$ family and for $L^{1,3,1}/\mathbb{Z}_2$, 
previously derived results in \cite{Eager:2010dk} and \cite{Bao:2024nyu} allow for a direct comparison and we confirm the agreement. 
Furthermore, we verify complete agreement with the Hilbert series computations for a wider range of examples, including complex cones over Pseudo del Pezzo surfaces. 
The corresponding Hilbert series expressions are collected in Appendix \ref{app_HS}.
To the best of our knowledge, no such expressions are currently available for 
$L^{a,b,c}$, $X^{\mathsf{p},\mathsf{q}}$ and $Z^{\mathsf{p},\mathsf{q}}$ families in the literature. 
Our combinatorial formula thus furnishes a novel prediction in these cases. 

The general form of the proposal applies beyond the toric Calabi-Yau setting. 
We confirm it for $\mathbb{RP}^5$, whose K\"{a}hler cone admits no toric diagram, 
finding perfect agreement with the geometric expectation. 

Apart from the residue analysis discussed above, 
special values of the supersymmetric zeta function can provide complementary data of the anomaly coefficients. 
In particular, it is proposed \cite{Nakayama:2025hzr} that 
the supersymmetric zeta values with $s=-1$ and $z=0$ for the supersymmetric field theories in 2d, 4d and 6d encode the supersymmetric Casimir energies, 
which are conjecturally determined by the anomaly coefficients. 
For the 4d SCFTs the anomaly coefficients are characterized by the central charges $a$ and $c$. 
In theories with a gravity dual description as Type IIB string theory on $AdS_5\times Y$ via the AdS/CFT correspondence, 
these central charges agree with each other at the leading term in the large $N$ limit \cite{Henningson:1998gx}, 
where $N$ denotes the quantized 5-form flux through the Sasaki-Einstein 5-manifold $Y$. 
It is controlled by the inverse volume of $Y$ so that the minimization of the volume \cite{Martelli:2005tp} with the critical Reeb vector field 
corresponds to the field theoretical $a$-maximization \cite{Intriligator:2003jj} with the exact R-symmetry in the IR fixed point. 
Our analysis of the gravitational supersymmetric zeta functions for the holographic dual Sasaki-Einstein 5-manifolds 
shows that their special values with $s=-1$ and $z=0$ capture the subleading holographic anomaly data, 
which are independent of the $AdS$ radius. 
Such subleading terms may break the equality between $a$ and $c$ in the large $N$ limit, 
generating non-trivial contribution to the difference of the central charges $c-a$. 
While such subleading anomaly data may arise from higher-derivative and loop corrections in the effective bulk action holographically 
\cite{Brigante:2007nu,Kats:2007mq,Brigante:2008gz,Cremonini:2008tw,Buchel:2008vz,Liu:2010gz}, 
they typically depend on the combinatorial data of the associated toric diagrams for the toric Sasaki-Einstein manifolds. 

\subsection{Structure}
The structure of this paper is organized as follows. 
In section \ref{sec_SE} we review Sasaki-Einstein manifolds 
and summarize their key properties, with particular emphasis on the toric Sasaki-Einstein 5-manifolds $Y$. 
In section \ref{sec_spectral} we discuss the holographic spectral functions for Type IIB string theory on $AdS_5\times Y$, 
which is holographically dual to 4d SCFTs. 
Along the way, we assemble the fundamental formulae and principal results of this work, 
including combinatorial formulae for the supersymmetric indices, supersymmetric zeta functions and supersymmetric determinants and for the integrated curvature-squared invariants, 
which are formulated in terms of the toric data reviewed in section \ref{sec_SE}. 
In section \ref{sec_symSE} we analyze the Sasaki-Einstein manifolds $S^5$, $\mathbb{RP}^5$ and $(S^5/\mathbb{Z}_k)_{\textrm{ALE}}$, 
which are holographically dual to gauge theories with extended supersymmetry. 
In these examples, we focus primarily on theories with gauge groups given by products of $U(N)$, associated with the Lie algebras $A_{N-1}$. 
For $\mathcal{N}=4$ super Yang-Mills theory, however, we also consider theories with BCD-type gauge groups, 
which are holographically dual to $\mathbb{RP}^5$.
In sections \ref{sec_orb}, \ref{sec_Ypq}, \ref{sec_Labc}, \ref{sec_orbLabc} and \ref{sec_dP} 
we study the Sasaki-Einstein manifolds that are dual to 4d $\mathcal{N}=1$ quiver gauge theories, 
namely orbifold $S^5/\mathbb{Z}_k$, $Y^{\mathsf{p},\mathsf{q}}$, $L^{a,b,c}$, $L^{a,b,c}/\mathbb{Z}_2$, 
and geometries associated with the (Pseudo) del Pezzo surfaces, respectively. 
In all these cases, the gauge group is taken to be a product of $SU(N)$ factors. 
In section \ref{sec_XZ} we analyze the supersymmetric spectral functions using the proposed formulae 
and the combinatorial data encoded in the toric diagram for the Sasaki-Einstein manifolds $X^{\mathsf{p},\mathsf{q}}$ and $Z^{\mathsf{p},\mathsf{q}}$ 
without referring to the detailed structures of the gauge theories. 
In Appendix \ref{app_HS}, we collect the Hilbert series for various Calabi-Yau cones. 
From these series, we extract the corresponding volumes and the curvature-squared integrals. 
In Appendix \ref{app_Riem2}, we summarize the volumes and curvature-squared integrals of Sasaki-Einstein manifolds. 
In Appendix \ref{app_proof}, we indicate how the combinatorial formula for the curvature-squared invariant can be established directly for smooth toric Sasaki-Einstein manifolds, 
without appealing to the conjectured identification of the leading Cardy coefficient.

\subsection{Future directions}

\begin{itemize}

\item 
It would be interesting to explore further physical and geometrical significance of the supersymmetric zeta values, 
corresponding to special values of supersymmetric zeta functions at negative integers, and more generally its special values at complex arguments. 
Moreover, investigating other analytic properties of the supersymmetric zeta function, 
such as its zeros or functional relations, may also provide valuable insights into the underlying holographic and geometric interpretations.

\item 
In \cite{Eager:2012hx}, it was observed that 
the index can be expressed as the Euler characteristic of the cyclic homology of the Ginzburg dg algebra \cite{Ginzburg:2006fu}, 
and equivalently represented on the gravity side in terms of the Kohn-Rossi cohomology \cite{MR177135,MR604043} of the Sasaki-Einstein 5-manifold. 
While this agreement provides compelling structural evidence, it does not yet furnish an explicit operator-level identification. 
Clarifying a direct operator-theoretic interpretation of the supersymmetric determinant on the gravity side may therefore 
shed light on the underlying mechanism and potentially provide a more structural justification of this claim. 
Holographically, extremal BPS mesons will correspond to the Kaluza-Klein modes of the scalar Laplacian on the smooth Sasaki-Einstein manifold, 
associated with holomorphic functions on the Calabi-Yau cone.
It is argued in \cite{Kihara:2005nt,Oota:2005mr} that 
the Laplacian on the Sasaki-Einstein 5-manifolds $L^{a,b,c}$ leads to Heun's equations upon the separation of variables. 
These equations can be embedded into the framework of isomonodromic deformations \cite{Jimbo:1981tov} of Fuchsian systems. 
In this formulation, the accessory parameter corresponding to the eigenvalue is generated by a Painlev\'{e} VI $\tau$-function \cite{MR642657,MR4300140}.  
For the BPS modes polynomial solutions arise and the accessory parameter  takes discrete values. 
In such a case, the $\tau$-function may effectively reduce to a finite polynomial multiplied by a product of gamma function factors \cite{Gamayun:2012ma}, 
which resembles the structure of the supersymmetric determinants.

\item 
The supersymmetric index for a stack of $N$ branes is conjectured to admit a giant graviton expansion \cite{Arai:2019xmp,Arai:2020qaj,Gaiotto:2021xce}. 
Although the supersymmetric zeta function is a potential tool for an analytic derivation of the giant graviton expansion, 
a structural subtlety of the expansion arises from the fact that a sum over the wrapping number appears inside the logarithm, 
which prevents a direct factorization of the supersymmetric zeta function into multiplicative Dirichlet components. 
Nevertheless, if the dominant contribution appears for a large wrapping number, 
the sum over the wrapping number can be viewed as the Laplace transform and 
it can be systematically analyzed by means of the saddle-point method. 
We hope to present the detailed result in a future work. 

\item 
The extension of the present analysis to the $AdS_4\times Y$ with $Y$ being the Sasaki-Einstein 7-manifold 
dual to 3d SCFTs arising from a stack of M2-branes would be interesting (see e.g. \cite{Oh:1998qi,Fabbri:1999hw,Ceresole:1999zg}). 
A systematic study of the holographic spectral functions and the associated toric combinatorics may uncover new geometric data encoded in the 3d SCFT. 
A promising avenue is to develop the asymptotic spectral analysis in \cite{Hayashi:2023txz} in terms of the holographic spectral functions.

\end{itemize}

\section{Sasaki-Einstein manifolds}
\label{sec_SE}

\subsection{Sasaki-Einstein manifolds}
A Sasakian $(2n+1)$-manifold is a Riemannian manifold $(Y^{2n+1},g_Y)$ 
such that its metric cone $(C(Y),g_C)$ $=$ $(\mathbb{R}_{>0}\times Y, dr^2+r^2 g_Y)$ is a K\"{a}hler manifold. 
It is equipped with a contact 1-form $\eta$ obeying the condition $\eta\wedge (d\eta)^n$ $\neq$ $0$. 
At each point $p\in Y$, 
\begin{align}
\label{ker_eta}
\ker\eta_p&=\{v\in T_pY\mid \eta_p(v)=0\}
\end{align}
defines a $2n$-dimensional hyperplane (codimension-1 subspace) in the tangent space. 
It carries a symplectic structure given by $d\eta$ on the contact hyperplane (\ref{ker_eta}). 
Given the contact 1-form $\eta$, the Reeb vector $\xi$ is defined by
\begin{align}
\iota_\xi d\eta=0, \qquad 
\eta(\xi)&=1. 
\end{align}
The first condition forces $\xi$ to be transverse to the $2n$-dimensional symplectic hyperplane and the second fixes the normalization. 

If the metric $g_Y$ is Einstein, i.e. $\mathrm{Ric}_{g_Y}$ $=$ $2ng_Y$, it is called a Sasaki-Einstein $(2n+1)$-manifold. 
It then follows that its metric cone $C(Y)$ is Ricci-flat K\"{a}hler, i.e. a Calabi-Yau manifold (see e.g. \cite{MR2382957}). 
On the Calabi-Yau cone $X$, the Reeb vector lifts to a holomorphic Killing vector 
\begin{align}
\xi=J(r\partial_r). 
\end{align} 
Together with the radial vector $r\partial_r$, it generates a complexified action on the cone, 
where the real flows define an $\mathbb{R}_{>0}$ dilation of the cone 
and an isometric $\mathbb{R}$-action generated by $\xi$. 
When the Sasaki structure is (quasi-)regular, the latter closes to a $U(1)$ action, 
while in the irregular case, the closure of the flow yields a torus action in the isometry group. 
If $X$ is toric, there is a maximal torus action of $U(1)^{n+1}$. 
Let $\phi_i$ $\sim$ $\phi_i+2\pi$ be angular coordinates on $X$. 
Then the Reeb vector is expressed as a real linear combination
\begin{align}
\xi&=\sum_{i=1}^{n+1}b_i\partial_{\phi_i}, 
\end{align}
where $\partial_{\phi_i}$ are the commuting Killing vectors which generate the $i$-th $U(1)$ action. 
Here the coefficients define the Reeb vector 
\begin{align}
\label{Reeb_1}
b&=(b_1,\cdots,b_{n+1})\in \mathbb{R}^{n+1}. 
\end{align}

\subsection{Toric Sasaki-Einstein 5-manifolds}

Let us focus on the Sasaki-Einstein 5-manifold $Y$ with $n=2$ whose metric cone $X=C(Y)$ is a toric Calabi-Yau 3-fold. 
The toric cone $X$ is completely encoded by a set of lattice vectors
\begin{align}
\label{toric_vertices}
v_I&=(1,x_I,y_I)\in \mathbb{Z}^3, \qquad I=1,\cdots, d. 
\end{align}
The first component is set to $1$ according to the Calabi-Yau condition. 
It defines the toric diagram, a convex lattice polygon in $\mathbb{Z}^2$. 
The 1-cone (ray)
\begin{align}
\sigma^{(1)}&=\mathbb{R}_{\ge0} v_I
\end{align}
is generated by a single vector $v_I$. 
A vector $v_I$ is called primitive if it is the shortest lattice vector along its ray. 
In such a case the vectors $v_I$ are identified with the primitive generators of the 1-cones of the toric fan. 
Each vector $v_I$ corresponds to a toric divisor $D_I$, a codimension-1 subvariety (non-compact 4-cycle). 
It also controls the BPS baryonic operator in the 4d $\mathcal{N}=1$ quiver gauge theory. 
The external lattice points $v_I$ in the toric diagram correspond to the baryonic operators built from the chiral multiplets $\Phi_I$. 
They are described by D3-branes wrapped on supersymmetric 3-cycles $\Sigma_I$ $=$ $D_I\cap Y$ in the Sasaki-Einstein base $Y$. 
The multiplicity of the field $\Phi_I$ is given by \cite{Franco:2005sm}\footnote{Note that extra lattice points lying strictly on edges of the toric diagram do not introduce new independent baryonic charges. }
\begin{align}
\#(\Phi_I)
&=\det (v_{I-1},v_I,v_{I+1}). 
\end{align}

The 2-cone 
\begin{align}
\sigma^{(2)}&=\mathbb{R}_{\ge0}v_I+\mathbb{R}_{\ge0}v_J
\end{align}
generated by two linearly independent rays corresponds to a codimension-2 subvariety (2-cycle) 
as an intersection $D_I\cap D_J$ of two toric divisors. 
On the toric diagram it corresponds to an edge. 
The 2-cones control which divisors intersect and how baryons are related and charged. 

The 3-cone 
\begin{align}
\sigma^{(3)}&=\mathbb{R}_{\ge0}v_I+\mathbb{R}_{\ge0}v_J+\mathbb{R}_{\ge0}v_K
\end{align}
generated by three linearly independent rays corresponds to an intersection point 
as a triple intersection $D_I\cap D_J\cap D_K$. 
On the toric diagram it corresponds to a triangle in the triangulation. 
It can be viewed as a local patch of the geometry as it locally looks like $\mathbb{C}^3$ up to orbifolds. 
In the 4d $\mathcal{N}=1$ quiver gauge theory it corresponds to a superpotential term. 
To summarize, the polyhedral cone 
\begin{align}
\mathcal{C}&=\left\{
\sum_{I}c_I v_I \;\middle|\; c_I\ge0 \right\}\subset \mathbb{R}^3
\end{align}
encodes the combinatorial and topological data of $X$. 
For an affine toric variety the image of the moment map is a convex polyhedral cone. 
Thus $y$ $\in$ $\mathcal{C}$ provide the natural radial coordinates on the toric Calabi-Yau cone. 

The dual cone $\mathcal{C}^{\vee}$ is defined by
\begin{align}
\mathcal{C}^{\vee}&=\left\{m \in \mathbb{R}^3\mid (m,v_J) \ge 0\  \forall J \right\}. 
\end{align}
The dual cone encodes the set of linear functions that are non-negative on the toric cone. 
An extremal ray of $\mathcal{C}^{\vee}$ is a one-dimensional cone 
\begin{align}
\rho^{\vee}&=\mathbb{R}_{\ge 0}m_I, 
\end{align}
where $m_I$ generate the edges of the dual cone. 
Physically, they correspond to the BPS mesons in the 4d $\mathcal{N}=1$ quiver gauge theory 
which are gauge invariant operators without epsilon symbols and correspond to holomorphic functions on $X$. 
The vector $m_I$ $\in$ $\mathbb{Z}^3$ is called the primitive generator if it is the shortest lattice vector along its ray. 
For the toric Calabi-Yau 3-fold cone, 
each facet generated by two adjacent rays $v_I$ and $v_{I+1}$ 
of the primal cone determines a unique primitive dual vector $m_{I}$ such that\footnote{
The two independent linear constraints fix the solution of the three-dimensional vector $m_I$ up to an overall real scale. 
As $m$ is a primitive lattice vector, there are only two primitive lattice solutions. 
Exactly one of them can satisfy the inequality $(m_I, v)>0$ 
for all other rays $v$ since the fan is strongly convex. 
} 
\begin{align}
\label{primitive_m}
(m_I, v_I)&=(m_{I}, v_{I+1})=0, \nonumber\\
(m_I, v)&>0 \quad \textrm{for all other $v$}. 
\end{align}
Note that the above definition does not guarantee that the vectors $m_I$ are distinct. 
Indeed, if several consecutive vertices of the toric diagram belong to the same face of the cone, 
the corresponding primitive normal vector annihilates all of them, leading to coincident generators. 
The primitive generator $m_I$ obeying (\ref{primitive_m}) cannot be written as a positive sum of other lattice points. 
Hence every other BPS meson in that direction can be realized as a power of the BPS meson corresponding to the primitive generator $m_I$. 
On the other hand, in the quiver gauge theory dual to the toric Sasaki-Einstein manifold 
the \textit{extremal BPS meson} is defined as a chiral primary operator that maximizes the flavor charges for fixed R-charge \cite{Benvenuti:2005ja,Benvenuti:2005cz}. 
Additivity of global charges then implies that any other BPS meson is realized as a power of the extremal BPS meson. 
This definition coincides precisely with the primitive generators $m_I$ of the dual cone described above. 
So we denote the extremal BPS meson by $\mathcal{M}(m_I)$. 
While the non-extremal BPS mesons correspond to extended semiclassical string configurations exploring the Sasaki-Einstein manifold, 
the extremal BPS mesons are a finite subset distinguished by the property that 
they correspond to the point-like BPS geodesics on the Sasaki-Einstein manifold\cite{Benvenuti:2005ja,Benvenuti:2005cz}.

The Reeb vector must lie in the interior of the dual cone since the canonical pairing between 
the positive vector $y$ $\in$ $\mathcal{C}$ and the Reeb vector $b$ provides the non-negative cone radius. 
For the toric variety the canonical bundle $K_X$ of $X$ is
\begin{align}
\label{toric_prop1}
K_X=-\sum_I D_I, 
\end{align}
where $D_I$ are the toric divisors corresponding to the vertices $v_I$. 
According to the Calabi-Yau condition $c_1(X)=0$, the canonical bundle $K_X$ of $X$ is trivial. 
This implies that there exists some $m_0$ $\in$ $\mathcal{C}^{\vee}$ such that 
$K_X$ $=$ $\mathrm{div}(f_{m_0})$, 
where $\mathrm{div}(f)$ $=$ $\sum_{I}\mathrm{ord}_{D_I}(f)D_I$ is the divisor of a rational function $f$ on $X$ 
and $f_{m}$ $\sim$ $u_I^{\mathrm{ord}_{D_I}(f_m)}$ is some monomial in the local holomorphic coordinate $u_I$ that vanishes on the toric divisor $D_I$. 
Since we have $\mathrm{ord}_{D_I}(f_m)$ $=$ $(m,v_I)$ for the toric variety, 
it follows from (\ref{toric_prop1}) and the Calabi-Yau condition that 
\begin{align}
(m_0,v_I)&=1\quad \forall I. 
\end{align}
This condition ensures that  
there is a nowhere-vanishing holomorphic $(3,0)$-form $\Omega$ $=$ $f_{m_0}\Omega_0$, 
where a simple pole along each $D_I$ of the logarithmic form 
$\Omega_0$ $=$ $\frac{dz_1}{z_1}$ $\wedge$ $\frac{dz_2}{z_2}$ $\wedge$ $\frac{dz_3}{z_3}$ 
of the torus exactly cancels a zero of $f_{m_0}$ on each $D_I$. 
Since the vertices $v_I$ are given by (\ref{toric_vertices}), we have $m_0$ $=$ $(1,0,0)$. 
As the holomorphic $(3,0)$-form $\Omega$ has Reeb charge $3$ generated by $b=J(r\partial_r)$, 
we can write 
\begin{align}
\label{Reeb_2}
b&=(3,b_1,b_2). 
\end{align}

Given a Reeb vector $b$, we define the polytope 
\begin{align}
\label{vol_polytope}
\Delta_b&=\left\{y\in \mathcal{C}|(b,y)\le \frac12\right\}. 
\end{align}
Here $(b,y)$ $=$ $r^2/2$ measures the radial coordinate $r$. 
The hyperplane defined by $(b,y)=1/2$ corresponds to fixing $r=1$, i.e. the Sasaki-Einstein manifold $Y$ 
and the region $(b,y)\le 1/2$ to the cone truncated at $r=1$. 
Making use of Stokes' theorem, we can evaluate the volume of $\Delta_b$ as \cite{Martelli:2005tp}
\begin{align}
\mathrm{Vol}(\Delta_b)
&=\frac{1}{18}
\sum_{I}
\frac{1}{|v_I|}
\mathrm{Vol}(\mathcal{F}_I), 
\end{align}
where 
\begin{align}
\frac{1}{|v_I|}
\mathrm{Vol}(\mathcal{F}_I)
&=
\frac{1}{8}
\frac{\det(v_{I-1},v_I,v_{I+1})}{\det(b,v_{I-1},v_I)\det(b,v_I,v_{I+1})}
\end{align}
is the volume of the $I$-th facet 
\begin{align}
\mathcal{F}_I&=\{y\in \Delta_b|(y,v_I)=0\}. 
\end{align}
The volume (\ref{vol_polytope}) of the polytope $\Delta_b$ and that of the facet $\mathcal{F}_I$ 
are related to the volume of the Sasaki-Einstein manifold $Y$ 
and that of the supersymmetric 3-cycle $\Sigma_I$ $\subset$ $Y$ respectively. 
Martelli, Sparks, and Yau \cite{Martelli:2005tp} show that 
\begin{align}
\mathrm{Vol}(Y)&=48\pi^3 \mathrm{Vol}(\Delta_b), \\
\mathrm{Vol}(\Sigma_I)&=16\pi^2 \frac{1}{|v_I|}\mathrm{Vol}(\mathcal{F}_I), 
\end{align}
where $C(\Sigma_I)$ $=$ $D_I$. 
Explicitly, the volume of the Sasaki-Einstein manifold $Y$ is computed as \cite{Martelli:2005tp}
\begin{align}
\label{vol_Y1}
\mathrm{Vol}(Y)&=
\frac{\pi^3}{3}
\sum_{I}\frac{\det(v_{I-1},v_I,v_{I+1})}{\det(b,v_{I-1},v_I)\det(b,v_I,v_{I+1})}, \\
\mathrm{Vol}(\Sigma_I)&=
2\pi^2 \frac{\det(v_{I-1},v_I,v_{I+1})}{\det(b,v_{I-1},v_I)\det(b,v_I,v_{I+1})}. 
\end{align}

Alternatively, the volume can be obtained from the Duistermaat-Heckman localization \cite{MR674406,MR696693}, 
as demonstrated by Martelli, Sparks, and Yau in their subsequent work \cite{Martelli:2006yb}. 
Let us consider a triangulation of the toric diagram that induces a decomposition of the cone $\mathcal{C}$ into simplicial cones 
\begin{align}
\mathcal{C}&=\bigcup_{i} \mathcal{C}_{i}, \qquad 
\mathcal{C}_{i}=\mathbb{R}_{\ge0}\langle v_{i,1},v_{i,2},v_{i,3}\rangle. 
\end{align}
Here $v_{i,1}$, $v_{i,2}$ and $v_{i,3}$ are three vectors in the toric diagram 
and the index $i$ labels the simplex, i.e. triangle in the toric diagram. 
The index $i$ runs over $i=1,\cdots, 2A$, 
where $A$ is the lattice area of the toric diagram. 
Applying Pick's theorem, it is evaluated as
\begin{align}
\label{Area_toric}
A&=n_I+\frac{n_B}{2}-1
=n_{\textrm{total}}-\frac{n_B}{2}-1, 
\end{align}
where $n_I$, $n_B$ and $n_{\textrm{total}}$ are 
the number of interior lattice points, the number of boundary lattice points, and the total number of lattice points respectively. 
For the simplicial cone $\mathcal{C}_{i}$, the dual cone 
\begin{align}
\mathcal{C}_{i}^{\vee}&=\mathbb{R}_{\ge0}\langle m_{i,1}, m_{i,2}, m_{i,3}\rangle
\end{align}
is also simplicial. 
$\{m_{i,1},m_{i,2},m_{i,3}\}$ is a set of three primitive inward-pointing normal vectors to the three facets of $\mathcal{C}_{i}$. 
Equivalently, each of them annihilates two of the three generators of $\mathcal{C}_{i}$ 
and the vector $m_{i,1}$ can be uniquely defined by the conditions
\begin{align}
(m_{i,1},v_{i,1})&=(m_{i,1},v_{i,2})=0, \\
(m_{i,1},v_{i,3})&>0. 
\end{align}
Similarly for $m_{i,2}$ and $m_{i,3}$ by cyclic permutation. 
Then the volume is expressible as a sum over the dual simplicial cones \cite{Martelli:2006yb}
\begin{align}
\label{vol_Y2}
\mathrm{Vol}(Y)&=\pi^3 \sum_{i}\frac{1}{(b,m_{i,1}) (b,m_{i,2}) (b,m_{i,3})}. 
\end{align}
Intuitively, the simplicial cone labeled by the index $i$ in the triangulation corresponds to a local $\mathbb{C}^3$-type patch 
and therefore the volume is obtained by summing these local contributions.

Among all Reeb vectors compatible with the Calabi-Yau structure, 
the Sasaki-Einstein metric corresponds to the unique minimizer of the volume \cite{Martelli:2005tp}. 
Therefore the Reeb vector is determined by solving
\begin{align}
\frac{\partial}{\partial b_i}\mathrm{Vol}(Y)&=0, \qquad i=1,2. 
\end{align}
The volume of the Sasaki-Einstein manifold is related to the coefficients of the $N^2$ terms in the 4d central charges of the dual 4d $\mathcal{N}=1$ SCFT
\begin{align}
a_{\mathcal{O}(N^2)}&=c_{\mathcal{O}(N^2)}
=\frac{\pi^3}{4\mathrm{Vol}(Y)}N^2
\end{align}
and the volume minimization procedure corresponds to the $a$-maximization \cite{Intriligator:2003jj}, 
for which the local maximum of the trial central charge determines the exact R-symmetry of the theory at its superconformal fixed point.
Given the Reeb vector $b$, 
the R-charge of the chiral multiplet $\Phi_I$ that forms the baryonic operator is given by \cite{Gubser:1998fp}
\begin{align}
\label{Rch_baryon}
R(\Phi_I)&=\frac{\pi}{3} \frac{\mathrm{Vol}(\Sigma_I)}{\mathrm{Vol}(Y)}. 
\end{align}
For the extremal BPS mesonic operators $\mathcal{M}_I$ 
associated with the primitive generators $m_I$ of the dual cone satisfying the condition (\ref{primitive_m}), 
the R-charges and conformal dimensions are given by
\begin{align}
\label{Rch_meson}
R(\mathcal{M}_I)&=\frac23 (b, m_I), \\
\Delta(\mathcal{M}_I)&=\frac32R(\mathcal{M}_I)=(b,m_I). 
\end{align}
Besides, the operator length, i.e. the total number of the chiral multiplet fields $\Phi_J$, in the mesonic operator $\mathcal{M}_I$ can be measured by
\begin{align}
\label{L_meson}
L_J(\mathcal{M}_I)&=(m_I,v_J). 
\end{align}

\section{Supersymmetric spectral functions}
\label{sec_spectral}

\subsection{Supersymmetric indices}
\label{sec_ind}
We begin with the superconformal index for the 4d SCFT \cite{Kinney:2005ej,Romelsberger:2005eg}.  
It can be defined as a weighted trace
\begin{align}
\label{DEF_ind}
\mathcal{I}(p;q)
&={\Tr}_{\mathcal{H}}(-1)^F p^{j_3+\overline{j}_3+\frac{R}{2}}q^{j_3-\overline{j}_3+\frac{R}{2}}
\prod_{\alpha}x_{\alpha}^{f_{\alpha}}
\end{align}
over the Hilbert space $\mathcal{H}$ of the radially quantized theory on $S^3$. 
Here $F$ is the Fermion number operator, 
$j_3$ and $\overline{j}_3$ stand for the Cartan generators of the rotation group $SU(2)_1$ $\times$ $SU(2)_2$ on $S^3$, 
$R$ is the $U(1)_R$ R-charge, 
and $f_{\alpha}$ are the Cartan generators of the other global symmetries. 
The index is invariant under the RG flow and it encodes protected operator spectra. 

For the Lagrangian quiver gauge theories the index can be expressed as a matrix integral 
over the gauge holonomy variables with contributions from vector and chiral multiplets. 
The single-particle index for a vector multiplet is given by
\begin{align}
i^{\textrm{vec}}(s;p;q)
&=i_v(p;q)\chi_{\textrm{adj}}(s)
\end{align}
where 
\begin{align}
\label{sind_vec}
i_v(p;q)&=-\frac{p}{1-p}-\frac{q}{1-q}
\end{align}
and $\chi_{\textrm{adj}}(s)$ is the character of the adjoint representation of the gauge group. 
For a chiral multiplet with R-charge $R=r$ transforming in a representation $\mathcal{R}$ of gauge group 
the single-particle index reads
\begin{align}
i^{\textrm{chiral}_{\mathcal{R},r}}(s;p;q)
&=i_{\textrm{cm}_{r}}(p;q)\chi_{\mathcal{R}}(s)
+\tilde{i}_{\textrm{cm}_{r}}(p;q)\chi_{\overline{\mathcal{R}}}(s), 
\end{align}
where 
\begin{align}
\label{sind_cm1}
i_{\textrm{cm}_{r}}(p;q)&=\frac{p^{\frac{r}{2}}q^{\frac{r}{2}}}{(1-p)(1-q)}, \\
\label{sind_cm2}
\tilde{i}_{\textrm{cm}_{r}}(p;q)&=-\frac{p^{1-\frac{r}{2}}q^{1-\frac{r}{2}}}{(1-p)(1-q)}, 
\end{align}
and $\chi_{\mathcal{R}}(s)$ is the character of the representation $\mathcal{R}$. 
The multi-particle states are generated by the plethystic exponential (PE) 
of the sum of all single-particle indices $i(s,p;q)$
\begin{align}
\mathrm{PE}[i(s;p;q)]&:=\exp\left[
\sum_{k=1}^{\infty}\frac{1}{k}i(s^k;p^k;q^k)
\right]. 
\end{align}
Projection onto the gauge invariant operators can be achieved by integrating over the gauge holonomy variables. 
Hence the index of the quiver gauge theory is evaluated as a matrix integral of the form 
\begin{align}
\label{ind_matrix1}
\mathcal{I}(p;q)
&=\int \prod_i \left[ds^{(i)}\right] 
\mathrm{PE}
\Biggl[
\sum_{i}i^{\textrm{vec}}\left(s^{(i)};p;q\right)
+\sum_{\mathcal{R}}i^{\textrm{chiral}_{\mathcal{R},r}}\left(s^{(i)};p;q\right)
\Biggr], 
\end{align}
where the indices $i$ labelb the gauge nodes of the quiver. 

The 4d gauge theories which are holographically dual to the Sasaki-Einstein manifolds $Y$ arise 
as the low-energy effective theories of $N$ coincident D3-branes probing the Calabi-Yau cone over $Y$. 
In the case of toric Calabi-Yau cones, there exists a well-established and systematic framework 
based on brane tilings (dimer models) that allows one to extract the quiver gauge theory data, 
i.e. the gauge groups, matter content and superpotential from the toric data \cite{Feng:2000mi,Feng:2002zw,Hanany:2005ve,Franco:2005rj,Franco:2005sm}. 
Open strings ending on a stack of D3-branes carry the Chan-Paton labels $a,b=1,\cdots, N$, 
which generates $U(N)$ vector multiplets as the gauge degrees of freedom. 
So one may associate to each node the $U(N)$ gauge group. 
Alternatively, one may work with $SU(N)$ gauge grop, effectively removing the $U(1)$ gauge factors, 
which typically decouple from the interacting sector. 
The Haar measure for the $U(N)$ gauge node is given by
\begin{align}
\left[ds^{(i)}\right]
&=\frac{1}{N!}\prod_{a=1}^{N}
\frac{ds_a^{(i)}}{2\pi is_a^{(i)}}
\prod_{1\le a<b\le N}
\left(
1-s_a^{(i)\pm}s_b^{(i)\mp}
\right)
\end{align}
and we impose the condition $\prod_{a=1}^N s_a^{(i)}=1$ for the $SU(N)$ gauge group. 
Setting $s_a^{(i)}$ $=$ $e^{i\theta_a^{(i)}}$, 
we can rewrite it as 
\begin{align}
\left[ds^{(i)}\right]
&=\frac{1}{N!}\prod_{a=1}^{N}
\frac{d\theta_a^{(i)}}{2\pi}
\Delta\left(\theta^{(i)}\right)^2, 
\end{align}
where 
\begin{align}
\label{vandermonde}
\Delta\left(\theta^{(i)}\right)^2
&=
\prod_{1\le a<b\le N}
\left|1-e^{i \left(\theta_a^{(i)}-\theta_b^{(i)}\right) }\right|^2. 
\end{align}
For the $SU(N)$ gauge group we insert the factor $\delta\left(\sum_{a=1}^N \theta_a^{(i)} \right)$. 

The chiral multiplet transforming in the representations $(\mathbf{N}_i,\overline{\mathbf{N}}_j)$ 
are described as an arrow from $i$-th node to $j$-th node in the quiver. 
The analysis requires special care in the theories with adjoint chiral multiplets corresponding to the loops in the quiver. 
The loop at $i$-th node in the quiver can be viewed as $(\mathbf{N}_i,\overline{\mathbf{N}}_i)$, 
i.e. the bifundamental chiral from $i$-th node to itself. 
Since $\mathbf{N}_i\otimes \overline{\mathbf{N}}_i$ $=$ $\mathrm{adj}$ $\oplus$ $\mathbf{1}$ for $SU(N)_i$, 
it decomposes as the traceless part of dimension $N_i^2-1$ and the trace part as a gauge neutral chiral multiplet.\footnote{We note that there exists an alternative convention in which neutral chiral multiplets are not introduced; instead, the $i$-th loop is treated as an $SU(N)_i$ adjoint field (see e.g. \cite{Agarwal:2013pba,ArabiArdehali:2014otj,ArabiArdehali:2015iow} for the corresponding indices). } 
As discussed in \cite{Gadde:2010en} and further elaborated later in the following sections, 
the contributions appearing in the index admit an interpretation in terms of single-trace operators 
associated with extremal BPS mesons \cite{Benvenuti:2005ja,Benvenuti:2005cz} corresponding to independent loops in the quiver. 
In the presence of adjoint chiral multiplets, these operators can be realized as loops of adjoint fields.
If no neutral chiral multiplets are present, however, the interpretation in terms of extremal BPS mesons is no longer compatible with this operator realization. 
Thus, in most of our analysis of $SU(N)$ gauge theories with adjoint chiral multiplets below, 
we will introduce these neutral chiral multiplets unless otherwise stated. 

There is a further practical advantage to retaining the trace mode. 
If the trace mode were excluded, $a$-maximization would require a modification of order $\mathcal{O}(N^0)$ 
to determine the superconformal R-charge in the presence of adjoint chiral fields, 
leading to a discrepancy with volume minimization. 
While string theory analysis provides no definitive evidence for or against this phenomenon, 
we find it more consistent to include the trace mode to maintain the exact agreement between $a$-maximization and volume minimization. 

For $\mathcal{N}\ge2$ enhanced supersymmetric gauge theories, 
the adjoint chiral multiplets at $i$-th node that belong to the $\mathcal{N}\ge2$ vector multiplets 
are required to transform as the adjoint representation of the associated gauge group to form the complete supermultiplet. 
We thus work with $\prod_i U(N)_i$ gauge group instead of $\prod_i SU(N)_i$ gauge group for $\mathcal{N}\ge 2$ quiver gauge theories dual to Sasaki-Einstein manifolds. 
We should add that since the index does not depend on the coupling, this gauging is not inconsistent with the fact that the $U(1)$ can be IR free. 

For generic $\mathcal{N}=1$ quiver gauge theories, 
the chiral multiplets are not part of the gauge degrees of freedom and supersymmetry does not relate the adjoint chiral multiplets to vector multiplets. 
Accordingly, we can simply regard the trace component as the legitimate gauge invariant operator 
and we identify the loop at $i$-th node as the chiral multiplet transforming as $\mathrm{adj}$ $\oplus$ $\mathbf{1}$ of $SU(N)_i$ gauge group. 

While the variables $\{\theta_a^{(i)}\}$ describe the configuration of eigenvalues for a unitary matrix, 
it is convenient to change the basis to study the large $N$ limit of the matrix integral (\ref{ind_matrix1}) \cite{Kinney:2005ej}. 
Let us introduce the empirical eigenvalue density
\begin{align}
\rho^N(\theta^{(i)})&=\frac{1}{N}\sum_{a=1}^N \delta(\theta^{(i)}-\theta_a^{(i)})
\end{align}
and expand it in Fourier modes 
\begin{align}
\label{dens_F}
\rho_k^{(i)}&:=\int d\theta^{(i)} \rho^N(\theta^{(i)}) e^{-ik\theta^{(i)}}
=\frac{1}{N}\sum_{a=1}^N e^{-ik\theta_a^{(i)}}, 
\end{align}
with ${\rho_k^{(i)}}^*$ $=$ $\rho_{-k}^{(i)}$. 
While the variables $\rho_k^{(i)}$ are not independent at finite $N$, 
they can form a complete set of collective coordinates for smooth eigenvalue distributions of $\theta_a^{(i)}$ in the large $N$ limit. 
In terms of the Fourier modes (\ref{dens_F}) we can write the $\mathfrak{su}(N)$ characters as
\begin{align}
\label{suNch_adj}
\chi_{\textrm{adj}}^{\mathfrak{su}(N)}(s^k)
&=\sum_{a,b}e^{ik\left(\theta_a^{(i)}-\theta_b^{(i)}\right)}-1
=N^2\left|\rho_k^{(i)}\right|^2-1, \\
\label{suNch_bifund}
\chi_{\textrm{bifund: $i\rightarrow j$}}^{\mathfrak{su}(N)}(s^k)
&=\sum_{a,b}e^{ik\left(\theta_a^{(i)} - \theta_{b}^{(j)} \right)}
=N^2 \rho_{-k}^{(i)} \rho_{k}^{(j)}. 
\end{align}
The measure factor (\ref{vandermonde}) can be rewritten as 
\begin{align}
\label{vandermonde2}
\Delta\left(\theta^{(i)}\right)^2
&=\exp\left[
-\sum_{k=1}^{\infty}
\frac{1}{k}\sum_{a,b}e^{ik\left(\theta_a^{(i)}-\theta_b^{(i)}\right)}
\right]
\nonumber\\
&=\exp\left[
-N^2\sum_{k=1}^{\infty}\frac{1}{k} \left| \rho_k^{(i)} \right|^2
\right]
\end{align}
up to the terms which are independent of eigenvalues. 
Plugging (\ref{suNch_adj}), (\ref{suNch_bifund}) and (\ref{vandermonde2}) into (\ref{ind_matrix1}), 
the matrix integral (\ref{ind_matrix1}) reduces to
\begin{align}
\label{ind_largeN2}
&\mathcal{I}(p;q)
=\int 
\prod_{i}
\prod_{k=1}^{\infty}
d\rho_{-k}^{(i)}d\rho_{k}^{(i)}
\exp
\Biggl[
-N^2 \sum_{k=1}^{\infty}
\frac{1}{k}\sum_{i,j}\rho_{-k}^{(i)}
\left(\delta_{ij}-M_{ij}(p^k;q^k)\right)
\rho_{k}^{(j)}
\Biggr] 
\end{align}
for quiver gauge theories with unitary gauge group $\prod_i U(N)_i$ and to 
\begin{align}
\label{ind_largeN1}
&\mathcal{I}^{\mathcal{N}=1}(p;q)
=\int 
\prod_{i}
\prod_{k=1}^{\infty}
d\rho_{-k}^{(i)}d\rho_{k}^{(i)}
\nonumber\\
&\times \exp
\Biggl[
-N^2 \sum_{k=1}^{\infty}
\frac{1}{k}\sum_{i,j}\rho_{-k}^{(i)}
\left(\delta_{ij}-M_{ij}(p^k;q^k)\right)
\rho_{k}^{(j)}
-
\sum_{i} 
\sum_{k=1}^{\infty}
\frac{1}{k}
i_v(p^k;q^k)
\Biggr]
\end{align}
for $\mathcal{N}=1$ supersymmetric quiver gauge theories with gauge group $\prod_i SU(N)_i$ 
with the loop at $i$-th node as the chiral multiplet transforming as $\mathrm{adj}$ $\oplus$ $\mathbf{1}$ of $SU(N)_i$ gauge group. 
The integrals (\ref{ind_largeN2}) and (\ref{ind_largeN1}) are now performed over the Fourier modes (\ref{dens_F}). 
The matrix $M_{ij}(p;q)$ is the weighted adjacency matrix of the quiver
whose entries are expressed in terms of the single-particle indices (\ref{sind_vec}), (\ref{sind_cm1}) and (\ref{sind_cm2}). 
In the exponent of the integrand the dominant terms in the large $N$ limit are quadratic in $\{\rho_{-k}^{(i)},\rho_{k}^{(i)}\}$. 
Therefore we find the exact large $N$ results by performing the Gaussian integrals (\ref{ind_largeN2}) and (\ref{ind_largeN1})
\begin{align}
\label{ind_multi_gravity2}
\mathcal{I}(p;q)
&=\left(
\prod_{k=1}^{\infty}
\frac{1}{\det\left(1-M(p^k;q^k)\right)}
\right), \\
\label{ind_multi_gravity1}
\mathcal{I}^{\mathcal{N}=1}(p;q)
&=\left(
\prod_{k=1}^{\infty}
\frac{1}{\det\left(1-M(p^k;q^k)\right)}
\right)
e^{-\sum_i \sum_{k=1}^{\infty}\frac{1}{k} i_v(p^k;q^k)}. 
\end{align}
Here we have dropped the factor that is independent of the fugacities $\{p,q\}$. 
For the SCFT dual to Type IIB string theory on $AdS_5\times Y$, 
the large $N$ index (\ref{ind_multi_gravity2}) or (\ref{ind_multi_gravity1}) can be interpreted as the gravity index for 
the multi-particle BPS states of Type IIB supergravity reduced on the Sasaki-Einstein manifold $Y$. 
Making use of the identities 
\begin{align}
\sum_{d|n}\frac{\mu(d)}{d}&=\frac{\varphi(n)}{n}, \\
\sum_{d|n}\mu(d)&=\delta_{n,1}, 
\end{align}
where $\mu(d)$ is the M\"{o}bius function 
and $\varphi(n)$ is Euler’s totient function, 
we get the single-particle gravity index as the plethystic logarithm (PL) of the large $N$ indices (\ref{ind_multi_gravity2}) and (\ref{ind_multi_gravity1}) 
\cite{Gadde:2010en}
\begin{align}
\label{ind_gravity}
i(p;q)&=\mathrm{PL}\left[\mathcal{I}(p;q)\right]
=-\sum_{k=1}^{\infty}\frac{\varphi(k)}{k}\log\left[\det (1-M(p^k;q^k))\right], \\
\label{ind_gravity1}
i^{\mathcal{N}=1}(p;q)&=\mathrm{PL}\left[\mathcal{I}^{\mathcal{N}=1}(p;q)\right]
=-\sum_{k=1}^{\infty}\frac{\varphi(k)}{k}\log\left[\det (1-M(p^k;q^k))\right]-\sum_i i_v(p;q), 
\end{align}
which are expected to be equivalent to the single-particle gravity index for the Sasaki-Einstein 5-manifold $Y$ defined by 
\begin{align}
\label{ind_gravityDEF}
i(p;q)&={\Tr}_{\mathcal{H}_{\textrm{sp}}} (-1)^F p^{\frac{E+j_2}{3}+j_1}q^{\frac{E+j_2}{3}-j_1}. 
\end{align}
Here $\mathcal{H}_{\textrm{sp}}$ is the single-particle Hilbert space and $E$ is the $AdS$ energy dual to the conformal dimension. 
The quantum numbers $j_1$ and $j_2$ are the spins, i.e. the Cartan generators of $SU(2)_L\times SU(2)_R$ $\cong$ $SO(4)$ isometry group of the spatial slice of the $AdS_5$. 
They correspond to the Cartan generators $\overline{j}_3$ and $j_3$ of the rotational group on $S^3$ for the dual 4d SCFT. 
The BPS states which contribute to the index (\ref{ind_gravityDEF}) satisfy the shortening condition $E=\frac32r+2j_2$, 
where $r$ is the moment along the Reeb direction, corresponding to the $U(1)_R$ charge in the dual 4d SCFT. 
Accordingly, when one expands the single-particle gravity index (\ref{ind_gravityDEF}) as
\begin{align}
\label{ind_gravityExp}
i(p;q)&=\sum_{n,m\ge0} \delta(n,m)p^n q^m, 
\end{align}
the expansion coefficients $\delta(n,m)$ enumerate the single-particle BPS states 
of Type IIB supergravity geometry $AdS_5\times Y$ reduced on the Sasaki-Einstein manifold $Y$. 

In the following sections, we explicitly compute the gravity indices from the gauge theory side for a variety of examples. 
These computations suggest that 
when we consider $\prod_i U(N)_i$ gauge group, 
the single-particle gravity indices (\ref{ind_gravityDEF}) for the toric Sasaki-Einstein manifolds can be universally computed by the following formula:
\begin{align}
\label{ind_gravity_conj0}
i(p;q)&=\sum_{I=1}^{d}\frac{(pq)^{\frac{(b,m_I)}{3}}}{1-(pq)^{\frac{(b,m_I)}{3}}}+2A i_v(p;q), 
\end{align}
where the pairing $(b,m_I)$ of the Reeb vector $b$ and the generator $m_I$ of the dual cone 
amounts to one half of the R-charge (\ref{Rch_meson}) of the $I$-th extremal BPS mesonic operators \cite{Benvenuti:2005ja,Benvenuti:2005cz}. 
$A$ is the area of the lattice polygon of the toric diagram. 
In other words, the index can be evaluated from the combinatorial data of the associated toric diagram! 
In particular, for general toric Sasaki-Einstein geometries dual to the generic $\mathcal{N}=1$ SCFTs described by $\prod_i SU(N)_i$ quiver gauge theories, 
the single-particle gravity indices (\ref{ind_gravityDEF}) can be written as
\begin{align}
\label{ind_gravity_conj}
i^{\mathcal{N}=1}(p;q)&=\sum_{I=1}^{d}\frac{(pq)^{\frac{(b,m_I)}{3}}}{1-(pq)^{\frac{(b,m_I)}{3}}}. 
\end{align}
Compared to the $U(N)$ description, 
the contributions from the decoupled $U(1)$ vector multiplets have been excluded.\footnote{Note that the corresponding multi-particle indices satisfy the angular momentum bound
proposed in \cite{Nakayama:2007jy}, where it is conjectured that the $q \to 0$
limit with $p$ fixed makes the superconformal indices $1$ in holographic $SU(N)$ gauge
theories (but not in the $U(1)$ part).}
This is implemented by the term $-\sum_i i_v(p,q)$ in (\ref{ind_gravity1}). 
The structure of our formula (\ref{ind_gravity_conj}) is in agreement with the general framework established in \cite{Eager:2012hx}. 
However, we further provide an explicit representation that can be computed directly from the toric diagram and the associated dual cone. 
This formulation makes the connection between geometric data and the formula more transparent. 
As we show below, this enables explicit evaluation of the indices for more intricate toric Sasaki-Einstein manifolds.
Also it should be noted that although \cite{Eager:2012hx} assumed the absence of adjoint chiral multiplets, 
in our current setup the inclusion of neutral chiral multiplets ensures that the formula (\ref{ind_gravity_conj}) remains valid even when adjoint chirals are present. 
Of course, contributions from such neutral chiral multiplets are optional and can be simply subtracted at the final stage if one wishes to exclude them. 

\subsection{Supersymmetric zeta functions}
So far, we have focused on the supersymmetric indices. 
Let us now turn to the supersymmetric zeta functions and supersymmetric determinants recently introduced in \cite{Nakayama:2025hzr}, 
which serve as new classes of spectral functions in supersymmetric theories. 

We define the supersymmetric zeta function for the Sasaki-Einstein manifold by
\begin{align}
\label{DEF_zeta}
\mathfrak{Z}(s,z;\omega_1,\omega_2)
&={\Tr}_{\mathcal{H}_{\textrm{sp}}} (-1)^F \frac{1}{\left( \left(\frac{E+j_2}{3}+j_1\right)\omega_1+\left(\frac{E+j_2}{3}-j_1\right)\omega_2+z \right)^s}
\nonumber\\
&=\sum_{n,m\ge0}\frac{\delta(n,m)}{(n\omega_1+m\omega_2+z)^s}. 
\end{align}
Here $E$, $j_1$ and $j_2$ are the quantum numbers introduced in the definition (\ref{ind_gravityDEF}) of the single-particle gravity index. 
$\delta(n,m)$ are the BPS degeneracies as the expansion coefficients of the single-particle gravity index (\ref{ind_gravityExp}). 
The supersymmetric zeta function can be obtained from the Mellin transform of the gravity index 
\begin{align}
\mathfrak{Z}(s,z;\omega_1,\omega_2)
&=
\frac{1}{\Gamma(s)}\int_{0}^{\infty}
d\beta \, i(p;q) e^{-\beta z}\beta^{s-1}, 
\end{align}
with $p=e^{-\beta \omega_1}$ and $q=e^{-\beta\omega_2}$. 

By performing the Mellin transform to the combinatorial formula (\ref{ind_gravity_conj0}) for the index, 
we get the supersymmetric zeta function for the toric Sasaki-Einstein manifold 
dual to quiver gauge theory with unitary gauge group $\prod_i U(N)_i$ of the form
\begin{align}
\label{zeta_gravity_conj0}
\mathfrak{Z}(s,z;\omega_1,\omega_2)
&=\sum_{I=1}^{d}
\left(\frac{(b,m_I)}{3} (\omega_1+\omega_2)\right)^{-s}
\zeta\left(s,1+\frac{3z}{(b,m_I)(\omega_1+\omega_2)}\right)
\nonumber\\
&-2A\left[\omega_1^{-s}\zeta\left(s,1+\frac{z}{\omega_1}\right)+\omega_2^{-s}\zeta\left(s,1+\frac{z}{\omega_2}\right)\right], 
\end{align}
where
\begin{align} 
\zeta(s,z)&=\sum_{n=0}^{\infty}\frac{1}{(n+z)^s}
\end{align}
is the Hurwitz zeta function. 
For the toric Sasaki-Einstein manifold dual to $\mathcal{N}=1$ $\prod_{i}SU(N)_i$ quiver gauge theory, 
it can be written as 
\begin{align}
\label{zeta_gravity_conj}
\mathfrak{Z}^{\mathcal{N}=1}(s,z;\omega_1,\omega_2)
&=\sum_{I=1}^{d}
\left(\frac{(b,m_I)}{3} (\omega_1+\omega_2)\right)^{-s}
\zeta\left(s,1+\frac{3z}{(b,m_I)(\omega_1+\omega_2)}\right). 
\end{align}
In comparison with the $U(N)$ gauge theory, 
the contributions from the decoupled $U(1)$ vector multiplets are removed in the $SU(N)$ theory. 

\subsection{Supersymmetric determinants}
Furthermore, we define the supersymmetric determinant for the Sasaki-Einstein geometry that is holographically dual to the 4d SCFT by
\begin{align}
\label{DEF_det}
\mathfrak{D}(z;\omega_1,\omega_2)
&=\exp\left[
\frac{\partial}{\partial s}\mathfrak{Z}(s,z;\omega_1,\omega_2)
\right]\Biggl|_{s=0}, 
\end{align}
where $\mathfrak{Z}(s,z;\omega_1,\omega_2)$ is the supersymmetric zeta function defined by (\ref{DEF_zeta}). 
It can be written as the product form
\begin{align}
\mathfrak{D}(z;\omega_1,\omega_2)
&=\prod_{n,m}(n\omega_1+m\omega_2+z)^{-\delta(n,m)}. 
\end{align}
One can compute it from the supersymmetric zeta function 
by using the formula 
\begin{align}
\exp\left[
\frac{\partial}{\partial s}
\left(a^{-s}\zeta\left(s,1+\frac{z}{a}\right)\right)
\right]\Biggl|_{s=0}
&=\frac{a^{\frac12+\frac{z}{a}} \Gamma\left(1+\frac{z}{a}\right)}
{\sqrt{2\pi}}. 
\end{align}

For the toric Sasaki-Einstein geometry dual to 
the quiver gauge theory with gauge group $\prod_i U(N)_i$, 
the supersymmetric determinant is computed from (\ref{zeta_gravity_conj0}) as follows: 
\begin{align}
\label{det_gravity_conj0}
\mathfrak{D}(z;\omega_1,\omega_2)
&=
\left(
\frac{2\pi \omega_1^{-\frac12-\frac{z}{\omega_1}} \omega_2^{-\frac12-\frac{z}{\omega_2}}}
{\Gamma\left(1+\frac{z}{\omega_1}\right)\Gamma\left(1+\frac{z}{\omega_2}\right)}
\right)^{2A}
\nonumber\\
&\times 
\prod_{I=1}^{d}
\left(\frac{(b,m_I) (\omega_1+\omega_2)}{3}\right)^{\frac12+\frac{3z}{(b,m_I) (\omega_1+\omega_2)}}
\frac{\Gamma\left(1+\frac{3z}{(b,m_I)(\omega_1+\omega_2)}\right)}{\sqrt{2\pi}}. 
\end{align}
For the toric Sasaki-Einstein dual to $\mathcal{N}=1$ quiver gauge theory with gauge group $\prod_i SU(N)_i$, 
the supersymmetric determinant takes a simpler form
\begin{align}
\label{det_gravity_conj}
\mathfrak{D}^{\mathcal{N}=1}(z;\omega_1,\omega_2)
&=\prod_{I=1}^{d}
\left(\frac{(b,m_I) (\omega_1+\omega_2)}{3}\right)^{\frac12+\frac{3z}{(b,m_I) (\omega_1+\omega_2)}}
\frac{\Gamma\left(1+\frac{3z}{(b,m_I)(\omega_1+\omega_2)}\right)}{\sqrt{2\pi}}. 
\end{align}
The factors associated with the decoupled $U(1)$ vector multiplets, which appear in the $U(N)$ gauge theory, 
are absent in the $SU(N)$ gauge theory. 

Although we do not pursue the question of whether the supersymmetric determinants (\ref{det_gravity_conj0}) and (\ref{det_gravity_conj}) 
admit a direct operator-theoretic interpretation from the gravity side in the present work, 
it would be highly intriguing to understand the gravitational origins of the eigenvalue problems underlying the supersymmetric determinants. 

\subsection{Cardy-like limit}
By means of the systematic analysis of the effective field theory (EFT) of the superconformal index, 
Di Pietro and Komargodski \cite{DiPietro:2014bca} argued 
that the leading behavior of the Cardy-like limit of the superconformal index for 4d SCFT is universally captured by the mixed $U(1)$-gravitational anomalies. 
The analytic properties of the supersymmetric zeta functions and supersymmetric determinants play a crucial role 
in uncovering not just the leading term but also the subleading terms. 
In \cite{Nakayama:2025hzr} it is shown that 
\begin{align}
\label{Cardy_lim}
&
\log \mathcal{I}(p;q)
\nonumber\\
&\sim 
\sum_{i}
\frac{\mathrm{Res}_{s=\alpha_i}\mathfrak{Z}(s,0;\omega_1,\omega_2)\Gamma(\alpha_i)\zeta(\alpha_i+1)}
{\beta^{\alpha_i}}
-\mathfrak{Z}(0,0;\omega_1,\omega_2)\log\beta
+\log \mathfrak{D}(0;\omega_1,\omega_2)
\end{align}
as $\beta$ $\rightarrow$ $0$. 
Importantly, there exist a logarithmic correction and a constant term in the asymptotic expansion. 
The logarithmic term is captured by the specialization $\mathfrak{Z}(0,0,;\omega_1,\omega_2)$ of the supersymmetric zeta function with $s=0$ and $z=0$. 
It is independent of the parameters $\omega_1$ and $\omega_2$ 
and it can be regarded as the analog of the Witten index, which we refer to as the \textit{Zeta-index} \cite{Nakayama:2025hzr}. 
The constant term is controlled by the specialization $\mathfrak{D}(0;\omega_1,\omega_2)$ of the supersymmetric determinant with $z=0$. 
At $\omega_1=\omega_2$ $=$ $1$, the resulting quantity captures 
the asymptotic degeneracy as a constant term, as we will see in (\ref{asymp_dn}), 
which we call the \textit{vacuum exponent} \cite{Nakayama:2025hzr}. 
The asymptotic formula (\ref{Cardy_lim}) maintains its validity regardless of the existence of a well-defined EFT. 

Let us now derive general expressions for the analytic data of the supersymmetric zeta functions and the supersymmetric determinants 
which characterize the Cardy-like limit (\ref{Cardy_lim}). 
The supersymmetric zeta function (\ref{zeta_gravity_conj0}) for the toric Sasaki-Einstein geometry dual to the quiver gauge theory of gauge group $\prod_i U(N)_i$
has a simple pole at $s=1$ and the corresponding residue is computed as
\begin{align}
\label{Res_formula0}
\mathrm{Res}_{s=1}\mathfrak{Z}(s,0;\omega_1,\omega_2)
&=3\sum_{I=1}^d \frac{1}{(b,m_I)}\frac{1}{\omega_1+\omega_2}
-2A\left(\frac{1}{\omega_1}+\frac{1}{\omega_2}\right). 
\end{align}
For the toric Sasaki-Einstein geometry dual to the $\mathcal{N}=1$ quiver gauge theory of gauge group $\prod_i SU(N)_i$ 
we have 
\begin{align}
\label{Res_formula}
\mathrm{Res}_{s=1}\mathfrak{Z}^{\mathcal{N}=1}(s,0;\omega_1,\omega_2)
&=3\sum_{I=1}^d \frac{1}{(b,m_I)}\frac{1}{\omega_1+\omega_2}. 
\end{align}
The leading Cardy-like asymptotics are captured by these residues of the corresponding supersymmetric zeta functions. 

Next consider the Zeta-indices which control the logarithmic terms in the Cardy-like asymptotics. 
For the toric Sasaki-Einstein manifold dual to the quiver gauge theory with unitary gauge group $\prod_i U(N)_i$ 
we obtain 
\begin{align}
\label{0_formula0}
\mathfrak{Z}(0,0;\omega_1,\omega_2)&=-\frac{d}{2}+2A.
\end{align}
For the toric Sasaki-Einstein dual to $\mathcal{N}=1$ quiver gauge theory of gauge group $\prod_i SU(N)_i$ 
the Zeta-index is equal to $-1/2$ times the number $d$ of the vertices of the toric diagram
\begin{align}
\label{0_formula}
\mathfrak{Z}^{\mathcal{N}=1}(0,0;\omega_1,\omega_2)&=-\frac{d}{2}.
\end{align}

Finally, let us consider the special values of the supersymmetric determinants at $z=0$ 
which determine the constant terms in the Cardy-like asymptotics. 
For the toric Sasaki-Einstein manifold dual to the quiver gauge theory of gauge group $\prod_i U(N)_i$ 
it is evaluated from (\ref{det_gravity_conj0}) as
\begin{align}
\label{0det_formula0}
\mathfrak{D}(0;\omega_1,\omega_2)
&=
\left(\frac{2\pi}{\sqrt{\omega_1\omega_2}}\right)^{2A}
\prod_{I=1}^d \sqrt{\frac{(b,m_I)(\omega_1+\omega_2)}{6\pi}}. 
\end{align}
The vacuum exponent is 
\begin{align}
\label{vac_formula0}
\mathfrak{D}(0;1,1)
&=
(2\pi)^{2A}
\prod_{I=1}^d \sqrt{\frac{(b,m_I)}{3\pi}}. 
\end{align}
In the case of the toric Sasaki-Einstein manifold dual to $\mathcal{N}=1$ quiver gauge theory of gauge group $\prod_i SU(N)_i$ we have 
\begin{align}
\label{0det_formula}
\mathfrak{D}^{\mathcal{N}=1}(0;\omega_1,\omega_2)
&=\prod_{I=1}^d \sqrt{\frac{(b,m_I)(\omega_1+\omega_2)}{6\pi}}. 
\end{align}
So the vacuum exponent is given by
\begin{align}
\label{vac_formula}
\mathfrak{D}^{\mathcal{N}=1}(0;1,1)
&=\prod_{I=1}^d \sqrt{\frac{(b,m_I)}{3\pi}}. 
\end{align}

Plugging (\ref{Res_formula0}), (\ref{0_formula0}) and (\ref{0det_formula0}) into (\ref{Cardy_lim}), 
we find that the Cardy-like limit of the multi-particle gravity index for the toric Sasaki-Einstein manifold 
dual to $\mathcal{N}=1$ quiver gauge theory of gauge group $\prod_{i}U(N)_i$ is given by
\begin{align}
\label{Cardy_lim_gravity0}
\log \mathcal{I}(p;q)
&\sim \frac{\pi^2}{2\beta}
\left[
\sum_{I=1}^{d}\frac{1}{(b,m_I)}\frac{1}{\omega_1+\omega_2}
-\frac{2A}{3}\left(\frac{1}{\omega_1}+\frac{1}{\omega_2}\right)
\right]
+\left(\frac{d}{2}-2A\right) \log\beta
\nonumber\\
&+\frac12 \sum_{I=1}^{d}
\log\left[
\frac{(b,m_I)(\omega_1+\omega_2)}{6\pi}
\right]
+2A\log \left(\frac{2\pi}{\sqrt{\omega_1\omega_2}}\right). 
\end{align}
In the case of the $\mathcal{N}=1$ quiver gauge theory of gauge group $\prod_i SU(N)_i$ it is given by 
\begin{align}
\label{Cardy_lim_gravity}
\log \mathcal{I}^{\mathcal{N}=1}(p;q)
&\sim \frac{\pi^2}{2\beta}
\left[
\sum_{I=1}^{d}\frac{1}{(b,m_I)}\frac{1}{\omega_1+\omega_2}
\right]
+\frac{d}{2}\log\beta
+\frac12 \sum_{I=1}^{d}
\log\left[
\frac{(b,m_I)(\omega_1+\omega_2)}{6\pi}
\right]. 
\end{align}
One should note that 
the order in which the Cardy-like limit and the large $N$ limit are taken is delicate and they do not commute 
(see section \ref{sec_finiteNlargeN} for more details). 
In the large $N$ limit the matrix integral is evaluated by a saddle-point method first in a way that it keeps geometric degrees of freedom. 
Consequently, the subsequent Cardy-like limit $\beta$ $\rightarrow$ $0$ reorganizes the effective action for the gravity dual that will contain the volume and curvature squared terms of the Sasaki-Einstein manifold. In the next subsection, we argue that 
the residues of the supersymmetric zeta functions, which govern the leading term in the Cardy-like limit, encode geometric invariants of the Sasaki-Einstein manifold as its volume and curvature squared term. 

\subsection{Curvature-squared integrals}
\label{sec_Riem2}
The gravity index for the toric Sasaki-Einstein manifold $Y$ counts the BPS operators in a way very similar to 
how one studies the spectral density of the Laplace-type operator on a compact manifold. 
In the context of spectral geometry, a central question is how the spectrum of differential operators encodes the underlying manifold. 
The foundational results show that the asymptotic expansion of such operators 
contains coefficients that are given by geometric invariants such as volume, scalar curvature, and integrated curvature-squared terms (see e.g. \cite{MR31145,MR217739,MR400315,MR861271,MR1064867,MR1616718,MR1749048}). 
It is conjectured by Arabi Ardehali, Liu and Szepietowski \cite{ArabiArdehali:2015iow} 
that the leading coefficient of the term $\frac{1}{\beta(\omega_1+\omega_2)}$ in the Cardy-like limit of the multi-particle gravity index is given by\footnote{The proposal (\ref{ALS_conj}) was formulated in \cite{ArabiArdehali:2014otj,ArabiArdehali:2015iow} on the basis of the asymptotic expansion of the Hilbert series obtained in \cite{Eager:2010dk}; see also Appendix \ref{app_HS}. }
\begin{align}
\label{ALS_conj}
\frac{1}{16\pi}\left(19\mathrm{Vol}(Y)+\frac18 \int_Y \mathrm{Riem}^2\right). 
\end{align}

According to (\ref{Cardy_lim}), the leading coefficient in the Cardy-like limit
is controlled by the residue of the supersymmetric zeta function at $s=1$. 
Reading (\ref{ALS_conj}) in this way, we are led to propose that 
for a Sasaki-Einstein 5-manifold $Y$ the curvature-squared invariant is determined by 
the analytic structure of $\mathfrak{Z}(s,0;\omega_1,\omega_2)$ at $s=1$ through
\begin{align}
\label{Riem2_conj2}
\int_Y \mathrm{Riem}^2&= 
\frac{64\pi^3}{3} \mathrm{Res}_{s=1}\mathfrak{Z}(s,0;\omega_1,\omega_2)\Bigl|_{\frac{1}{\omega_1+\omega_2}}-152\,\mathrm{Vol}(Y). 
\end{align}
Here it is understood that one extracts the coefficient of $\frac{1}{\omega_1+\omega_2}$ from the residue, 
in accordance with the formulation of (\ref{ALS_conj}). 
For the $\prod_i SU(N)_i$ quiver gauge theories this is automatic, 
the residue (\ref{Res_formula}) being proportional to $\frac{1}{\omega_1+\omega_2}$. 
Terms with a different dependence on $\omega_1$ and $\omega_2$ arise once the decoupled $U(1)$ vector multiplets are retained,
as in (\ref{Res_formula0}), and also for $\mathbb{RP}^5$ in (\ref{Res_RP5}). 
The essential point about (\ref{Riem2_conj2}) is that its right-hand side is determined 
by the single-particle BPS spectrum $\mathcal{H}_{\textrm{sp}}$ entering (\ref{DEF_zeta}) and requires no combinatorial input, 
so that the relation is not tied to toric geometry. 
We should add that when $Y$ is singular this spectrum is not that of Kaluza-Klein modes on the smooth locus alone, 
and correspondingly $\int_Y \mathrm{Riem}^2$ receives localized contributions. 
We return to this in sections \ref{sec_S5Z2} and \ref{sec_S5Zk}.

For a toric Sasaki-Einstein manifold the residue was evaluated in (\ref{Res_formula}) 
purely in terms of the toric data, and (\ref{Riem2_conj2}) then specializes to 
a closed combinatorial formula for the integral of the square of the Riemann tensor:
\begin{align}
\label{Riem2_conj}
\int_Y \mathrm{Riem}^2&=64\pi^3 \sum_{I=1}^d \frac{1}{(b,m_I)}
-\frac{152\pi^3}{3}\sum_{I} \frac{(m_I,v_{I-1})}{(b,m_{I-1}) (b,m_I)}, 
\end{align}
where we have rewritten the Martelli-Sparks-Yau volume formula (\ref{vol_Y1}) for the toric Sasaki-Einstein manifold $Y$ in terms of the primitive generators $m_I$ of the dual cone 
according to the relation $\det(v_{I-1},v_I,v_{I+1})$ $=$ $(m_I,v_{I-1})$ and $\det(b,v_{I-1},v_I)$ $=$ $(b,m_{I-1})$. 

In this toric specialization an equivalent relation is implicit in \cite{ArabiArdehali:2015iow}. 
Combining (\ref{ALS_conj}) with the large $N$ Cardy expansion of the index of an $SU(N)$ toric quiver, 
eq.(4.2) of that reference expresses $\int_Y \mathrm{Riem}^2$ through $\sum_I 1/r_I$, 
where $r_I$ are the R-charges of the extremal BPS mesons \cite{Benvenuti:2005ja,Benvenuti:2005cz}, 
and this is equivalent to (\ref{Riem2_conj}) by (\ref{Rch_meson}) and (\ref{vol_Y1}).
The description in terms of $\{r_I\}$, however, is available only in the toric case. 
Outside the toric Calabi-Yau setting the set $\{r_I\}$ is not merely hard to determine but undefined. 
The notion of an extremal BPS meson presupposes a mesonic flavor symmetry with respect to which a chiral primary of given R-charge can be extremal, 
together with the dictionary between such operators and the primitive generators of the dual cone, 
and neither survives once the toric Calabi-Yau structure is given up. 
The residue of $\mathfrak{Z}$, by contrast, is defined whenever the single-particle BPS spectrum is. 
It is in this sense that we regard (\ref{Riem2_conj2}), rather than its toric form (\ref{Riem2_conj}), as the basic statement. 

From the holographic perspective, the volume appears from the Einstein-Hilbert term in the effective action 
that is obtained from the two-derivative Type IIB supergravity classical action, 
whereas the curvature-squared term does not. 
It is an interesting problem to elucidate the origin of these terms in the Cardy-like limit, including coefficients from the gravity side. 
For instance, one plausible candidate is provided by the one-loop correction 
by quantizing the quadratic fluctuations of the supergravity fields around the classical background. 
Hence $\int_{Y} \mathrm{Riem}^2$ provides a geometric probe that goes beyond the classical Einstein data in the context of holography. 
Despite its significance, the direct evaluation of the integral of the Riemann square on a generic Sasaki-Einstein manifold is highly non-trivial even when an explicit metric is known. 
The formula (\ref{Riem2_conj}) shows that for toric geometries the invariant is nevertheless determined by the combinatorial data of the toric diagram alone, 
and in the following sections we evaluate it for numerous examples, 
checking at the same time the index formula (\ref{ind_gravity_conj}) on which the derivation rests.

The scope of (\ref{Riem2_conj2}) is genuinely wider.
In section \ref{sec_RP5} we consider $\mathbb{RP}^5$, whose K\"{a}hler cone $\mathbb{C}^3/\mathbb{Z}_2$ is again a toric variety but is not Gorenstein. 
The quotient lattice admits no vector $m_0$ with $(m_0,v_I)=1$, equivalently the $\mathbb{Z}_2$ action does not preserve the holomorphic 3-form, 
so that no toric diagram in the sense of section \ref{sec_SE} exists and with it neither the dual cone generators $m_I$ nor the extremal BPS mesons. 
Accordingly the single-particle index (\ref{ind_RP5}) is not of the form (\ref{ind_gravity_conj0}) or (\ref{ind_gravity_conj}), 
and the supersymmetric zeta function (\ref{zeta_RP5}) is not a finite sum of Hurwitz zeta functions 
but involves the sign-twisted Barnes multiple zeta functions (\ref{DEF_twistmultiplezeta}), with possible poles at $s=1,\dots,5$. 
Only the pole at $s=1$ survives, and (\ref{Riem2_conj2}) correctly reproduces $\int_{\mathbb{RP}^5}\mathrm{Riem}^2=20\pi^3$.
The relation (\ref{ALS_conj}) was formulated for toric Sasaki-Einstein manifolds, 
where the leading Cardy coefficient is controlled by the R-charges of the extremal BPS mesons. 
Our proposal (\ref{Riem2_conj2}) asserts that it is the residue of the supersymmetric zeta function, 
rather than that combinatorial datum, which carries the geometric information, and $\mathbb{RP}^5$ provides a test of this outside the toric Calabi-Yau setting.

\subsection{Asymptotic growth}
The above analytic properties of the supersymmetric zeta functions and supersymmetric determinants also reveal the asymptotic behavior of the BPS degeneracy. 
According to the gravity index formula (\ref{ind_gravity_conj}) for the toric Sasaki-Einstein manifold 
dual to the $\mathcal{N}=1$ quiver gauge theory with gauge group $\prod_i SU(N)_i$, 
the fugacities $p$ and $q$ only appear through the combination $pq$. 
So let us set $p=q$ for simplicity. 
When we expand the multi-particle gravity index as
\begin{align}
\mathcal{I}^{\mathcal{N}=1}(q;q)&=\sum_{n\ge0}d(n)q^n, 
\end{align}
the degeneracy $d(n)$ exhibits the Cardy-like growth at large $n$.\footnote{The index (\ref{ind_gravity_conj0}) for the quiver gauge theories of gauge group $\prod_i U(N)_i$ 
receives additional contributions from the decoupled $U(1)$ vector multiplets. 
The expansion coefficients of the index for the $U(1)$ vector multiplet do not show exponential growth 
but rather they exhibit oscillatory behavior. 
Consequently, the degeneracy does not grow exponentially in general. } 
Using the analysis of \cite{Nakayama:2025hzr}, 
the asymptotic degeneracy for the toric Sasaki-Einstein manifold dual to an $\mathcal{N}=1$ SCFT is given by
\begin{align}
\label{asymp_dn}
d(n)&\sim 
\frac{\prod_{I=1}^{d}\sqrt{(b,m_I)}}
{2^{\frac{d+3}{2}} 3^{\frac{d}{2}}n^{\frac{d+3}{4}}}
\left(\sum_{I=1}^{d}\frac{1}{(b,m_I)}\right)^{\frac{d+1}{4}}
\exp\left[
\pi\sqrt{\sum_{I=1}^{d}\frac{1}{(b,m_I)}}n^{\frac12}
\right]. 
\end{align}

\subsection{Supersymmetric zeta values}
In addition to the analytic structure at $s=1$ and $s=0$, which capture the asymptotic behavior in the Cardy-like limit, 
the supersymmetric zeta function (\ref{DEF_zeta}) admits a family of weighted moments defined by its evaluation at special negative values of $s$. 
Our analysis of a wide range of examples strongly suggests that this quantity captures the subleading holographic anomaly data which is independent of the $AdS$ radius. 

\subsubsection{$U(N)$ and $SU(N)$ quiver gauge theories}
When the associated gauge theory has a unitary gauge group $\prod_{i} U(N)_i$, 
we find that the special supersymmetric zeta value 
obtained from the supersymmetric zeta function (\ref{zeta_gravity_conj0}) with $s=-1$ and $z=0$ exactly vanishes
\begin{align}
\label{-1_formula0}
\mathfrak{Z}(-1,0;\omega_1,\omega_2)
&=0.  
\end{align}

For the toric Sasaki-Einstein manifold dual to the $\mathcal{N}=1$ quiver gauge theory of gauge group $\prod_i SU(N)_i$, 
we obtain from (\ref{zeta_gravity_conj}) the special supersymmetric zeta value for $s=-1$ and $z=0$ with the form 
\begin{align}
\label{-1_formula}
\mathfrak{Z}^{\mathcal{N}=1}(-1,0;\omega_1,\omega_2)
&=-\frac{A}{6}(\omega_1+\omega_2), 
\end{align}
where $A$ is the area of a lattice polygon in the toric diagram. 

Here we observe that the supersymmetric zeta values (\ref{-1_formula0}) and (\ref{-1_formula})  
encode the subleading terms of the 4d central charges $c$ and $a$ in large $N$ limit, i.e. the $\mathcal{O}(N^0)$ terms. 
In the $\prod_i U(N)_i$ quiver gauge theories the $\mathcal{O}(N^0)$ contributions $c_{\mathcal{O}(N^0)}$ and $a_{\mathcal{O}(N^0)}$ vanish. 
In the $\mathcal{N}=1$ $\prod_i SU(N)_i$ quiver gauge theories, 
there are $\mathcal{O}(N^0)$ contributions to the 4d central charges from the vector multiplets, 
whereas there is no contribution from the chiral multiplets. 
Physically, the area $A$ is equals one half of the number $n_{\textrm{vec}}$ of $\mathcal{N}=1$ vector multiplets  
or equivalently the number of gauge nodes in the quiver gauge theory
\begin{align}
A&=\frac{n_{\textrm{vec}}}{2}. 
\end{align}
Thus we have the $\mathcal{O}(N^0)$ contributions 
\begin{align}
\label{c0_formula}
c_{\mathcal{O}(N^0)}&=-\frac{1}{4}A, \\
\label{a0_formula}
a_{\mathcal{O}(N^0)}&=-\frac{3}{8}A, 
\end{align}
and the difference of the central charges 
\begin{align}
c_{\textrm{4d}}-a_{\textrm{4d}}&=\frac{1}{8}A. 
\end{align}

Accordingly, the two expressions (\ref{-1_formula0}) and (\ref{-1_formula}) 
for the special supersymmetric zeta value with $s=-1$ and $z=0$ can be unified into the following single expression: 
\begin{align}
\label{-1_formula2}
\mathfrak{Z}(-1,0;\omega_1,\omega_2)
&=-\frac43\left(c_{\mathcal{O}(N^0)}-a_{\mathcal{O}(N^0)}\right)(\omega_1+\omega_2)
\end{align}
from the field theoretical perspective. 
In other words, it encodes the non-trivial difference $c-a$. 
Holographically, it captures higher-derivative interactions and quantum effects in the bulk effective action. 
It is associated with a deviation of the shear-viscosity-to-entropy ration $\eta/s$ 
from its universal Einstein value $1/4\pi$ 
(see e.g. \cite{Brigante:2007nu,Kats:2007mq,Brigante:2008gz,Cremonini:2008tw,Buchel:2008vz,Liu:2010gz}).

\subsubsection{General cases}
So far we have focused on the $\mathcal{N}\ge 2$ $\prod_i U(N)_i$ quiver gauge theories 
and $\mathcal{N}=1$ $\prod_i SU(N)_i$ quiver gauge theories 
in which loops in the quiver are identified with both the $SU(N)$ adjoint and neutral chiral multiplets. 
However, the difference of central charges has been extensively studied holographically in different theories, 
e.g. $\mathcal{N}\ge 2$ gauge theories with gauge group $SU(N)$. 
It is therefore worth noting that 
the supersymmetric zeta values at $s=-1$ and $z=0$ for more general cases 
can encode \textit{subleading holographic anomalies}, i.e. subleading contributions to central charges of the dual SCFTs. 

We propose that the supersymmetric zeta values with $s=-1$ and $z=0$ 
for more general cases are given by the following expression: 
\begin{align}
\label{-1_formula3}
&
\mathfrak{Z}^{\textrm{general}}(-1,0;\omega_1,\omega_2)
\nonumber\\
&=-\frac43\left(c_{\mathcal{O}(N_{\textrm{eff}}^0)}-a_{\mathcal{O}(N_{\textrm{eff}}^0)}\right)(\omega_1+\omega_2)
+\frac{4}{27}\left(3c_{\mathcal{O}(N_{\textrm{eff}}^0)}-2a_{\mathcal{O}(N_{\textrm{eff}}^0)}\right)\frac{(\omega_1+\omega_2)^3}{\omega_1\omega_2}. 
\end{align}
When we consider $\mathcal{N}\ge 2$ quiver gauge theories of $\prod_i U(N)_i$ gauge theories 
and $\mathcal{N}=1$ $\prod_i SU(N)_i$ quiver gauge theories, 
the expression (\ref{-1_formula3}) reduces to (\ref{-1_formula2}) with $N_{\textrm{eff}}$ $=$ $N$. 
By extracting the coefficients of the supersymmetric zeta values (\ref{-1_formula3}), 
one can directly determine the $\mathcal{O}(N_{\textrm{eff}}^0)$ contributions to the 4d central charges. 

For example, when we consider $SU(N)$ gauge groups instead of $U(N)$ gauge groups for $\mathcal{N}\ge 2$ supersymmetric gauge theories, 
the term with $\frac{(\omega_1+\omega_2)^3}{\omega_1\omega_2}$ appears. 
In such cases $N_{\textrm{eff}}$ $=$ $N$ and 
we get $c_{\mathcal{O}(N^0)}$ $=$ $a_{\mathcal{O}(N^0)}$ $=$ $-\frac14$ 
for $\mathcal{N}=4$ $SU(N)$ gauge theory 
(see \cite{Bilal:1999ph,Bilal:1999ty,Mansfield:2000zw,Mansfield:2002pa,Mansfield:2003gs} for the holographic analysis), 
$c_{\mathcal{O}(N^0)}$ $=$ $-\frac13$, $a_{\mathcal{O}(N^0)}$ $=$ $-\frac{5}{12}$  
for $\mathcal{N}=2$ $SU(N)\times SU(N)$ quiver gauge theory describing a stack of $N$ D3-branes probing 
$\mathbb{C}\times \mathbb{C}^2/\mathbb{Z}_2$ (see \cite{ArabiArdehali:2013jiu} for the holographic analysis) 
and $c_{\mathcal{O}(N^0)}$ $=$ $-\frac{k}{6}$, $a_{\mathcal{O}(N^0)}$ $=$ $-\frac{5k}{24}$  
for $\mathcal{N}=2$ $\prod_{i=1}^k SU(N)_i$ quiver gauge theory for the low-energy description of $N$ D3-branes probing 
$\mathbb{C}\times \mathbb{C}^2/\mathbb{Z}_k$. 

For $\mathcal{N}=4$ super Yang-Mills theory with orthogonal and symplectic gauge groups, 
the subleading contributions should not be identified with a naive $\mathcal{O}(N^0)$ terms. 
Rather, the relevant expansion parameter is the effective flux
\begin{align}
\label{Neff}
N_{\textrm{eff}}&=\begin{cases}
N+\frac14&\textrm{for $SO(2N+1)$ and $USp(2N)$} \cr
N-\frac14&\textrm{for $O(2N)$} \cr
\end{cases}, 
\end{align}
which determines the $AdS$ curvature scale and is shifted by the presence of the O3-plane. 
The $+1/4$ shift in (\ref{Neff}) originates from the D3-brane charge of the $\widetilde{\textrm{O3}}^-$- and O3$^+$-planes, 
corresponding respectively to the $SO(2N+1)$ and $USp(2N)$ gauge theories, 
By contrast, the $-1/4$ shift is associated with the D3-brane charge of the O3$^-$-plane, which gives rise to the $O(2N)$ gauge theory. 
The subleading holographic anomaly information is encoded in the $\mathcal{O}(N_{\textrm{eff}}^0)$ terms, 
i.e., the pieces independent of the shifted $AdS$ radius, rather than in naive $\mathcal{O}(N^0)$ contributions. 
The latter do not correctly capture the intrinsic holographic quantities for orthogonal or symplectic gauge theories realized by introducing an orientifold plane. 
We substantiate this point in section \ref{sec_RP5}.

Our proposed formula (\ref{-1_formula3}) for the supersymmetric zeta values parallels the observation made in \cite{ArabiArdehali:2014otj,ArabiArdehali:2014cey}, 
where the relevant information is extracted by applying a differential operator to a reparametrized index and taking an appropriate fugacity limit. 
Despite its technical appearance, this operation is nothing but the supersymmetric zeta value evaluated at $s=-1$ and $z=0$. 

A further natural question is whether other supersymmetric zeta values admit an intrinsic geometric interpretation.
In the class of examples considered in this work, i.e. the toric Sasaki-Einstein manifolds, 
the special supersymmetric zeta values with $s=-2$ and $z=0$ are found to vanish
\begin{align}
\mathfrak{Z}(-2,0;\omega_1,\omega_2)
&=0.  
\end{align}
On the other hand, the special supersymmetric zeta values with $s=-3$ and $z=0$ are non-trivial. 
One may ask whether it can be written as a higher intrinsic functional of the toric polytope, such as a weighted sum over edges or vertices, 
extending the convex-geometric dictionary between spectral data and toric combinatorics.

\subsection{Finite-$N$ field theory vs large-$N$ geometry}
\label{sec_finiteNlargeN}
A crucial distinction must be made between 
the finite-$N$ field-theoretic supersymmetric spectral functions studied in \cite{Nakayama:2025hzr} 
and the holographic supersymmetric spectral functions that appears as their large-$N$ geometric limits discussed here. 
Consequently, the large $N$ limit taken prior to the asymptotic expansion differs from the large $N$ limit of the asymptotic expansion for finite-$N$ gauge theory result, 
including the Di Pietro and Komargodski formula \cite{DiPietro:2014bca}. 
For the field theoretic supersymmetric zeta function at fixed $N$ 
the residue at a simple pole $s=1$ that captures the leading term in the Cardy-like limit (\ref{Cardy_lim}) of the index 
can be expressed as the difference of the central charges \cite{Nakayama:2025hzr}
\begin{align}
\label{Cardy_4dSCFT}
\mathrm{Res}_{s=1}\mathfrak{Z}^{\textrm{4d $\mathcal{N}=1$ field}}(s,z;\omega_1,\omega_2)
&=16(c-a) \frac{\omega_1+\omega_2}{\omega_1\omega_2}. 
\end{align}
Obviously, the residue (\ref{Res_formula0}) or (\ref{Res_formula}) is not simply reproduced by taking its large $N$ limit. 
A noteworthy feature concerns the supersymmetric zeta value at $s=-1$ and $z=0$, which determines the supersymmetric Casimir energy at finite $N$ 
\cite{Kim:2009wb,Kim:2012ava,Assel:2014paa,Assel:2014tba,Cassani:2014zwa,Lorenzen:2014pna,Assel:2015nca,Bobev:2015kza,Martelli:2015kuk}. 
It can be also determined by the central charges as \cite{Nakayama:2025hzr}
\begin{align}
\label{Cas_4dSCFT}
&
\mathfrak{Z}^{\textrm{4d $\mathcal{N}=1$ field}}(-1,0;\omega_1,\omega_2)
\nonumber\\
&=-\frac43\left(c-a\right)(\omega_1+\omega_2)
+\frac{4}{27}\left( 3c-2a \right)\frac{(\omega_1+\omega_2)^3}{\omega_1\omega_2}. 
\end{align}
Remarkably, the holographic zeta value (\ref{-1_formula3}) exhibits precisely the same functional dependence on 
the parameters $\omega_1$, $\omega_2$ and on the central charges. 
However, the $\mathcal{O}(N_{\textrm{eff}}^2)$ terms of the central charges do not contribute to the holographic supersymmetric zeta value (\ref{-1_formula3}). 
In other words, the large-$N$ geometric limit effectively projects onto the subleading $\mathcal{O}(N_{\textrm{eff}}^0)$ terms of the central charges. 
This should be contrasted with the supersymmetric Casimir energy obtained in \cite{BenettiGenolini:2016qwm}. 
It is identified with the large $N$ limit of the expression (\ref{Cas_4dSCFT}), 
for which the leading $\mathcal{O}(N_{\textrm{eff}}^2)$ contributions survive. 
The difference between the two expressions (\ref{-1_formula3}) and (\ref{Cas_4dSCFT}) for the supersymmetric zeta values can also be traced to the non-commutativity of the large $N$ limit and the low-temperature limit. 

\subsection{Summary of the spectral data}
\label{sec_summary}
Before turning to examples we summarize, in Table \ref{tab_spectral}, the
analytic data extracted in this section for all the geometries treated in this paper. 
Each of the toric entries is obtained in closed form from the toric diagram alone, 
by way of the index (\ref{ind_gravity_conj}), 
the supersymmetric zeta function (\ref{zeta_gravity_conj}) 
and the supersymmetric determinant (\ref{det_gravity_conj}). 
Reading the table by columns exhibits three patterns which are not visible example by example. 
The Zeta-index equals $-d/2$ for every $\prod_i SU(N)_i$ quiver, so that it counts the lattice points of the toric diagram rather than any metric quantity. 
The supersymmetric zeta value at $s=-1$ is $-\frac{A}{6}(\omega_1+\omega_2)$, with $A$ the area of the lattice polygon,
equivalently one half the number of gauge nodes; and the vacuum exponent is a product over the extremal BPS mesons. 
The corresponding volumes and curvature-squared integrals are collected in Appendix \ref{app_Riem2}.

\begin{table}[H]
\centering
\begin{tabular}{|c|c|c|c|c|c|}
\hline
$Y$ & $d$ & $A$ & $\mathfrak{Z}(0,0)$ 
& $-\frac{\mathfrak{Z}(-1,0)}{\omega_1+\omega_2}$ & $\mathfrak{D}(0;1,1)$ \\ 
\hline\hline
$S^5$ $^\dagger$ & $3$ & $\frac12$ & $-\frac12$ & $0$ & $\frac{2}{3\sqrt{3\pi}}$ \\ \hline
$\mathbb{RP}^5$ $^\ddagger$ & --- & --- & $-\frac12$ & (\ref{-1_RP5}) & --- \\ \hline
$S^5/\mathbb{Z}_2$ $^\dagger$ & $4$ & $1$ & $0$ & $0$ & $\frac89$ \\ \hline
$(S^5/\mathbb{Z}_k)_{\textrm{ALE}}$ $^\dagger$ & $k+2$ & $\frac k2$ & $\frac k2-1$ & $0$ 
& $\frac{2^{k}k\pi^{\frac k2-1}}{3^{\frac k2+1}}$ \\ \hline
$S^5/\mathbb{Z}_{2n-1}$ & $3$ & $\frac{2n-1}{2}$ & $-\frac32$ & $\frac{2n-1}{12}$ 
& $\frac{(2n-1)^{\frac32}}{(3\pi)^{\frac32}}$ \\ \hline
$S^5/\mathbb{Z}_{2n}$ & $4$ & $n$ & $-2$ & $\frac n6$ & $\frac{2n^2}{9\pi^2}$ \\ \hline
$T^{1,1}$ & $4$ & $1$ & $-2$ & $\frac16$ & $\frac{1}{4\pi^2}$ \\ \hline
$T^{2,2}$ & $4$ & $2$ & $-2$ & $\frac13$ & $\frac{1}{\pi^2}$ \\ \hline
$T^{k,k}$ & $4$ & $k$ & $-2$ & $\frac k6$ & $\frac{k^2}{4\pi^2}$ \\ \hline
$Y^{2,1}$ & $4$ & $2$ & $-2$ & $\frac13$ & $\frac{4(2\sqrt{13}-5)}{9\pi^2}$ \\ \hline
$Y^{\mathsf{p},\mathsf{q}}$ & $4$ & $\mathsf{p}$ & $-2$ & $\frac{\mathsf{p}}{6}$ 
& (\ref{det_Ypq}) \\ \hline
$L^{1,2,1}$ & $5$ & $\frac32$ & $-\frac52$ & $\frac14$ 
& $\frac{1}{3^{\frac54}\pi^{\frac52}}$ \\ \hline
$L^{a,b,a}$ & $a+b+2$ & $\frac{a+b}{2}$ & $-\frac{a+b+2}{2}$ & $\frac{a+b}{12}$ 
& (\ref{vac_Laba}) \\ \hline
$L^{1,5,2}$ & $4$ & $3$ & $-2$ & $\frac12$ & (\ref{vac_L152}) \\ \hline
$L^{1,7,3}$ & $4$ & $4$ & $-2$ & $\frac23$ & (\ref{vac_L173}) \\ \hline
$L^{2,6,3}$ & $6$ & $4$ & $-3$ & $\frac23$ & (\ref{vac_L263}) \\ \hline
$L^{a,b,c}$ & $4$ & $\frac{a+b}{2}$ & $-2$ & $\frac{a+b}{12}$ & (\ref{vac_Labc}) \\ \hline
$L^{1,2,1}/\mathbb{Z}_2$ & $6$ & $3$ & $-3$ & $\frac12$ 
& $\frac{2\sqrt{2(1+\sqrt3)}}{3\sqrt3\pi^3}$ \\ \hline
$L^{1,3,1}/\mathbb{Z}_2$ & $8$ & $4$ & $-4$ & $\frac23$ 
& $\frac{\sqrt{11290+4501\sqrt7}}{162\pi^4}$ \\ \hline
$Y_{dP_2}$ & $5$ & $\frac52$ & $-\frac52$ & $\frac{5}{12}$ 
& $\frac{3\sqrt{558-78\sqrt{33}}}{32\pi^{\frac52}}$ \\ \hline
$Y_{dP_3}$ & $6$ & $3$ & $-3$ & $\frac12$ & $\frac{1}{\pi^3}$ \\ \hline
$Y_{PdP_{3a}}$ & $6$ & $3$ & $-3$ & $\frac12$ & $\frac{4}{3\sqrt3\pi^3}$ \\ \hline
$Y_{PdP_{4a}}$ & $7$ & $\frac72$ & $-\frac72$ & $\frac{7}{12}$ & (\ref{vac_PdP4a}) \\ \hline
$Y_{PdP_{4b}}$ & $7$ & $\frac72$ & $-\frac72$ & $\frac{7}{12}$ & (\ref{vac_PdP4b}) \\ \hline
$Y_{PdP_5}$ & $8$ & $4$ & $-4$ & $\frac23$ & $\frac{1}{\pi^4}$ \\ \hline
$X^{\mathsf{p},\mathsf{q}}$ & $5$ & $\frac{2\mathsf{p}+1}{2}$ & $-\frac52$ 
& $\frac{2\mathsf{p}+1}{12}$ & (\ref{vac_Xpq}) \\ \hline
$Z^{\mathsf{p},\mathsf{q}}$ & $6$ & $\mathsf{p}+1$ & $-3$ & $\frac{\mathsf{p}+1}{6}$ 
& (\ref{vac_Zpq}) \\ \hline
\end{tabular}
\caption{Spectral data for the Sasaki-Einstein manifolds treated in this paper.
$d$ is the number of lattice points of the toric diagram, $A$ the area of the
lattice polygon. Entries marked $\dagger$ refer to $\prod_i U(N)_i$ quiver gauge
theories, for which $\mathfrak{Z}(0,0)=-\frac d2+2A$ and
$\mathfrak{Z}(-1,0)=0$; all others are $\prod_i SU(N)_i$ theories, for which
$\mathfrak{Z}(0,0)=-\frac d2$ and
$\mathfrak{Z}(-1,0)=-\frac{A}{6}(\omega_1+\omega_2)$. The entry marked $\ddagger$
admits no toric diagram. In all $SU(N)$ cases the difference of central charges
is $c_{\mathcal{O}(N^0)}-a_{\mathcal{O}(N^0)}=A/8$.}
\label{tab_spectral}
\end{table}

\section{Highly symmetric Sasaki-Einstein manifolds}
\label{sec_symSE}
The following sections are organized in a uniform way. 
For each geometry we first record the gauge theory and the toric data, then the single-particle index and the extremal BPS mesons, 
and finally the supersymmetric zeta function, the supersymmetric determinant and the spectral data extracted from them. 
The point of treating so many cases is not to tabulate formulas but to establish which features of these spectral functions are universal. 
The reader who wishes to see the outcome of this comparison may consult Table \ref{tab_spectral} of section \ref{sec_summary}, 
where the results are collected and the regularities are stated.

In this section we discuss the supersymmetric spectral functions of the Sasaki-Einstein manifolds that 
are holographically dual to supersymmetric gauge theories with at least $\mathcal{N}=2$ supersymmetry. 
As we discussed in section \ref{sec_ind}, we take the gauge group to be $U(N)$ for the extended supersymmetric gauge theories based on the Lie algebra $A_{N-1}$. 
This choice allows us to include the trace component of the adjoint chiral multiplet, 
which plays a role in the construction of the protected mesonic operators. 
We further study, as a case falling outside the toric Calabi-Yau setting, 
$\mathcal{N}=4$ supersymmetric gauge theories with BCD-type gauge groups dual to $AdS_5\times \mathbb{RP}^5$.
\subsection{$AdS_5\times S^5$}
Consider 4d $\mathcal{N}=4$ super Yang-Mills theory with gauge group $U(N)$.  
It is a low-energy effective theory on a stack of $N$ D3-branes propagating in flat space $\mathbb{C}^3$. 
In $\mathcal{N}=1$ language there is a vector multiplet of $U(N)$ gauge group and three adjoint chiral multiplets $X_{11}^{1}$, $X_{11}^{2}$ and $X_{11}^{3}$ of R-charge $R=2/3$. 
The quiver diagram is shown as follows: 
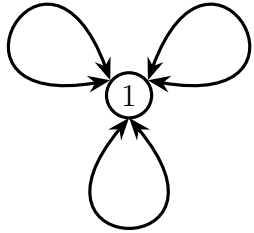
\begin{align}
\label{quiver_S5}
\begin{tikzpicture}[<->]
    \draw[very thick] (0,0) circle (3mm);
    \node at (0,0) {1};
\draw[>=Stealth,scale=5,very thick] (0,-0.06)  to[in=-50,out=-130,loop] (0,0);
\draw[>=Stealth,scale=5,very thick] (-0.05,0.04)  to[in=-170,out=-250,loop] (0,0);
\draw[>=Stealth,scale=5,very thick] (0.05,0.04)  to[in=70,out=-10,loop] (0,0);
\end{tikzpicture}. 
\end{align}

It is holographically dual to Type IIB string theory on $AdS_5\times S^5$, 
where $S^5$ is the base of the Calabi-Yau cone $\mathbb{C}^3$. 
The toric diagram of $\mathbb{C}^3$ is generated by
\begin{align}
\label{vertex_C3}
v_1&=(1,0,0), \qquad 
v_2=(1,1,0), \qquad 
v_3=(1,0,1). 
\end{align}
Projecting these onto the two-dimensional plane, we obtain 
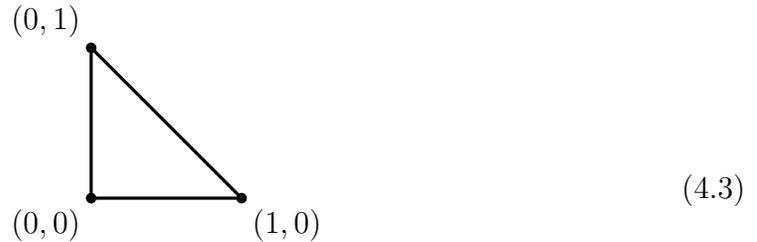
\begin{align}
\label{toric_C3}
\begin{tikzpicture}[scale=2, baseline={(0,0)}]
  \fill (0,0) circle (1pt) node[below left] {$(0,0)$};
  \fill (1,0) circle (1pt) node[below right] {$(1,0)$};
  \fill (0,1) circle (1pt) node[above left] {$(0,1)$};
  \draw[very thick] (0,0) -- (1,0) -- (0,1) -- cycle;
\end{tikzpicture}
\end{align}
We have 
\begin{align}
\det (v_{I-1},v_I,v_{I+1})&=1, \qquad I=1,2,3, 
\end{align}
which enumerates each type of chiral multiplets $X_{11}^I$ for $I=1,2,3$.  
The area of the polygon is 
\begin{align}
A&=\frac12. 
\end{align}
The primitive lattice generators of the dual cone are
\begin{align}
\label{dualG_C3}
m_1&=(0,0,1), \qquad 
m_2=(1,-1,-1), \qquad 
m_3=(0,1,0). 
\end{align}
The Reeb vector is
\begin{align}
b&=(3,1,1). 
\end{align}
The volume of $S^5$ is 
\begin{align}
\label{vol_S5}
\mathrm{Vol}(S^5)&=\pi^3. 
\end{align}
The R-charges of the chiral multiplets $X_{11}^I$ are evaluated as
\begin{align}
R(X_{11}^I)&=\frac{\pi}{3}\frac{\mathrm{Vol}(\Sigma_I)}{\mathrm{Vol}(S^5)}
=\frac23, \qquad I=1,2,3, 
\end{align}
where for $I=1,2,3$, we find 
\begin{align}
\mathrm{Vol}(\Sigma_I)&=2\pi^2 \frac{\det (v_{I-1},v_I,v_{I+1})}{\det(b,v_{I-1},v_I)\det(b,v_I,v_{I+1})}=2\pi^2. 
\end{align}
This correctly matches the volume of $S^3$. 

The weighted adjacency matrix is simply 
\begin{align}
\label{M_S5}
M^{AdS_5\times S^5}
&=i_v+3i_{\textrm{cm}_{\frac23}}+3\tilde{i}_{\textrm{cm}_{\frac23}}. 
\end{align}
The single-particle gravity index is evaluated as \cite{Kinney:2005ej}
\begin{align}
\label{ind_S5}
i^{AdS_5\times S^5}(p;q)
&=-\sum_{k=1}^{\infty}\frac{\varphi(k)}{k}\log\left[\det (1-M(p^k;q^k))\right]
\nonumber\\
&=3\frac{(pq)^{\frac13}}{1-(pq)^{\frac13}}-\frac{p}{1-p}-\frac{q}{1-q}. 
\end{align}
Here we have no subtractive terms as we consider $U(N)$ gauge group. 
The first term of (\ref{ind_S5}) is contributed by three extremal BPS mesonic operators as single-trace operators of the three adjoint chirals
\begin{align}
\mathcal{M}_1&=\Tr (X_{11}^{3}), & 
\mathcal{M}_2&=\Tr (X_{11}^{1}), &
\mathcal{M}_3&=\Tr (X_{11}^{2}). 
\end{align}
The three extremal BPS mesons $\mathcal{M}_I$ correspond to the three primitive generators $m_I$ of the dual cone in (\ref{dualG_C3}). 
In fact, we obtain from the formula (\ref{Rch_meson}) the R-charges
\begin{align}
\label{Rmeson_S5}
R(\mathcal{M}_1)&=\frac23 (b,m_1)=\frac23, \nonumber\\
R(\mathcal{M}_2)&=\frac23 (b,m_2)=\frac23, \nonumber\\
R(\mathcal{M}_3)&=\frac23 (b,m_3)=\frac23. 
\end{align}
The operator lengths of the extremal BPS mesons $\mathcal{M}_I$ are
\begin{align}
L_J(\mathcal{M}_1)&=(m_1,v_J)=\delta_{J,3}, \\
L_J(\mathcal{M}_2)&=(m_2,v_J)=\delta_{J,1}, \\
L_J(\mathcal{M}_3)&=(m_3,v_J)=\delta_{J,2}, 
\end{align}
which implies that the theory has the three adjoint chiral multiplets. 

The supersymmetric zeta function is evaluated as
\begin{align}
\label{zeta_S5}
&
\mathfrak{Z}^{AdS_5\times S^5}(s,z;\omega_1,\omega_2)
\nonumber\\
&=3\left(\frac{\omega_1+\omega_2}{3}\right)^{-s}\zeta\left(s,1+\frac{3z}{\omega_1+\omega_2}\right)
-\omega_1^{-s}\zeta\left(s,1+\frac{z}{\omega_1}\right)
-\omega_2^{-s}\zeta\left(s,1+\frac{z}{\omega_2}\right). 
\end{align}
The residue at a simple pole $s=1$ is
\begin{align}
\label{Res_S5}
\mathrm{Res}_{s=1}\mathfrak{Z}^{AdS_5\times S^5}(s,z;\omega_1,\omega_2)
&=\frac{9}{\omega_1+\omega_2}-\frac{1}{\omega_1}-\frac{1}{\omega_2}. 
\end{align}
Substituting the coefficient of the term $\frac{1}{\omega_1+\omega_2}$ 
and the volume (\ref{vol_S5}) into (\ref{Riem2_conj}), 
we get the integral of the Riemann square over $S^5$
\begin{align}
\label{Riem2_S5}
\int_{S^5} \mathrm{Riem}^2&=40\pi^3=1240.25\ldots. 
\end{align}
The Zeta-index is evaluated as
\begin{align}
\label{0_S5}
\mathfrak{Z}^{AdS_5\times S^5}(0,0;\omega_1,\omega_2)&=-\frac12. 
\end{align}
The supersymmetric zeta value at $s=-1$ and $z=0$ vanishes\footnote{
If one considers $\mathcal{N}=4$ $SU(N)$ SYM theory, 
the supersymmetric zeta function is
\begin{align}
&
\mathfrak{Z}^{AdS_5\times S^5}(s,z;\omega_1,\omega_2)
\nonumber\\
&=3\left(\frac{\omega_1+\omega_2}{3}\right)^{-s}\zeta\left(s,1+\frac{3z}{\omega_1+\omega_2}\right)
\nonumber\\
&-3\zeta_2\left(s,\frac{\omega_1+\omega_2}{3}+z;\omega_1,\omega_2\right)
+3\zeta_2\left(s,\frac{2(\omega_1+\omega_2)}{3}+z;\omega_1,\omega_2\right)
\end{align}
and the supersymmetric zeta value at $s=-1$ and $z=0$ is evaluated as $-\frac{(\omega_1+\omega_2)^3}{27\omega_1\omega_2}$. 
Then we find from the formula (\ref{-1_formula3})
\begin{align}
c_{\mathcal{O}(N^0)}&=a_{\mathcal{O}(N^0)}=-\frac14. 
\end{align}
}
\begin{align}
\label{-1_S5}
\mathfrak{Z}^{AdS_5\times S^5}(-1,0;\omega_1,\omega_2)
&=0. 
\end{align}
According to the formulae (\ref{c0_formula}) and (\ref{a0_formula}), we obtain 
\begin{align}
c_{\mathcal{O}(N^0)}&=0, \\
a_{\mathcal{O}(N^0)}&=0, 
\end{align}
and 
\begin{align}
c_{\mathcal{O}(N^0)}-a_{\mathcal{O}(N^0)}&=0, 
\end{align}
as expected. 

The supersymmetric determinant is evaluated as
\begin{align}
\label{det_S5}
\mathfrak{D}^{AdS_5\times S^5}(z;\omega_1,\omega_2)
&=\frac{
(\omega_1+\omega_2)^{\frac32+\frac{9z}{\omega_1+\omega_2}}
}
{2^{\frac12}3^{3\left(\frac12+\frac{3z}{\omega_1+\omega_2}\right)}\pi^{\frac12}\omega_1^{\frac12+\frac{z}{\omega_1}} \omega_2^{\frac12+\frac{z}{\omega_2}}}
\frac{\Gamma\left(1+\frac{3z}{\omega_1+\omega_2}\right)^3}
{\Gamma\left(1+\frac{z}{\omega_1}\right)\Gamma\left(1+\frac{z}{\omega_2}\right)}. 
\end{align}
The special value at $z=0$ is 
\begin{align}
\mathfrak{D}^{AdS_5\times S^5}(0;\omega_1,\omega_2)
&=\frac{1}{3\sqrt{6\pi}}\frac{(\omega_1+\omega_2)^{\frac32}}{(\omega_1\omega_2)^{\frac12}}. 
\end{align}
With $\omega_1=\omega_2=1$ it reduces to the vacuum exponent 
\begin{align}
\label{vac_S5}
\mathfrak{D}^{AdS_5\times S^5}(0;1,1)
&=\frac{2}{3\sqrt{3\pi}}=0.217157\ldots. 
\end{align}

\subsection{$AdS_5\times \mathbb{RP}^5$}
\label{sec_RP5}
Next consider 4d $\mathcal{N}=4$ super Yang-Mills theory with orthogonal and symplectic gauge groups. 
It is holographically dual to Type IIB string theory on $AdS_5\times \mathbb{RP}^5$ \cite{Witten:1998xy}. 
The dual geometry $\mathbb{RP}^5$ arises from the quotient of $S^5$, corresponding to the identification of points under the $\mathbb{Z}_2$ involution. 
The volume of $\mathbb{RP}^5$ is exactly one half of that of $S^5$
\begin{align}
\label{vol_RP5}
\mathrm{Vol}(\mathbb{RP}^5)&=\frac{\pi^3}{2}. 
\end{align}
Although the quotient $\mathbb{C}^3/\mathbb{Z}_2$ is again a toric variety, 
the quotient lattice admits no vector $m_0$ with $(m_0,v_I)=1$, equivalently 
the $\mathbb{Z}_2$ action does not preserve the holomorphic 3-form. 
The cone is therefore not Gorenstein, no toric diagram in the sense of section 
\ref{sec_SE} exists, and with it neither the dual cone generators $m_I$ nor the 
extremal BPS mesons are available. 

The single-particle gravity index is obtained from the large $N$ limit of 
the superconformal indices of $\mathcal{N}=4$ super Yang-Mills theory with gauge groups $SO(2N+1)$, $USp(2N)$ and $O(2N)$ 
(see e.g. \cite{Imamura:2016abe}). 
It is obtained by implementing the $\mathbb{Z}_2$ gauging procedure as follows:  
\begin{align}
\label{ind_RP5}
i^{AdS_5\times \mathbb{RP}^5}(p;q)
&=\frac12
\left(
i^{AdS_5\times S^5}(p;q)+i^{AdS_5\times S^5_{\textrm{odd}}}(p;q)
\right), 
\end{align}
where $i^{AdS_5\times S^5}(p;q)$ is the single particle gravity index (\ref{ind_S5}) for $AdS_5\times S^5$ and 
\begin{align}
i^{AdS_5\times S^5_{\textrm{odd}}}(p;q)
&=\frac12
\left(
-1+\frac{(1+p)(1+q)(1-(pq)^{\frac13})^3}{(1-p)(1-q)(1+(pq)^{\frac13})^3}
\right)
\end{align}
is the index for the $\mathbb{Z}_2$-odd KK modes. 

Analogously to the single-particle gravity index (\ref{ind_RP5}), 
the supersymmetric zeta function for $AdS_5\times \mathbb{RP}^5$ involves contributions from KK modes even and odd under the $\mathbb{Z}_2$ action
\begin{align}
\label{zeta_RP5}
\mathfrak{Z}^{AdS_5\times \mathbb{RP}^5}(s,z;\omega_1,\omega_2)
&=\frac12
\left(
\mathfrak{Z}^{AdS_5\times S^5}(s,z;\omega_1,\omega_2)+\mathfrak{Z}^{AdS_5\times S^5_{\textrm{odd}}}(s,z;\omega_1,\omega_2)
\right). 
\end{align}
Here the contributions from 
the $\mathbb{Z}_2$-even KK modes are given by the supersymmetric zeta function (\ref{zeta_S5}) for $AdS_5\times S^5$, 
whereas those from the $\mathbb{Z}_2$-odd KK modes are given by
\begin{align}
&
\mathfrak{Z}^{AdS_5\times S^5_{\textrm{odd}}}(s,z;\omega_1,\omega_2)
\nonumber\\
&=\omega_1^{-s}\zeta\left(s,1+\frac{z}{\omega_1}\right)
+\omega_2^{-s}\zeta\left(s,1+\frac{z}{\omega_2}\right)
-3\zeta_1\left(s,\frac{\omega_1+\omega_2}{3}+z ; \frac{\omega_1+\omega_2}{3}  \;\middle|\;  -1\right)
\nonumber\\
&+2\zeta_2\left(s,\omega_1+\omega_2+z;\omega_1,\omega_2\right)
-6\sum_{i=1}^2\zeta_2\left(s,\omega_i+\frac{\omega_1+\omega_2}{3}+z;\omega_i,\frac{\omega_1+\omega_2}{3} \;\middle|\; 1,-1 \right)
\nonumber\\
&+6\zeta_2\left(s,\frac{2(\omega_1+\omega_2)}{3}+z; \left\{\frac{\omega_1+\omega_2}{3}\right\}^2 \;\middle|\; \{-1\}^2 \right)
\nonumber\\
&-12\zeta_3\left(s,\frac{4(\omega_1+\omega_2)}{3}+z; \omega_1,\omega_2,\frac{\omega_1+\omega_2}{3} \;\middle|\; \{-1\}^3 \right)
\nonumber\\
&+12\sum_{i=1}^2\zeta_3\left(s,\omega_i+\frac{2(\omega_1+\omega_2)}{3}+z;\omega_i,\left\{\frac{\omega_1+\omega_2}{3}\right\}^2 \;\middle|\; 1,\{-1\}^2 \right)
\nonumber\\
&+24\zeta_4\left(s,\frac{5(\omega_1+\omega_2)}{3}+z;\omega_1,\omega_2,\left\{\frac{\omega_1+\omega_2}{3}\right\}^2 \;\middle|\; \{1\}^2,\{-1\}^2 \right)
\nonumber\\
&-8\sum_{i=1}^2 \zeta_4\left(s,\omega_i+\omega_1+\omega_2+z;\omega_i,\left\{\frac{\omega_1+\omega_2}{3}\right\}^3 \;\middle|\; 1,\{-1\}^3 \right)
\nonumber\\
&-16\zeta_5\left(s,2(\omega_1+\omega_2)+z;\omega_1,\omega_2,\left\{\frac{\omega_1+\omega_2}{3}\right\}^3 \;\middle|\; \{1\}^2,\{-1\}^3 \right), 
\end{align}
where we have defined the \textit{sign-twisted Barnes multiple zeta function} as
\begin{align}
\label{DEF_twistmultiplezeta}
\zeta_k\left(s,z;\omega_1,\cdots,\omega_k|\epsilon_1,\cdots,\epsilon_k\right)
&=\sum_{n_1,\cdots,n_k\ge0}
\frac{\prod_{j=1}^k \epsilon_j^{n_j}}{(\omega_1 n_1+\cdots +\omega_k n_k+z)^{s}}, 
\end{align}
with $\epsilon$ $=$ $(\epsilon_1,\cdots,\epsilon_k)$ $\in$ $\{\pm 1\}^k$. 
We see that $\zeta_k\left(s,z;\omega_1,\cdots,\omega_k|\epsilon_1,\cdots,\epsilon_k\right)$ has an integral representation
\begin{align}
&
\zeta_k\left(s,z;\omega_1,\cdots,\omega_k|\epsilon_1,\cdots,\epsilon_k\right)
\nonumber\\
&=\frac{1}{\Gamma(s)}\sum_{n_1,\cdots,n_k\ge0}
\int_0^{\infty} t^{s-1}
\prod_{j=1}^{k}\epsilon_j^{n_j}
e^{-(\omega_1n_1+\cdots+\omega_k n_k+z)t}dt
\nonumber\\
&=\frac{1}{\Gamma(s)}\int_0^{\infty}t^{s-1}
\frac{e^{-zt}}{(1-\epsilon_1 e^{-\omega_1 t}) \cdots (1-\epsilon_k e^{-\omega_k t})}dt. 
\end{align}
The analytic properties of the sign-twisted Barnes multiple zeta function (\ref{DEF_twistmultiplezeta}), 
including location of poles, their residues and values at negative integers come from the asymptotic expansion of the heat kernel 
(see e.g. \cite{MR1396308})
\begin{align}
\frac{e^{-zt}}{(1-\epsilon_1 e^{-\omega_1 t}) \cdots (1-\epsilon_k e^{-\omega_k t})}. 
\end{align}
The sign-twisted Barnes multiple zeta function (\ref{DEF_twistmultiplezeta}) has poles at $s=k-j$, with $j=0,\cdots, k-1$. 
So each term of the supersymmetric zeta function (\ref{zeta_RP5}) may admit poles at $s=1,\cdots, 5$. 
However, one finds that the residues at poles with $s>1$ vanish, while the residue at the simple pole $s=1$ is given by
\begin{align}
\label{Res_RP5}
\mathrm{Res}_{s=1}\mathfrak{Z}^{AdS_5\times \mathbb{RP}^5}(s,z;\omega_1,\omega_2)
&=\frac92\frac{1}{\omega_1+\omega_2}
-\frac12\left(\frac{1}{\omega_1}+\frac{1}{\omega_2}\right). 
\end{align}
Upon inserting the residue coefficient of (\ref{Res_RP5}) and the volume (\ref{vol_RP5}) into the formula (\ref{Riem2_conj2}), 
we then obtain the Riemann-squared integral 
\begin{align}
\label{Riem2_RP5}
\int_{\mathbb{RP}^5} \mathrm{Riem}^2&=20\pi^3, 
\end{align}
which is one half of the Riemann-squared integral (\ref{Riem2_S5}) of $S^5$. 
From a geometric viewpoint, $\mathbb{RP}^5$ inherits the constant curvature of the sphere under the free $\mathbb{Z}_2$ identification. 
The uniformity of the Riemann tensor implies that the curvature-squared integral scales linearly with the volume. 
The value of the supersymmetric zeta function at $s=0$, $z=0$ is given by
\begin{align}
\label{0_RP5}
\mathfrak{Z}^{AdS_5\times \mathbb{RP}^5}(0,0;\omega_1,\omega_2)
&=-\frac12. 
\end{align}
The supersymmetric zeta function evaluated at $s=-1$ and $z=0$ takes the following value:
\begin{align}
\label{-1_RP5}
\mathfrak{Z}^{AdS_5\times \mathbb{RP}^5}(-1,0;\omega_1,\omega_2)
&=-\frac{1}{216}\frac{(\omega_1+\omega_2)^3}{\omega_1\omega_2}. 
\end{align}
According to the formula (\ref{-1_formula3}) we obtain 
\begin{align}
c_{\mathcal{O}(N_{\textrm{eff}}^0)}&=-\frac{1}{32}, \\
a_{\mathcal{O}(N_{\textrm{eff}}^0)}&=-\frac{1}{32}, 
\end{align}
and 
\begin{align}
c_{\mathcal{O}(N_{\textrm{eff}}^0)}-
a_{\mathcal{O}(N_{\textrm{eff}}^0)}&=0. 
\end{align}
These values are consistent with the 
central charges of $\mathcal{N}=4$ super Yang-Mills theories of orthogonal and symplectic gauge groups 
\begin{align}
c=\frac{N_{\textrm{eff}}^2}{2}-\frac{1}{32}, \\
a=\frac{N_{\textrm{eff}}^2}{2}-\frac{1}{32}, 
\end{align}
where $N_{\textrm{eff}}$ is given by the effective flux (\ref{Neff}) in the presence of the orientifold plane. 

\subsection{$AdS_5\times S^5/\mathbb{Z}_2$}
\label{sec_S5Z2}
Let us consider the 4d quiver gauge theory that is realized on a stack of $N$ D3-branes in the background of the $\mathbb{Z}_2$ orbifold of $\mathbb{C}^3$. 
In this case there is only one Calabi-Yau $\mathbb{Z}_2$ orbifold of $\mathbb{C}^3$ with the following ALE-type action on the three complex coordinates on $\mathbb{C}^3$: 
\begin{align}
\label{Z2_action}
\theta: (z_1,z_2,z_3)\mapsto (-z_1,-z_2,z_3). 
\end{align}
The theory on the D3-branes has $\mathcal{N}=2$ supersymmetry. 
We identify it with the $\mathcal{N}=2$ $U(N)\times U(N)$ quiver gauge theory, 
where each $\mathcal{N}=2$ $U(N)$ vector multiplet contains an adjoint chiral multiplet $X_{ii}$, with $i=1,2$. 
Also the theory has a pair of chiral multiplets $X_{i\ i+1}^{\alpha}$, with $\alpha=1,2$ transforming in the bifundamental representation. 
These chiral multiplets carry R-charge $R=2/3$. 
The quiver diagram is illustrated as
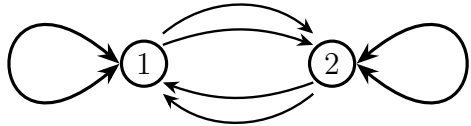
\begin{align}
\begin{tikzpicture}[<->]
    \draw[very thick] (-1.25,0) circle (3mm);
    \node at (-1.25,0) {1};
     \draw[very thick] (1.25,0) circle (3mm);
    \node at (1.25,0) {2};
\draw[>=Stealth,scale=5,very thick] (-0.315,0)  to[in=-140,out=-220,loop] (0,0);
\draw[>=Stealth,scale=5,very thick] (0.315,0)  to[in=40,out=-40,loop] (0,0);
    \draw[-{Stealth[length=6pt,width=5pt]}, line width=0.9pt, bend left=40]
         (-1,0.4) to node[]{} (1,0.4);
    \draw[-{Stealth[length=6pt,width=5pt]}, line width=0.9pt, bend left=20]
         (-1,0.25) to node[]{} (1,0.25);
     \draw[-{Stealth[length=6pt,width=5pt]}, line width=0.9pt, bend left=40]
         (1,-0.4) to node[]{} (-1,-0.4);
    \draw[-{Stealth[length=6pt,width=5pt]}, line width=0.9pt, bend left=20]
         (1,-0.25) to node[]{} (-1,-0.25);
\end{tikzpicture}. 
\end{align}
The superpotential takes the form
\begin{align}
\mathcal{W}&=\sum_{\alpha=1}^2 \Tr (X_{11}X_{12}^{\alpha}X_{21}^{\alpha}-X_{22}X_{21}^{\alpha}X_{12}^{\alpha}). 
\end{align}
It is holographically dual to Type IIB string theory on $AdS_5\times S^5/\mathbb{Z}_2$. 
The fixed locus of the group action (\ref{Z2_action}) is the set of points left invariant by all elements of the stabilizer. 
The fixed set is the entire complex line $\{z_1=z_2=0\}$ $\cong$ $\mathbb{C}$ parameterized by $z_3$. 
Hence there are infinitely many fixed points as non-isolated singularity 
and locally we have $\mathbb{C}^3/\mathbb{Z}_2$ $\cong$ $\mathbb{C}^2/\mathbb{Z}_2\times \mathbb{C}$. 
The toric diagram is generated by 
\begin{align}
v_1&=(1,0,0), \qquad 
v_2=(1,1,0), \qquad 
v_3=(1,1,1), \qquad 
v_4=(1,-1,0). 
\end{align}
It lives in the two-dimensional plane as
\begin{align}
\label{toric_C3/Z2}
\begin{tikzpicture}[scale=1.5, baseline={(0,0)}]
  \fill (0,0) circle (2pt) node[below] {$(0,0)$};
  \fill (1,0) circle (2pt) node[below right] {$(1,0)$};
  \fill (1,1) circle (2pt) node[above right] {$(1,1)$};
  \fill (-1,0) circle (2pt) node[below left] {$(-1,0)$};
  \draw[very thick] (0,0) -- (1,0) -- (1,1) -- (-1,0) -- cycle;
\end{tikzpicture}
\end{align}
This is a triangle with an integer point on its boundary edge. 
Although the diagram contains four vertices, only three primitive cones contribute to independent baryonic sectors; 
the vertex $v_1$ does not generate an additional chiral multiplet. 
The corresponding multiplicities are therefore given by the determinants
\begin{align}
\det (v_4,v_2,v_3)&=2, \\
\det (v_2,v_3,v_4)&=2, \\
\det (v_3,v_4,v_2)&=2, 
\end{align}
all equal to two.
They coincide with the multiplicities of three types of chiral multiplets $X_{ii}$, $X_{i\ i+1}^{1}$ and $X_{i\ i+1}^{2}$, 
where the indices $i=1,2$ are understood modulo $2$. 
The area of the polygon is 
\begin{align}
\label{A_C3/Z2}
A&=1. 
\end{align}
The dual vectors are
\begin{align}
\label{dualG_C3/Z2}
m_1&=(0,0,1),\qquad 
m_2=(1,-1,0),\qquad 
m_3=(1,1,-2), \qquad 
m_4=(0,0,1). 
\end{align}
The Reeb vector is
\begin{align}
b&=(3,1,1). 
\end{align}
The volume of $S^5/\mathbb{Z}_2$ is evaluated from (\ref{vol_Y1}) or (\ref{vol_Y2}) as
\begin{align}
\label{vol_S5/Z2}
\mathrm{Vol}(S^5/\mathbb{Z}_2)&=\frac{\pi^3}{2}. 
\end{align}
The R-charges of the chiral multiplets are evaluated as
\begin{align}
R(X_{i\ i+1}^{1})&=\frac{\pi}{3}\frac{\mathrm{Vol}(\Sigma_2)}{\mathrm{Vol}(S^5/\mathbb{Z}_2)}=\frac23, \\
R(X_{ii})&=\frac{\pi}{3}\frac{\mathrm{Vol}(\Sigma_3)}{\mathrm{Vol}(S^5/\mathbb{Z}_2)}=\frac23, \\
R(X_{i\ i+1}^{2})&=\frac{\pi}{3}\frac{\mathrm{Vol}(\Sigma_4)}{\mathrm{Vol}(S^5/\mathbb{Z}_2)}=\frac23, 
\end{align}
where 
\begin{align}
\mathrm{Vol}(\Sigma_2)&=\frac{\det(v_4,v_2,v_3)}{\det(b,v_4,v_2)\det(b,v_2,v_3)}=\pi^2, \\
\mathrm{Vol}(\Sigma_3)&=\frac{\det(v_2,v_3,v_4)}{\det(b,v_2,v_3)\det(b,v_3,v_4)}=\pi^2, \\
\mathrm{Vol}(\Sigma_4)&=\frac{\det(v_3,v_4,v_2)}{\det(b,v_3,v_4)\det(b,v_4,v_2)}=\pi^2, 
\end{align}
are equal to the volume of $S^3/\mathbb{Z}_2$. 

The weighted adjacency matrix is given by
\begin{align}
\label{M_S5/Z2}
M^{AdS_5\times S^{5}/\mathbb{Z}_2}
&=\left(
\begin{matrix}
i_v+i_{\textrm{cm}_{\frac23}}+\tilde{i}_{\textrm{cm}_{\frac23}}
&2\left(i_{\textrm{cm}_{\frac23}}+\tilde{i}_{\textrm{cm}_{\frac23}}\right)\\
2\left(i_{\textrm{cm}_{\frac23}}+\tilde{i}_{\textrm{cm}_{\frac23}}\right)&i_v+i_{\textrm{cm}_{\frac23}}+\tilde{i}_{\textrm{cm}_{\frac23}}\\
\end{matrix}
\right). 
\end{align}
The single-particle gravity index is evaluated from (\ref{M_S5/Z2}) as \cite{Nakayama:2005mf}
\begin{align}
i^{AdS_5\times S^{5}/\mathbb{Z}_2}(p;q)
&=2\frac{p^{\frac13}q^{\frac13}}{1-p^{\frac13}q^{\frac13}}
+2\frac{p^{\frac23}q^{\frac23}}{1-p^{\frac23}q^{\frac23}}
-2\left(\frac{p}{1-p}+\frac{q}{1-q}\right). 
\end{align}
There are four extremal BPS mesonic operators. 
They correspond to the four vectors (\ref{dualG_C3/Z2}) of the dual cone. 
Two of them carry R-charge $R=2/3$, whereas the other two have R-charge $R=4/3$. 
These R-charges can be evaluated from (\ref{Rch_meson}) as
\begin{align}
R(\mathcal{M}_1)&=\frac23(b,m_1)=\frac23, \nonumber\\
R(\mathcal{M}_2)&=\frac23(b,m_2)=\frac43, \nonumber\\
R(\mathcal{M}_3)&=\frac23(b,m_3)=\frac43, \nonumber\\
R(\mathcal{M}_4)&=\frac23(b,m_4)=\frac23. 
\end{align}
The operator lengths of the extremal BPS mesons $\mathcal{M}_I$ are
\begin{align}
L_J(\mathcal{M}_1)&=(m_1,v_J)=\delta_{J,3}, \\
L_J(\mathcal{M}_2)&=(m_2,v_J)=2\delta_{J,4}, \\
L_J(\mathcal{M}_3)&=(m_3,v_J)=2\delta_{J,2}, \\
L_J(\mathcal{M}_4)&=(m_4,v_J)=\delta_{J,3}, 
\end{align}
with $J=2,3,4$. 
Hence the two extremal BPS mesons $\mathcal{M}_1$ and $\mathcal{M}_4$ are built from the two adjoint chiral multiplets
\begin{align}
\mathcal{M}_1=\Tr (X_{11}),\qquad 
\mathcal{M}_4=\Tr (X_{22}). 
\end{align}
The other two $\mathcal{M}_2$ and $\mathcal{M}_3$ consist of the two bifundamentals
\begin{align}
\mathcal{M}_2=\Tr (X_{12}^{2}X_{21}^{2}), \qquad 
\mathcal{M}_3=\Tr (X_{12}^{1}X_{21}^{1}). 
\end{align}

The supersymmetric zeta function is
\begin{align}
&
\mathfrak{Z}^{AdS_5\times S^{5}/\mathbb{Z}_2}(s,z;\omega_1,\omega_2)
\nonumber\\
&=2\left(\frac{\omega_1+\omega_2}{3}\right)^{-s}
\zeta\left(s,1+\frac{3z}{\omega_1+\omega_2}\right)
+2\left(\frac{2(\omega_1+\omega_2)}{3}\right)^{-s}
\zeta\left(s,1+\frac{\frac32z}{\omega_1+\omega_2}\right)
\nonumber\\
&-2\omega_1^{-s}\zeta\left(s,1+\frac{z}{\omega_1}\right)
-2\omega_2^{-s}\zeta\left(s,1+\frac{z}{\omega_2}\right). 
\end{align}
The residue at a simple pole $s=1$ is evaluated as
\begin{align}
\label{Res_S5/Z2}
\mathrm{Res}_{s=1}\mathfrak{Z}^{AdS_5\times S^{5}/\mathbb{Z}_2}(s,z;\omega_1,\omega_2)
&=\frac{9}{\omega_1+\omega_2}-2\left(\frac{1}{\omega_1}+\frac{1}{\omega_2}\right). 
\end{align}
The coefficient of the term $\frac{1}{\omega_1+\omega_2}$ 
is expected to encode the integral of the Riemann square according to the formula (\ref{Riem2_conj}). 
We find
\begin{align}
\label{Riem2_S5/Z2}
\int_{S^5/\mathbb{Z}_2} \mathrm{Riem}^2 
&=116\pi^3=3596.73\ldots. 
\end{align}
If we quotient $Y$ by a freely acting finite group $\Gamma$, the local curvature invariant is given by
\begin{align}
\frac{1}{|\Gamma|} \int_{Y} \mathrm{Riem}^2. 
\end{align}
However, if the geometry has fixed points and it is not smooth, 
while the local curvature away from the fixed loci behaves as above, 
there are additional localized contributions according to the Kawasaki orbifold index theorem \cite{MR527023,MR641150}. 
For $S^5/\mathbb{Z}_2$ there exists a $\mathbb{Z}_2$ which acts trivially on one complex coordinate. 
The fixed set is a copy of $S^1$ so that the geometry locally looks like 
the $A_1$ ALE space and $\mathbb{C}^2/\mathbb{Z}_2\times S^1$ near the fixed locus. 
We observe that 
the Chern-Gauss-Bonnet formula, 
\begin{align}
\label{CGB_formula}
\chi(M)
&=\frac{1}{32\pi^2}\int_M 
(\mathrm{Riem}^2-4\mathrm{Ric}^2+R^2)
+\int_{\partial M}B, 
\end{align}
where $\chi(M)$ is the Euler characteristic of a 4-manifold $M$ and $B$ is the Chern-Simons boundary contribution 
constructed from the boundary curvature invariant \cite{MR1763657,MR1784799}, 
allows us to compute the integral of the squared Riemann tensor over $\mathbb{C}^2/\mathbb{Z}_2$
\begin{align}
\int_{\mathbb{C}^2/\mathbb{Z}_2} \mathrm{Riem}^2&=48\pi^2. 
\end{align}
Here we have used that 
the ALE space is Ricci-flat ($\mathrm{Ric}^2$ $=$ $R^2$ $=$ $0$), 
that its Euler characteristic is $2$ \cite{MR992334} and that 
the boundary at infinity of the $A_1$ ALE space is $S^3/\mathbb{Z}_2$, 
yielding a boundary contribution $1/2$ equal to one half of that of $S^3$. 
Then the localized contribution is given by
\begin{align}
\mathrm{Vol}(S^1)\times \int_{\mathbb{C}^2/\mathbb{Z}_2} \mathrm{Riem}^2
=96\pi^3
\end{align}
so that we obtain the result (\ref{Riem2_S5/Z2})
\begin{align}
\underbrace{\frac{40\pi^3}{2}}_{\frac{1}{2}\int_{S^5}\mathrm{Riem}^2}
+\underbrace{96\pi^3}_{\mathrm{Vol}(S^1)\times \int_{\mathbb{C}^2/\mathbb{Z}_2} \mathrm{Riem}^2}
=116\pi^3. 
\end{align}
In fact, this value is in perfect agreement with the one computed from the Hilbert series (see Appendix \ref{app_HS}). 
The Zeta-index is
\begin{align}
\label{0_S5/Z2}
\mathfrak{Z}^{AdS_5\times S^{5}/\mathbb{Z}_2}(0,0;\omega_1,\omega_2)&=0. 
\end{align}
The supersymmetric zeta value for $s=-1$ and $z=0$ is\footnote{
For $SU(N)\times SU(N)$ quiver gauge theory, 
the supersymmetric zeta function is
\begin{align}
&
\mathfrak{Z}^{AdS_5\times S^{5}/\mathbb{Z}_2}(s,z;\omega_1,\omega_2)
\nonumber\\
&=2\left(\frac{\omega_1+\omega_2}{3}\right)^{-s}
\zeta\left(s,1+\frac{3z}{\omega_1+\omega_2}\right)
+2\left(\frac{2(\omega_1+\omega_2)}{3}\right)^{-s}
\zeta\left(s,1+\frac{\frac32z}{\omega_1+\omega_2}\right)
\nonumber\\
&-2\zeta_2\left(s,\frac{\omega_1+\omega_2}{3}+z;\omega_1,\omega_2\right)
+2\zeta_2\left(s,\frac{2(\omega_1+\omega_2)}{3}+z;\omega_1,\omega_2\right)
\end{align}
and the supersymmetric zeta value at $s=-1$ and $z=0$ is $-\frac{\omega_1+\omega_2}{9}-\frac{2(\omega_1+\omega_2)^3}{81\omega_1\omega_2}$. 
In this case, the formula (\ref{-1_formula3}) leads to
\begin{align}
c_{\mathcal{O}(N^0)}&=-\frac13, \qquad 
a_{\mathcal{O}(N^0)}=-\frac{5}{12}. 
\end{align}
}
\begin{align}
\label{-1_S5/Z2}
\mathfrak{Z}^{AdS_5\times S^{5}/\mathbb{Z}_2}(-1,0;\omega_1,\omega_2)
&=0.  
\end{align}
Correspondingly, the $\mathcal{O}(N^0)$ contributions $c_{\mathcal{O}(N^0)}$ and $a_{\mathcal{O}(N^0)}$ vanish
\begin{align}
c_{\mathcal{O}(N^0)}&=0, \\
a_{\mathcal{O}(N^0)}&=0. 
\end{align}

The supersymmetric determinant is
\begin{align}
\label{det_S5/Z2}
&
\mathfrak{D}^{AdS_5\times S^{5}/\mathbb{Z}_2}(z;\omega_1,\omega_2)
\nonumber\\
&=\frac{2^{1+\frac{3z}{\omega_1+\omega_2}} (\omega_1+\omega_2)^{2+\frac{9z}{\omega_1+\omega_2}}}
{3^{2+\frac{9z}{\omega_1+\omega_2}} \omega_1^{1+\frac{2z}{\omega_1}} \omega_2^{1+\frac{2z}{\omega_2}}}
\frac{\Gamma\left(1+\frac{3z}{2(\omega_1+\omega_2)}\right)^2 \Gamma\left(1+\frac{3z}{\omega_1+\omega_2}\right)^2}
{\Gamma\left(1+\frac{z}{\omega_1}\right)^2\Gamma\left(1+\frac{z}{\omega_2}\right)^2}. 
\end{align}
The special value with $z=0$ is
\begin{align}
\mathfrak{D}^{AdS_5\times S^{5}/\mathbb{Z}_2}(0;\omega_1,\omega_2)
&=\frac{2(\omega_1+\omega_2)^2}{9\omega_1\omega_2}. 
\end{align}
The vacuum exponent is
\begin{align}
\label{vac_S5/Z2}
\mathfrak{D}^{AdS_5\times S^{5}/\mathbb{Z}_2}(0;1,1)
&=\frac89=0.888889\ldots.
\end{align}

\subsection{$AdS_5\times (S^5/\mathbb{Z}_k)_{\textrm{ALE}}$}
\label{sec_S5Zk}
For a stack of $N$ D3-branes in the background of the $\mathbb{Z}_k$ orbifold of $\mathbb{C}^3$ with $k\ge3$, 
 we need to specify the orbifold action on the three complex coordinates on $\mathbb{C}^3$: 
\begin{align}
\label{orb1}
\theta: (z_1,z_2,z_3)\mapsto (e^{\frac{2\pi i a_1}{k}}z_1,e^{\frac{2\pi i a_2}{k}}z_2,e^{\frac{2\pi i a_3}{k}}z_3), 
\end{align}
where the weights $a_i$ of the action defined modulo $k$ obey the Calabi-Yau condition
\begin{align}
\label{orb2}
a_1+a_2+a_3\equiv 0\quad \mod k. 
\end{align}
Let us consider the case with 
\begin{align}
\label{orbifold_C2/Zk}
(a_1,a_2,a_3)&\equiv (a,-a,0). 
\end{align}
In this case, the geometry is the ALE orbifold of the form $\mathbb{C}^2/\mathbb{Z}_k\times \mathbb{C}$. 
The effective theory on the D3-branes is identified with the 4d $\mathcal{N}=2$ necklace quiver gauge theory with $k$ $U(N)$ gauge nodes. 
The theory has bifundamental chiral multiplets $(X_{i\ i+1},X_{i+1\ i})$ between the adjacent gauge nodes. 
In this case, each gauge node corresponds to the $\mathcal{N}=2$ vector multiplet including an adjoint chiral multiplet. 
The chiral multiplets have R-charge $R=2/3$. 
The quiver diagram is 
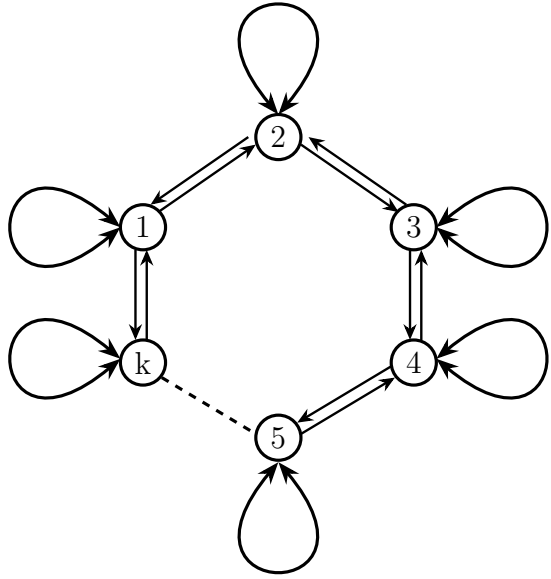
\begin{align}
\begin{tikzpicture}
    \draw[very thick] (-1.8,0) circle (3mm);
    \node at (-1.8,0) {1};
        \draw[very thick] (0,1.2) circle (3mm);
    \node at (0,1.2) {2};
     \draw[very thick] (1.8,0) circle (3mm);
    \node at (1.8,0) {3};
         \draw[very thick] (1.8,-1.8) circle (3mm);
    \node at (1.8,-1.8) {4};
             \draw[very thick] (0,-2.8) circle (3mm);
    \node at (0,-2.8) {5};
             \draw[very thick] (-1.8,-1.8) circle (3mm);
    \node at (-1.8,-1.8) {k};
\draw[<->,>=Stealth,scale=5,very thick] (-0.42,0)  to[in=-140,out=-220,loop] (0,0);
\draw[<->,>=Stealth,scale=5,very thick] (0,0.3)  to[in=-230,out=-310,loop] (0,0);
\draw[<->,>=Stealth,scale=5,very thick] (0.42,0)  to[in=40,out=-40,loop] (0,0);
\draw[<->,>=Stealth,scale=5,very thick] (0.42,-0.35)  to[in=40,out=-40,loop] (0,0);
\draw[<->,>=Stealth,scale=5,very thick] (0,-0.625)  to[in=-50,out=-130,loop] (0,0);
\draw[<->,>=Stealth,scale=5,very thick] (-0.42,-0.35)  to[in=-140,out=-220,loop] (0,0);
    \draw[<->,-{Stealth[length=6pt,width=5pt]}, line width=0.9pt]
         (-1.6,0.2) to node[]{} (-0.3,1.1);
    \draw[-{Stealth[length=6pt,width=5pt]}, line width=0.9pt]
         (-0.4,1.2) to node[]{} (-1.7,0.3);
    \draw[-{Stealth[length=6pt,width=5pt]}, line width=0.9pt]
         (0.3,1.1) to node[]{} (1.6,0.2);
    \draw[-{Stealth[length=6pt,width=5pt]}, line width=0.9pt]
         (1.7,0.3) to node[]{} (0.4,1.2);
    \draw[-{Stealth[length=6pt,width=5pt]}, line width=0.9pt]
         (1.75,-0.3) to node[]{} (1.75,-1.5);
    \draw[-{Stealth[length=6pt,width=5pt]}, line width=0.9pt]
           (1.9,-1.5) to node[]{} (1.9,-0.3);
    \draw[-{Stealth[length=6pt,width=5pt]}, line width=0.9pt]
         (1.5,-1.85) to node[]{} (0.25,-2.6);
    \draw[-{Stealth[length=6pt,width=5pt]}, line width=0.9pt]
         (0.3,-2.75) to node[]{} (1.55,-2);
    \draw[-{Stealth[length=6pt,width=5pt]}, line width=0.9pt]
         (-1.75,-1.5) to node[]{} (-1.75,-0.3);
    \draw[-{Stealth[length=6pt,width=5pt]}, line width=0.9pt]
         (-1.9,-0.3) to node[]{} (-1.9,-1.5);
     \draw [dashed,very thick] (-1.55,-2) to node[]{} (-0.3,-2.75);
\end{tikzpicture}
\end{align}
The superpotential is
\begin{align}
\mathcal{W}&=\sum_{i=1}^{k} \Tr (X_{ii}X_{i\ i+1}X_{i+1\ i}-X_{i+1 i+1}X_{i+1\ i}X_{i\ i+1}). 
\end{align}
The theory is holographically dual to Type IIB string theory on $AdS_5\times (S^5/\mathbb{Z}_k)_{\textrm{ALE}}$, 
where $(S^5/\mathbb{Z}_k)_{\textrm{ALE}}$ is the Sasaki-Einstein base 
of the Calabi-Yau 3-fold $\mathbb{C}^2/\mathbb{Z}_k\times \mathbb{C}$ 
where the orbifold $\mathbb{C}^2/\mathbb{Z}_k$ defines the $A_{k-1}$ surface singularity. 
As a toric variety its toric diagram consists of $k+1$ colinear points \cite{MR1234037,MR2810322}. 
Combining an additional vertex associated with the trivial $\mathbb{C}$ factor, 
we take the toric diagram for $\mathbb{C}^2/\mathbb{Z}_k\times \mathbb{C}$ 
to be generated by the following $k+2$ vectors:\footnote{For example, the toric diagram for $k=3$ can be found in \cite{Bianchi:2014qma}. } 
\begin{align}
v_1&=(1,0,0),\qquad v_2=(1,1,0),\qquad v_3=(1,1,1), \nonumber\\
v_I&=(1,-k+I-3,0), \qquad I=4,5,\cdots, k+2. 
\end{align}
It is drawn as
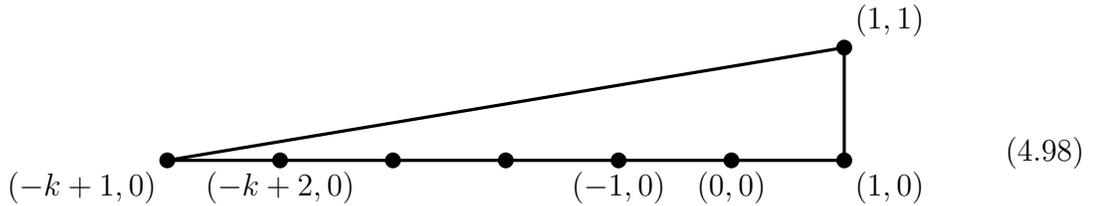
\begin{align}
\label{toric_ALEZk}
\begin{tikzpicture}[scale=1.5, baseline={(0,0)}]
  \fill (0,0) circle (2pt) node[below] {$(0,0)$};
  \fill (1,0) circle (2pt) node[below right] {$(1,0)$};
  \fill (1,1) circle (2pt) node[above right] {$(1,1)$};
  \fill (-5,0) circle (2pt) node[below left] {$(-k+1,0)$};
  \fill (-4,0) circle (2pt) node[below] {$(-k+2,0)$};
  \fill (-3,0) circle (2pt) node[below left] {};  
  \fill (-2,0) circle (2pt) node[below left] {};  
  \fill (-1,0) circle (2pt) node[below] {$(-1,0)$};
  \draw[very thick] (0,0) -- (1,0) -- (1,1) -- (-5,0) -- cycle;
\end{tikzpicture}
\end{align}
For $k=2$ it reduces to the toric diagram (\ref{toric_C3/Z2}) of $\mathbb{C}^2/\mathbb{Z}_2$. 
Here the vertices $v_1$, $v_5$, $\cdots$, $v_{k+2}$ give rise to no independent chiral multiplet 
as they correspond to non-vertex lattice points on the boundary of the toric diagram. 
The independent chiral multiplets are associated with primitive three-cones generated by adjacent external vertices $v_2$, $v_3$ and $v_4$ 
and the following three determinants 
\begin{align}
\det (v_4,v_2,v_3)&=k, \\
\det (v_2,v_3,v_4)&=k, \\
\det (v_3,v_4,v_2)&=k, 
\end{align}
give rise to the multiplicities of three types of chiral multiplets $X_{i\ i+1}$, $X_{ii}$ and $X_{i\ i-1}, $with $i=1,2,\cdots, k$. 
The area of the lattice polygon is 
\begin{align}
\label{A_ALEZk}
A&=\frac{k}{2}. 
\end{align}
The lattice generators of the dual cone are
\begin{align}
\label{dualG_ALEZk}
m_1&=(0,0,1),\qquad 
m_2=(1,-1,0),\qquad 
m_3=(k-1,1,-k), \qquad \nonumber\\
m_{I}&=(0,0,1),\qquad I=4,5,\cdots, k+2. 
\end{align}
We note that since the set of vertices $v_I$ includes internal lattice points along edges, 
several consecutive pairs lie on the same supporting hyperplane. 
The corresponding dual generators $m_I$ therefore coincide, without implying any geometric inconsistency. 
The Reeb vector is
\begin{align}
b&=(3,3-k,1). 
\end{align}
From (\ref{vol_Y1}) or (\ref{vol_Y2}) we obtain the volume of $S^5/\mathbb{Z}_k$
\begin{align}
\label{vol_ALEZk}
\mathrm{Vol}(S^5/\mathbb{Z}_k)&=\frac{\pi^3}{k}. 
\end{align}
The R-charges of the chiral multiplets are evaluated as
\begin{align}
R(X_{i\ i+1})&=\frac{\pi}{3}\frac{\mathrm{Vol}(\Sigma_2)}{\mathrm{Vol}(S^5/\mathbb{Z}_k)}=\frac23, \\
R(X_{ii})&=\frac{\pi}{3}\frac{\mathrm{Vol}(\Sigma_3)}{\mathrm{Vol}(S^5/\mathbb{Z}_k)}=\frac23, \\
R(X_{i\ i-1})&=\frac{\pi}{3}\frac{\mathrm{Vol}(\Sigma_4)}{\mathrm{Vol}(S^5/\mathbb{Z}_k)}=\frac23, 
\end{align}
where 
\begin{align}
\mathrm{Vol}(\Sigma_2)&=\frac{\det(v_4,v_2,v_3)}{\det(b,v_4,v_2)\det(b,v_2,v_3)}=\frac{2\pi^2}{k}, \\
\mathrm{Vol}(\Sigma_3)&=\frac{\det(v_2,v_3,v_4)}{\det(b,v_2,v_3)\det(b,v_3,v_4)}=\frac{2\pi^2}{k}, \\
\mathrm{Vol}(\Sigma_4)&=\frac{\det(v_3,v_4,v_2)}{\det(b,v_3,v_4)\det(b,v_4,v_2)}=\frac{2\pi^2}{k}. 
\end{align}
are the volume of $S^3/\mathbb{Z}_k$. 

We have the weighted adjacency matrix of the form 
\begin{align}
&
M^{AdS_5\times (S^5/\mathbb{Z}_k)_{\textrm{ALE}}}
\nonumber\\
&=\left(
\begin{smallmatrix}
i_v+i_{\textrm{cm}_{\frac23}}+\tilde{i}_{\textrm{cm}_{\frac23}}&i_{\textrm{cm}_{\frac23}}+\tilde{i}_{\textrm{cm}_{\frac23}}&0&\cdots&0&i_{\textrm{cm}_{\frac23}}+\tilde{i}_{\textrm{cm}_{\frac23}}\\
i_{\textrm{cm}_{\frac23}}+\tilde{i}_{\textrm{cm}_{\frac23}}&i_v+i_{\textrm{cm}_{\frac23}}+\tilde{i}_{\textrm{cm}_{\frac23}}&i_{\textrm{cm}_{\frac23}}+\tilde{i}_{\textrm{cm}_{\frac23}}
&\cdots&0&0\\
0&i_{\textrm{cm}_{\frac23}}+\tilde{i}_{\textrm{cm}_{\frac23}}&i_v+i_{\textrm{cm}_{\frac23}}+\tilde{i}_{\textrm{cm}_{\frac23}}&\cdots&0&0\\
\vdots&\vdots&\vdots&\ddots&&\vdots\\
\vdots&\vdots&\vdots&\vdots&\ddots&\vdots\\
i_{\textrm{cm}_{\frac23}}+\tilde{i}_{\textrm{cm}_{\frac23}}&0&0&\cdots&i_{\textrm{cm}_{\frac23}}+\tilde{i}_{\textrm{cm}_{\frac23}}&i_v+i_{\textrm{cm}_{\frac23}}+\tilde{i}_{\textrm{cm}_{\frac23}}\\
\end{smallmatrix}
\right). 
\end{align}
As we consider the $\mathcal{N}=2$ $U(N)$ gauge nodes, there is no subtractive trace term in the formula (\ref{ind_gravity}). 
We find the single-particle gravity index 
\begin{align}
\label{ind_ALEZk}
i^{AdS_5\times (S^5/\mathbb{Z}_k)_{\textrm{ALE}}}(p;q)
&=k\frac{p^{\frac13}q^{\frac13}}{1-p^{\frac13}q^{\frac13}}
+2\frac{p^{\frac{k}{3}}q^{\frac{k}{3}}}{1-p^{\frac{k}{3}}q^{\frac{k}{3}}}
-k\left(\frac{p}{1-p}+\frac{q}{1-q}\right). 
\end{align}
It enumerates two types of the extremal BPS mesonic operators. 
The first type involves $k$ extremal BPS mesonic operators 
\begin{align}
\mathcal{M}_{i+3}=\Tr X_{ii}, 
\end{align}
with $i=1,\cdots, k$ and $\mathcal{M}_{k+3}$ $\equiv$ $\mathcal{M}_1$, 
built from the adjoint chirals $X_{ii}$ in the $\mathcal{N}=2$ vector multiplets. 
They correspond to the $k$ vectors $m_{i+3}$ of the dual cone in (\ref{dualG_ALEZk}). 
Their R-charges are computed from the formula (\ref{Rch_meson})
\begin{align}
R(\mathcal{M}_{i+3})&=\frac23 (b,m_{i+3})=\frac23. 
\end{align}
The operator lengths are reproduced from the pairings
\begin{align}
L_J(\mathcal{M}_{i+3})&=(m_{i+3},v_J)=\delta_{J,3}. 
\end{align}
The other type is given by
\begin{align}
\mathcal{M}_{2}&=\Tr(X_{1 k}\cdots X_{21}), \nonumber\\
\mathcal{M}_{3}&=\Tr(X_{12}\cdots X_{k1}), 
\end{align}
which come from the two independent long mesonic loops 
consisting of the bifundamental chiral matter fields of the $\mathcal{N}=2$ hypermultiplets. 
They correspond to the primitive generators $m_2$ and $m_3$ of the dual cone. 
Again the R-charges are obtained from the formula (\ref{Rch_meson})
\begin{align}
R(\mathcal{M}_{2})&=\frac23 (b,m_{2})=\frac{2k}{3}, \\
R(\mathcal{M}_{3})&=\frac23 (b,m_{3})=\frac{2k}{3}. 
\end{align}
The operator lengths are
\begin{align}
L_J(\mathcal{M}_{2})&=(m_{2},v_J)=k\delta_{J,4}, \\
L_J(\mathcal{M}_{3})&=(m_{3},v_J)=k\delta_{J,2}. 
\end{align}

The supersymmetric zeta function is
\begin{align}
&
\mathfrak{Z}^{AdS_5\times (S^5/\mathbb{Z}_k)_{\textrm{ALE}}}(s,z;\omega_1,\omega_2)
\nonumber\\
&=k\left(\frac{\omega_1+\omega_2}{3}\right)^{-s}
\zeta\left(s,\frac{\omega_1+\omega_2+3z}{\omega_1+\omega_2}\right)
+2\left(\frac{k(\omega_1+\omega_2)}{3}\right)^{-s}
\zeta\left(s,\frac{\omega_1+\omega_2+\frac{3}{k}z}{\omega_1+\omega_2}\right)
\nonumber\\
&-k\omega_1^{-s} \zeta\left(s,1+\frac{z}{\omega_1}\right)
-k\omega_2^{-s} \zeta\left(s,1+\frac{z}{\omega_2}\right). 
\end{align}
We find the residue at a simple pole $s=1$
\begin{align}
\mathrm{Res}_{s=1}\mathfrak{Z}^{AdS_5\times (S^5/\mathbb{Z}_k)_{\textrm{ALE}}}(s,z;\omega_1,\omega_2)
&=\frac{3(k^2+2)}{k}\frac{1}{\omega_1+\omega_2}
-k\left(\frac{1}{\omega_1}+\frac{1}{\omega_2}\right). 
\end{align}
Substituting the coefficient of the term with $\frac{1}{\omega_1+\omega_2}$ and the volume (\ref{vol_ALEZk}) 
into the formula (\ref{Riem2_conj}), we find
\begin{align}
\label{Riem2_ALEZk}
\int_{ (S^5/\mathbb{Z}_k)_{\textrm{ALE}} } \mathrm{Riem}^2
&=\frac{8(8k^2-3)}{k}\pi^3. 
\end{align}
When $k=1$ and $2$ it reduces to (\ref{Riem2_S5}) and (\ref{Riem2_S5/Z2}). 
The geometry $(S^5/\mathbb{Z}_k)_{\textrm{ALE}}$ does not possess isolated fixed points. 
Instead, the fixed set forms a circle $S^1$ and the local contribution is governed by the geometry of $S^1\times \mathbb{C}^2/\mathbb{Z}_k$. 
Hence there are extra contributions to the integral $\int \mathrm{Riem}^2$. 
Applying the Chern-Gauss-Bonnet formula (\ref{CGB_formula}) to the $A_{k-1}$ ALE space, 
and using the Ricci-flatness together with $\chi$ $=$ $k$ \cite{MR992334}, 
we obtain
\begin{align}
\int_{\mathbb{C}^2/\mathbb{Z}_k}\mathrm{Riem}^2
&=32\pi^2 \left(k-\frac{1}{k}\right), 
\end{align}
where the subtraction $1/k$ arises from the Gauss-Bonnet boundary term of the asymptotic lens space $S^3/\mathbb{Z}_k$. 
Therefore, the result (\ref{Riem2_ALEZk}) is consistently reproduced as follows: 
\begin{align}
\underbrace{
\frac{40\pi^3}{k}}_{\frac{1}{k}\int_{S^5}\mathrm{Riem}^2}
+
\underbrace{
64\pi^3\left(
k-\frac{1}{k}
\right)}_{\mathrm{Vol}(S^1)\times \int_{\mathbb{C}^2/\mathbb{Z}_k}\mathrm{Riem}^2}
=\frac{8(8k^2-3)}{k}\pi^3. 
\end{align}
Also this value exactly agrees with the result obtained from the Hilbert series (see Appendix \ref{app_HS}). 
The Zeta-index is
\begin{align}
\mathfrak{Z}^{AdS_5\times (S^5/\mathbb{Z}_k)_{\textrm{ALE}}}(0,0;\omega_1,\omega_2)
&=\frac{k}{2}-1. 
\end{align}
The supersymmetric zeta value with $s=-1$ and $z=0$ vanishes
\begin{align}
\mathfrak{Z}^{AdS_5\times (S^5/\mathbb{Z}_k)_{\textrm{ALE}}}(-1,z;\omega_1,\omega_2)
&=0. 
\end{align}
This is consistent with the formula (\ref{-1_formula2}) 
as neither $c_{\mathcal{O}(N^0)}$ nor $a_{\mathcal{O}(N^0)}$ exists for the theory
\begin{align}
c_{\mathcal{O}(N^0)}&=0, \\
a_{\mathcal{O}(N^0)}&=0.  
\end{align}

The supersymmetric determinant is
\begin{align}
&
\mathfrak{D}^{AdS_5\times (S^5/\mathbb{Z}_k)_{\textrm{ALE}}}(z;\omega_1,\omega_2)
\nonumber\\
&=\frac{(2\pi)^{\frac{k}{2}-1} k^{1+\frac{6z}{k(\omega_1+\omega_2)}} (\omega_1+\omega_2)^{1+\frac{k}{2}+\frac{3(2+k^2)z}{k(\omega_1+\omega_2)}}}
{3^{1+\frac{k}{2}+\frac{3(2+k^2)z}{k(\omega_1+\omega_2)}} \omega_1^{k(\frac12+\frac{z}{\omega_1})} \omega_2^{k(\frac12+\frac{z}{\omega_2})}}
\frac{\Gamma\left(1+\frac{3z}{\omega_1+\omega_2}\right)^k \Gamma\left(1+\frac{3z}{k(\omega_1+\omega_2)}\right)^2}
{\Gamma\left(1+\frac{z}{\omega_1}\right)^k\Gamma\left(1+\frac{z}{\omega_2}\right)^k}. 
\end{align}
Setting $z=0$, we obtain
\begin{align}
\mathfrak{D}^{AdS_5\times (S^5/\mathbb{Z}_k)_{\textrm{ALE}}}(0;\omega_1,\omega_2)
&=\frac{(2\pi)^{\frac{k}{2}-1}}{3^{\frac{k}{2}+1}}
\frac{(\omega_1+\omega_2)^{\frac{k}{2}+1}}{\omega_1^{\frac{k}{2}} \omega_2^{\frac{k}{2}}}. 
\end{align}
The vacuum exponent is
\begin{align}
\mathfrak{D}^{AdS_5\times (S^5/\mathbb{Z}_k)_{\textrm{ALE}}}(0;1,1)
&=\frac{2^{k}k\pi^{\frac{k}{2}-1}}{3^{\frac{k}{2}+1}}. 
\end{align}

\section{Weighted projective Sasaki-Einstein orbifolds}
\label{sec_orb}

In section \ref{sec_S5Zk} we focused on the $\mathbb{Z}_k$ actions (\ref{orb1}) with $k\ge3$ 
obeying the condition (\ref{orb2}) with $\mathcal{N}=2$ supersymmetry. 
We now turn to the actions (\ref{orb1}) that preserve only $\mathcal{N}=1$ supersymmetry. 
It is realized by 
\begin{align}
(a_1,a_2,a_3)&\equiv (a,a,-2a). 
\end{align}
This is a toric Calabi-Yau 3-fold orbifold singularity $\mathbb{C}^3/\mathbb{Z}_k (1,1,-2)$. 
It defines an isolated toric Calabi-Yau cone 
whose quasi-regular Sasaki-Einstein base is a $U(1)$ fibration over a weighted projective surface. 
This contrasts with the geometry $\mathbb{C}^3/\mathbb{Z}_k (1,-1,0)$ $\cong$ $(\mathbb{C}^2/\mathbb{Z}_k)\times \mathbb{C}$ in section \ref{sec_S5Zk}, 
which is non-isolated and of product type. 
We refer to the Sasaki-Einstein base of $\mathbb{C}^3/\mathbb{Z}_k (1,1,-2)$ as the \textit{weighted projective Sasaki-Einstein orbifold}. 

The theory on the D3-branes preserves $\mathcal{N}=1$ supersymmetry. 
In the following sections, we adopt the convention that 
the gauge group of the $\mathcal{N}=1$ quiver gauge theory is taken to be $SU(N)$ rather than $U(N)$.
The 4d $\mathcal{N}=1$ quiver gauge theory associated with this orbifold has $k$ $SU(N)$ gauge nodes. 
It contains three types of bifundamental chiral multiplets $X_{i\ i+1}^{1}$, $X_{i\ i+1}^{2}$ and $X_{i\ i-2}$, 
where the indices $i=1,\cdots, k$ are understood modulo $k$. 
The chiral multiplets have R-charge $R=2/3$. 
The quiver diagram is given by
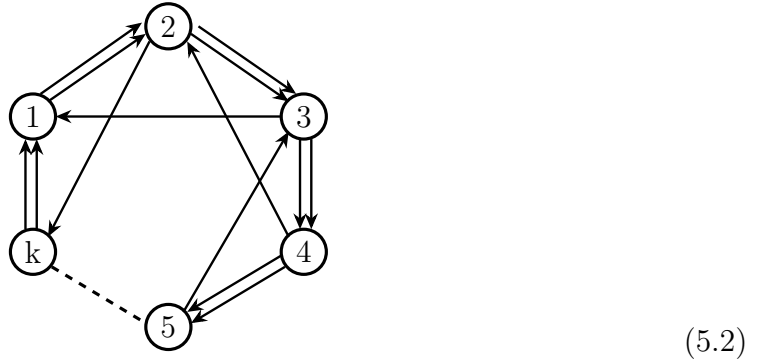
\begin{align}
\begin{tikzpicture}
    \draw[very thick] (-1.8,0) circle (3mm);
    \node at (-1.8,0) {1};
        \draw[very thick] (0,1.2) circle (3mm);
    \node at (0,1.2) {2};
     \draw[very thick] (1.8,0) circle (3mm);
    \node at (1.8,0) {3};
         \draw[very thick] (1.8,-1.8) circle (3mm);
    \node at (1.8,-1.8) {4};
             \draw[very thick] (0,-2.8) circle (3mm);
    \node at (0,-2.8) {5};
             \draw[very thick] (-1.8,-1.8) circle (3mm);
    \node at (-1.8,-1.8) {k};
    \draw[-{Stealth[length=6pt,width=5pt]}, line width=0.9pt]
         (-1.6,0.2) to node[]{} (-0.3,1.1);
    \draw[-{Stealth[length=6pt,width=5pt]}, line width=0.9pt]
         (-1.7,0.3) to node[]{} (-0.35,1.25);
    \draw[-{Stealth[length=6pt,width=5pt]}, line width=0.9pt]
         (0.3,1.1) to node[]{} (1.6,0.2);
    \draw[-{Stealth[length=6pt,width=5pt]}, line width=0.9pt]
         (0.4,1.2) to node[]{} (1.7,0.3);
    \draw[-{Stealth[length=6pt,width=5pt]}, line width=0.9pt]
         (1.75,-0.3) to node[]{} (1.75,-1.5);
    \draw[-{Stealth[length=6pt,width=5pt]}, line width=0.9pt]
           (1.9,-0.3) to node[]{} (1.9,-1.5);
    \draw[-{Stealth[length=6pt,width=5pt]}, line width=0.9pt]
         (1.5,-1.85) to node[]{} (0.25,-2.6);
    \draw[-{Stealth[length=6pt,width=5pt]}, line width=0.9pt]
         (1.55,-2) to node[]{} (0.3,-2.75);
    \draw[-{Stealth[length=6pt,width=5pt]}, line width=0.9pt]
         (-1.75,-1.5) to node[]{} (-1.75,-0.3);
    \draw[-{Stealth[length=6pt,width=5pt]}, line width=0.9pt]
         (-1.9,-1.5) to node[]{} (-1.9,-0.3);
     \draw [dashed,very thick] (-1.55,-2) to node[]{} (-0.3,-2.75);
    \draw[-{Stealth[length=6pt,width=5pt]}, line width=0.9pt]
         (-0.25,1.0) to node[]{} (-1.6,-1.6);
    \draw[-{Stealth[length=6pt,width=5pt]}, line width=0.9pt]
         (1.5,0) to node[]{} (-1.5,0);
    \draw[-{Stealth[length=6pt,width=5pt]}, line width=0.9pt]
         (1.6,-1.6) to node[]{} (0.25,1.0);
    \draw[-{Stealth[length=6pt,width=5pt]}, line width=0.9pt]
         (0.2,-2.6) to node[]{} (1.6,-0.2);
\end{tikzpicture}
\end{align}
The superpotential is
\begin{align}
\mathcal{W}&=\sum_{i=1}^{k} \Tr (X_{i\ i+1}^1X_{i+1\ i+2}^2X_{i+2\ i}-X_{i\ i+1}^2X_{i+1\ i+2}^1X_{i+2\ i}). 
\end{align}
The theory is holographically dual to Type IIB string theory on $AdS_5\times S^5/\mathbb{Z}_k$. 

Since the situation differs for even and odd $k$, we analyze the two cases separately. 

\subsection{$AdS_5\times S^5/\mathbb{Z}_{2n-1}$}
For odd $k=2n-1$ with $n>1$ the only fixed point is the origin and the singularity is isolated. 
Namely, there are no orbifold singularities of any codimension. 
The toric diagram for $\mathbb{C}^3/\mathbb{Z}_{2n-1}$ has three external vertices\footnote{
There are $(n-1)$ interior lattice points $(1,m-1,m-1)$ with $m=1,\cdots, n-1$ 
which are related to exceptional divisors in the crepant resolution (see e.g. \cite{DelaOssa:2001blj}). } 
\begin{align}
v_1&=(1,-n+1,-n+1), \qquad 
v_2=(1,1,0), \qquad
v_3=(1,0,1). 
\end{align}
It is depicted as 
\begin{align}
\label{toric_C3/Z2k-1}
\begin{tikzpicture}[scale=1, baseline={(0,0)}]
  \fill (-2,-2) circle (2pt) node[below left] {$(-n+1,-n+1)$};
  \fill (1,0) circle (2pt) node[below right] {$(1,0)$};
  \fill (0,1) circle (2pt) node[above left] {$(0,1)$};
  \draw[very thick] (-2,-2) -- (1,0) -- (0,1) -- cycle;
\end{tikzpicture}
\end{align}
When $n=1$, it reduces to the toric diagram (\ref{toric_C3}) for $\mathbb{C}^3$. 
We have 
\begin{align}
\det (v_{I-1},v_{I},v_{I+1})&=2n-1, 
\end{align}
where the indices $I=1,2,3$ are understood modulo $3$. 
These three determinants encode the multiplicities of three types of the chiral multiplets $X_{i\ i+1}^{1}$, $X_{i\ i+1}^{2}$ and $X_{i\ i-2}$ forming the baryonic operators.  
The area of the polygon is 
\begin{align}
A&=\frac{2n-1}{2}.  
\end{align}
The primitive lattice generators of the dual cone are
\begin{align}
\label{dualG_C3/Z2k-1}
m_1&=(n-1,-n+1,n), \qquad 
m_2=(1,-1,-1), \qquad 
m_3=(n-1,n,-n+1). 
\end{align}
The Reeb vector is
\begin{align}
b&=(3,-n+2,-n+2). 
\end{align}
The volume of $S^5/\mathbb{Z}_{2n-1}$ is evaluated from (\ref{vol_Y1}) or (\ref{vol_Y2}) as 
\begin{align}
\label{vol_S5/Z2k-1}
\mathrm{Vol}(S^5/\mathbb{Z}_{2n-1})&=\frac{\pi^3}{2n-1}. 
\end{align}
The R-charges of the chiral multiplet fields are given by
\begin{align}
R(X_{i\ i+1}^{1})&=\frac{\pi}{3}\frac{\mathrm{Vol}(\Sigma_1)}{\mathrm{Vol}(S^5/\mathbb{Z}_{2n-1})}=\frac23, \\
R(X_{i\ i+1}^{2})&=\frac{\pi}{3}\frac{\mathrm{Vol}(\Sigma_2)}{\mathrm{Vol}(S^5/\mathbb{Z}_{2n-1})}=\frac23, \\
R(X_{i\ i-2})&=\frac{\pi}{3}\frac{\mathrm{Vol}(\Sigma_3)}{\mathrm{Vol}(S^5/\mathbb{Z}_{2n-1})}=\frac23, 
\end{align}
where 
\begin{align}
\mathrm{Vol}(\Sigma_1)&=2\pi^2 \frac{\det(v_3,v_1,v_2)}{\det(b,v_3,v_1)\det(b,v_1,v_2)}=\frac{2\pi^2}{2n-1}, \\
\mathrm{Vol}(\Sigma_2)&=2\pi^2 \frac{\det(v_1,v_2,v_3)}{\det(b,v_1,v_2)\det(b,v_2,v_3)}=\frac{2\pi^2}{2n-1}, \\
\mathrm{Vol}(\Sigma_3)&=2\pi^2 \frac{\det(v_2,v_3,v_1)}{\det(b,v_2,v_3)\det(b,v_3,v_1)}=\frac{2\pi^2}{2n-1}. 
\end{align}

The weighted adjacency matrix for $S^5/\mathbb{Z}_k$ takes the following form:
\begin{align}
\label{M_S5/Zk}
M^{AdS_5\times S^5/\mathbb{Z}_k}
&=\left(
\begin{smallmatrix}
i_v&2i_{\textrm{cm}_{\frac23}}&\tilde{i}_{\textrm{cm}_{\frac23}}&0&\cdots&0&i_{\textrm{cm}_{\frac23}}&2\tilde{i}_{\textrm{cm}_{\frac23}} \\
2\tilde{i}_{\textrm{cm}_{\frac23}}&i_v&2i_{\textrm{cm}_{\frac23}}&\tilde{i}_{\textrm{cm}_{\frac23}}&0&\cdots&0&i_{\textrm{cm}_{\frac23}} \\
i_{\textrm{cm}_{\frac23}}&2\tilde{i}_{\textrm{cm}_{\frac23}}&i_v&2i_{\textrm{cm}_{\frac23}}&\tilde{i}_{\textrm{cm}_{\frac23}}&0&\cdots&0 \\
\vdots&\vdots&\vdots&\ddots&\vdots&\vdots&\vdots&\vdots \\
\vdots&\vdots&\vdots&\vdots&\ddots&\vdots&\vdots&\vdots \\
0&\cdots&0&i_{\textrm{cm}_{\frac23}}&2\tilde{i}_{\textrm{cm}_{\frac23}}&i_v&2i_{\textrm{cm}_{\frac23}}&\tilde{i}_{\textrm{cm}_{\frac23}} \\
\tilde{i}_{\textrm{cm}_{\frac23}}&0&\cdots&0&i_{\textrm{cm}_{\frac23}}&2\tilde{i}_{\textrm{cm}_{\frac23}}&i_v&2i_{\textrm{cm}_{\frac23}} \\
2i_{\textrm{cm}_{\frac23}}&\tilde{i}_{\textrm{cm}_{\frac23}}&0&\cdots&0&i_{\textrm{cm}_{\frac23}}&2\tilde{i}_{\textrm{cm}_{\frac23}}&i_v \\
\end{smallmatrix}
\right). 
\end{align}
For odd $k=2n-1$ we find from (\ref{M_S5/Zk}) the gravity index 
\begin{align}
i^{AdS_5\times S^5/\mathbb{Z}_{2n-1}}(p;q)
&=
3\frac{(pq)^{\frac{2n-1}{3}}}{1-(pq)^{\frac{2n-1}{3}}}. 
\end{align}
There are three extremal BPS mesonic operators
\begin{align}
\mathcal{M}_1^{\textrm{odd}}&=\Tr(X_{1\ 2n-2}X_{2n-2\ 2n-4}\cdots X_{42}X_{2\ 2n-1}X_{2n-1\ 2n-3}\cdots X_{53}X_{31}), \\
\mathcal{M}_2^{\textrm{odd}}&=\Tr(X_{12}^1X_{23}^1\cdots X_{2n-1\ 1}^1), \\
\mathcal{M}_3^{\textrm{odd}}&=\Tr(X_{12}^2X_{23}^2\cdots X_{2n-1\ 1}^2). 
\end{align}
They correspond to the three primitive generators (\ref{dualG_C3/Z2k-1}) of the dual cone 
such that their R-charges are evaluated as
\begin{align}
R(\mathcal{M}_I^{\textrm{odd}})&=\frac23(b,m_I)=\frac{2(2n-1)}{3}, \qquad I=1,2,3. 
\end{align}
The operator lengths of the extremal BPS mesons $\mathcal{M}_I$ are
\begin{align}
L_J(\mathcal{M}_1^{\textrm{odd}})&=(m_1,v_J)=(2n-1)\delta_{J,3}, \\
L_J(\mathcal{M}_2^{\textrm{odd}})&=(m_2,v_J)=(2n-1)\delta_{J,1}, \\
L_J(\mathcal{M}_3^{\textrm{odd}})&=(m_3,v_J)=(2n-1)\delta_{J,2}. 
\end{align}

The supersymmetric zeta function is given by
\begin{align}
\label{zeta_S5/Z2k-1}
&
\mathfrak{Z}^{AdS_5\times S^5/\mathbb{Z}_{2n-1}}(s,z;\omega_1,\omega_2)
\nonumber\\
&=3\left(\frac{2n-1}{3}(\omega_1+\omega_2)\right)^{-s}\zeta\left(s,1+\frac{3z}{(2n-1)(\omega_1+\omega_2)}\right). 
\end{align}
The residue of the supersymmetric zeta function (\ref{zeta_S5/Z2k-1}) at a simple pole $s=1$ is
\begin{align}
\label{Res_S5/Z2k-1}
\mathrm{Res}_{s=1}\mathfrak{Z}^{AdS_5\times S^5/\mathbb{Z}_{2n-1}}(s,z;\omega_1,\omega_2)
&=\frac{9}{2n-1}\frac{1}{\omega_1+\omega_2}. 
\end{align}
From (\ref{Riem2_conj}), (\ref{vol_S5/Z2k-1}) and (\ref{Res_S5/Z2k-1}) 
we get the integral of the Riemann square
\begin{align}
\label{Riem2_S5/Z2k-1}
\int_{S^5/\mathbb{Z}_{2n-1}} \mathrm{Riem}^2
&=\frac{40\pi^3}{2n-1}. 
\end{align}
In this case, the quotient is smooth and there are no localized curvature contributions analogous to those which arise in $S^5/\mathbb{Z}_2$. 
For $n=2$, i.e. $S^5/\mathbb{Z}_3$ this also reproduces the result in \cite{Eager:2010dk} evaluated from the Hilbert series 
(see Appendix \ref{app_HS}). 
The Zeta-index is
\begin{align}
\mathfrak{Z}^{AdS_5\times S^5/\mathbb{Z}_{2n-1}}(0,0;\omega_1,\omega_2)
&=-\frac32, 
\end{align}
which is equal to minus half the number of the lattice points of the toric diagram. 
The special supersymmetric zeta value with $s=-1$ and $z=0$ is given by
\begin{align}
\label{-1_S5/Z2k-1}
\mathfrak{Z}^{AdS_5\times S^5/\mathbb{Z}_{2n-1}}(-1,0;\omega_1,\omega_2)
&=-\frac{2n-1}{12}(\omega_1+\omega_2). 
\end{align}

The supersymmetric determinant is given by
\begin{align}
&
\mathfrak{D}^{AdS_5\times S^5/\mathbb{Z}_{2n-1}}(z;\omega_1,\omega_2)
\nonumber\\
&=\frac{\left((2n-1)(\omega_1+\omega_2)\right)^{\frac32+\frac{9z}{(2n-1)(\omega_1+\omega_2)}} \Gamma\left(1+\frac{3z}{(2n-1)(\omega_1+\omega_2)}\right)^3}
{3^{\frac32+\frac{9z}{(2n-1)(\omega_1+\omega_2)}} 2^{\frac32}\pi^{\frac32}}. 
\end{align}
We have the vacuum exponent
\begin{align}
\mathfrak{D}^{AdS_5\times S^5/\mathbb{Z}_{2n-1}}(0;1,1)
&=
\frac{(2n-1)^{\frac32}}{3^{\frac32}\pi^{\frac32}}. 
\end{align}

\subsection{$AdS_5\times S^5/\mathbb{Z}_{2n}$}
\label{sec_S5/Z2k}
For even $k=2n$ with $n>1$ the cyclic group $\mathbb{Z}_k$ contains an element $\theta^{\frac{k}{2}}$ $\in$ $\mathbb{Z}_k$ of order $2$ that acts as
\begin{align}
(z_1,z_2,z_3)&\mapsto (-z_1,-z_2,z_3). 
\end{align}
The fixed locus is a complex line with $z_1=z_2=0$, $z_3\neq 0$. 
Taking the Sasaki-Einstein base of the Calabi-Yau cone, 
we have the intersection with $S^5$ defined by $|z_1|^2+|z_2|^2+|z_3|^2$ $=$ $1$. 
Thus the fixed locus becomes a circle $|z_3|=1$ and the Sasaki-Einstein base has the non-isolated orbifold singularity. 
The toric diagram of $\mathbb{C}^3/\mathbb{Z}_{2n}$ is generated by 
\begin{align}
v_1&=(1,0,0), \qquad 
v_2=(1,1,0), \qquad 
v_3=(1,n,n), \qquad 
v_4=(1,-1,0). 
\end{align}
It is shown as 
\begin{align}
\label{toric_C3/Z2k}
\begin{tikzpicture}[scale=1, baseline={(0,0)}]
  \fill (0,0) circle (2pt) node[below] {$(0,0)$};
  \fill (1,0) circle (2pt) node[below right] {$(1,0)$};
  \fill (3,3) circle (2pt) node[above right] {$(n,n)$};
  \fill (-1,0) circle (2pt) node[below left] {$(-1,0)$};
  \draw[very thick] (0,0) -- (1,0) -- (3,3) -- (-1,0) -- cycle;
\end{tikzpicture}
\end{align}
For $k=1$ it reduces to the toric diagram (\ref{toric_C3/Z2}) for $\mathbb{C}^2/\mathbb{Z}_2$ of the degenerate quadrilateral. 
We have 
\begin{align}
\det (v_{4},v_{2},v_{3})&=2n, \\
\det (v_{2},v_{3},v_{4})&=2n, \\
\det (v_{3},v_{4},v_{2})&=2n, 
\end{align}
which encode the multiplicities of three types of the chiral multiplets $X_{i\ i+1}^{1}$, $X_{i\ i-2}$ and $X_{i\ i+1}^{2}$ forming the baryons. 
The area of the polygon is 
\begin{align}
A&=n. 
\end{align}
The primitive lattice generators of the dual cone are
\begin{align}
\label{dualG_C3/Z2k}
m_1&=(0,0,1),\qquad 
m_2=(n,-n,n-1), \qquad 
m_3=(n,n,-n-1), \qquad 
m_4=(0,0,1). 
\end{align}
The coincidence of dual vectors $m_1$ and $m_4$ arises from the presence of collinear boundary lattice points in the toric diagram (\ref{toric_C3/Z2k}).
The Reeb vector is
\begin{align}
b&=(3,n,n). 
\end{align}
From (\ref{vol_Y1}) or (\ref{vol_Y2}) we get the volume of $S^5/\mathbb{Z}_{2n}$
\begin{align}
\label{vol_S5/Z2k}
\mathrm{Vol}(S^5/\mathbb{Z}_{2n})&=\frac{\pi^3}{2n}. 
\end{align}
The R-charges of the chiral multiplet fields are 
\begin{align}
R(X_{i\ i+1}^{1})&=\frac{\pi}{3}\frac{\mathrm{Vol}(\Sigma_2)}{\mathrm{Vol}(S^5/\mathbb{Z}_{2n})}=\frac23, \\
R(X_{i\ i-2})&=\frac{\pi}{3}\frac{\mathrm{Vol}(\Sigma_3)}{\mathrm{Vol}(S^5/\mathbb{Z}_{2n})}=\frac23, \\
R(X_{i\ i+1}^{2})&=\frac{\pi}{3}\frac{\mathrm{Vol}(\Sigma_4)}{\mathrm{Vol}(S^5/\mathbb{Z}_{2n})}=\frac23, 
\end{align}
where 
\begin{align}
\mathrm{Vol}(\Sigma_2)&=2\pi^2 \frac{\det(v_4,v_2,v_3)}{\det(b,v_4,v_2)\det(b,v_2,v_3)}=\frac{\pi^2}{n}, \\
\mathrm{Vol}(\Sigma_3)&=2\pi^2 \frac{\det(v_2,v_3,v_4)}{\det(b,v_2,v_3)\det(b,v_3,v_4)}=\frac{\pi^2}{n}, \\
\mathrm{Vol}(\Sigma_4)&=2\pi^2 \frac{\det(v_3,v_4,v_2)}{\det(b,v_3,v_4)\det(b,v_4,v_2)}=\frac{\pi^2}{n}. 
\end{align}

The gravity index is evaluated from (\ref{M_S5/Zk}) as
\begin{align}
i^{AdS_5\times S^5/\mathbb{Z}_{2n}}(p;q)
&=
2\frac{(pq)^{\frac{n}{3}}}{1-(pq)^{\frac{n}{3}}}+2\frac{(pq)^{\frac{2n}{3}}}{1-(pq)^{\frac{2n}{3}}}. 
\end{align}
There are four extremal BPS mesonic operators
\begin{align}
\mathcal{M}_1^{\textrm{even}}&=\Tr (X_{1\ 2n-1}X_{2n-1\ 2n-3}\cdots X_{53}X_{31}), \\
\mathcal{M}_2^{\textrm{even}}&=\Tr (X_{12}^2X_{23}^2 \cdots X_{2n-1\ 2n}^2X_{2n\ 1}^2), \\
\mathcal{M}_3^{\textrm{even}}&=\Tr (X_{12}^1X_{23}^1 \cdots X_{2n-1\ 2n}^1X_{2n\ 1}^1), \\
\mathcal{M}_4^{\textrm{even}}&=\Tr (X_{2\ 2n}X_{2n\ 2n-2} \cdots X_{64}X_{42}). 
\end{align}
Here $\mathcal{M}_1^{\textrm{even}}$ and $\mathcal{M}_4^{\textrm{even}}$ have R-charge $R=2n/3$ 
and the other two carry R-charge $R=4n/3$.  
These R-charges are computed from the corresponding primitive generators (\ref{dualG_C3/Z2k}) of the dual cone as
\begin{align}
R(\mathcal{M}_1^{\textrm{even}})&=\frac23(b, m_1)=\frac{2n}{3}, \nonumber\\
R(\mathcal{M}_2^{\textrm{even}})&=\frac23(b, m_2)=\frac{4n}{3}, \nonumber\\
R(\mathcal{M}_3^{\textrm{even}})&=\frac23(b, m_3)=\frac{4n}{3}, \nonumber\\
R(\mathcal{M}_4^{\textrm{even}})&=\frac23(b, m_4)=\frac{2n}{3}. 
\end{align}
The operator lengths of the extremal BPS mesons $\mathcal{M}_I$ are
\begin{align}
L_J(\mathcal{M}_1^{\textrm{even}})&=(m_1,v_J)=n\delta_{J,3}, \\
L_J(\mathcal{M}_2^{\textrm{even}})&=(m_2,v_J)=2n\delta_{J,4}, \\
L_J(\mathcal{M}_3^{\textrm{even}})&=(m_3,v_J)=2n\delta_{J,2}, \\
L_J(\mathcal{M}_4^{\textrm{even}})&=(m_4,v_J)=n\delta_{J,3}. 
\end{align}

We have the supersymmetric zeta function 
\begin{align}
\label{zeta_S5/Z2k}
\mathfrak{Z}^{AdS_5\times S^5/\mathbb{Z}_{2n}}(s,z;\omega_1,\omega_2)
&=2\left(\frac{n}{3}(\omega_1+\omega_2)\right)^{-s}
\zeta\left(s,1+\frac{3z}{n(\omega_1+\omega_2)}\right)
\nonumber\\
&+2\left(\frac{2n}{3}(\omega_1+\omega_2)\right)^{-s}
\zeta\left(s,1+\frac{3z}{{2n}(\omega_1+\omega_2)}\right). 
\end{align}
The residue at a simple pole $s=1$ is
\begin{align}
\label{Res_S5/Z2k}
\mathrm{Res}_{s=1}\mathfrak{Z}^{AdS_5\times S^5/\mathbb{Z}_{2n}}(s,z;\omega_1,\omega_2)
&=
\frac{9}{n}\frac{1}{\omega_1+\omega_2}. 
\end{align}
From (\ref{Riem2_conj}), (\ref{vol_S5/Z2k}) and (\ref{Res_S5/Z2k}) 
we find
\begin{align}
\label{Riem2_S5/Z2k}
\int_{S^5/\mathbb{Z}_{2n}} \mathrm{Riem}^2
&=\frac{116\pi^3}{n}. 
\end{align}
Since $\mathbb{Z}_{2n}$ $\cong$ $\mathbb{Z}_{n}\ltimes \mathbb{Z}_2$, 
we can view $S^5/\mathbb{Z}_{2n}$ as $(S^5/\mathbb{Z}_2)/\mathbb{Z}_n$, 
where the first quotient by $\mathbb{Z}_2$ creates non-trivial fixed locus 
and the $\mathbb{Z}_n$ action is free. 
Therefore the result (\ref{Riem2_S5/Z2k}) is simply obtained 
by dividing the integrated curvature-square (\ref{Riem2_S5/Z2}) for $S^5/\mathbb{Z}_2$ by $n$. 
For example, for $n=2$, i.e. $S^3/\mathbb{Z}_4$ 
the expression (\ref{Riem2_S5/Z2k}) perfectly matches the result obtained from the Hilbert series (see Appendix \ref{app_HS}). 
The Zeta-index reads
\begin{align}
\mathfrak{Z}^{AdS_5\times S^5/\mathbb{Z}_{2n}}(0,0;\omega_1,\omega_2)
&=-2. 
\end{align}
The special supersymmetric zeta value with $s=-1$ and $z=0$ is given by
\begin{align}
\label{-1_S5/Z2k}
\mathfrak{Z}^{AdS_5\times S^5/\mathbb{Z}_{2n}}(-1,0;\omega_1,\omega_2)
&=-\frac{n}{6}(\omega_1+\omega_2). 
\end{align}
Note that 
the special supersymmetric zeta values (\ref{-1_S5/Z2k-1}) and (\ref{-1_S5/Z2k}) can be written as
\begin{align}
\label{-1_S5/Zk}
\mathfrak{Z}^{AdS_5\times S^5/\mathbb{Z}_{k}}(-1,0;\omega_1,\omega_2)
&=-\frac{k}{12}(\omega_1+\omega_2). 
\end{align}
According to the formulae (\ref{c0_formula}) and (\ref{a0_formula}), we get
\begin{align}
c_{\mathcal{O}(N^0)}&=-\frac{k}{8}, \\
a_{\mathcal{O}(N^0)}&=-\frac{3k}{16}. 
\end{align}
Hence we have 
\begin{align}
\label{c-a_S5/Zk}
c_{\mathcal{O}(N^0)}-a_{\mathcal{O}(N^0)}&=\frac{k}{16}. 
\end{align}
The gravitational derivation of the difference (\ref{c-a_S5/Zk}) of the central charges was reported in \cite{ArabiArdehali:2013bac,ArabiArdehali:2013jiu}. 

The supersymmetric determinant is 
\begin{align}
\mathfrak{D}^{AdS_5\times S^5/\mathbb{Z}_{2n}}(z;\omega_1,\omega_2)
&=\frac{
\left(n(\omega_1+\omega_2)\right)^{\frac{9z}{n(\omega_1+\omega_2)}}z^2 
\Gamma\left(\frac{3z}{n(\omega_1+\omega_2)}\right)^2 \Gamma\left(1+\frac{3z}{2n(\omega_1+\omega_2)}\right)^2
}{2^{1-\frac{3z}{n(\omega_1+\omega_2)}} 3^{\frac{9z}{n(\omega_1+\omega_2)}}\pi^2}. 
\end{align}
The vacuum exponent is given by
\begin{align}
\mathfrak{D}^{AdS_5\times S^5/\mathbb{Z}_{2n}}(z;1,1)
&=\frac{2n^2}{9\pi^2}
\end{align}

\section{The $Y^{\mathsf{p},\mathsf{q}}$ Sasaki-Einstein manifolds}
\label{sec_Ypq}

The $Y^{\mathsf{p},\mathsf{q}}$ family of toric Sasaki-Einstein manifolds 
is characterized by two relatively prime positive integers $\mathsf{p}$, $\mathsf{q}$ with $\mathsf{q}<\mathsf{p}$. 
They are all topologically $S^2\times S^3$. 
There are infinitely many quasi-regular and irregular examples. 
The corresponding Sasaki-Einstein metrics are known \cite{Gauntlett:2004zh,Gauntlett:2004yd,Gauntlett:2004hh} 
and the toric descriptions are found \cite{Martelli:2004wu}. 
These notions refer to the orbit structure of the Reeb vector field, not to metric regularity, as all metrics are smooth on $S^2\times S^3$. 
The quasi-regularity occurs when $4\mathsf{p}^2 - 3\mathsf{q}^2$ is a square, 
while otherwise the manifold is irregular, where the corresponding Killing vector field does not integrate to a $U(1)$ action. 
Rather, it generates an $\mathbb{R}$-action with generic orbits. 

\subsection{$AdS_5\times T^{1,1}$}
As the simplest example in the $Y^{\mathsf{p},\mathsf{q}}$ family, let us consider $Y^{1,0}$. 
The corresponding SCFT is the Klebanov-Witten theory \cite{Klebanov:1998hh}, 
which is a quiver gauge theory with $\mathcal{N}=1$ vector multiplet of gauge group $SU(N)_1\times SU(N)_2$ 
coupled to four chiral multiplets, $X_{12}^1$, $X_{12}^2$, $X_{21}^1$ and $X_{21}^2$ with R-charge $R=1/2$. 
Under $SU(N)_1\times SU(N)_2$ the fields $X_{12}^{\alpha}$ transform as $(\bf{N},\overline{\bf{N}})$ and $X_{21}^{\beta}$ as $(\overline{\bf{N}},\bf{N})$. 
The theory has $SU(2)_a$ and $SU(2)_b$ flavor symmetries rotating $X_{12}^{\alpha}$ and $X_{21}^{\beta}$ respectively, 
as well as the $U(1)_B$ baryonic symmetry under which $X_{12}^{\alpha}$ have charge $+1$ and $X_{21}^{\beta}$ have charge $-1$. 
It describes a stack of $N$ D3-branes probing the conifold. 
The field content is summarized as 
\begin{align}
\begin{array}{c|cccccc}
&SU(N)_1&SU(N)_2&SU(2)_a&SU(2)_b&U(1)_B&U(1)_R\\ \hline 
X_{12}^{\alpha}&\bf{N}&\overline{\bf{N}}&\bf{2}&\bf{1}&+1&\frac12 \\
X_{21}^{\beta}&\overline{\bf{N}}&\bf{N}&\bf{1}&\bf{2}&-1&\frac12 \\
\end{array}
\end{align}
The quiver diagram is
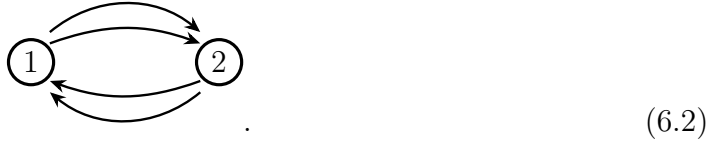
\begin{align}
\label{quiver_T11}
\begin{tikzpicture}[<->]
    \draw[very thick] (-1.25,0) circle (3mm);
    \node at (-1.25,0) {1};
     \draw[very thick] (1.25,0) circle (3mm);
    \node at (1.25,0) {2};
    \draw[-{Stealth[length=6pt,width=5pt]}, line width=0.9pt, bend left=40]
         (-1,0.4) to node[]{} (1,0.4);
    \draw[-{Stealth[length=6pt,width=5pt]}, line width=0.9pt, bend left=20]
         (-1,0.25) to node[]{} (1,0.25);
     \draw[-{Stealth[length=6pt,width=5pt]}, line width=0.9pt, bend left=40]
         (1,-0.4) to node[]{} (-1,-0.4);
    \draw[-{Stealth[length=6pt,width=5pt]}, line width=0.9pt, bend left=20]
         (1,-0.25) to node[]{} (-1,-0.25);
\end{tikzpicture}. 
\end{align}
The superpotential is
\begin{align}
\mathcal{W}&=\Tr (X_{12}^1X_{21}^1X_{12}^2X_{21}^2-X_{12}^1X_{21}^2X_{12}^2X_{21}^1). 
\end{align}
The $Y^{1,0}$ geometry is the homogeneous Sasaki-Einstein manifold $T^{1,1}$, 
which is topologically $S^2\times S^3$ and may be viewed as the coset defined by
\begin{align}
T^{l,m}&=\frac{SU(2)\times SU(2)}{U(1)_{l,m}}, 
\end{align}
where $U(1)_{l,m}$ is embedded in the two $SU(2)$ factors with weights $(l,m)$ being $(1,1)$.\footnote{The $U(1)$ is embedded as 
\begin{align}
e^{i\theta}:\qquad (g_1,g_2)\mapsto 
(g_1 e^{\frac{il\theta\sigma_3}{2}},g_2 e^{\frac{im\theta\sigma_3}{2}}), 
\end{align}
where $g_1$ and $g_2$ are elements of the two $SU(2)$ factors and $\sigma_3$ is the Pauli matrix. } 
The Calabi-Yau cone over $T^{1,1}$ is the conifold whose toric diagram is generated by
\begin{align}
v_1&=(1,0,0),\qquad
v_2=(1,1,0),\qquad
v_3=(1,1,1),\qquad
v_4=(1,0,1).  
\end{align}
It is illustrated as
\begin{align}
\label{toric_conifold}
\begin{tikzpicture}[scale=2, baseline={(0,0)}]
  \fill (0,0) circle (1pt) node[below left] {$(0,0)$};
  \fill (1,0) circle (1pt) node[below right] {$(1,0)$};
  \fill (1,1) circle (1pt) node[above right] {$(1,1)$};
  \fill (0,1) circle (1pt) node[above left] {$(0,1)$};
  \draw[very thick] (0,0) -- (1,0) -- (1,1) -- (0,1) -- cycle;
\end{tikzpicture}
\end{align}
We have 
\begin{align}
\det (v_{I-1},v_{I},v_{I+1})&=1, \qquad I=1,2,3,4, 
\end{align}
which encode the multiplicities of four types of the chiral multiplets forming the baryons.  
The area of the polygon is
\begin{align}
A&=1. 
\end{align}
The primitive generators of the dual cone are
\begin{align}
\label{dualG_conifold}
m_1&=(0,0,1), \qquad 
m_2=(1,-1,0), \qquad 
m_3=(1,0,-1), \qquad 
m_4=(0,1,0). 
\end{align}
The Reeb vector is fixed as
\begin{align}
b&=\left(3,\frac32,\frac32\right). 
\end{align}
The volume of $T^{1,1}$ is evaluated from (\ref{vol_Y1}) and (\ref{vol_Y2}) 
\begin{align}
\label{vol_T11}
\mathrm{Vol}(T^{1,1})&=\frac{16\pi^3}{27}. 
\end{align}
The R-charges of the chiral multiplet fields are given by
\begin{align}
R(X_{12}^1)&=\frac{\pi}{3}\frac{\mathrm{Vol}(\Sigma_1)}{\mathrm{Vol}(T^{1,1})}=\frac12, \\
R(X_{21}^1)&=\frac{\pi}{3}\frac{\mathrm{Vol}(\Sigma_2)}{\mathrm{Vol}(T^{1,1})}=\frac12, \\
R(X_{12}^2)&=\frac{\pi}{3}\frac{\mathrm{Vol}(\Sigma_3)}{\mathrm{Vol}(T^{1,1})}=\frac12, \\
R(X_{21}^2)&=\frac{\pi}{3}\frac{\mathrm{Vol}(\Sigma_4)}{\mathrm{Vol}(T^{1,1})}=\frac12, 
\end{align}
where 
\begin{align}
\mathrm{Vol}(\Sigma_I)&=2\pi^2 \frac{\det(v_{I-1},v_I,v_{I+1})}{\det(b,v_{I-1},v_I)\det(b,v_{I},v_{I+1})}=\frac{8\pi^2}{9}, 
\end{align}
with $I=1,2,3,4$, is the volume of the supersymmetric cycle $\Sigma_I$ wrapped by D3-branes, 
corresponding to the associated baryonic operators. 

The weighted adjacency matrix is given by
\begin{align}
\label{M_T11}
M^{AdS_5\times T^{1,1}}
&=\left(
\begin{matrix}
i_v&2\left(i_{\textrm{cm}_{\frac12}}+\tilde{i}_{\textrm{cm}_{\frac12}}\right)\\
2\left(i_{\textrm{cm}_{\frac12}}+\tilde{i}_{\textrm{cm}_{\frac12}}\right)&i_v\\
\end{matrix}
\right). 
\end{align}
The gravity index is evaluated as \cite{Nakayama:2006ur,Gadde:2010en}
\begin{align}
\label{ind_T11}
i^{AdS_5\times T^{1,1}}(p;q)
&=\frac{4p^{\frac12}q^{\frac12}}{1-p^{\frac12}q^{\frac12}}. 
\end{align}
This enumerates four extremal BPS mesons of R-charge $R=1$ as the single-trace operators  
\begin{align}
\mathcal{M}_1&=\Tr (X_{12}^2X_{21}^2),&
\mathcal{M}_2&=\Tr (X_{12}^1X_{21}^2),
\nonumber\\
\mathcal{M}_3&=\Tr (X_{12}^1X_{21}^1), &
\mathcal{M}_4&=\Tr (X_{12}^2X_{21}^1). 
\end{align}
They correspond to the four primitive generators (\ref{dualG_conifold}) of the dual cone. 
In fact, the R-charges are evaluated as
\begin{align}
R(\mathcal{M}_I)&=\frac23(b,m_I)=1, \qquad I=1,2,3,4
\end{align}
and the operator lengths are given by
\begin{align}
L_J(\mathcal{M}_1)&=(m_1,v_J)=\delta_{J,3}+\delta_{J,4}, \\
L_J(\mathcal{M}_2)&=(m_2,v_J)=\delta_{J,4}+\delta_{J,1}, \\
L_J(\mathcal{M}_3)&=(m_3,v_J)=\delta_{J,1}+\delta_{J,2}, \\
L_J(\mathcal{M}_4)&=(m_4,v_J)=\delta_{J,2}+\delta_{J,3}. 
\end{align}

The supersymmetric zeta function is
\begin{align}
\label{zeta_T11}
\mathfrak{Z}^{AdS_5\times T^{1,1}}(s,z;\omega_1,\omega_2)
&=4\left(\frac{\omega_1+\omega_2}{2}\right)^{-s}
\zeta\left(s,\frac{\omega_1+\omega_2+2z}{\omega_1+\omega_2}\right). 
\end{align}
The residue at a simple pole $s=1$ is
\begin{align}
\label{Res_T11}
\mathrm{Res}_{s=1}\mathfrak{Z}^{AdS_5\times T^{1,1}}(s,z;\omega_1,\omega_2)
&=\frac{8}{\omega_1+\omega_2}. 
\end{align}
It follows from (\ref{Riem2_conj}), (\ref{vol_T11}) and (\ref{Res_T11}) that 
\begin{align}
\label{Riem2_T11}
\int_{T^{1,1}}\mathrm{Riem}^2
&=\frac{2176}{27}\pi^3=2498.88\ldots. 
\end{align}
This value agrees exactly with that computed from the Hilbert series (see Appendix \ref{app_HS}).
The Zeta-index is
\begin{align}
\mathfrak{Z}^{AdS_5\times T^{1,1}}(0,0;\omega_1,\omega_2)
&=-2. 
\end{align}
The supersymmetric zeta value with $s=-1$ and $z=0$ is
\begin{align}
\mathfrak{Z}^{AdS_5\times T^{1,1}}(-1,0;\omega_1,\omega_2)
&=-\frac{\omega_1+\omega_2}{6}. 
\end{align}
From (\ref{c0_formula}) and (\ref{a0_formula}) we obtain
\begin{align}
c_{\mathcal{O}(N^0)}&=-\frac14, \\
a_{\mathcal{O}(N^0)}&=-\frac{3}{8}. 
\end{align}
Hence
\begin{align}
c_{\mathcal{O}(N^0)}-a_{\mathcal{O}(N^0)}&=\frac18. 
\end{align}

The supersymmetric determinant is
\begin{align}
\mathfrak{D}^{AdS_5\times T^{1,1}}(z;\omega_1,\omega_2)
&=\frac{z^4 (\omega_1+\omega_2)^{\frac{8z}{\omega_1+\omega_2}-2}\Gamma\left(\frac{2z}{\omega_1+\omega_2}\right)^4}
{\pi^2 4^{\frac{4z}{\omega_1+\omega_2}}}. 
\end{align}
The vacuum exponent is
\begin{align}
\mathfrak{D}^{AdS_5\times T^{1,1}}(0;1,1)&=\frac{1}{4\pi^2}=0.0253303\ldots. 
\end{align}

\subsection{$AdS_5\times T^{2,2}$}
As a further example, we examine $T^{2,2}$, that is $Y^{2,0}$. 
The corresponding field theory is realized on a stack of $N$ D3-branes probing the $\mathbb{Z}_2$ orbifold of the conifold. 
It is the 4d $\mathcal{N}=1$ quiver gauge theory of a $4$-node necklace with eight chiral multiplets $X_{i\ i+1}^{1,2}$ with $i=1,2,3,4$, 
which are described by two arrows between each adjacent pair of gauge nodes. 
The field content is 
\begin{align}
\begin{array}{c|cc}
&\prod_{i=1}^4SU(N)_{i}&U(1)_R\\ \hline 
X_{12}^{1},X_{12}^{2}&(\bf{N},\overline{\bf{N}},\bf{1},\bf{1})&\frac12 \\
X_{23}^{1},X_{23}^{2}&(\bf{1},\bf{N},\overline{\bf{N}},\bf{1})&\frac12 \\
X_{34}^{1},X_{34}^{2}&(\bf{1},\bf{1},\bf{N},\overline{\bf{N}})&\frac12 \\
X_{41}^{1},X_{41}^{2}&(\overline{\bf{N}},\bf{1},\bf{1},\bf{N})&\frac12 \\
\end{array}
\end{align}
The quiver diagram is
\begin{align}
\begin{tikzpicture}[<->]
    \draw[very thick] (-1,0) circle (3mm);
    \node at (-1,0) {1};
    \draw[very thick] (1,0) circle (3mm);
    \node at (1,0) {2};
    \draw[very thick] (1,-2) circle (3mm);
    \node at (1,-2) {3};
    \draw[very thick] (-1,-2) circle (3mm);
    \node at (-1,-2) {4};
    \draw[-{Stealth[length=6pt,width=5pt]}, line width=0.9pt]
          (-0.65,0.1) to node[]{} (0.65,0.1);
    \draw[-{Stealth[length=6pt,width=5pt]}, line width=0.9pt]
          (-0.65,-0.1) to node[]{} (0.65,-0.1);
    \draw[-{Stealth[length=6pt,width=5pt]}, line width=0.9pt]
         (0.9,-0.35) to node[]{} (0.9,-1.6);
    \draw[-{Stealth[length=6pt,width=5pt]}, line width=0.9pt]
          (1.1,-0.35) to node[]{} (1.1,-1.6);
    \draw[-{Stealth[length=6pt,width=5pt]}, line width=0.9pt]
         (-0.9,-1.6) to node[]{} (-0.9,-0.35);
    \draw[-{Stealth[length=6pt,width=5pt]}, line width=0.9pt]
          (-1.1,-1.6) to node[]{} (-1.1,-0.35);
    \draw[-{Stealth[length=6pt,width=5pt]}, line width=0.9pt]
          (0.65,-1.9) to node[]{} (-0.65,-1.9);
    \draw[-{Stealth[length=6pt,width=5pt]}, line width=0.9pt]
          (0.65,-2.1) to node[]{} (-0.65,-2.1);
\end{tikzpicture}. 
\end{align}
The superpotential takes the form
\begin{align}
\mathcal{W}&=
\epsilon_{\alpha\beta} \epsilon_{\gamma\delta} \Tr 
(
X_{12}^{\alpha}X_{23}^{\gamma}X_{34}^{\beta}X_{41}^{\delta}
). 
\end{align}
The theory is gravity dual to Type IIB string theory on $AdS_5\times T^{2,2}$. 
The Calabi-Yau cone whose Sasaki-Einstein base is $T^{2,2}$ is the $\mathbb{Z}_2$ orbifold of the conifold, 
which is also known as the complex cone over the Hirzebruch surface $F_0$ $=$ $\mathbb{P}^1\times \mathbb{P}^1$. 
The toric diagram is generated by 
\begin{align}
v_1&=(1,0,0),\qquad 
v_2=(1,1,0),\qquad 
v_3=(1,2,2),\qquad 
v_4=(1,1,2). 
\end{align}
\begin{align}
\label{toric_dconifold}
\begin{tikzpicture}[scale=1.5, baseline={(0,0)}]
  \fill (0,0) circle (2pt) node[below] {$(0,0)$};
  \fill (1,0) circle (2pt) node[below] {$(1,0)$};
  \fill (2,2) circle (2pt) node[above] {$(2,2)$};
  \fill (1,2) circle (2pt) node[above] {$(1,2)$};
  \draw[very thick] (0,0) -- (1,0) -- (2,2) -- (1,2) -- cycle;
\end{tikzpicture}
\end{align}
The multiplicities of four types of the chiral multiplets 
$X^{(1)}$ $=$ $\{X_{12}^1,X_{34}^1\}$, 
$X^{(2)}$ $=$ $\{X_{23}^1,X_{41}^1\}$, 
$X^{(3)}$ $=$ $\{X_{12}^{2}, X_{34}^2 \}$ 
and $X^{(4)}$ $=$ $\{X_{23}^{2},X_{41}^2\}$ are given by the determinants 
\begin{align}
\det (v_{I-1},v_{I},v_{I+1})&=2, \qquad I=1,2,3,4. 
\end{align}
The area of the polygon is
\begin{align}
A&=2. 
\end{align}
The primitive generators of the dual cone are
\begin{align}
\label{dualG_dconifold}
m_1&=(0,0,1), \qquad 
m_2=(2,-2,1), \qquad 
m_3=(2,0,-1), \qquad 
m_4=(0,2,-1). 
\end{align}
The Reeb vector is fixed as
\begin{align}
b&=\left(3,3,3\right). 
\end{align}
The volume of $T^{2,2}$ is evaluated from (\ref{vol_Y1}) and (\ref{vol_Y2}) 
\begin{align}
\label{vol_T22}
\mathrm{Vol}(T^{2,2})&=\frac{8\pi^3}{27}. 
\end{align}
The R-charges of the chiral multiplet fields are given by
\begin{align}
R(X^{(1)})&=\frac{\pi}{3}\frac{\mathrm{Vol}(\Sigma_1)}{\mathrm{Vol}(T^{2,2})}=\frac12, \\
R(X^{(2)})&=\frac{\pi}{3}\frac{\mathrm{Vol}(\Sigma_2)}{\mathrm{Vol}(T^{2,2})}=\frac12, \\
R(X^{(3)})&=\frac{\pi}{3}\frac{\mathrm{Vol}(\Sigma_3)}{\mathrm{Vol}(T^{2,2})}=\frac12, \\
R(X^{(4)})&=\frac{\pi}{3}\frac{\mathrm{Vol}(\Sigma_4)}{\mathrm{Vol}(T^{2,2})}=\frac12, 
\end{align}
where 
\begin{align}
\mathrm{Vol}(\Sigma_1)&=2\pi^2 \frac{\det(v_{4},v_{1},v_{2})}{\det(b,v_{4},v_1)\det(b,v_{1},v_{2})}=\frac{4\pi^2}{9}, \\
\mathrm{Vol}(\Sigma_2)&=2\pi^2 \frac{\det(v_{1},v_{2},v_{3})}{\det(b,v_{1},v_2)\det(b,v_{2},v_{3})}=\frac{4\pi^2}{9}, \\
\mathrm{Vol}(\Sigma_3)&=2\pi^2 \frac{\det(v_{2},v_{3},v_{4})}{\det(b,v_{2},v_3)\det(b,v_{3},v_{4})}=\frac{4\pi^2}{9}, \\
\mathrm{Vol}(\Sigma_4)&=2\pi^2 \frac{\det(v_{3},v_{4},v_{1})}{\det(b,v_{3},v_4)\det(b,v_{4},v_{1})}=\frac{4\pi^2}{9}. 
\end{align}

The weighted adjacency matrix is 
\begin{align}
\label{M_T22}
M^{AdS_5\times T^{2,2}}
&=\left(
\begin{matrix}
i_v&2i_{\textrm{cm}_{\frac12}}&0&2\tilde{i}_{\textrm{cm}_{\frac12}}\\
2\tilde{i}_{\textrm{cm}_{\frac12}}&i_v&2i_{\textrm{cm}_{\frac12}}&0\\
0&2\tilde{i}_{\textrm{cm}_{\frac12}}&i_v&2i_{\textrm{cm}_{\frac12}}\\
2i_{\textrm{cm}_{\frac12}}&0&2\tilde{i}_{\textrm{cm}_{\frac12}}&i_v\\
\end{matrix}
\right). 
\end{align}
The gravity index is given by
\begin{align}
\label{ind_T22}
i^{AdS_5\times T^{2,2}}(p;q)
&=4\frac{pq}{1-pq}. 
\end{align}
There are four independent extremal BPS mesonic operators of R-charge $2$ as single-trace operators
\begin{align}
\mathcal{M}_1&=\Tr(X_{12}^2X_{23}^2X_{34}^2X_{41}^2), & 
\mathcal{M}_2&=\Tr(X_{12}^1X_{23}^2X_{34}^1X_{41}^2), \\
\mathcal{M}_3&=\Tr(X_{12}^1X_{23}^1X_{34}^1X_{41}^1), &
\mathcal{M}_4&=\Tr(X_{12}^2X_{23}^1X_{34}^2X_{41}^1). 
\end{align}
They correspond to the primitive generators (\ref{dualG_dconifold}) and the R-charge is given by
\begin{align}
\frac23(b,m_I)&=2, \qquad I=1,2,3,4. 
\end{align}
The operator lengths are
\begin{align}
L_J(\mathcal{M}_1)&=(m_1,v_J)=2\delta_{J,3}+2\delta_{J,4}, \\
L_J(\mathcal{M}_2)&=(m_2,v_J)=2\delta_{J,4}+2\delta_{J,1}, \\
L_J(\mathcal{M}_3)&=(m_3,v_J)=2\delta_{J,1}+2\delta_{J,2}, \\
L_J(\mathcal{M}_4)&=(m_4,v_J)=2\delta_{J,2}+2\delta_{J,3}. 
\end{align}

The supersymmetric zeta function is
\begin{align}
\mathfrak{Z}^{AdS_5\times T^{2,2}}(s,z;\omega_1,\omega_2)
&=4(\omega_1+\omega_2)^{-s}
\zeta\left(s,\frac{\omega_1+\omega_2+z}{\omega_1+\omega_2}\right). 
\end{align}
The residue at a simple pole $s=1$ is
\begin{align}
\mathrm{Res}_{s=1}\mathfrak{Z}^{AdS_5\times T^{2,2}}(s,z;\omega_1,\omega_2)
&=\frac{4}{\omega_1+\omega_2}. 
\end{align}
Applying the formula (\ref{Riem2_conj}), we arrive at
\begin{align}
\label{Riem2_T22}
\int_{T^{2,2}}\mathrm{Riem}^2
&=\frac{1088}{27}\pi^3=1249.44\ldots. 
\end{align}
The curvature-squared integral does not receive any contributions from the fixed point 
and it is simply one half of (\ref{Riem2_T11}) since the Sasaki-Einstein manifold $T^{2,2}$ is smooth. 
The value (\ref{Riem2_T22}) coincides with the result from the Hilbert series (see Appendix \ref{app_HS}). 
The Zeta-index is
\begin{align}
\mathfrak{Z}^{AdS_5\times T^{2,2}}(0,0;\omega_1,\omega_2)
&=-2. 
\end{align}
The special supersymmetric zeta value with $s=-1$ and $z=0$ is
\begin{align}
\mathfrak{Z}^{AdS_5\times T^{2,2}}(-1,0;\omega_1,\omega_2)
&=-\frac{\omega_1+\omega_2}{3}. 
\end{align}
It follows from (\ref{c0_formula}) and (\ref{a0_formula}) that 
\begin{align}
c_{\mathcal{O}(N^0)}&=-\frac12, \\
a_{\mathcal{O}(N^0)}&=-\frac{3}{4}. 
\end{align}
Hence
\begin{align}
c_{\mathcal{O}(N^0)}-a_{\mathcal{O}(N^0)}&=\frac14. 
\end{align}

The supersymmetric determinant is
\begin{align}
\mathfrak{D}^{AdS_5\times T^{2,2}}(z;\omega_1,\omega_2)
&=\frac{(\omega_1+\omega_2)^{2+\frac{4z}{\omega_1+\omega_2}} \Gamma\left(1+\frac{z}{\omega_1+\omega_2}\right)^4}
{4\pi^2}. 
\end{align}
The vacuum exponent is
\begin{align}
\mathfrak{D}^{AdS_5\times T^{2,2}}(0;1,1)
&=\frac{1}{\pi^2}=0.101321\ldots. 
\end{align}

\subsection{$AdS_5\times T^{k,k}$}
\label{sec_Tkk}
More generally, let us consider $T^{k,k}$, i.e. $Y^{k,0}$. The theory describes 
a stack of $N$ D3-branes probing the $\mathbb{Z}_k$ orbifold of the conifold. It is given by the 4d $\mathcal{N}=1$ quiver gauge theory with $2k$ $SU(N)$ gauge nodes 
and four types of the bifundamental chiral multiplets $X_{2i-1\ 2i}^{1,2}$, $X_{2i\ 2i+1}$ and $X_{2i\ 2i-3}$, with $i=1,\cdots, k$. 
The chiral multiplets share the same R-charge $R=1/2$. 
The quiver diagram is depicted as
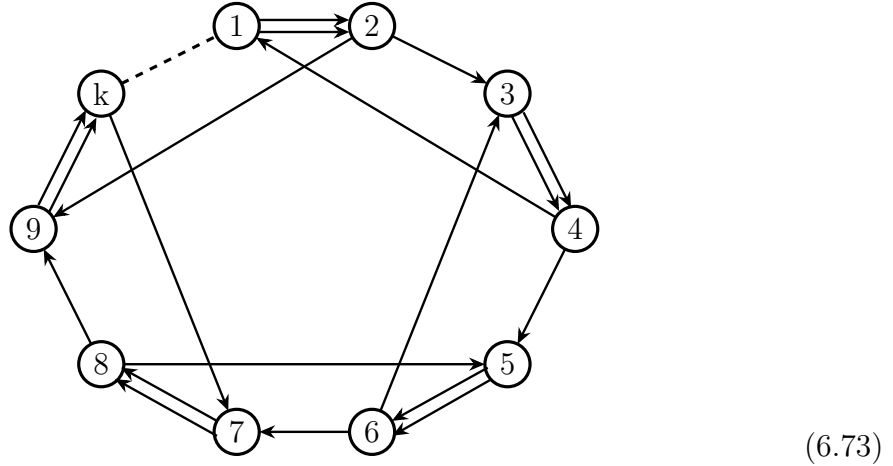
\begin{align}
\begin{tikzpicture}
\def\S{1.8}\def\R{3mm}
\coordinate (v1) at (0,0);
\coordinate (v2) at (\S*1,0);
\coordinate (v3) at (\S*2,\S*-0.5);
\coordinate (v4) at (\S*2.5,\S*-1.5);
\coordinate (v5) at (\S*2,\S*-2.5);
\coordinate (v6) at (\S*1,\S*-3);
\coordinate (v7) at (0,\S*-3);
\coordinate (v8) at (\S*-1,\S*-2.5);
\coordinate (v9) at (\S*-1.5,\S*-1.5);
\coordinate (vk) at (\S*-1,\S*-0.5);
\foreach \v/\lab in {v1/1,v2/2,v3/3,v4/4,v5/5,v6/6,v7/7,v8/8,v9/9,vk/k}{
  \draw[very thick] (\v) circle (\R);\node at (\v) {\lab};
  }
\tikzset{qarrow/.style={-{Stealth[length=6pt,width=5pt]},line width=0.9pt,shorten <=\R,shorten >=\R}}
\draw[qarrow] ($(v1)+(0,0.08)$)--($(v2)+(0,0.08)$);
\draw[qarrow] ($(v1)+(0,-0.08)$)--($(v2)+(0,-0.08)$);
\draw[qarrow] (v2)--(v3);
\draw[qarrow] (v2)--(v9);
\draw[qarrow] ($(v3)+(0.08,0.02)$)--($(v4)+(0.08,0.01)$);
\draw[qarrow] ($(v3)+(-0.08,-0.02)$)--($(v4)+(-0.08,-0.03)$);
\draw[qarrow] (v4)--(v5);
\draw[qarrow] (v4)--(v1);
\draw[qarrow] ($(v5)+(0,0.1)$)--($(v6)+(0,0)$);
\draw[qarrow] ($(v5)+(0.05,-0.05)$)--($(v6)+(0.01,-0.2)$);
\draw[qarrow] (v6)--(v7);
\draw[qarrow] (v6)--(v3);
\draw[qarrow] ($(v7)+(0,0)$)--($(v8)+(0,0.1)$);
\draw[qarrow] ($(v7)+(0,-0.2)$)--($(v8)+(-0.08,-0.05)$);
\draw[qarrow] (v8)--(v9);
\draw[qarrow] (v8)--(v5);
\draw[qarrow] ($(v9)+(0.07,-0.05)$)--($(vk)+(0.07,-0.05)$);
\draw[qarrow] ($(v9)+(-0.07,0.05)$)--($(vk)+(-0.07,0.05)$);
\draw[qarrow] (vk)--(v7);
\draw[dashed,very thick,shorten <=\R,shorten >=\R] (vk)--(v1);
\end{tikzpicture}
\end{align}
The superpotential takes the form
\begin{align}
\mathcal{W}&=\sum_{i=1}^k \epsilon^{\alpha\beta}
\Tr (X_{2i-1\ 2i}^{\alpha}X_{2i\ 2i-3}X_{2i-3\ 2i-2}^{\beta}X_{2i-2\ 2i-1}). 
\end{align}
The gravity dual is Type IIB string theory on $AdS_5\times T^{k,k}$. 
The Calabi-Yau cone whose Sasaki-Einstein base is $T^{k,k}$ is the $\mathbb{Z}_k$ orbifold of the conifold. 
The toric diagram is generated by 
\begin{align}
v_1&=(1,0,0),\qquad 
v_2=(1,1,0),\qquad 
v_3=(1,k,k),\qquad 
v_4=(1,k-1,k). 
\end{align}
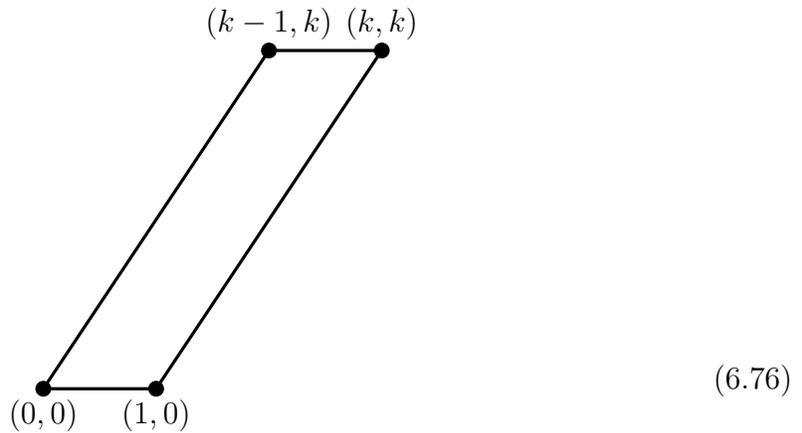
\begin{align}
\label{toric_conifoldZk}
\begin{tikzpicture}[scale=1.5, baseline={(0,0)}]
  \fill (0,0) circle (2pt) node[below] {$(0,0)$};
  \fill (1,0) circle (2pt) node[below] {$(1,0)$};
  \fill (3,3) circle (2pt) node[above] {$(k,k)$};
  \fill (2,3) circle (2pt) node[above] {$(k-1,k)$};
  \draw[very thick] (0,0) -- (1,0) -- (3,3) -- (2,3) -- cycle;
\end{tikzpicture}
\end{align}
The multiplicities of four types of the chiral multiplets 
$X_{2i-1\ 2i}^1$, $X_{2i\ 2i+1}$, $X_{2i-1\ 2i}^2$ and $X_{2i\ 2i-3}$ are given by the determinants 
\begin{align}
\det (v_{I-1},v_{I},v_{I+1})&=k, \qquad I=1,2,3,4. 
\end{align}
The area of the polygon is
\begin{align}
\label{A_conifoldZk}
A&=k. 
\end{align}
The primitive generators of the dual cone are
\begin{align}
\label{dualG_conifoldZk}
m_1&=(0,0,1), \qquad 
m_2=(k,-k,k-1), \qquad 
m_3=(k,0,-1), \qquad 
m_4=(0,k,-k+1). 
\end{align}
The Reeb vector is fixed as
\begin{align}
b&=\left(3,\frac{3k}{2},\frac{3k}{2}\right). 
\end{align}
The volume of $T^{k,k}$ is evaluated from (\ref{vol_Y1}) and (\ref{vol_Y2}) 
\begin{align}
\mathrm{Vol}(T^{k,k})&=\frac{16\pi^3}{27k}. 
\end{align}
The R-charges of the chiral multiplet fields are given by
\begin{align}
R(X_{2i-1\ 2i}^1)&=\frac{\pi}{3}\frac{\mathrm{Vol}(\Sigma_1)}{\mathrm{Vol}(T^{k,k})}=\frac12, \\
R(X_{2i\ 2i+1})&=\frac{\pi}{3}\frac{\mathrm{Vol}(\Sigma_2)}{\mathrm{Vol}(T^{k,k})}=\frac12, \\
R(X_{2i-1\ 2i}^2)&=\frac{\pi}{3}\frac{\mathrm{Vol}(\Sigma_3)}{\mathrm{Vol}(T^{k,k})}=\frac12, \\
R(X_{2i\ 2i-3})&=\frac{\pi}{3}\frac{\mathrm{Vol}(\Sigma_4)}{\mathrm{Vol}(T^{k,k})}=\frac12, 
\end{align}
where 
\begin{align}
\mathrm{Vol}(\Sigma_1)&=2\pi^2 \frac{\det(v_{4},v_{1},v_{2})}{\det(b,v_{4},v_1)\det(b,v_{1},v_{2})}=\frac{8\pi^2}{9k}, \\
\mathrm{Vol}(\Sigma_2)&=2\pi^2 \frac{\det(v_{1},v_{2},v_{3})}{\det(b,v_{1},v_2)\det(b,v_{2},v_{3})}=\frac{8\pi^2}{9k}, \\
\mathrm{Vol}(\Sigma_3)&=2\pi^2 \frac{\det(v_{2},v_{3},v_{4})}{\det(b,v_{2},v_3)\det(b,v_{3},v_{4})}=\frac{8\pi^2}{9k}, \\
\mathrm{Vol}(\Sigma_4)&=2\pi^2 \frac{\det(v_{3},v_{4},v_{1})}{\det(b,v_{3},v_4)\det(b,v_{4},v_{1})}=\frac{8\pi^2}{9k}. 
\end{align}

The weighted adjacency matrix takes the form
\begin{align}
&
M^{AdS_5\times T^{k,k}}
\nonumber\\
&=\left(
\begin{matrix}
i_v&2i_{\textrm{cm}_{\frac12}}&0&\tilde{i}_{\textrm{cm}_{\frac12}}&0&0&\cdots&0&0&\tilde{i}_{\textrm{cm}_{\frac12}} \\
2\tilde{i}_{\textrm{cm}_{\frac12}}&i_v&i_{\textrm{cm}_{\frac12}}&0&0&\cdots&0&0&i_{\textrm{cm}_{\frac12}}&0\\
0&\tilde{i}_{\textrm{cm}_{\frac12}}&i_v&2i_{\textrm{cm}_{\frac12}}&0&\tilde{i}_{\textrm{cm}_{\frac12}}&0&\cdots&\cdots&0 \\
i_{\textrm{cm}_{\frac12}}&0&2\tilde{i}_{\textrm{cm}_{\frac12}}&i_v&i_{\textrm{cm}_{\frac12}}&0&0&\cdots&0&0\\
\vdots&\vdots&\vdots&\vdots&\ddots&\vdots&\vdots&\vdots&\vdots&\vdots \\
\vdots&\vdots&\vdots&\vdots&\vdots&\ddots&\vdots&\vdots&\vdots&\vdots \\
0&0&\cdots&0&0&\tilde{i}_{\textrm{cm}_{\frac12}}&i_v&2i_{\textrm{cm}_{\frac12}}&0&\tilde{i}_{\textrm{cm}_{\frac12}} \\
0&\cdots&0&0&i_{\textrm{cm}_{\frac12}}&0&2\tilde{i}_{\textrm{cm}_{\frac12}}&i_v&i_{\textrm{cm}_{\frac12}}&0\\
0&\tilde{i}_{\textrm{cm}_{\frac12}}&0&0&\cdots&0&0&\tilde{i}_{\textrm{cm}_{\frac12}}&i_v&2i_{\textrm{cm}_{\frac12}} \\
i_{\textrm{cm}_{\frac12}}&0&0&\cdots&0&0&i_{\textrm{cm}_{\frac12}}&0&2\tilde{i}_{\textrm{cm}_{\frac12}}&i_v\\
\end{matrix}
\right). 
\end{align}
The single-particle gravity index is evaluated as
\begin{align}
i^{AdS_5\times T^{k,k}}(p;q)
&=4\frac{(pq)^{\frac{k}{2}}}{1-(pq)^{\frac{k}{2}}}. 
\end{align}
There are four extremal BPS mesons as single-trace operators of R-charge $R=k$
\begin{align}
\mathcal{M}_1&=\Tr (X_{12}^2 X_{2\ 2k-1} X_{2k-1\ 2k}^2 \cdots X_{34}^2 X_{41}), \\
\mathcal{M}_2&=\Tr (X_{12}^1 X_{2\ 2k-1} X_{2k-1\ 2k}^1 \cdots X_{34}^1 X_{41}), \\
\mathcal{M}_3&=\Tr (X_{12}^1 X_{23} X_{34}^1 \cdots X_{2k-1\ 2k}^{1} X_{2k\ 1}), \\
\mathcal{M}_4&=\Tr (X_{12}^2 X_{23} X_{34}^2 \cdots X_{2k-1\ 2k}^{2} X_{2k\ 1}).  
\end{align}
They correspond to the primitive generators (\ref{dualG_conifoldZk}) of the dual cone. 
The R-charges are reproduced from the formula (\ref{Rch_meson})
\begin{align}
\frac23(b,m_I)&=k, \qquad I=1,2,3,4. 
\end{align}
The operator lengths are given by
\begin{align}
L_J(\mathcal{M}_1)&=(m_1,v_J)=k\delta_{J,3}+k\delta_{J,4}, \\
L_J(\mathcal{M}_2)&=(m_2,v_J)=k\delta_{J,4}+k\delta_{J,1}, \\
L_J(\mathcal{M}_3)&=(m_3,v_J)=k\delta_{J,1}+k\delta_{J,2}, \\
L_J(\mathcal{M}_4)&=(m_4,v_J)=k\delta_{J,2}+k\delta_{J,3}. 
\end{align}

The supersymmetric zeta function is 
\begin{align}
\mathfrak{Z}^{AdS_5\times T^{k,k}}(s,z;\omega_1,\omega_2)
&=4\left(\frac{k(\omega_1+\omega_2)}{2}\right)^{-s}
\zeta\left(s,1+\frac{2z}{k(\omega_1+\omega_2)}\right). 
\end{align}
The residue at a simple pole $s=1$ is
\begin{align}
\mathrm{Res}_{s=1}\mathfrak{Z}^{AdS_5\times T^{k,k}}(s,z;\omega_1,\omega_2)
&=\frac{8}{k}\frac{1}{\omega_1+\omega_2}. 
\end{align}
From the formula (\ref{Riem2_conj}) we find the integral of the Riemann square
\begin{align}
\int_{T^{k,k}} \mathrm{Riem}^2&=\frac{2176}{27k}\pi^3. 
\end{align}
While the Sasaki-Einstein manifold $T^{k,k}$ can be viewed as $T^{1,1}/\mathbb{Z}_k$, 
it is smooth so that the curvature invariants do not receive contributions from localized singularities. 
The Zeta-index is
\begin{align}
\mathfrak{Z}^{AdS_5\times T^{k,k}}(0,0;\omega_1,\omega_2)&=-2. 
\end{align}
The supersymmetric zeta value for $s=-1$ and $z=0$ is
\begin{align}
\mathfrak{Z}^{AdS_5\times T^{k,k}}(-1,0;\omega_1,\omega_2)
&=-\frac{k}{6}(\omega_1+\omega_2). 
\end{align}
From the formula (\ref{c0_formula}) and (\ref{a0_formula}) we obtain
\begin{align}
c_{\mathcal{O}(N^0)}&=-\frac{k}{4}, \\
a_{\mathcal{O}(N^0)}&=-\frac{3k}{8}
\end{align}
and 
\begin{align}
c_{\mathcal{O}(N^0)}-a_{\mathcal{O}(N^0)}&=\frac{k}{8}. 
\end{align}

The supersymmetric determinant is
\begin{align}
\mathfrak{D}^{AdS_5\times T^{k,k}}(z;\omega_1,\omega_2)
&=\frac{
\left(k(\omega_1+\omega_2)\right)^{2+\frac{8z}{k(\omega_1+\omega_2)}}\Gamma\left(1+\frac{2z}{k(\omega_1+\omega_2)}\right)^4
}
{2^{4+\frac{8z}{k(\omega_1+\omega_2)}} \pi^2}. 
\end{align}
The vacuum exponent is
\begin{align}
\mathfrak{D}^{AdS_5\times T^{k,k}}(0;1,1)
&=\frac{k^2}{4\pi^2}. 
\end{align}

\subsection{$AdS_5\times Y^{2,1}$}
Let us study the 4d $\mathcal{N}=1$ quiver gauge theory describing a stack of $N$ D3-branes 
probing the complex cone over the first del Pezzo surface $dP_1$ that is the blow-up of $\mathbb{P}^2$ at one point. 
The probed geometry is isomorphic to the complex cone over the first Hirzebruch surface $F_1$. 
The theory has four $SU(N)$ gauge nodes and ten chiral multiplets \cite{Feng:2001xr,Feng:2002zw,Feng:2004uq,Bertolini:2004xf,Franco:2005rj,Forcella:2008bb,Davey:2009bp,Hanany:2012hi}. 
The field content and charges are summarized as
\begin{align}
\begin{array}{c|ccccc}
&\prod_{i=1}^4SU(N)_{i}&SU(2)&U(1)_R\\ \hline 
X_{12}&(\bf{N},\overline{\bf{N}},\bf{1},\bf{1})&\bf{1}&\frac{-17+5\sqrt{13}}{3} \\
X_{23}^{\alpha}&(\bf{1},\bf{N},\overline{\bf{N}},\bf{1})&\bf{2}&\frac{4(4-\sqrt{13})}{3} \\
X_{34}^{\alpha}&(\bf{1},\bf{1},\bf{N},\overline{\bf{N}})&\bf{2}&\frac{-1+\sqrt{13}}{3} \\
X_{34}^3&(\bf{1},\bf{1},\bf{N},\overline{\bf{N}})&\bf{1}&-3+\sqrt{13} \\
X_{41}^{\alpha}&(\overline{\bf{N}},\bf{1},\bf{1},\bf{N})&\bf{2}&\frac{4(4-\sqrt{13})}{3} \\
X_{13}&(\bf{N},\bf{1},\overline{\bf{N}},\bf{1})&\bf{1}&-3+\sqrt{13} \\
X_{42}&(\bf{1},\overline{\bf{N}},\bf{1},\bf{N})&\bf{1}&-3+\sqrt{13} \\
\end{array}. 
\end{align}
The quiver diagram is 
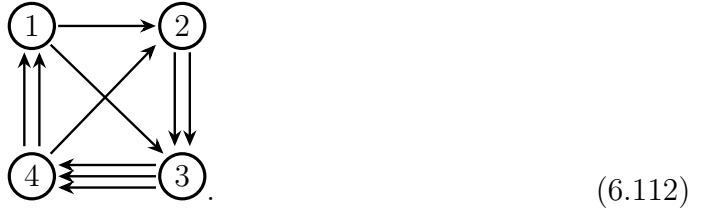
\begin{align}
\begin{tikzpicture}[<->]
    \draw[very thick] (-1,0) circle (3mm);
    \node at (-1,0) {1};
    \draw[very thick] (1,0) circle (3mm);
    \node at (1,0) {2};
    \draw[very thick] (1,-2) circle (3mm);
    \node at (1,-2) {3};
    \draw[very thick] (-1,-2) circle (3mm);
    \node at (-1,-2) {4};
    \draw[-{Stealth[length=6pt,width=5pt]}, line width=0.9pt]
          (-0.65,0) to node[]{} (0.65,0);
    \draw[-{Stealth[length=6pt,width=5pt]}, line width=0.9pt]
          (-0.75,-0.25) to node[]{} (0.75,-1.7);
    \draw[-{Stealth[length=6pt,width=5pt]}, line width=0.9pt]
          (-0.75,-1.7) to node[]{} (0.65,-0.25);
    \draw[-{Stealth[length=6pt,width=5pt]}, line width=0.9pt]
         (0.9,-0.35) to node[]{} (0.9,-1.6);
    \draw[-{Stealth[length=6pt,width=5pt]}, line width=0.9pt]
          (1.1,-0.35) to node[]{} (1.1,-1.6);
    \draw[-{Stealth[length=6pt,width=5pt]}, line width=0.9pt]
         (-0.9,-1.6) to node[]{} (-0.9,-0.35);
    \draw[-{Stealth[length=6pt,width=5pt]}, line width=0.9pt]
          (-1.1,-1.6) to node[]{} (-1.1,-0.35);
    \draw[-{Stealth[length=6pt,width=5pt]}, line width=0.9pt]
          (0.65,-1.85) to node[]{} (-0.65,-1.85);
    \draw[-{Stealth[length=6pt,width=5pt]}, line width=0.9pt]
          (0.65,-2) to node[]{} (-0.65,-2);
    \draw[-{Stealth[length=6pt,width=5pt]}, line width=0.9pt]
          (0.65,-2.15) to node[]{} (-0.65,-2.15);
\end{tikzpicture}. 
\end{align}
The superpotential is
\begin{align}
\mathcal{W}&=\Tr(X_{34}^1X_{42}X_{23}^1+X_{34}^2X_{41}^2X_{13}+X_{23}^2X_{34}^3X_{41}^1X_{12}
\nonumber\\
&-X_{41}^1X_{13}X_{34}^1-X_{42}X_{23}^2X_{34}^2-X_{34}^3X_{41}^2X_{12}X_{23}^1). 
\end{align}

The theory is holographically dual to Type IIB string theory on $AdS_5\times Y^{2,1}$ \cite{Feng:2000mi}. 
The corresponding Calabi-Yau is the complex cone over $dP_1$ whose toric diagram is generated by
\begin{align}
v_1&=(1,0,0),\qquad 
v_2=(1,1,0),\qquad 
v_3=(1,2,2),\qquad 
v_4=(1,0,1). 
\end{align}
It is shown as
\begin{align}
\label{toric_Y21}
\begin{tikzpicture}[scale=1.5, baseline={(0,0)}]
  \fill (0,0) circle (2pt) node[below left] {$(0,0)$};
  \fill (1,0) circle (2pt) node[below right] {$(1,0)$};
  \fill (2,2) circle (2pt) node[above right] {$(2,2)$};
  \fill (0,1) circle (2pt) node[above left] {$(0,1)$};
  \draw[very thick] (0,0) -- (1,0) -- (2,2) -- (0,1) -- cycle;
\end{tikzpicture}
\end{align}
The multiplicities of four types of the chiral multiplets 
$Z$ $=$ $X_{12}$, 
$U_1$ $=$ $\{X_{23}^{1},X_{41}^1\}$, 
$Y$ $=$ $\{X_{34}^3,X_{13},X_{42}\}$ 
and $U_2$ $=$ $\{X_{23}^{2}, X_{41}^2\}$ forming the baryonic operators are 
\begin{align}
\det (v_{4},v_{1},v_{2})&=1, \\
\det (v_{1},v_{2},v_{3})&=2, \\
\det (v_{2},v_{3},v_{4})&=3, \\
\det (v_{3},v_{4},v_{1})&=2. 
\end{align}
The area of the polygon is
\begin{align}
\label{A_Y21}
A&=2. 
\end{align}
The primitive generators of the dual cone are
\begin{align}
\label{dualG_Y21}
m_1&=(0,0,1), \qquad 
m_2=(2,-2,1), \qquad 
m_3=(2,1,-2), \qquad 
m_4=(0,1,0). 
\end{align}
The Reeb vector is fixed as
\begin{align}
b&=\left(3,\sqrt{13}-1,\sqrt{13}-1\right). 
\end{align}
The volume of $Y^{2,1}$ is evaluated from (\ref{vol_Y1}) and (\ref{vol_Y2}) 
\begin{align}
\label{vol_Y21}
\mathrm{Vol}(Y^{2,1})&=\frac{46+13\sqrt{13}}{324}\pi^3. 
\end{align}
The R-charges of the chiral multiplet fields are given by
\begin{align}
R(Z)&=\frac{\pi}{3}\frac{\mathrm{Vol}(\Sigma_1)}{\mathrm{Vol}(Y^{2,1})}=\frac{-17+5\sqrt{13}}{3}, \\
R(U_1)&=\frac{\pi}{3}\frac{\mathrm{Vol}(\Sigma_2)}{\mathrm{Vol}(Y^{2,1})}=\frac{4(4-\sqrt{13})}{3}, \\
R(Y)&=\frac{\pi}{3}\frac{\mathrm{Vol}(\Sigma_3)}{\mathrm{Vol}(Y^{2,1})}=-3+\sqrt{13}, \\
R(U_2)&=\frac{\pi}{3}\frac{\mathrm{Vol}(\Sigma_4)}{\mathrm{Vol}(Y^{2,1})}=\frac{4(4-\sqrt{13})}{3}, \\
R(V_{1})&=\frac{\pi}{3}\frac{\mathrm{Vol}(\Sigma_4+\Sigma_1)}{\mathrm{Vol}(Y^{2,1})}=\frac{-1+\sqrt{13}}{3}, \\
R(V_{2})&=\frac{\pi}{3}\frac{\mathrm{Vol}(\Sigma_1+\Sigma_2)}{\mathrm{Vol}(Y^{2,1})}=\frac{-1+\sqrt{13}}{3}, 
\end{align}
where 
\begin{align}
\mathrm{Vol}(\Sigma_1)&=2\pi^2 \frac{\det(v_{4},v_{1},v_{2})}{\det(b,v_{4},v_1)\det(b,v_{1},v_{2})}=\frac{2\pi^2}{(-1+\sqrt{13})^2}, \\
\mathrm{Vol}(\Sigma_2)&=2\pi^2 \frac{\det(v_{1},v_{2},v_{3})}{\det(b,v_{1},v_2)\det(b,v_{2},v_{3})}=\frac{\pi^2}{-5+2\sqrt{13}}, \\
\mathrm{Vol}(\Sigma_3)&=2\pi^2 \frac{\det(v_{2},v_{3},v_{4})}{\det(b,v_{2},v_3)\det(b,v_{3},v_{4})}=\frac{6\pi^2}{(-7+\sqrt{13})^2}, \\
\mathrm{Vol}(\Sigma_4)&=2\pi^2 \frac{\det(v_{3},v_{4},v_{1})}{\det(b,v_{3},v_4)\det(b,v_{4},v_{1})}=\frac{\pi^2}{-5+2\sqrt{13}}, 
\end{align}
and $V_{\alpha}=\{X_{34}^{\alpha}\}$ can be viewed as the composites, 
corresponding to unions of adjacent toric divisors. 

We have the weighted adjacency matrix
\begin{align}
\label{M_Y21}
&
M^{AdS_5\times Y^{2,1}}
\nonumber\\
&=\left(
\begin{matrix}
i_v&i_{\textrm{cm}_{\frac13(-17+5\sqrt{13})}}&i_{\textrm{cm}_{-3+\sqrt{13}}}&2\tilde{i}_{\textrm{cm}_{\frac43(4-\sqrt{13})}}\\
\tilde{i}_{\textrm{cm}_{\frac13(-17+\sqrt{13})}}&i_v&2i_{\textrm{cm}_{\frac43(4-\sqrt{13})}}&\tilde{i}_{\textrm{cm}_{-3+\sqrt{13}}}\\
\tilde{i}_{\textrm{cm}_{-3+\sqrt{13}}}&2\tilde{i}_{\textrm{cm}_{\frac43(4-\sqrt{13})}}&i_v&2i_{\textrm{cm}_{\frac13(-1+\sqrt{13})}}+i_{\textrm{cm}_{-3+\sqrt{13}}}\\
2i_{\textrm{cm}_{\frac43(4-\sqrt{13})}}&i_{\textrm{cm}_{-3+\sqrt{13}}}&2\tilde{i}_{\textrm{cm}_{\frac13(-1+\sqrt{13})}}+\tilde{i}_{\textrm{cm}_{-3+\sqrt{13}}}&i_v\\
\end{matrix}
\right). 
\end{align}
The gravity index is
\begin{align}
\label{ind_Y21}
i^{AdS_5\times Y^{2,1}}(p;q)
&=2\frac{(pq)^{\frac{7-\sqrt{13}}{3}}}{1-(pq)^{\frac{7-\sqrt{13}}{3}}}
+2\frac{(pq)^{\frac{-1+\sqrt{13}}{3}}}{1-(pq)^{\frac{-1+\sqrt{13}}{3}}}. 
\end{align}
It counts two extremal BPS mesonic operators of R-charge $\frac{2(7-\sqrt{13})}{3}$
\begin{align}
\mathcal{M}_1&=\Tr (X_{34}^3X_{41}^1X_{13}), \\
\mathcal{M}_4&=\Tr(X_{23}^1X_{34}^3X_{42}), 
\end{align}
and two extremal BPS mesons of R-charge $\frac{2(-1+\sqrt{13})}{3}$
\begin{align}
\mathcal{M}_2&=\Tr (X_{12}X_{23}^2X_{34}^1X_{41}^2), \\
\mathcal{M}_3&=\Tr (X_{12}X_{23}^1X_{34}^2X_{41}^1). 
\end{align}
These extremal BPS mesons correspond to the primitive generators (\ref{dualG_Y21}) of the dual cone. 
According to the formula (\ref{Rch_meson}) 
the R-charges are evaluated from the toric diagram as
\begin{align}
R(\mathcal{M}_1)&=\frac23(b,m_1)=\frac{2(\sqrt{13}-1)}{3}, \\
R(\mathcal{M}_2)&=\frac23(b,m_2)=\frac{2(7-\sqrt{13})}{3}, \\
R(\mathcal{M}_3)&=\frac23(b,m_3)=\frac{2(7-\sqrt{13})}{3}, \\
R(\mathcal{M}_4)&=\frac23(b,m_4)=\frac{2(\sqrt{13}-1)}{3}. 
\end{align}

The supersymmetric zeta function is
\begin{align}
\label{zeta_Y21}
\mathfrak{Z}^{AdS_5\times Y^{2,1}}(s,z;\omega_1,\omega_2)
&=2\left(\frac{7-\sqrt{13}}{3}(\omega_1+\omega_2)\right)^{-s}
\zeta\left(s,\frac{\omega_1+\omega_2+\frac{3z}{7-\sqrt{13}}}{\omega_1+\omega_2}\right)
\nonumber\\
&+2\left(\frac{-1+\sqrt{13}}{3}(\omega_1+\omega_2)\right)^{-s}
\zeta\left(s,\frac{\omega_1+\omega_2+\frac{3z}{-1+\sqrt{13}}}{\omega_1+\omega_2}\right). 
\end{align}
The residue at a simple pole $s=1$ is
\begin{align}
\label{Res_Y21}
\mathrm{Res}_{s=1}\mathfrak{Z}^{AdS_5\times Y^{2,1}}(s,z;\omega_1,\omega_2)
&=\frac{5+2\sqrt{13}}{3}\frac{1}{\omega_1+\omega_2}. 
\end{align}
Plugging (\ref{Res_Y21}) and (\ref{vol_Y21}) into the formula (\ref{Riem2_conj}), 
we get
\begin{align}
\int_{Y^{2,1}}\mathrm{Riem}^2
&=\frac{2(566+329\sqrt{13})}{81}\pi^3. 
\end{align}
This is in exact agreement with the Hilbert series computation (see Appendix \ref{app_HS}).
The Zeta-index is
\begin{align}
\mathfrak{Z}^{AdS_5\times Y^{2,1}}(0,0;\omega_1,\omega_2)
&=-2. 
\end{align}
The supersymmetric zeta value with $s=-1$ and $z=0$ is
\begin{align}
\mathfrak{Z}^{AdS_5\times Y^{2,1}}(-1,0;\omega_1,\omega_2)
&=-\frac{\omega_1+\omega_2}{3}. 
\end{align}
From the formula (\ref{c0_formula}) and (\ref{a0_formula}) we find
\begin{align}
c_{\mathcal{O}(N^0)}&=-\frac12, \\
a_{\mathcal{O}(N^0)}&=-\frac{3}{4}. 
\end{align}
Hence
\begin{align}
c_{\mathcal{O}(N^0)}-a_{\mathcal{O}(N^0)}&=\frac14. 
\end{align}

We find the supersymmetric determinant
\begin{align}
&
\mathfrak{D}^{AdS_5\times Y^{2,1}}(z;\omega_1,\omega_2)
\nonumber\\
&=\frac{
(7-\sqrt{13})^{\frac{7+\sqrt{13}}{6(\omega_1+\omega_2)}z}
(-1+\sqrt{13})^{\frac{1+\sqrt{13}}{2(\omega_1+\omega_2)}z}
(-5+2\sqrt{13})
(\omega_1+\omega_2)^{2+\frac{5+2\sqrt{13}}{3(\omega_1+\omega_2)}z}
}
{3^{2+\frac{(5+2\sqrt{13})}{3(\omega_1+\omega_2)}z} \pi^2 }
\nonumber\\
&\times 
\Gamma\left(1+\frac{1+\sqrt{13}}{4(\omega_1+\omega_2)}z\right)^2
\Gamma\left(1+\frac{7+\sqrt{13}}{12(\omega_1+\omega_2)}z\right)^2. 
\end{align}
The vacuum exponent is
\begin{align}
\mathfrak{D}^{AdS_5\times Y^{2,1}}(0;1,1)
&=\frac{4(-5+2\sqrt{13})}{9\pi^2}
\nonumber\\
&=0.0995696\ldots. 
\end{align}

\subsection{$AdS_5\times Y^{\mathsf{p},\mathsf{q}}$}
Now we consider the 4d $\mathcal{N}=1$ quiver gauge theory 
for a stack of $N$ D3-branes probing the Calabi-Yau cone over the Sasaki-Einstein manifold $Y^{\mathsf{p},\mathsf{q}}$ 
with $\mathsf{p}>\mathsf{q}>0$.\footnote{$Y^{\mathsf{p},0}$ and $Y^{\mathsf{p},\mathsf{p}}$ correspond to $T^{\mathsf{p},\mathsf{p}}$ in section \ref{sec_Tkk} 
and $S^5/\mathbb{Z}_\mathsf{2p}$ in section \ref{sec_S5/Z2k} respectively. }
It contains $2\mathsf{p}$ $SU(N)$ gauge nodes and $(4\mathsf{p}+2\mathsf{q})$ bifundamental chiral multiplets composed of four types, 
the $\mathsf{p}$ $SU(2)$ doublets $U_{\alpha}$, 
$\mathsf{q}$ $SU(2)$ doublets $V_{\alpha}$, 
$(\mathsf{p}+\mathsf{q})$ $SU(2)$ singlets $Y$ 
and $(\mathsf{p}-\mathsf{q})$ $SU(2)$ singlets $Z$ with $\alpha=1,2$. 
The field content is summarized as \cite{Benvenuti:2004dy}
\begin{align}
\begin{array}{c|ccccc}
&\textrm{number}&SU(2)&U(1)_F&U(1)_B&U(1)_R\\ \hline 
U_{\alpha}&\mathsf{p}&\bf{2}&0&-\mathsf{p}&1-\frac{x+y}{2} \\
V_{\alpha}&\mathsf{q}&\bf{2}&+1&\mathsf{q}&1+\frac{x-y}{2} \\
Y&\mathsf{p}+\mathsf{q}&\bf{1}&-1&\mathsf{p}-\mathsf{q}&y \\
Z&\mathsf{p}-\mathsf{q}&\bf{1}&+1&\mathsf{p}+\mathsf{q}&x \\
\end{array}
\end{align}
Here $x$ and $y$ are given by 
\begin{align}
x&=\frac{-4\mathsf{p}^2+3\mathsf{q}^2-2\mathsf{p}\mathsf{q}+(2\mathsf{p}+\mathsf{q})\sqrt{4\mathsf{p}^2-3\mathsf{q}^2}}{3\mathsf{q}^2}, \\
y&=\frac{-4\mathsf{p}^2+3\mathsf{q}^2+2\mathsf{p}\mathsf{q}+(2\mathsf{p}-\mathsf{q})\sqrt{4\mathsf{p}^2-3\mathsf{q}^2}}{3\mathsf{q}^2}. 
\end{align}
The superpotential takes the form \cite{Benvenuti:2004dy}
\begin{align}
\mathcal{W}&=\sum 
\epsilon^{\alpha\beta}\Tr (U_{\alpha}V_{\beta}Y+V_{\alpha}U_{\beta}Y)
+\epsilon^{\alpha\beta}\Tr (ZU_{\alpha}YU_{\beta}). 
\end{align}
The sum is taken over the indices of gauge nodes. 
The first terms include cubic terms, corresponding to the $2\mathsf{q}$ triangles, 
and the second contain quartic terms, corresponding to the $(\mathsf{p}-\mathsf{q})$ rectangles.  
The theory is holographically dual to Type IIB string theory on $AdS_5\times Y^{\mathsf{p},\mathsf{q}}$. 
The toric diagram of the Calabi-Yau cone over $Y^{\mathsf{p},\mathsf{q}}$ is generated by
\begin{align}
v_1&=(1,0,0), \qquad 
v_2=(1,1,0), \qquad 
v_3=(1,\mathsf{p},\mathsf{p}), \qquad 
v_4=(1,\mathsf{p}-\mathsf{q}-1,\mathsf{p}-\mathsf{q}). 
\end{align}
It is illustrated as
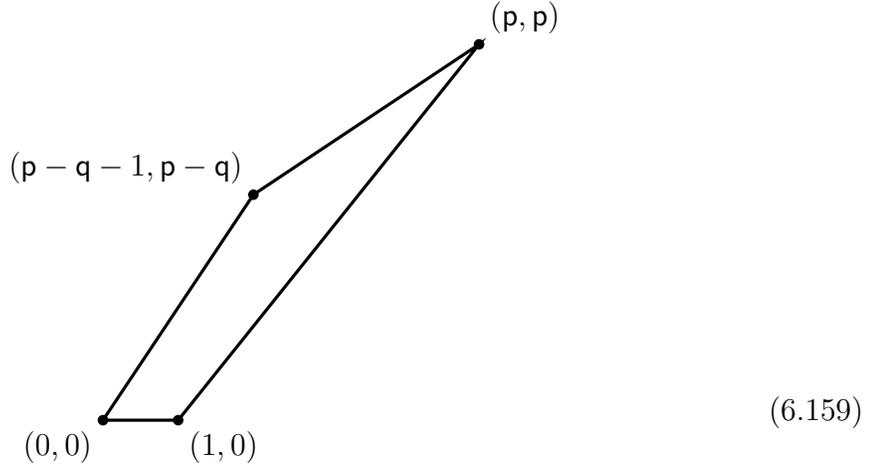
\begin{align}
\label{toric_Ypq}
\begin{tikzpicture}[scale=1, baseline={(0,0)}]
  \fill (0,0) circle (2pt) node[below left] {$(0,0)$};
  \fill (1,0) circle (2pt) node[below right] {$(1,0)$};
  \fill (5,5) circle (2pt) node[above right] {$(\mathsf{p},\mathsf{p})$};
  \fill (2,3) circle (2pt) node[above left] {$(\mathsf{p}-\mathsf{q}-1,\mathsf{p}-\mathsf{q})$};
  \draw[very thick] (0,0) -- (1,0) -- (5,5) -- (2,3) -- cycle;
\end{tikzpicture}
\end{align}
We obtain the multiplicities of four types of the chiral multiplets 
$Z$, $U_1$, $Y$ and $U_2$ forming the baryonic operators 
\begin{align}
\det (v_{4},v_{1},v_{2})&=\mathsf{p}-\mathsf{q}, \\
\det (v_{1},v_{2},v_{3})&=\mathsf{p}, \\
\det (v_{2},v_{3},v_{4})&=\mathsf{p}+\mathsf{q}, \\
\det (v_{3},v_{4},v_{1})&=\mathsf{p}. 
\end{align}
The area of the polygon is
\begin{align}
\label{A_Ypq}
A&=\mathsf{p}. 
\end{align}
The primitive generators of the dual cone are
\begin{align}
\label{dualG_Ypq}
m_1&=(0,0,1), \qquad 
m_2=(\mathsf{p},-\mathsf{p},\mathsf{p}-1), \nonumber\\
m_3&=(\mathsf{p},\mathsf{q},-\mathsf{q}-1), \qquad 
m_4=(0,\mathsf{p}-\mathsf{q},-\mathsf{p}+\mathsf{q}+1). 
\end{align}
We have the Reeb vector
\begin{align}
b&=\left(3,
\frac{-2\mathsf{p}^2+3\mathsf{p}\mathsf{q}+\mathsf{p}
\sqrt{4\mathsf{p}^2-3\mathsf{q}^2}}{2\mathsf{q}},
\frac{-2\mathsf{p}^2+3\mathsf{p}\mathsf{q}+\mathsf{p}
\sqrt{4\mathsf{p}^2-3\mathsf{q}^2}}{2\mathsf{q}}
\right). 
\end{align}
The volume of $Y^{\mathsf{p},\mathsf{q}}$ is evaluated from (\ref{vol_Y1}) and (\ref{vol_Y2})
\begin{align}
\label{vol_Ypq}
\mathrm{Vol}(Y^{\mathsf{p},\mathsf{q}})
&=\frac{\mathsf{q}^2(2\mathsf{p}+\sqrt{4\mathsf{p}^2-3\mathsf{q}^2})}
{3\mathsf{p}^2(-2\mathsf{p}^2+3\mathsf{q}^2+\mathsf{p}\sqrt{4\mathsf{p}^2-3\mathsf{q}^2})}
\pi^3. 
\end{align}
The R-charges of the chiral multiplet fields are 
\begin{align}
R(Z)&=\frac{\pi}{3}\frac{\mathrm{Vol}(\Sigma_1)}{\mathrm{Vol}(Y^{\mathsf{p},\mathsf{q}})}=x, \\
R(U_1)&=\frac{\pi}{3}\frac{\mathrm{Vol}(\Sigma_2)}{\mathrm{Vol}(Y^{\mathsf{p},\mathsf{q}})}=1-\frac{x+y}{2}, \\
R(Y)&=\frac{\pi}{3}\frac{\mathrm{Vol}(\Sigma_3)}{\mathrm{Vol}(Y^{\mathsf{p},\mathsf{q}})}=y, \\
R(U_2)&=\frac{\pi}{3}\frac{\mathrm{Vol}(\Sigma_4)}{\mathrm{Vol}(Y^{\mathsf{p},\mathsf{q}})}=1-\frac{x+y}{2}, \\
R(V_{1})&=\frac{\pi}{3}\frac{\mathrm{Vol}(\Sigma_4+\Sigma_1)}{\mathrm{Vol}(Y^{\mathsf{p},\mathsf{q}})}=1+\frac{x-y}{2}, \\
R(V_{2})&=\frac{\pi}{3}\frac{\mathrm{Vol}(\Sigma_1+\Sigma_2)}{\mathrm{Vol}(Y^{\mathsf{p},\mathsf{q}})}=1+\frac{x-y}{2}, 
\end{align}
where 
\begin{align}
\mathrm{Vol}(\Sigma_1)&=2\pi^2 \frac{\det(v_{4},v_{1},v_{2})}{\det(b,v_{4},v_1)\det(b,v_{1},v_{2})}
=\frac{8(\mathsf{p}-\mathsf{q})\mathsf{q}^2\pi^2}{\mathsf{p}^2(-2\mathsf{p}+3\mathsf{q}+\sqrt{4\mathsf{p}^2-3\mathsf{q}^2})^2}, \\
\mathrm{Vol}(\Sigma_2)&=2\pi^2 \frac{\det(v_{1},v_{2},v_{3})}{\det(b,v_{1},v_2)\det(b,v_{2},v_{3})}
=\frac{2\mathsf{q}^2\pi^2}{\mathsf{p}(-2\mathsf{p}^2+3\mathsf{q}^2+\mathsf{p}\sqrt{4\mathsf{p}^2-3\mathsf{q}^2})}, \\
\mathrm{Vol}(\Sigma_3)&=2\pi^2 \frac{\det(v_{2},v_{3},v_{4})}{\det(b,v_{2},v_3)\det(b,v_{3},v_{4})}
=\frac{8(\mathsf{p}+\mathsf{q})\mathsf{q}^2\pi^2}{\mathsf{p}^2(-2\mathsf{p}-3\mathsf{q}+\sqrt{4\mathsf{p}^2-3\mathsf{q}^2})^2}, \\
\mathrm{Vol}(\Sigma_4)&=2\pi^2 \frac{\det(v_{3},v_{4},v_{1})}{\det(b,v_{3},v_4)\det(b,v_{4},v_{1})}
=\frac{2\mathsf{q}^2\pi^2}{\mathsf{p}(-2\mathsf{p}^2+3\mathsf{q}^2+\mathsf{p}\sqrt{4\mathsf{p}^2-3\mathsf{q}^2})}. 
\end{align}

It is conjectured in \cite{Gadde:2010en} 
that the single-particle gravity index for the Sasaki-Einstein manifold $Y^{\mathsf{p},\mathsf{q}}$ 
can be written in the following form: 
\begin{align}
\label{ind_Ypq}
i^{AdS_5\times Y^{\mathsf{p},\mathsf{q}}}(p;q)
&=2\frac{(pq)^{\frac{\mathsf{p}(3\mathsf{q}+2\mathsf{p}-\sqrt{4\mathsf{p}^2-3\mathsf{q}^2})}{6\mathsf{q}}}}
{1-(pq)^{\frac{\mathsf{p}(3\mathsf{q}+2\mathsf{p}-\sqrt{4\mathsf{p}^2-3\mathsf{q}^2})}{6\mathsf{q}}}}
+2\frac{(pq)^{\frac{\mathsf{p}(3\mathsf{q}-2\mathsf{p}+\sqrt{4\mathsf{p}^2-3\mathsf{q}^2})}{6\mathsf{q}}}}
{1-(pq)^{\frac{\mathsf{p}(3\mathsf{q}-2\mathsf{p}+\sqrt{4\mathsf{p}^2-3\mathsf{q}^2})}{6\mathsf{q}}}}. 
\end{align}
This conjectured expression is consistently reproduced from the combinatorial formula (\ref{ind_gravity_conj}).
There are two types of extremal BPS mesons. 
The two extremal BPS mesons of the first type carry R-charge
\begin{align}
\frac{2\mathsf{p^2}+3\mathsf{p}\mathsf{q}-\mathsf{p}\sqrt{4\mathsf{p}^2-3\mathsf{q}^2}}{3\mathsf{q}}. 
\end{align}
The other two extremal BPS mesons have R-charge 
\begin{align}
\frac{-2\mathsf{p^2}+3\mathsf{p}\mathsf{q}+\mathsf{p}\sqrt{4\mathsf{p}^2-3\mathsf{q}^2}}{3\mathsf{q}}. 
\end{align}
These extremal BPS mesons correspond to the four primitive generators (\ref{dualG_Ypq}) of the dual cone. 
The above R-charges are obtained from the formula (\ref{Rch_meson})
\begin{align}
R(\mathcal{M}_1)&=\frac23(b,m_1)=\frac{-2\mathsf{p^2}+3\mathsf{p}\mathsf{q}+\mathsf{p}\sqrt{4\mathsf{p}^2-3\mathsf{q}^2}}{3\mathsf{q}}, \\
R(\mathcal{M}_2)&=\frac23(b,m_2)=\frac{2\mathsf{p^2}+3\mathsf{p}\mathsf{q}-\mathsf{p}\sqrt{4\mathsf{p}^2-3\mathsf{q}^2}}{3\mathsf{q}}, \\
R(\mathcal{M}_3)&=\frac23(b,m_3)=\frac{2\mathsf{p^2}+3\mathsf{p}\mathsf{q}-\mathsf{p}\sqrt{4\mathsf{p}^2-3\mathsf{q}^2}}{3\mathsf{q}}, \\
R(\mathcal{M}_4)&=\frac23(b,m_4)=\frac{-2\mathsf{p^2}+3\mathsf{p}\mathsf{q}+\mathsf{p}\sqrt{4\mathsf{p}^2-3\mathsf{q}^2}}{3\mathsf{q}}. 
\end{align}
The effective operator lengths are given by
\begin{align}
L_J(\mathcal{M}_1)&=(m_1,v_J)=\mathsf{p}\delta_{J,3}+(\mathsf{p}-\mathsf{q})\delta_{J,4}, \\
L_J(\mathcal{M}_2)&=(m_2,v_J)=(\mathsf{p}+\mathsf{q})\delta_{J,4}+\mathsf{p}\delta_{J,1}, \\
L_J(\mathcal{M}_3)&=(m_3,v_J)=\mathsf{p}\delta_{J,1}+(\mathsf{p}+\mathsf{q})\delta_{J,2}, \\
L_J(\mathcal{M}_4)&=(m_4,v_J)=(\mathsf{p}-\mathsf{q})\delta_{J,2}+\mathsf{p}\delta_{J,3}. 
\end{align}
The extremal BPS meson $\mathcal{M}_1$ can be built from $\mathsf{p}$ $Y$ fields and $\mathsf{p}-\mathsf{q}$ $U_2$ fields 
and $\mathcal{M}_4$ can be realized from $\mathsf{p}-\mathsf{q}$ $U_1$ fields and $\mathsf{p}$ $Y$ fields. 
The other types of the mesons $\mathcal{M}_2$ and $\mathcal{M}_3$ cannot be constructed 
only from the chiral multiplets $Z$, $U_1$, $Y$, $U_2$, forming the baryons but also from the composite chiral multiplets $V_1$ and $V_2$. 

The supersymmetric zeta function is evaluated as
\begin{align}
\label{zeta_Ypq}
&
\mathfrak{Z}^{AdS_5\times Y^{\mathsf{p},\mathsf{q}}}(s,z;\omega_1,\omega_2)
\nonumber\\
&=2\left(
\frac{\mathsf{p}(3\mathsf{q}+2\mathsf{p}-\sqrt{4\mathsf{p}^2-3\mathsf{q}^2})}{6\mathsf{q}}
(\omega_1+\omega_2)
\right)^{-s}
\zeta\left(
s,1+\frac{6\mathsf{q}z}{\mathsf{p}(3\mathsf{q}+2\mathsf{p}-\sqrt{4\mathsf{p}^2-3\mathsf{q}^2})}
\frac{1}{\omega_1+\omega_2}
\right)
\nonumber\\
&+2\left(
\frac{\mathsf{p}(3\mathsf{q}-2\mathsf{p}+\sqrt{4\mathsf{p}^2-3\mathsf{q}^2})}{6\mathsf{q}}
(\omega_1+\omega_2)
\right)^{-s}
\zeta\left(
s,1+\frac{6\mathsf{q}z}{\mathsf{p}(3\mathsf{q}-2\mathsf{p}+\sqrt{4\mathsf{p}^2-3\mathsf{q}^2})}
\frac{1}{\omega_1+\omega_2}
\right). 
\end{align}
The residue at a simple pole $s=1$ is
\begin{align}
\label{Res_Ypq}
\mathrm{Res}_{s=1}\mathfrak{Z}^{AdS_5\times Y^{\mathsf{p},\mathsf{q}}}(s,z;\omega_1,\omega_2)
&=\frac{18\mathsf{q}^2}{\mathsf{p}(3\mathsf{q}^2-2\mathsf{p}^2+\mathsf{p}\sqrt{4\mathsf{p}^2-3\mathsf{q}^2})}
\frac{1}{\omega_1+\omega_2}. 
\end{align}
According to the formula (\ref{Riem2_conj}), the integral of the Riemann square is given by
\begin{align}
\label{Riem2_Ypq}
\int_{Y^{\mathsf{p},\mathsf{q}}}\mathrm{Riem}^2
&=\frac{8\mathsf{q}^2(106\mathsf{p}-19\sqrt{4\mathsf{p}^2-3\mathsf{q}^2})}
{3\mathsf{p}^2(3\mathsf{q}^2-2\mathsf{p}^2+\mathsf{p}\sqrt{4\mathsf{p}^2-3\mathsf{q}^2})}\pi^3. 
\end{align}
This agrees with the integrated curvature evaluated from the Hilbert series \cite{Eager:2010dk}. 
As discussed in section \ref{sec_Riem2}, this is equivalent to the claim of \cite{ArabiArdehali:2015iow} in the toric case.
The Zeta-index is
\begin{align}
\label{0_Ypq}
\mathfrak{Z}^{AdS_5\times Y^{\mathsf{p},\mathsf{q}}}(0,0;1,1)
&=-2. 
\end{align}
The supersymmetric zeta value with $s=-1$ and $z=0$ is
\begin{align}
\label{-1_Ypq}
\mathfrak{Z}^{AdS_5\times Y^{\mathsf{p},\mathsf{q}}}(-1,0;\omega_1,\omega_2)
&=-\frac{\mathsf{p}}{6}(\omega_1+\omega_2). 
\end{align}
According to the formulae (\ref{c0_formula}) and (\ref{a0_formula}), we get
\begin{align}
c_{\mathcal{O}(N^0)}&=-\frac{\mathsf{p}}{4}, \\
a_{\mathcal{O}(N^0)}&=-\frac{3\mathsf{p}}{8}, 
\end{align}
and 
\begin{align}
\label{c-a_Ypq}
c_{\mathcal{O}(N^0)}-a_{\mathcal{O}(N^0)}&=\frac{\mathsf{p}}{8}. 
\end{align}
The holographic computation of the central charge difference (\ref{c-a_Ypq}) was addressed in \cite{ArabiArdehali:2013vyp}. 

The supersymmetric determinant is
\begin{align}
\label{det_Ypq}
&
\mathfrak{D}^{AdS_5\times Y^{\mathsf{p},\mathsf{q}}}(z;\omega_1,\omega_2)
\nonumber\\
&=\frac{\mathsf{p}^2 (3\mathsf{q}^2-2\mathsf{p}^2+\mathsf{p}\sqrt{4\mathsf{p}^2-3\mathsf{q}^2})}
{\pi^2 \mathsf{q}^2 6^{2+\frac{18\mathsf{q}^2z}{\mathsf{p}(3\mathsf{q}^2-2\mathsf{p}^2+\mathsf{p}\sqrt{4\mathsf{p}^2-3\mathsf{q}^2})}}}
(\omega_1+\omega_2)^2
\nonumber\\
&\times 
\left(
\frac{\mathsf{p}(3\mathsf{q}+2\mathsf{p}-\sqrt{4\mathsf{p}^2-3\mathsf{q}^2})}{\mathsf{q}}
(\omega_1+\omega_2)
\right)^{\frac{3\mathsf{q}+2\mathsf{p}+\sqrt{4\mathsf{p}^2-3\mathsf{q}^2}}{\mathsf{p}(\mathsf{p}+\mathsf{q}) (\omega_1+\omega_2)}z}
\nonumber\\
&\times 
\left(
\frac{\mathsf{p}(3\mathsf{q}-2\mathsf{p}+\sqrt{4\mathsf{p}^2-3\mathsf{q}^2})}{\mathsf{q}}
(\omega_1+\omega_2)
\right)^{\frac{12\mathsf{q}}{\mathsf{p}(3\mathsf{q}-2\mathsf{p}+\sqrt{4\mathsf{p}^2-3\mathsf{q}^2})(\omega_1+\omega_2)} z}
\nonumber\\
&\times 
\Gamma\left(1+\frac{6\mathsf{q}}{\mathsf{p}(3\mathsf{q}-2\mathsf{p}+\sqrt{4\mathsf{p}^2-3\mathsf{q}^2}) (\omega_1+\omega_2)}z\right)^2
\nonumber\\
&\times 
\Gamma\left(1+\frac{3\mathsf{q}+2\mathsf{p}+\sqrt{4\mathsf{p}^2-3\mathsf{q}^2}}{2\mathsf{p}(\mathsf{p}+\mathsf{q}) (\omega_1+\omega_2)}z\right)^2. 
\end{align}
The vacuum exponent is 
\begin{align}
\mathfrak{D}^{AdS_5\times Y^{\mathsf{p},\mathsf{q}}}(0;1,1)
&=\frac{\mathsf{p}^2(3\mathsf{q}^2-2\mathsf{p}^2+\mathsf{p}\sqrt{4\mathsf{p}^2-3\mathsf{q}^2})}
{9\pi^2\mathsf{q}^2}. 
\end{align}

\section{The $L^{a,b,c}$ Sasaki-Einstein manifolds}
\label{sec_Labc}
The $L^{a,b,c}$ manifolds constitute a three-parameter family of toric Sasaki-Einstein 5-manifolds labeled by three positive integers $(a,b,c)$ \cite{Cvetic:2005ft,Cvetic:2005vk}. 
They are topologically $S^2\times S^3$ and generalize the earlier infinite family of toric Sasaki-Einstein manifolds $Y^{\mathsf{p},\mathsf{q}}$ 
in such a way that it coincides with the $Y^{\mathsf{p},\mathsf{q}}$ for $(a,b,c)$ $=$ $(\mathsf{p}-\mathsf{q},\mathsf{p}+\mathsf{q},\mathsf{p})$. 
For generic integer parameters, the resulting space may be smooth or may possess orbifold or more general toric singularities, 
depending on the lattice structure underlying the toric data. 

\subsection{$AdS_5\times L^{1,2,1}$}
\label{sec_L121}
Let us examine the 4d $\mathcal{N}=1$ quiver gauge theory on a stack of $N$ D3-branes probing the Suspended Pinch Point (SPP) \cite{Morrison:1998cs,Franco:2005sm,Forcella:2008bb}. 
It contains three $SU(N)$ gauge nodes and bifundamental chiral multiplets connecting the adjacent nodes as well as an adjoint chiral multiplet at one of the nodes. 
The field content is
\begin{align}
\begin{array}{c|ccccc}
&\prod_{i=1}^3 SU(N)_i&U(1)_{F_1}&U(1)_{F_2}&U(1)_B&U(1)_R\\ \hline 
X_{31}&(\overline{\bf{N}},\bf{1},\bf{N})&+1&0&+1&\frac{1}{\sqrt{3}} \\
X_{13}&(\bf{N},\bf{1},\overline{\bf{N}})&+1&0&-1&\frac{1}{\sqrt{3}} \\
X_{12}&(\bf{N},\overline{\bf{N}},\bf{1})&+1&-1&1&1-\frac{1}{\sqrt{3}} \\
X_{21}&(\overline{\bf{N}},\bf{N},\bf{1})&-1&-1&-1&1-\frac{1}{\sqrt{3}} \\
X_{23}&(\bf{1},\bf{N},\overline{\bf{N}})&0&+1&0&\frac{1}{\sqrt{3}} \\
X_{32}&(\bf{1},\overline{\bf{N}},\bf{N})&0&-1&0&\frac{1}{\sqrt{3}} \\
X_{33}&(\bf{1},\bf{1},\bf{N^2})&0&0&0&2(1-\frac{1}{\sqrt{3}}) \\
\end{array}
\end{align}
Here we note that the loop at the third node in the quiver is identified with $X_{33}$ 
as an adjoint chiral multiplet together with a neutral chiral multiplet. 
Accordingly, the theory contains both the traceless part and the trace part. 
The quiver diagram is 
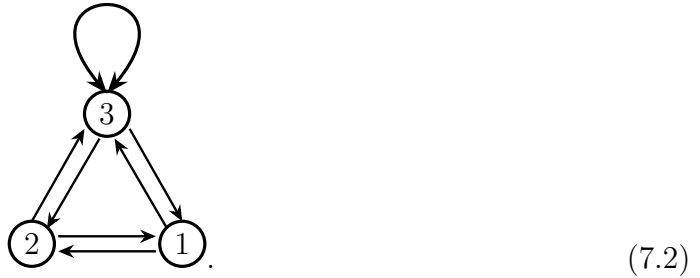
\begin{align}
\begin{tikzpicture}[<->]
    \draw[very thick] (0,1.73) circle (3mm);
    \node at (0,1.73) {3};
    \draw[very thick] (1,0) circle (3mm);
    \node at (1,0) {1};
    \draw[very thick] (-1,0) circle (3mm);
    \node at (-1,0) {2};
\draw[<->,>=Stealth,scale=4,very thick] (0,0.5)  to[in=-230,out=-310,loop] (0,0);
    \draw[-{Stealth[length=6pt,width=5pt]}, line width=0.9pt]
          (0.3,1.53) to node[]{} (1,0.3);
     \draw[-{Stealth[length=6pt,width=5pt]}, line width=0.9pt]
          (0.8,0.2) to node[]{} (0.1,1.4);
    \draw[-{Stealth[length=6pt,width=5pt]}, line width=0.9pt]
          (-0.65,0.1) to node[]{} (0.65,0.1);
    \draw[-{Stealth[length=6pt,width=5pt]}, line width=0.9pt]
          (0.65,-0.1) to node[]{} (-0.65,-0.1);
    \draw[-{Stealth[length=6pt,width=5pt]}, line width=0.9pt]
          (-1,0.3) to node[]{} (-0.3,1.53);
     \draw[-{Stealth[length=6pt,width=5pt]}, line width=0.9pt]
          (-0.1,1.4) to node[]{} (-0.8,0.2) ;
\end{tikzpicture}. 
\end{align}
The superpotential is \cite{Forcella:2008bb}
\begin{align}
\mathcal{W}&=\Tr (X_{33}(X_{31}X_{13}-X_{32}X_{23})+X_{23}X_{32}X_{21}X_{12}-X_{13}X_{31}X_{12}X_{21}). 
\end{align}
The theory is holographically dual to Type IIB string theory on $AdS_5\times L^{1,2,1}$, where $L^{1,2,1}$ is the Sasaki-Einstein base of the SPP. 
The toric diagram of the SPP is generated by five primitive vectors 
\begin{align}
v_1&=(1,0,0), \qquad 
v_2=(1,1,0), \qquad 
v_3=(1,1,1), \nonumber\\
v_4&=(1,1,2), \qquad 
v_5=(1,0,1). 
\end{align}
It is shown as
\begin{align}
\label{toric_SPP}
\begin{tikzpicture}[scale=1.25, baseline={(0,0)}]
  \fill (0,0) circle (2pt) node[below left] {$(0,0)$};
  \fill (1,0) circle (2pt) node[below right] {$(1,0)$};
  \fill (1,1) circle (2pt) node[right] {$(1,1)$};
  \fill (1,2) circle (2pt) node[above] {$(1,2)$};
  \fill (0,1) circle (2pt) node[left] {$(0,1)$};
  \draw[very thick] (0,0) -- (1,0) -- (1,1) -- (1,2) -- (0,1) -- cycle;
\end{tikzpicture}
\end{align}
We have the multiplicities of four types of the chiral multiplets 
$U_2$ $=$ $\{X_{12}\}$, 
$Y$ $=$ $\{X_{32},X_{13}\}$, 
$U_1$ $=$ $\{X_{23},X_{31}\}$ and 
$Z$ $=$ $\{X_{21}\}$ forming the baryonic operators
\begin{align}
\det (v_{5},v_{1},v_{2})&=1, \\
\det (v_{1},v_{2},v_{4})&=2, \\
\det (v_{2},v_{4},v_{5})&=2, \\
\det (v_{4},v_{5},v_{1})&=1.
\end{align}
The area of the polygon is
\begin{align}
\label{A_SPP}
A&=\frac32. 
\end{align}
The primitive generators of the dual cone are
\begin{align}
\label{dualG_SPP}
m_1&=(0,0,1), &
m_2&=(1,-1,0), & 
m_3&=(1,-1,0), \nonumber\\
m_4&=(1,1,-1), & 
m_5&=(0,1,0). 
\end{align}
The Reeb vector is given by
\begin{align}
b&=\left(3,\sqrt{3},\frac{3+\sqrt{3}}{2}\right). 
\end{align}
The volume of $L^{1,2,1}$ is evaluated from (\ref{vol_Y1}) and (\ref{vol_Y2}) 
\begin{align}
\label{vol_L121}
\mathrm{Vol}(L^{1,2,1})
&=\frac{2\sqrt{3}}{9}\pi^3. 
\end{align}
The R-charges of the chiral multiplet fields are 
\begin{align}
R(U_2)&=\frac{\pi}{3}\frac{\mathrm{Vol}(\Sigma_1)}{\mathrm{Vol}(L^{1,2,1})}=1-\frac{1}{\sqrt{3}}, \\
R(Y)&=\frac{\pi}{3}\frac{\mathrm{Vol}(\Sigma_2)}{\mathrm{Vol}(L^{1,2,1})}=\frac{1}{\sqrt{3}}, \\
R(U_{1})&=\frac{\pi}{3}\frac{\mathrm{Vol}(\Sigma_4)}{\mathrm{Vol}(L^{1,2,1})}=\frac{1}{\sqrt{3}}, \\
R(Z)&=\frac{\pi}{3}\frac{\mathrm{Vol}(\Sigma_5)}{\mathrm{Vol}(L^{1,2,1})}=1-\frac{1}{\sqrt{3}},
\end{align}
where 
\begin{align}
\mathrm{Vol}(\Sigma_1)&=2\pi^2 \frac{\det(v_{5},v_{1},v_{2})}{\det(b,v_{5},v_1)\det(b,v_{1},v_{2})}
=\frac{2(\sqrt{3}-1)\pi^2}{3}, \\
\mathrm{Vol}(\Sigma_2)&=2\pi^2 \frac{\det(v_{1},v_{2},v_{4})}{\det(b,v_{1},v_2)\det(b,v_{2},v_{4})}
=\frac{2\pi^2}{3}, \\
\mathrm{Vol}(\Sigma_4)&=2\pi^2 \frac{\det(v_{2},v_{4},v_{5})}{\det(b,v_{2},v_4)\det(b,v_{4},v_{5})}
=\frac{2\pi^2}{3}, \\
\mathrm{Vol}(\Sigma_5)&=2\pi^2 \frac{\det(v_{4},v_{5},v_{1})}{\det(b,v_{4},v_5)\det(b,v_{5},v_{1})}
=\frac{2(\sqrt{3}-1)\pi^2}{3}.
\end{align}

We have the weighted adjacency matrix
\begin{align}
\label{M_L121}
&
M^{AdS_5\times L^{1,2,1}}
\nonumber\\
&=\left(
\begin{matrix}
i_v
&i_{\mathrm{cm}_{1-\frac{1}{\sqrt{3}}}}+\tilde{i}_{\mathrm{cm}_{1-\frac{1}{\sqrt{3}}}}
&i_{\mathrm{cm}_{\frac{1}{\sqrt{3}}}}+\tilde{i}_{\mathrm{cm}_{\frac{1}{\sqrt{3}}}}\\
i_{\mathrm{cm}_{1-\frac{1}{\sqrt{3}}}}+\tilde{i}_{\mathrm{cm}_{1-\frac{1}{\sqrt{3}}}}
&i_v&i_{\mathrm{cm}_{\frac{1}{\sqrt{3}}}}+\tilde{i}_{\mathrm{cm}_{\frac{1}{\sqrt{3}}}} \\
i_{\mathrm{cm}_{\frac{1}{\sqrt{3}}}}+\tilde{i}_{\mathrm{cm}_{\frac{1}{\sqrt{3}}}}
&i_{\mathrm{cm}_{\frac{1}{\sqrt{3}}}}+\tilde{i}_{\mathrm{cm}_{\frac{1}{\sqrt{3}}}}
&i_v+i_{\mathrm{cm}_{2(1-\frac{1}{\sqrt{3}})}}+\tilde{i}_{\mathrm{cm}_{2(1-\frac{1}{\sqrt{3}})}}\\
\end{matrix}
\right). 
\end{align}
The gravity index is evaluated as
\begin{align}
\label{ind_L121}
i^{AdS_5\times L^{1,2,1}}(p;q)
&=\frac{(pq)^{\frac{1}{\sqrt{3}}}}{1-(pq)^{\frac{1}{\sqrt{3}}}}
+2\frac{(pq)^{\frac{3-\sqrt{3}}{3}}}{1-(pq)^{\frac{3-\sqrt{3}}{3}}}
+2\frac{(pq)^{\frac{3+\sqrt{3}}{6}}}{1-(pq)^{\frac{3+\sqrt{3}}{6}}}. 
\end{align}
The first term corresponds to the single-trace operator of a length-$2$ loop $\mathcal{M}_5$ $=$ $\Tr(X_{31}X_{13})$ with R-charge $\frac{2}{\sqrt{3}}$. 
The second corresponds to the other length-$2$ loop $\mathcal{M}_2$ $=$ $\Tr(X_{12}X_{21})$ 
and the length-$1$ loop $\mathcal{M}_3$ $=$ $\Tr(X_{33})$ with R-charge $\frac{2(3-\sqrt{3})}{3}$. 
The third corresponds to the two triangle mesons $\mathcal{M}_1$ $=$ $\Tr(X_{13}X_{32}X_{21})$ 
and $\mathcal{M}_4$ $=$ $\Tr(X_{31}X_{12}X_{23})$ with R-charge $\frac{3+\sqrt{3}}{3}$. 
They correspond to the five primitive generators (\ref{dualG_SPP}) of the dual cone. 
In fact, the R-charges are evaluated from (\ref{Rch_meson}) as
\begin{align}
R(\mathcal{M}_1)&=\frac23(b,m_1)=\frac{3+\sqrt{3}}{3}, \\
R(\mathcal{M}_2)&=\frac23(b,m_2)=\frac{2(3-\sqrt{3})}{3}, \\
R(\mathcal{M}_3)&=\frac23(b,m_3)=\frac{2(3-\sqrt{3})}{3}, \\
R(\mathcal{M}_4)&=\frac23(b,m_4)=\frac{3+\sqrt{3}}{3}, \\
R(\mathcal{M}_5)&=\frac23(b,m_5)=\frac{2}{\sqrt{3}}. 
\end{align}

The supersymmetric zeta function is given by
\begin{align}
\label{zeta_L121}
&
\mathfrak{Z}^{AdS_5\times L^{1,2,1}}(s,z;\omega_1,\omega_2)
\nonumber\\
&=\left(\frac{\omega_1+\omega_2}{\sqrt{3}}\right)^{-s}
\zeta\left(s,1+\frac{\sqrt{3}z}{\omega_1+\omega_2}\right)
+2\left(\frac{3-\sqrt{3}}{3}(\omega_1+\omega_2)\right)^{-s}
\zeta\left(s,1+\frac{\frac{3}{3-\sqrt{3}}z}{\omega_1+\omega_2}\right)
\nonumber\\
&+2\left(\frac{3+\sqrt{3}}{6}(\omega_1+\omega_2)\right)^{-s}
\zeta\left(s,1+\frac{\frac{6}{3+\sqrt{3}}z}{\omega_1+\omega_2}\right). 
\end{align}
The residue at a simple pole $s=1$ is
\begin{align}
\label{Res_L121}
\mathrm{Res}_{s=1}\mathfrak{Z}^{AdS_5\times L^{1,2,1}}(s,z;\omega_1,\omega_2)
&=\frac{9}{\omega_1+\omega_2}. 
\end{align}
By inserting into the formula (\ref{Riem2_conj}) 
the residue coefficient in (\ref{Res_L121}) and the volume (\ref{vol_L121}), 
we obtain the following expression for the integrated Riemann square
\begin{align}
\label{Riem2_L121}
\int_{L^{1,2,1}}\mathrm{Riem}^2
&=\frac{16(108-19\sqrt{3})}{9}\pi^3=4139.19\ldots. 
\end{align}
The Zeta-index is
\begin{align}
\mathfrak{Z}^{AdS_5\times L^{1,2,1}}(0,0;\omega_1,\omega_2)&=-\frac52. 
\end{align}
The special supersymmetric zeta value for $s=-1$ and $z=0$ is
\begin{align}
\mathfrak{Z}^{AdS_5\times L^{1,2,1}}(-1,0;\omega_1,\omega_2)
&=-\frac14 (\omega_1+\omega_2). 
\end{align}
From the formulae (\ref{c0_formula}) and (\ref{a0_formula}) we find
\begin{align}
\label{c0_L121}
c_{\mathcal{O}(N^0)}&=-\frac38, \\
\label{a0_L121}
a_{\mathcal{O}(N^0)}&=-\frac{9}{16}, 
\end{align}
and 
\begin{align}
c_{\mathcal{O}(N^0)}-a_{\mathcal{O}(N^0)}&=\frac{3}{16}. 
\end{align}

The supersymmetric determinant is
\begin{align}
&
\mathfrak{D}^{AdS_5\times L^{1,2,1}}(z;\omega_1,\omega_2)
\nonumber\\
&=\frac{(3+\sqrt{3})^{\frac{12z}{(3+\sqrt{3})(\omega_1+\omega_2)}}
\left(1-\frac{1}{\sqrt{3}}\right)^{\frac{6z}{(3-\sqrt{3})(\omega_1+\omega_2)}} 
(\omega_1+\omega_2)^{\frac52+\frac{9z}{\omega_1+\omega_2}}}
{2^{\frac52+\frac{12z}{(3+\sqrt{3})(\omega_1+\omega_2)}} 3^{\frac54+\frac{3(9+\sqrt{3})z}{2(3+\sqrt{3})(\omega_1+\omega_2)}} \pi^{\frac52} }
\nonumber\\
&\times 
\Gamma\left(1+\frac{\sqrt{3}z}{\omega_1+\omega_2}\right) 
\Gamma\left(1+\frac{3z}{(3-\sqrt{3})(\omega_1+\omega_2)}\right)^{2}
\Gamma\left(1+\frac{6z}{(3+\sqrt{3})(\omega_1+\omega_2)}\right)^2. 
\end{align}
The vacuum exponent is computed as
\begin{align}
\mathfrak{D}^{AdS_5\times L^{1,2,1}}(0;1,1)
&=\frac{1}{3^{\frac54}\pi^{\frac52}}=0.0144785\ldots.
\end{align}

\subsection{$AdS_5\times L^{a,b,a}$}
More generally, let us examine the gauge theory for a stack of $N$ D3-branes 
probing the toric Calabi-Yau 3-fold, which is a cone over the Sasaki-Einstein manifold $L^{a,b,a}$. 
These theories were first studied from Type IIA brane configurations of relatively rotated NS5-branes and D4-branes in \cite{Uranga:1998vf}. 
It has $(a+b)$ $SU(N)$ gauge nodes and $2(a+b)$ bifundamental chiral multiplets corresponding to the arrows between the adjacent gauge nodes 
and the $(b-a)$ adjoint chirals together with neutral chirals corresponding to the loops. 
The field content and charges are given by \cite{Franco:2005sm}
\begin{align}
\begin{array}{c|cc}
&\textrm{number}&U(1)_R\\ \hline 
X_{2i-1\ 2i}&a&x \\
X_{2i\ 2i-1}&a&x \\
X_{2i\ 2i+1}&a&1-x \\
X_{2i+1\ 2i}&a&1-x \\
X_{2a+j\ 2a+j+1}&b-a&1-x \\
X_{2a+j+1\ 2a+j}&b-a&1-x \\
X_{2a+j\ 2a+j}&b-a&2x \\
\end{array}
\end{align}
with $i=1,\cdots, a$ and $j=1,\cdots, b-a$. 
Here 
\begin{align}
x&=\frac{2b-a-\sqrt{a^2-ab+b^2}}{3(b-a)}. 
\end{align}
The quiver diagram is 
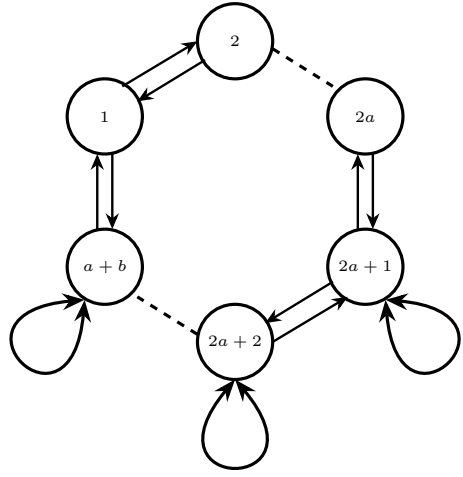
\begin{align}
\begin{tikzpicture}
    \draw[very thick] (0,2) circle (5mm);
    \node at (0,2) {\tiny 2};
    \draw[very thick] (1.732,1) circle (5mm);
    \node at (1.732,1) {\tiny $2a$};
    \draw[very thick] (1.732,-1) circle (5mm);
    \node at (1.732,-1) {\tiny $2a+1$};
    \draw[very thick] (0,-2) circle (5mm);
    \node at (0,-2) {\tiny $2a+2$};
    \draw[very thick] (-1.732,-1) circle (5mm);
    \node at (-1.732,-1) {\tiny $a+b$};
    \draw[very thick] (-1.732,1) circle (5mm);
    \node at (-1.732,1) {\tiny $1$};
\draw[<->,>=Stealth,scale=4,very thick] (0.5,-0.36)  to[in=-5,out=-85,loop] (0.5,-0.36);
\draw[<->,>=Stealth,scale=4,very thick] (-0.5,-0.36)  to[in=-95,out=-175,loop] (-0.5,-0.36);
\draw[<->,>=Stealth,scale=4,very thick] (0,-0.625)  to[in=-50,out=-130,loop] (0,-0.625);
    \draw[-{Stealth[length=6pt,width=5pt]}, line width=0.9pt]
          (-1.5,1.4) to node[]{} (-0.5,2);
    \draw[-{Stealth[length=6pt,width=5pt]}, line width=0.9pt]
          (-0.4,1.75) to node[]{} (-1.3,1.2);
     \draw [dashed,very thick] (0.5,1.9) to node[]{} (1.3,1.4);
    \draw[-{Stealth[length=6pt,width=5pt]}, line width=0.9pt]
          (1.832,0.5) to node[]{} (1.832,-0.5);
    \draw[-{Stealth[length=6pt,width=5pt]}, line width=0.9pt]
          (1.632,-0.5) to node[]{} (1.632,0.5);
    \draw[-{Stealth[length=6pt,width=5pt]}, line width=0.9pt]
          (0.5,-2) to node[]{} (1.5,-1.4);
    \draw[-{Stealth[length=6pt,width=5pt]}, line width=0.9pt]
          (1.3,-1.2) to node[]{} (0.4,-1.75);          
     \draw [dashed,very thick] (-0.5,-1.9) to node[]{} (-1.3,-1.4);
    \draw[-{Stealth[length=6pt,width=5pt]}, line width=0.9pt]
          (-1.832,-0.5) to node[]{} (-1.832,0.5);
    \draw[-{Stealth[length=6pt,width=5pt]}, line width=0.9pt]
          (-1.632,0.5) to node[]{} (-1.632,-0.5);
\end{tikzpicture}. 
\end{align}
The gravity dual is Type IIB string theory on $AdS_5\times L^{a,b,a}$. 
The toric diagram of the cone over $L^{a,b,a}$ is generated by $(a+b+2)$ vectors
\begin{align}
v_1&=(1,0,0),\nonumber\\
v_{I+2}&=(1,1,I), \qquad I=0,1,\cdots, b, \nonumber\\
v_{J+b+2}&=(1,0,a-J+1), \qquad J=1,\cdots, a. 
\end{align}
It is shown as
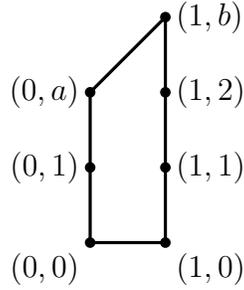
\begin{align}
\label{toric_Laba}
\begin{tikzpicture}[scale=1, baseline={(0,0)}]
  \fill (0,0) circle (2pt) node[below left] {$(0,0)$};
  \fill (1,0) circle (2pt) node[below right] {$(1,0)$};
  \fill (1,1) circle (2pt) node[right] {$(1,1)$};
  \fill (1,2) circle (2pt) node[right] {$(1,2)$};
  \fill (1,3) circle (2pt) node[right] {$(1,b)$};
  \fill (0,2) circle (2pt) node[left] {$(0,a)$};
  \fill (0,1) circle (2pt) node[left] {$(0,1)$};
  \draw[very thick] (0,0) -- (1,0) -- (1,1) -- (1,2) -- (1,3) -- (0,2) -- (0,1) -- cycle;
\end{tikzpicture}
\end{align}
It has $(a-1)$ lattice points on one external edge and $(b-1)$ lattice points on the opposite edge. 
When $a=1$ and $b=2$ it reduces to the toric diagram (\ref{toric_SPP}) for the SPP. 
The multiplicities of four types of the chiral multiplets 
$U_2$ $=$ $\{X_{2i-1\ 2i}\}$, 
$Y$ $=$ $\{X_{2i+1\ 2i}, X_{2a+j+1\ 2a+j}\}$, 
$U_1$ $=$ $\{X_{2i\ 2i+1}, X_{2a+j\ 2a+j+1}\}$ and 
$Z$ $=$ $\{X_{2i\ 2i-1}\}$, which form the baryonic operators, are given by
\begin{align}
\det (v_{b+3},v_{1},v_{2})&=a, \\
\det (v_{1},v_{2},v_{b+2})&=b, \\
\det (v_{2},v_{b+2},v_{b+3})&=b, \\
\det (v_{b+2},v_{b+3},v_{1})&=a, 
\end{align}
where $i=1,\cdots, a$ and $j=1,\cdots, b-a$. 
The area of the polygon is
\begin{align}
\label{A_Laba}
A&=\frac{a+b}{2}. 
\end{align}
The primitive generators of the dual cone are
\begin{align}
\label{dualG_Laba}
m_1&=(0,0,1), \nonumber\\
m_{I+1}&=(1,-1,0), \qquad I=1,\cdots, b, \nonumber\\
m_{b+2}&=(a,b-a,-1), \nonumber\\
m_{J+b+2}&=(0,1,0), \qquad J=1,\cdots, a. 
\end{align}
The Reeb vector is determined as
\begin{align}
b&=\left(3,\frac{b-2a+\sqrt{a^2-ab+b^2}}{b-a},\frac{a+b+\sqrt{a^2-ab+b^2}}{2}\right). 
\end{align}
The volume of $L^{a,b,a}$ is evaluated from (\ref{vol_Y1}) and (\ref{vol_Y2}). 
One finds \cite{Franco:2005sm}
\begin{align}
\label{vol_Laba}
\mathrm{Vol}(L^{a,b,a})
&=\frac{4\left(
(2a-b)(2b-a)(a+b)+2(a^2-ab+b^2)^{\frac32}
\right)}
{27a^2b^2}\pi^3. 
\end{align}
The R-charges of the chiral multiplet fields are 
\begin{align}
R(U_2)&=\frac{\pi}{3}\frac{\mathrm{Vol}(\Sigma_1)}{\mathrm{Vol}(L^{a,b,a})}=x, \\
R(Y)&=\frac{\pi}{3}\frac{\mathrm{Vol}(\Sigma_2)}{\mathrm{Vol}(L^{a,b,a})}=1-x, \\
R(U_1)&=\frac{\pi}{3}\frac{\mathrm{Vol}(\Sigma_{b+2})}{\mathrm{Vol}(L^{a,b,a})}=1-x, \\
R(Z)&=\frac{\pi}{3}\frac{\mathrm{Vol}(\Sigma_{b+3})}{\mathrm{Vol}(L^{a,b,a})}=x, 
\end{align}
where 
\begin{align}
\mathrm{Vol}(\Sigma_1)&=2\pi^2 \frac{\det(v_{b+3},v_{1},v_{2})}{\det(b,v_{b+3},v_1)\det(b,v_{1},v_{2})}
\nonumber\\
&=\frac{4(a^2-2b^2+2ab+(2b-a)\sqrt{a^2-ab+b^2})}{9a^2b}\pi^2, \\
\mathrm{Vol}(\Sigma_2)&=2\pi^2 \frac{\det(v_{1},v_{2},v_{b+2})}{\det(b,v_{1},v_2)\det(b,v_{2},v_{b+2})}
\nonumber\\
&=\frac{4(-2a^2+b^2+2ab+(2a-b)\sqrt{a^2-ab+b^2})}{9ab^2}\pi^2, \\
\mathrm{Vol}(\Sigma_{b+2})&=2\pi^2 \frac{\det(v_{2},v_{b+2},v_{b+3})}{\det(b,v_{2},v_{b+2})\det(b,v_{b+2},v_{b+3})}
\nonumber\\
&=\frac{4(-2a^2+b^2+2ab+(2a-b)\sqrt{a^2-ab+b^2})}{9ab^2}\pi^2, \\
\mathrm{Vol}(\Sigma_{b+3})&=2\pi^2 \frac{\det(v_{b+2},v_{b+3},v_{1})}{\det(b,v_{b+2},v_{b+3})\det(b,v_{b+3},v_{1})}
\nonumber\\
&=\frac{4(a^2-2b^2+2ab+(2b-a)\sqrt{a^2-ab+b^2})}{9a^2b}\pi^2. 
\end{align}

The weighted adjacency matrix takes the form
\begin{align}
&
M^{AdS_5\times L^{a,b,a}}
\nonumber\\
&=\left(
\begin{smallmatrix}
i_v&i_{\textrm{cm}_x}+\tilde{i}_{\textrm{cm}_x}&0&\cdots&\cdots&0&i_{\textrm{cm}_{1-x}}+\tilde{i}_{\textrm{cm}_{1-x}} \\
i_{\textrm{cm}_{x}}+\tilde{i}_{\textrm{cm}_{x}}&i_v&i_{\textrm{cm}_{1-x}}+\tilde{i}_{\textrm{cm}_{1-x}}&0&\cdots&\cdots&0 \\
\vdots&&\ddots&&&&\vdots \\
\cdots&0&i_{\textrm{cm}_{x}}+\tilde{i}_{\textrm{cm}_{x}}&i_v&i_{\textrm{cm}_{1-x}}+\tilde{i}_{\textrm{cm}_{1-x}}&0&\cdots \\
\cdots&\cdots&0&i_{\textrm{cm}_{1-x}}+\tilde{i}_{\textrm{cm}_{1-x}}&i_v+i_{2x}+\tilde{i}_{2x}&i_{\textrm{cm}_{1-x}}+\tilde{i}_{\textrm{cm}_{1-x}}&0 \\
\vdots&&&&\ddots&&\vdots \\
0&\cdots&\cdots&0&i_{\textrm{cm}_{1-x}}+\tilde{i}_{\textrm{cm}_{1-x}}&i_v+i_{2x}+\tilde{i}_{2x}&i_{\textrm{cm}_{1-x}}+\tilde{i}_{\textrm{cm}_{1-x}} \\
i_{\textrm{cm}_{1-x}}+\tilde{i}_{\textrm{cm}_{1-x}}&0&\cdots&\cdots&0&i_{\textrm{cm}_{1-x}}+\tilde{i}_{\textrm{cm}_{1-x}}&i_v+i_{2x}+\tilde{i}_{2x} \\
\end{smallmatrix}
\right). 
\end{align}
The single-particle gravity index can be written as
\begin{align}
\label{ind_Laba}
i^{AdS_5\times L^{a,b,a}}(p;q)
&=b\frac{(pq)^x}{1-(pq)^x}+a\frac{(pq)^{1-x}}{1-(pq)^{1-x}}
+2\frac{(pq)^{\frac{ax+b(1-x)}{2}}}{1-(pq)^{\frac{ax+b(1-x)}{2}}}. 
\end{align}
The same form was derived in \cite{Agarwal:2013pba} and is consistent with our combinatorial formula (\ref{ind_gravity_conj}).
The first term enumerates the following $a+(b-a)$ $=$ $b$ extremal BPS mesons of R-charge $R=2x$: 
\begin{align}
\mathcal{M}_i^{(I)}&=\Tr (X_{2i\ 2i+1}X_{2i+1\ 2i}),& i&=1,\cdots, a\\
\mathcal{M}_j^{(II)}&=\Tr (X_{2a+j\ 2a+j}),& j&=1,\cdots, b-a. 
\end{align}
The second term counts extremal BPS mesons of R-charge $R=2(1-x)$
\begin{align}
\mathcal{M}_i^{(III)}&=\Tr (X_{2i-1\ 2i}X_{2i\ 2i-1}),& i&=1,\cdots, a. 
\end{align}
The third term is contributed from two extremal BPS mesons as the single-trace operators of length $(a+b)$
\begin{align}
\mathcal{M}^{(IV)}&=\Tr (X_{12}X_{23}X_{34}\cdots X_{a+b\ 1}), \\
\mathcal{M}^{(V)}&=\Tr (X_{1\ a+b}X_{a+b\ a+b-1}X_{a+b-1\ a+b-2}\cdots X_{21}). 
\end{align}
They carry R-charge $R=ax+b(1-x)$. 
These extremal BPS mesons correspond to the primitive generators (\ref{dualG_Laba}). 
The R-charges can be evaluated as
\begin{align}
\frac23(b,m_1)&=ax+b(1-x)=\frac{a+b+\sqrt{a^2-ab+b^2}}{3}, \\
\frac23(b,m_{I+1})&=2x=\frac{2(2b-a-\sqrt{a^2-ab+b^2})}{3(b-a)}, \qquad I=1,\cdots, b\\
\frac23(b,m_{b+2})&=ax+b(1-x)=\frac{a+b+\sqrt{a^2-ab+b^2}}{3}, \\
\frac23(b,m_{J+b+2})&=2(1-x)=\frac{2(b-2a+\sqrt{a^2-ab+b^2})}{3(b-a)}, \qquad J=1,\cdots, a. 
\end{align}

We obtain the supersymmetric zeta function 
\begin{align}
\label{zeta_Laba}
&
\mathfrak{Z}^{AdS_5\times L^{a,b,a}}(s,z;\omega_1,\omega_2)
\nonumber\\
&=b\left(x(\omega_1+\omega_2)\right)^{-s}\zeta\left(s,1+\frac{z}{x(\omega_1+\omega_2)}\right)
\nonumber\\
&+a\left((1-x)(\omega_1+\omega_2)\right)^{-s}\zeta\left(s,1+\frac{z}{(1-x)(\omega_1+\omega_2)}\right)
\nonumber\\
&+2\left(\frac{ax+b(1-x)}{2}(\omega_1+\omega_2)\right)^{-s}\zeta\left(s,1+\frac{2z}{(ax+b(1-x))(\omega_1+\omega_2)}\right). 
\end{align}
The residue at a simple pole $s=1$ is
\begin{align}
\label{Res_Laba}
&
\mathrm{Res}_{s=1}\mathfrak{Z}^{AdS_5\times L^{a,b,a}}(s,z;\omega_1,\omega_2)
\nonumber\\
&=\frac{(ab+4)(a+b)+(2ab-4)\sqrt{a^2-ab+b^2}}{ab}\frac{1}{\omega_1+\omega_2}. 
\end{align}
Using the residue coefficient (\ref{Res_Laba}) and the volume in (\ref{vol_Laba}), 
we obtain the integrated Riemann square 
\begin{align}
\label{Riem2_Laba}
\int_{L^{a,b,a}}\mathrm{Riem}^2
&=\frac{32}{27a^2b^2}
\Biggl[
(a+b)(38a^2+38b^2+18a^2b^2-23ab)
\nonumber\\
&-2\sqrt{a^2-ab+b^2}(19a^2+19b^2-18a^2b^2+17ab)
\Biggr]\pi^3.
\end{align}
The Zeta-index is evaluated as
\begin{align}
\mathfrak{Z}^{AdS_5\times L^{a,b,a}}(0,0;\omega_1,\omega_2)
&=-\frac{a+b+2}{2}. 
\end{align}
The supersymmetric zeta value for $s=-1$ and $z=0$ is 
\begin{align}
\label{-1_Laba}
\mathfrak{Z}^{AdS_5\times L^{a,b,a}}(-1,0;\omega_1,\omega_2)
&=-\frac{a+b}{12}(\omega_1+\omega_2). 
\end{align}
From the formulae (\ref{c0_formula}) and (\ref{a0_formula}) 
we obtain
\begin{align}
c_{\mathcal{O}(N^0)}&=-\frac{a+b}{8}, \\
a_{\mathcal{O}(N^0)}&=-\frac{3(a+b)}{16}
\end{align}
and 
\begin{align}
c_{\mathcal{O}(N^0)}-a_{\mathcal{O}(N^0)}
&=\frac{a+b}{16}. 
\end{align}

The supersymmetric determinant is evaluated as
\begin{align}
\label{det_Laba}
&
\mathfrak{D}^{AdS_5\times L^{a,b,a}}(z;\omega_1,\omega_2)
\nonumber\\
&=
\frac{
(1-x)^{a\left(\frac12+\frac{z}{(1-x)(\omega_1+\omega_2)}\right)}
x^{b\left(\frac12+\frac{z}{x(\omega_1+\omega_2)}\right)}
(ax+b(1-x))^{1+\frac{4z}{(ax+b(1-x))(\omega_1+\omega_2)}}
}
{
2^{2+\frac{a+b}{2}+\frac{4z}{(ax+b(1-x))(\omega_1+\omega_2)}}
\pi^{\frac{a+b+2}{2}}
}
\nonumber\\
&\times 
(\omega_1+\omega_2)^{\frac12 \left[
2+a+b+\frac{2az}{(1-x)(\omega_1+\omega_2)}
+\frac{2\left(\frac{b}{x}+\frac{4}{ax+b(1-x)}\right)z}{\omega_1+\omega_2}
\right]}
\nonumber\\
&\times \Gamma\left(1+\frac{z}{(1-x)(\omega_1+\omega_2)}\right)^a
\Gamma\left(1+\frac{z}{x(\omega_1+\omega_2)}\right)^b
\Gamma\left(1+\frac{2z}{(ax+b(1-x))(\omega_1+\omega_2)}\right)^2. 
\end{align}
The vacuum exponent is
\begin{align}
\label{vac_Laba}
&
\mathfrak{D}^{AdS_5\times L^{a,b,a}}(0;1,1)
\nonumber\\
&=\frac{1}{2(3\pi)^{\frac{a}{2}+\frac{b}{2}+1}}
\frac{1}
{(b-a)^{\frac{a}{2}+\frac{b}{2}}}(a+b+\sqrt{a^2-ab+b^2})
\nonumber\\
&\times (2a-b-\sqrt{a^2-ab+b^2})^{\frac{a}{2}} 
(2b-a-\sqrt{a^2-ab+b^2})^{\frac{b}{2}}. 
\end{align}

\subsection{$AdS_5\times L^{a,b,c}$}
Consider the gauge theories for a stack of $N$ D3-branes 
probing the toric Calabi-Yau 3-fold, which is a cone over the non-singular Sasaki-Einstein manifold $L^{a,b,c}$ with $a\neq c$, 
for which each of the pair $a$ and $b$ must be coprime to each of $c$ and $a+b-c$. 

For $a=1$, $b=5$ and $c=2$, 
there are 6 gauge nodes and 16 chiral multiplets. 
The field content is summarized as
\begin{align}
\begin{array}{c|cc}
&\prod_{i=1}^6 SU(N)_i&U(1)_R\\ \hline 
X_{13}&(\bf{N},\bf{1},\overline{\bf{N}},\bf{1},\bf{1},\bf{1})&2-x_1-x_2\\
X_{15}&(\bf{N},\bf{1},\bf{1},\bf{1},\overline{\bf{N}},\bf{1})&x_1-x_3\\
X_{21}&(\overline{\bf{N}},\bf{N},\bf{1},\bf{1},\bf{1},\bf{1})&x_3\\
X_{23}&(\bf{1},\bf{N},\overline{\bf{N}},\bf{1},\bf{1},\bf{1})&x_2\\
X_{34}&(\bf{1},\bf{1},\bf{N},\overline{\bf{N}},\bf{1},\bf{1})&x_1\\
X_{36}^1&(\bf{1},\bf{1},\bf{N},\bf{1},\bf{1},\overline{\bf{N}})&2-x_1-x_2+x_3\\
X_{36}^2&(\bf{1},\bf{1},\bf{N},\bf{1},\bf{1},\overline{\bf{N}})&x_2\\
X_{41}&(\overline{\bf{N}},\bf{1},\bf{1},\bf{N},\bf{1},\bf{1})&x_2\\
X_{42}&(\bf{1},\overline{\bf{N}},\bf{1},\bf{N},\bf{1},\bf{1})&2-x_1-x_2\\
X_{46}&(\bf{1},\bf{1},\bf{1},\bf{N},\bf{1},\overline{\bf{N}})&x_1-x_3\\
X_{53}&(\bf{1},\bf{1},\overline{\bf{N}},\bf{1},\bf{N},\bf{1})&x_1-x_3\\
X_{54}^1&(\bf{1},\bf{1},\bf{1},\overline{\bf{N}},\bf{N},\bf{1})&2-x_1-x_2+x_3\\
X_{54}^2&(\bf{1},\bf{1},\bf{1},\overline{\bf{N}},\bf{N},\bf{1})&x_2\\
X_{62}&(\bf{1},\overline{\bf{N}},\bf{1},\bf{1},\bf{1},\bf{N})&x_1-x_3\\
X_{65}^1&(\bf{1},\bf{1},\bf{1},\bf{1},\overline{\bf{N}},\bf{N})&2-x_1-x_2+x_3\\
X_{65}^2&(\bf{1},\bf{1},\bf{1},\bf{1},\overline{\bf{N}},\bf{N})&x_2\\
\end{array}
\end{align}
The quiver diagram is depicted as
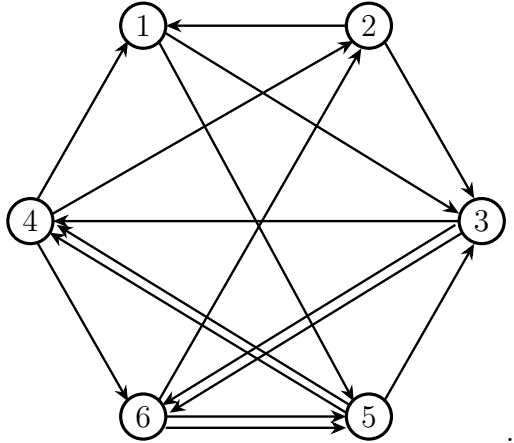
\begin{align}
\label{quiver_L152}
\begin{tikzpicture}[<->]
    \draw[very thick] (-1.5,2.6) circle (3mm);
    \node at (-1.5,2.6) {1};    
    \draw[very thick] (3,0) circle (3mm);
    \node at (3,0) {3};
    \draw[very thick] (1.5,2.6) circle (3mm);
    \node at (1.5,2.6) {2};
    \draw[very thick] (1.5,-2.6) circle (3mm);
    \node at (1.5,-2.6) {5};
    \draw[very thick] (-1.5,-2.6) circle (3mm);
    \node at (-1.5,-2.6) {6};
    \draw[very thick] (-3,0) circle (3mm);
    \node at (-3,0) {4};
    \draw[-{Stealth[length=6pt,width=5pt]}, line width=0.9pt]
          (1.2,2.6) to node[]{} (-1.2,2.6);
    \draw[-{Stealth[length=6pt,width=5pt]}, line width=0.9pt]
          (-1.2,2.5) to node[]{} (2.7,0.1);
    \draw[-{Stealth[length=6pt,width=5pt]}, line width=0.9pt]
          (-2.9,0.3) to node[]{} (-1.7,2.4);
    \draw[-{Stealth[length=6pt,width=5pt]}, line width=0.9pt]
          (-1.3,2.4) to node[]{} (1.3,-2.4);
    \draw[-{Stealth[length=6pt,width=5pt]}, line width=0.9pt]
          (1.7,2.4) to node[]{} (2.9,0.3);
    \draw[-{Stealth[length=6pt,width=5pt]}, line width=0.9pt]
          (-2.7,0.1) to node[]{} (1.3,2.4);
    \draw[-{Stealth[length=6pt,width=5pt]}, line width=0.9pt]
          (-1.3,-2.4) to node[]{} (1.4,2.3);
    \draw[-{Stealth[length=6pt,width=5pt]}, line width=0.9pt]
          (2.7,0) to node[]{} (-2.7,0);
    \draw[-{Stealth[length=6pt,width=5pt]}, line width=0.9pt]
          (1.7,-2.4) to node[]{} (2.9,-0.3);
    \draw[-{Stealth[length=6pt,width=5pt]}, line width=0.9pt]
          (2.65,-0.05) to node[]{} (-1.25,-2.45);
    \draw[-{Stealth[length=6pt,width=5pt]}, line width=0.9pt]
          (2.75,-0.15) to node[]{} (-1.15,-2.55);
    \draw[-{Stealth[length=6pt,width=5pt]}, line width=0.9pt]
          (1.25,-2.45) to node[]{} (-2.65,-0.05);
    \draw[-{Stealth[length=6pt,width=5pt]}, line width=0.9pt]
          (-2.9,-0.3) to node[]{} (-1.7,-2.4);
    \draw[-{Stealth[length=6pt,width=5pt]}, line width=0.9pt]
          (1.2,-2.55) to node[]{} (-2.75,-0.15);
    \draw[-{Stealth[length=6pt,width=5pt]}, line width=0.9pt]
          (-1.2,-2.6) to node[]{} (1.2,-2.6);
    \draw[-{Stealth[length=6pt,width=5pt]}, line width=0.9pt]
          (-1.2,-2.75) to node[]{} (1.2,-2.75);
\end{tikzpicture}. 
\end{align}
The theory is holographically dual to Type IIB string theory on $AdS_5\times L^{1,5,2}$. 
The toric diagram for the toric Calabi-Yau cone is generated by
\begin{align}
v_1&=(1,0,0),\qquad 
v_2=(1,1,0), \qquad 
v_3=(1,3,5), \qquad 
v_4=(1,1,2). 
\end{align}
It is illustrated as
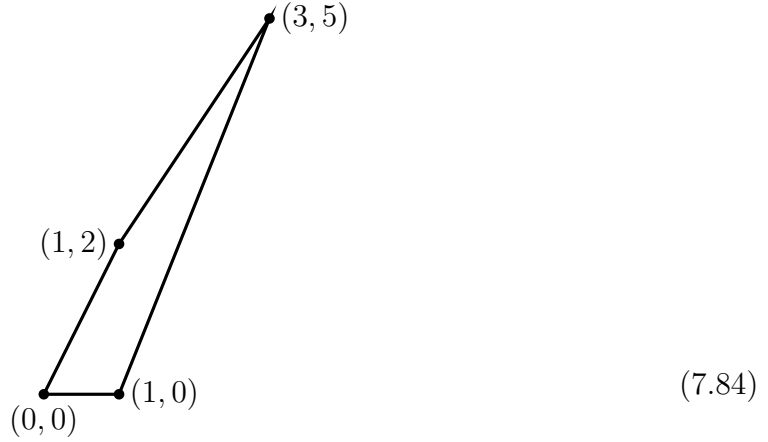
\begin{align}
\label{toric_L152}
\begin{tikzpicture}[scale=1, baseline={(0,0)}]
  \fill (0,0) circle (2pt) node[below] {$(0,0)$};
  \fill (1,0) circle (2pt) node[right] {$(1,0)$};
  \fill (3,5) circle (2pt) node[right] {$(3,5)$};
  \fill (1,2) circle (2pt) node[left] {$(1,2)$};
  \draw[very thick] (0,0) -- (1,0) -- (3,5) -- (1,2) -- cycle;
\end{tikzpicture}
\end{align}
The multiplicities of four types of the chiral multiplets 
$U_2$ $=$ $\{X_{13},X_{42}\}$, 
$Y$ $=$ $\{X_{23},$ $X_{36}^2,$ $X_{41},$ $X_{54}^2,$ $X_{65}^2\}$, 
$U_1$ $=$ $\{X_{15},$ $X_{46},$ $X_{53},$ $X_{62}\}$ 
and $Z$ $=$ $\{X_{21}\}$ forming the baryonic operators are
\begin{align}
\det (v_{4},v_{1},v_{2})&=2, \\
\det (v_{1},v_{2},v_{3})&=5, \\
\det (v_{2},v_{3},v_{4})&=4, \\
\det (v_{3},v_{4},v_{1})&=1.
\end{align}
The area of the polygon is
\begin{align}
\label{A_L152}
A&=3. 
\end{align}
The dual cone is generated by
\begin{align}
\label{dualG_L152}
m_1&=(0,0,1),\qquad 
m_2=(5,-5,2),\qquad 
m_3=(1,3,-2), \qquad 
m_4=(0,2,-1). 
\end{align}
In terms of the Reeb vector $b=(3,b_1,b_2)$ 
the volume of $L^{1,5,2}$ is given by
\begin{align}
\mathrm{Vol}(L^{1,5,2})
&=\frac{2(15+15b_1-8b_2)}
{b_2(15-5b_1+2b_2)(3+3b_1-2b_2)(2b_1-b_2)}\pi^3. 
\end{align}
According to the volume minimization \cite{Martelli:2005tp}, 
we numerically obtain the Reeb vector
\begin{align}
b_1&=4.04738\ldots, \qquad 
b_2=5.30826\ldots, 
\end{align}
and 
\begin{align}
\mathrm{Vol}(L^{1,5,2})
&=5.7248\ldots. 
\end{align}
The R-charges of the chiral multiplet fields are given by
\begin{align}
R(U_2)&=\frac{\pi}{3}\frac{\mathrm{Vol}(\Sigma_1)}{\mathrm{Vol}(L^{1,5,2})}
=\frac{2(3+3b_1-2b_2)(15-5b_1+2b_2)}{45+45b_1-24b_2}
=2-x_1-x_2, \\
R(Y)&=\frac{\pi}{3}\frac{\mathrm{Vol}(\Sigma_2)}{\mathrm{Vol}(L^{1,5,2})}
=\frac{5(3+3b_1-2b_2)(2b_1-b_2)}{45+45b_1-24b_2}
=x_2, \\
R(U_1)&=\frac{\pi}{3}\frac{\mathrm{Vol}(\Sigma_3)}{\mathrm{Vol}(L^{1,5,2})}
=\frac{4(2b_1-b_2)b_2}{45+45b_1-24b_2}
=x_1-x_3, \\
R(Z)&=\frac{\pi}{3}\frac{\mathrm{Vol}(\Sigma_4)}{\mathrm{Vol}(L^{1,5,2})}
=\frac{(15-5b_1+2b_2)b_2}{45+45b_1-24b_2}
=x_3, 
\end{align}
where 
\begin{align}
\mathrm{Vol}(\Sigma_1)&=2\pi^2 \frac{\det(v_{4},v_{1},v_{2})}{\det(b,v_{4},v_1)\det(b,v_{1},v_{2})}=\frac{4\pi^2}{2b_1b_2-b_2^2}, \\
\mathrm{Vol}(\Sigma_2)&=2\pi^2 \frac{\det(v_{1},v_{2},v_{3})}{\det(b,v_{1},v_2)\det(b,v_{2},v_{3})}=\frac{10\pi^2}{b_2(15-5b_1+2b_2)}, \\
\mathrm{Vol}(\Sigma_3)&=2\pi^2 \frac{\det(v_{2},v_{3},v_{4})}{\det(b,v_{2},v_3)\det(b,v_{3},v_{4})}=\frac{8\pi^2}{(3+3b_1-2b_2)(15-5b_1+2b_2)}, \\
\mathrm{Vol}(\Sigma_4)&=2\pi^2 \frac{\det(v_{3},v_{4},v_{1})}{\det(b,v_{3},v_4)\det(b,v_{4},v_{1})}=\frac{2\pi^2}{(3+3b_1-2b_2)(2b_1-b_2)}.
\end{align}

We have the weighted adjacency matrix 
\begin{align}
\label{M_L152}
&
M^{AdS_5\times L^{1,5,2}}
\nonumber\\
&=\left(
\begin{matrix}
i_v&\tilde{i}_{\textrm{cm}_{z}}&i_{\textrm{cm}_{u_2}}&\tilde{i}_{\textrm{cm}_{y}}&i_{\textrm{cm}_{u_1}}&0\\
i_{\textrm{cm}_{z}}&i_v&i_{\textrm{cm}_{y}}&\tilde{i}_{\textrm{cm}_{u_2}}&0&\tilde{i}_{\textrm{cm}_{u_1}} \\
\tilde{i}_{\textrm{cm}_{u_2}}&\tilde{i}_{\textrm{cm}_{y}}&i_v&i_{\textrm{cm}_{v_2}}&\tilde{i}_{\textrm{cm}_{u_1}}&i_{\textrm{cm}_{v_1}}+i_{\textrm{cm}_{y}} \\
i_{\textrm{cm}_{y}}&i_{\textrm{cm}_{u_2}}&\tilde{i}_{\textrm{cm}_{v_2}}&i_v&\tilde{i}_{\textrm{cm}_{v_1}}+\tilde{i}_{\textrm{cm}_{y}}&i_{\textrm{cm}_{u_1}} \\
\tilde{i}_{\textrm{cm}_{u_1}}&0&i_{\textrm{cm}_{u_1}}&i_{\textrm{cm}_{v_1}}+i_{\textrm{cm}_{y}} &i_v&\tilde{i}_{\textrm{cm}_{v_1}}+\tilde{i}_{\textrm{cm}_{y}}  \\
0&i_{\textrm{cm}_{u_1}}&\tilde{i}_{\textrm{cm}_{v_1}}+\tilde{i}_{\textrm{cm}_{y}}&\tilde{i}_{\textrm{cm}_{u_1}}&i_{\textrm{cm}_{v_1}}+i_{\textrm{cm}_{y}}&i_v \\
\end{matrix}
\right), 
\end{align}
where
\begin{align}
u_1&=x_1-x_3,& u_2&=2-x_1-x_2,& v_1&=2-x_1-x_2+x_3
\nonumber\\
v_2&=x_1,& y&=x_2,& z&=x_3. 
\end{align}
From (\ref{M_L152}) we obtain the single-particle gravity index 
\begin{align}
\label{ind_L152}
&
i^{AdS_5\times L^{1,5,2}}(p;q)
\nonumber\\
&=\frac{(pq)^{\frac{2-x_1+3x_2}{2}}}{1-(pq)^{\frac{2-x_1+3x_2}{2}}}
+\frac{(pq)^{\frac{x_1+2x_2-x_3}{2}}}{1-(pq)^{\frac{x_1+2x_2-x_3}{2}}}
+\frac{(pq)^{\frac{5x_1-3x_3}{2}}}{1-(pq)^{\frac{5x_1-3x_3}{2}}}
+\frac{(pq)^{5-\frac{5(x_1+x_2)}{2}+2x_3}}{1-(pq)^{5-\frac{5(x_1+x_2)}{2}+2x_3}}. 
\end{align}
There are four extremal BPS mesonic operators, which correspond to the four primitive generators (\ref{dualG_L152}). 
The R-charges are computed from the formula (\ref{Rch_meson}) as
\begin{align}
R(\mathcal{M}_1)&=\frac23(b,m_1)=\frac23b_2=5x_1-3x_3=3.53884\ldots, \\
R(\mathcal{M}_2)&=\frac23(b,m_2)=\frac23(15-5b_1+2b_2)=10-5(x_1+x_2)+4x_3=3.5864\ldots, \\
R(\mathcal{M}_3)&=\frac23(b,m_3)=\frac23(3+3b_1-2b_2)=2-x_1+3x_2=3.01708\ldots, \\
R(\mathcal{M}_4)&=\frac23(b,m_4)=\frac23 (2b_1-b_2)=x_1+2x_2-x_3=1.85767\ldots, 
\end{align}
with 
\begin{align}
b_1&=\frac32(3x_1+x_2-2x_3),\\
b_2&=\frac32(5x_1-3x_3). 
\end{align}

The supersymmetric zeta function is
\begin{align}
\label{zeta_L152}
&
\mathfrak{Z}^{AdS_5\times L^{1,5,2}}(s,z;\omega_1,\omega_2)
\nonumber\\
&=
\left(\left(\frac{2-x_1+3x_2}{2}\right)(\omega_1+\omega_2) \right)^{-s}\zeta\left(s,1+\frac{2z}{(2-x_1+3x_2) (\omega_1+\omega_2)}\right)
\nonumber\\
&+\left(\left(\frac{5x_1-3x_3}{2}\right)(\omega_1+\omega_2) \right)^{-s}\zeta\left(s,1+\frac{2z}{(5x_1-3x_3) (\omega_1+\omega_2)}\right)
\nonumber\\
&+\left(\left(\frac{x_1+2x_2-x_3}{2}\right)(\omega_1+\omega_2) \right)^{-s}\zeta\left(s,1+\frac{2z}{(x_1+2x_2-x_3) (\omega_1+\omega_2)}\right)
\nonumber\\
&+\left(\left(\frac{10-5(x_1+x_2)+4x_3}{2}\right)(\omega_1+\omega_2) \right)^{-s}\zeta\left(s,1+\frac{2z}{(10-5(x_1+x_2)+4x_3) (\omega_1+\omega_2)}\right). 
\end{align}
The residue at a simple pole $s=1$ is
\begin{align}
\label{Res_L152}
&
\mathrm{Res}_{s=1}\mathfrak{Z}^{AdS_5\times L^{1,5,2}}(s,z;\omega_1,\omega_2)
\nonumber\\
&=
\Biggl(
\frac{2}{2-x_1+3x_2}+\frac{2}{5x_1-3x_3}
+\frac{2}{x_1+2x_2-x_3}+\frac{2}{10-5(x_1+x_2)+4x_3}
\Biggr)
\frac{1}{\omega_1+\omega_2}
\nonumber\\
&=3\left(
\frac{1}{3+3b_1-2b_2}
+\frac{1}{2b_1-b_2}
+\frac{1}{b_2}
+\frac{1}{15-5b_1+2b_2}
\right)\frac{1}{\omega_1+\omega_2}. 
\end{align}
Substituting the residue coefficient (\ref{Res_L152}) and the volume into the formula (\ref{Riem2_conj}), 
we obtain the following expression for the integral of the squared Riemann tensor
\begin{align}
\int_{L^{1,5,2}} \mathrm{Riem}^2
&=\frac{16\pi^3}{b_2(15-5b_1+2b_2)(3+3b_1-2b_2)(2b_1-b_2)}
\nonumber\\
&\times 
\left(
-120b_1^3+112b_1^2b_2-24b_1b_2^2+240b_1^2-72b_2^2-48b_1b_2+75b_1+152b_2-285
\right)
\nonumber\\
&=1023.17\ldots. 
\end{align}
The Zeta-index is
\begin{align}
\label{0_L152}
\mathfrak{Z}^{AdS_5\times L^{1,5,2}}(0,0;\omega_1,\omega_2)
&=-2. 
\end{align}
The supersymmetric zeta value for $s=-1$ and $z=0$ is
\begin{align}
\label{-1_L152}
\mathfrak{Z}^{AdS_5\times L^{1,5,2}}(-1,0;\omega_1,\omega_2)
&=-\frac12 (\omega_1+\omega_2). 
\end{align}
According to the formulae (\ref{c0_formula}) and (\ref{a0_formula}) we find
\begin{align}
c_{\mathcal{O}(N^0)}&=-\frac34, \\
a_{\mathcal{O}(N^0)}&=-\frac98, 
\end{align}
and 
\begin{align}
c_{\mathcal{O}(N^0)}-a_{\mathcal{O}(N^0)}&=\frac38. 
\end{align}

The supersymmetric determinant takes the form
\begin{align}
\label{det_L152}
\mathfrak{D}^{AdS_5\times L^{1,5,2}}(z;\omega_1,\omega_2)
&=\prod_{I=1}^4 
\frac{\rho_I^{\frac12+\frac{z}{\rho_I}} \Gamma\left(1+\frac{z}{\rho_I}\right)}
{\sqrt{2\pi}}, 
\end{align}
where 
\begin{align}
\rho_I&=\frac13(b,m_I)(\omega_1+\omega_2). 
\end{align}
The vacuum exponent is evaluated as
\begin{align}
\label{vac_L152}
&
\mathfrak{D}^{AdS_5\times L^{1,5,2}}(0;1,1)
\nonumber\\
&=\frac{\sqrt{(2-x_1+3x_2)(5x_1-3x_3)(x_1+2x_2-x_3)(10-5(x_1+x_2)+4x_3)}}
{4\pi^2}. 
\end{align}

For $a=1$, $b=7$ and $c=3$, 
the theory contains 8 gauge nodes and 22 chiral multiplets. 
The field content is 
\begin{align}
\begin{array}{c|cc}
&\prod_{i=1}^8 SU(N)_i&U(1)_R\\ \hline 
X_{13}&(\bf{N},\bf{1},\overline{\bf{N}},\bf{1},\bf{1},\bf{1},\bf{1},\bf{1})&x_1-x_3 \\
X_{16}&(\bf{N},\bf{1},\bf{1},\bf{1},\bf{1},\overline{\bf{N}},\bf{1},\bf{1})&2-x_1-x_2\\
X_{21}&(\overline{\bf{N}},\bf{N},\bf{1},\bf{1},\bf{1},\bf{1},\bf{1},\bf{1})&x_3 \\
X_{25}&(\bf{1},\bf{N},\bf{1},\bf{1},\overline{\bf{N}},\bf{1},\bf{1},\bf{1})&x_2 \\
X_{35}&(\bf{1},\bf{1},\bf{N},\bf{1},\overline{\bf{N}},\bf{1},\bf{1},\bf{1})&x_1-x_3 \\
X_{37}&(\bf{1},\bf{1},\bf{N},\bf{1},\bf{1},\bf{1},\overline{\bf{N}},\bf{1})&x_2 \\
X_{38}&(\bf{1},\bf{1},\bf{N},\bf{1},\bf{1},\bf{1},\bf{1},\overline{\bf{N}})&2-x_1-x_2+x_3 \\
X_{42}&(\bf{1},\overline{\bf{N}},\bf{1},\bf{N},\bf{1},\bf{1},\bf{1},\bf{1})&x_1-x_3 \\
X_{43}&(\bf{1},\bf{1},\overline{\bf{N}},\bf{N},\bf{1},\bf{1},\bf{1},\bf{1})&x_2 \\
X_{47}&(\bf{1},\bf{1},\bf{1},\bf{N},\bf{1},\bf{1},\overline{\bf{N}},\bf{1})&2-x_1-x_2+x_3 \\
X_{54}&(\bf{1},\bf{1},\bf{1},\overline{\bf{N}},\bf{N},\bf{1},\bf{1},\bf{1})&2-x_1-x_2+x_3 \\
X_{56}&(\bf{1},\bf{1},\bf{1},\bf{1},\bf{N},\overline{\bf{N}},\bf{1},\bf{1})&x_2 \\
X_{57}&(\bf{1},\bf{1},\bf{1},\bf{1},\bf{N},\bf{1},\overline{\bf{N}},\bf{1})&x_1 \\
X_{63}&(\bf{1},\bf{1},\overline{\bf{N}},\bf{1},\bf{1},\bf{N},\bf{1},\bf{1})&2-x_1-x_2+x_3 \\
X_{64}&(\bf{1},\bf{1},\bf{1},\overline{\bf{N}},\bf{1},\bf{N},\bf{1},\bf{1})&x_2 \\
X_{68}&(\bf{1},\bf{1},\bf{1},\bf{1},\bf{1},\bf{N},\bf{1},\overline{\bf{N}})&x_1 \\
X_{72}&(\bf{1},\overline{\bf{N}},\bf{1},\bf{1},\bf{1},\bf{1},\bf{N},\bf{1})&2-x_1-x_2 \\
X_{76}&(\bf{1},\bf{1},\bf{1},\bf{1},\bf{1},\overline{\bf{N}},\bf{N},\bf{1})&x_1-x_3 \\
X_{78}&(\bf{1},\bf{1},\bf{1},\bf{1},\bf{1},\bf{1},\bf{N},\overline{\bf{N}})&x_2 \\
X_{81}&(\overline{\bf{N}},\bf{1},\bf{1},\bf{1},\bf{1},\bf{1},\bf{1},\bf{N})&x_2 \\
X_{84}&(\bf{1},\bf{1},\bf{1},\overline{\bf{N}},\bf{1},\bf{1},\bf{1},\bf{N})&x_1-x_3 \\
X_{85}&(\bf{1},\bf{1},\bf{1},\bf{1},\overline{\bf{N}},\bf{1},\bf{1},\bf{N})&2-x_1-x_2 \\
\end{array}
\end{align}
The quiver diagram is depicted as
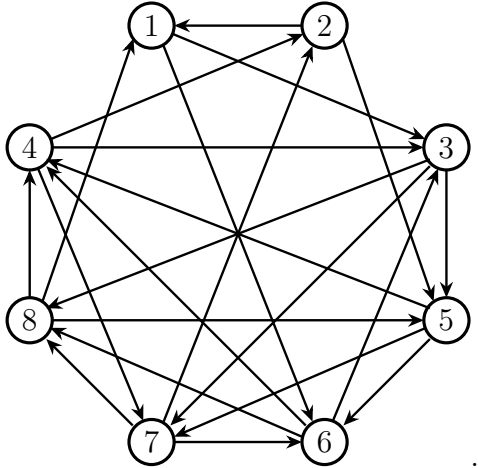
\begin{align}
\label{quiver_L173}
\begin{tikzpicture}[<->]
\draw[very thick] (-1.15,2.77) circle (3mm);
\node at (-1.15,2.77) {1};
\draw[very thick] (1.15,2.77) circle (3mm);
\node at (1.15,2.77) {2};
\draw[very thick] (2.77,1.15) circle (3mm);
\node at (2.77,1.15) {3};
\draw[very thick] (-2.77,1.15) circle (3mm);
\node at (-2.77,1.15) {4};
\draw[very thick] (-2.77,-1.15) circle (3mm);
\node at (-2.77,-1.15) {8};
\draw[very thick] (-1.15,-2.77) circle (3mm);
\node at (-1.15,-2.77) {7};
\draw[very thick] (1.15,-2.77) circle (3mm);
\node at (1.15,-2.77) {6};
\draw[very thick] (2.77,-1.15) circle (3mm);
\node at (2.77,-1.15) {5};
\draw[-{Stealth[length=6pt,width=5pt]}, line width=0.9pt]
(-0.87,2.66) to node[]{} (2.49,1.26);     
\draw[-{Stealth[length=6pt,width=5pt]}, line width=0.9pt]
(-0.99,2.5) to node[]{} (0.99,-2.5);      
\draw[-{Stealth[length=6pt,width=5pt]}, line width=0.9pt]
(0.85,2.77) to node[]{} (-0.85,2.77);       
\draw[-{Stealth[length=6pt,width=5pt]}, line width=0.9pt]
(1.39,2.6) to node[]{} (2.6,-0.9);       
\draw[-{Stealth[length=6pt,width=5pt]}, line width=0.9pt]
(2.77,0.85) to node[]{} (2.77,-0.85);       
\draw[-{Stealth[length=6pt,width=5pt]}, line width=0.9pt]
(2.55,0.9) to node[]{} (-0.9,-2.55);      
\draw[-{Stealth[length=6pt,width=5pt]}, line width=0.9pt]
(2.55,0.99) to node[]{} (-2.55,-0.99);      
\draw[-{Stealth[length=6pt,width=5pt]}, line width=0.9pt]
(-2.49,1.26) to node[]{} (0.87,2.66);       
\draw[-{Stealth[length=6pt,width=5pt]}, line width=0.9pt]
(-2.47,1.15) to node[]{} (2.47,1.15);       
\draw[-{Stealth[length=6pt,width=5pt]}, line width=0.9pt]
(-2.66,0.87) to node[]{} (-1.26,-2.49);     
\draw[-{Stealth[length=6pt,width=5pt]}, line width=0.9pt]
(2.55,-0.99) to node[]{} (-2.55,0.99);      
\draw[-{Stealth[length=6pt,width=5pt]}, line width=0.9pt]
(2.55,-1.4) to node[]{} (1.4,-2.55);      
\draw[-{Stealth[length=6pt,width=5pt]}, line width=0.9pt]
(2.5,-1.25) to node[]{} (-0.85,-2.7);     
\draw[-{Stealth[length=6pt,width=5pt]}, line width=0.9pt]
(1.26,-2.49) to node[]{} (2.66,0.87);       
\draw[-{Stealth[length=6pt,width=5pt]}, line width=0.9pt]
(0.9,-2.55) to node[]{} (-2.55,0.9);      
\draw[-{Stealth[length=6pt,width=5pt]}, line width=0.9pt]
(0.85,-2.7) to node[]{} (-2.5,-1.25);     
\draw[-{Stealth[length=6pt,width=5pt]}, line width=0.9pt]
(-0.99,-2.5) to node[]{} (0.99,2.5);      
\draw[-{Stealth[length=6pt,width=5pt]}, line width=0.9pt]
(-0.85,-2.77) to node[]{} (0.85,-2.77);     
\draw[-{Stealth[length=6pt,width=5pt]}, line width=0.9pt]
(-1.4,-2.55) to node[]{} (-2.55,-1.4);    
\draw[-{Stealth[length=6pt,width=5pt]}, line width=0.9pt]
(-2.6,-0.9) to node[]{} (-1.39,2.6);     
\draw[-{Stealth[length=6pt,width=5pt]}, line width=0.9pt]
(-2.77,-0.85) to node[]{} (-2.77,0.85);     
\draw[-{Stealth[length=6pt,width=5pt]}, line width=0.9pt]
(-2.47,-1.15) to node[]{} (2.47,-1.15);     
\end{tikzpicture}. 
\end{align}
The gravity dual is Type IIB string theory on $AdS_5\times L^{1,7,3}$. 
The primitive vectors of the toric diagram for the toric Calabi-Yau cone are chosen as
\begin{align}
v_1&=(1,0,0),\qquad 
v_2=(1,1,0), \qquad 
v_3=(1,5,7), \qquad 
v_4=(1,2,3). 
\end{align}
It is illustrated as
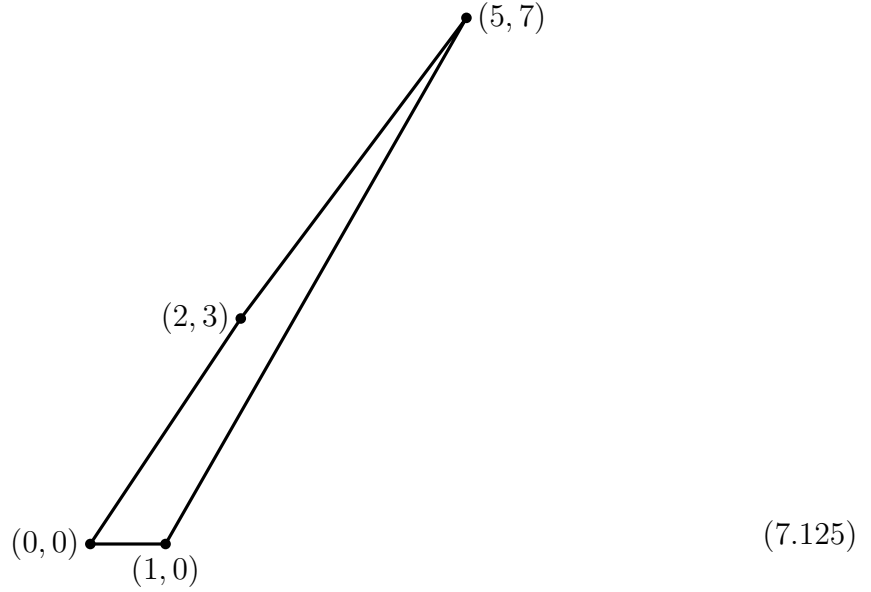
\begin{align}
\label{toric_L173}
\begin{tikzpicture}[scale=1, baseline={(0,0)}]
  \fill (0,0) circle (2pt) node[left] {$(0,0)$};
  \fill (1,0) circle (2pt) node[below] {$(1,0)$};
  \fill (5,7) circle (2pt) node[right] {$(5,7)$};
  \fill (2,3) circle (2pt) node[left] {$(2,3)$};
  \draw[very thick] (0,0) -- (1,0) -- (5,7) -- (2,3) -- cycle;
\end{tikzpicture}
\end{align}
The multiplicities of four types of the chiral multiplets 
$U_2$ $=$ $\{X_{16},X_{72},X_{85}\}$, 
$Y$ $=$ $\{X_{25},X_{37},X_{43},X_{56},X_{64},X_{78},X_{81}\}$, 
$U_1$ $=$ $\{X_{13},X_{35},X_{42},X_{76},X_{84}\}$ 
and $Z$ $=$ $\{X_{21}\}$ forming the baryonic operators are
\begin{align}
\det (v_{4},v_{1},v_{2})&=3, \\
\det (v_{1},v_{2},v_{3})&=7, \\
\det (v_{2},v_{3},v_{4})&=5, \\
\det (v_{3},v_{4},v_{1})&=1.
\end{align}
The area of the polygon is
\begin{align}
\label{A_L173}
A&=4. 
\end{align}
The dual cone is generated by
\begin{align}
\label{dualG_L173}
m_1&=(0,0,1),\qquad 
m_2=(7,-7,4),\qquad 
m_3=(1,4,-3), \qquad 
m_4=(0,3,-2). 
\end{align}
The volume of $L^{1,7,3}$ is 
\begin{align}
\label{vol_L173}
\mathrm{Vol}(L^{1,7,3})
&=\frac{63+84b_1-58b_2}
{b_2(21-7b_1+4b_2)(3+4b_1-3b_2)(3b_1-2b_2)}\pi^3, 
\end{align}
where $b=(3,b_1,b_2)$ is the Reeb vector. 
Applying the volume minimization \cite{Martelli:2005tp}, 
the Reeb vector is numerically determined as
\begin{align}
b_1&=6.11175\ldots, \qquad 
b_2=7.27183\ldots, 
\end{align}
and 
\begin{align}
\mathrm{Vol}(L^{1,7,3})
&=4.22672\ldots. 
\end{align}
The R-charges of the chiral multiplet fields are given by
\begin{align}
R(U_2)&=\frac{\pi}{3}\frac{\mathrm{Vol}(\Sigma_1)}{\mathrm{Vol}(L^{1,7,3})}
=\frac{2(3+4b_1-3b_2)(21-7b_1+4b_2)}{63+84b_1-58b_2}
=2-x_1-x_2, \\
R(Y)&=\frac{\pi}{3}\frac{\mathrm{Vol}(\Sigma_2)}{\mathrm{Vol}(L^{1,7,3})}
=\frac{14(3+4b_1-3b_2)(3b_1-2b_2)}{3(63+84b_1-58b_2)}
=x_2, \\
R(U_1)&=\frac{\pi}{3}\frac{\mathrm{Vol}(\Sigma_3)}{\mathrm{Vol}(L^{1,7,3})}
=\frac{10(3b_1-2b_2)b_2}{3(63+84b_1-58b_2)}
=x_1-x_3, \\
R(Z)&=\frac{\pi}{3}\frac{\mathrm{Vol}(\Sigma_4)}{\mathrm{Vol}(L^{1,7,3})}
=\frac{2(21-7b_1+4b_2)b_2}{3(63+84b_1-58b_2)}=x_3, 
\end{align}
where 
\begin{align}
\mathrm{Vol}(\Sigma_1)&=2\pi^2 \frac{\det(v_{4},v_{1},v_{2})}{\det(b,v_{4},v_1)\det(b,v_{1},v_{2})}=\frac{6\pi^2}{3b_1b_2-2b_2^2}, \\
\mathrm{Vol}(\Sigma_2)&=2\pi^2 \frac{\det(v_{1},v_{2},v_{3})}{\det(b,v_{1},v_2)\det(b,v_{2},v_{3})}=\frac{14\pi^2}{b_2(21-7b_1+4b_2)}, \\
\mathrm{Vol}(\Sigma_3)&=2\pi^2 \frac{\det(v_{2},v_{3},v_{4})}{\det(b,v_{2},v_3)\det(b,v_{3},v_{4})}=\frac{10\pi^2}{(3+4b_1-3b_2)(21-7b_1+4b_2)}, \\
\mathrm{Vol}(\Sigma_4)&=2\pi^2 \frac{\det(v_{3},v_{4},v_{1})}{\det(b,v_{3},v_4)\det(b,v_{4},v_{1})}=\frac{2\pi^2}{(3+4b_1-3b_2)(3b_1-2b_2)}.
\end{align}

The weighted adjacency matrix is 
\begin{align}
\label{M_L173}
&
M^{AdS_5\times L^{1,7,3}}
\nonumber\\
&=\left(
\begin{matrix}
i_v&\tilde{i}_{\textrm{cm}_{z}}&i_{\textrm{cm}_{u_1}}&0&0&i_{\textrm{cm}_{u_2}}&0&\tilde{i}_{\textrm{cm}_{y}} \\
i_{\textrm{cm}_{z}}&i_v&0&\tilde{i}_{\textrm{cm}_{u_1}}&i_{\textrm{cm}_{y}}&0&\tilde{i}_{\textrm{cm}_{u_2}}&0 \\
\tilde{i}_{\textrm{cm}_{u_1}}&0&i_v&\tilde{i}_{\textrm{cm}_{y}}&i_{\textrm{cm}_{u_1}}&\tilde{i}_{\textrm{cm}_{v_1}}&i_{\textrm{cm}_{y}}&i_{\textrm{cm}_{v_1}} \\
0&i_{\textrm{cm}_{u_1}}&i_{\textrm{cm}_{y}}&i_v&\tilde{i}_{\textrm{cm}_{v_1}}&\tilde{i}_{\textrm{cm}_{y}}&i_{\textrm{cm}_{v_1}}&\tilde{i}_{\textrm{cm}_{u_1}} \\
0&\tilde{i}_{\textrm{cm}_{y}}&\tilde{i}_{\textrm{cm}_{u_1}}&i_{\textrm{cm}_{v_1}}&i_v&i_{\textrm{cm}_{y}}&i_{\textrm{cm}_{v_2}}&\tilde{i}_{\textrm{cm}_{u_2}} \\
\tilde{i}_{\textrm{cm}_{u_2}}&0&i_{\textrm{cm}_{v_1}}&i_{\textrm{cm}_{y}}&\tilde{i}_{\textrm{cm}_{y}}&i_v&\tilde{i}_{\textrm{cm}_{u_1}}&i_{\textrm{cm}_{v_2}} \\
0&i_{\textrm{cm}_{u_2}}&\tilde{i}_{\textrm{cm}_{y}}&\tilde{i}_{\textrm{cm}_{v_1}}&\tilde{i}_{\textrm{cm}_{v_2}}&i_{\textrm{cm}_{u_1}}&i_v&i_{\textrm{cm}_{y}} \\
i_{\textrm{cm}_{y}}&0&\tilde{i}_{\textrm{cm}_{v_1}}&i_{\textrm{cm}_{u_1}}&i_{\textrm{cm}_{u_2}}&\tilde{i}_{\textrm{cm}_{v_2}}&\tilde{i}_{\textrm{cm}_{y}} &i_v \\
\end{matrix}
\right). 
\end{align}
From (\ref{M_L173}) we find the gravity index
\begin{align}
\label{ind_L173}
&
i^{AdS_5\times L^{1,7,3}}(p;q)
\nonumber\\
&=\frac{(pq)^{\frac{2-x_1+4x_2}{2}}}{1-(pq)^{\frac{2-x_1+4x_2}{2}}}
+\frac{(pq)^{\frac{x_1+3x_2-x_3}{2}}}{1-(pq)^{\frac{x_1+3x_2-x_3}{2}}}
+\frac{(pq)^{\frac{7x_1-4x_3}{2}}}{1-(pq)^{\frac{7x_1-4x_3}{2}}}
+\frac{(pq)^{7+\frac{-7(x_1+x_2)+5x_3}{2}}}{1-(pq)^{7+\frac{-7(x_1+x_2)+5x_3}{2}}}.
\end{align}
It enumerates four extremal BPS mesons. 
According to the formula (\ref{Rch_meson}) we get the R-charges of the extremal BPS mesons
\begin{align}
R(\mathcal{M}_1)&=\frac23(b,m_1)=\frac23b_2=7x_1-4x_3=4.84789\ldots, \\
R(\mathcal{M}_2)&=\frac23(b,m_2)=\frac23(21-7b_1+4b_2)=14-7(x_1+x_2)+5x_3=4.87002\ldots, \\
R(\mathcal{M}_3)&=\frac23(b,m_3)=\frac23(3+4b_1-3b_2)=2-x_1+4x_2=3.75436\ldots, \\
R(\mathcal{M}_4)&=\frac23(b,m_4)=\frac23(3b_1-2b_2)=x_1+3x_2-x_3=2.52774\ldots,
\end{align}
where 
\begin{align}
b_1&=\frac32(5x_1+x_2-3x_3), \\
b_2&=\frac32(7x_1-4x_3). 
\end{align}

We obtain the supersymmetric zeta function
\begin{align}
&
\mathfrak{Z}^{AdS_5\times L^{1,7,3}}(s,z;\omega_1,\omega_2)
\nonumber\\
&=
\left(\left(\frac{2-x_1+4x_2}{2}\right)(\omega_1+\omega_2) \right)^{-s}\zeta\left(s,1+\frac{2z}{(2-x_1+4x_2) (\omega_1+\omega_2)}\right)
\nonumber\\
&+\left(\left(\frac{x_1+3x_2-x_3}{2}\right)(\omega_1+\omega_2) \right)^{-s}\zeta\left(s,1+\frac{2z}{(x_1+3x_2-x_3) (\omega_1+\omega_2)}\right)
\nonumber\\
&+\left(\left(\frac{7x_1-4x_3}{2}\right)(\omega_1+\omega_2) \right)^{-s}\zeta\left(s,1+\frac{2z}{(7x_1-4x_3) (\omega_1+\omega_2)}\right)
\nonumber\\
&+\left(\left(\frac{14-7(x_1+x_2)+5x_3}{2}\right)(\omega_1+\omega_2) \right)^{-s}\zeta\left(s,1+\frac{2z}{(14-7(x_1+x_2)+5x_3) (\omega_1+\omega_2)}\right). 
\end{align}
It has the residue at a simple pole $s=1$
\begin{align}
\label{Res_L173}
&
\mathrm{Res}_{s=1}\mathfrak{Z}^{AdS_5\times L^{1,7,3}}(s,z;\omega_1,\omega_2)
\nonumber\\
&=
\Biggl(
\frac{2}{2-x_1+4x_2}+\frac{2}{7x_1-4x_3}
+\frac{2}{x_1+3x_2-x_3}+\frac{2}{14-7(x_1+x_2)+5x_3}
\Biggr)
\frac{1}{\omega_1+\omega_2}
\nonumber\\
&=3\left(
\frac{1}{3+4b_1-3b_2}+\frac{1}{3b_1-2b_2}+\frac{1}{b_2}+\frac{1}{21-7b_1+4b_2}
\right)\frac{1}{\omega_1+\omega_2}. 
\end{align}
Using the residue coefficient (\ref{Res_L173}) and the volume (\ref{vol_L173}) as inputs to the formula (\ref{Riem2_conj}), 
we compute the integral of the Riemann curvature squared of the form
\begin{align}
\int_{L^{1,7,3}}\mathrm{Riem}^2
&=\frac{8\pi^3}{b_2(21-7b_1+4b_2)(3+4b_1-3b_2)(3b_1-2b_2)}
\nonumber\\
&\times 
(-672b_1^3+80b_2^3+1040b_1^2b_2-512b_1b_2^2
\nonumber\\
&+1512b_1^2+24b_2^2
-1152b_1b_2-84b_1+598b_2-1197)
\nonumber\\
&=777.816\ldots. 
\end{align}
The Zeta-index is
\begin{align}
\label{0_L173}
\mathfrak{Z}^{AdS_5\times L^{1,7,3}}(0,0;\omega_1,\omega_2)
&=-2. 
\end{align}
The supersymmetric zeta value for $s=-1$ and $z=0$ is
\begin{align}
\label{-1_L173}
\mathfrak{Z}^{AdS_5\times L^{1,7,3}}(-1,0;\omega_1,\omega_2)
&=-\frac23 (\omega_1+\omega_2). 
\end{align}
From the formulae (\ref{c0_formula}) and (\ref{a0_formula}) we get
\begin{align}
c_{\mathcal{O}(N^0)}&=-1, \\
a_{\mathcal{O}(N^0)}&=-\frac32, 
\end{align}
and 
\begin{align}
c_{\mathcal{O}(N^0)}-a_{\mathcal{O}(N^0)}&=\frac12. 
\end{align}

The supersymmetric determinant is
\begin{align}
\label{det_L173}
\mathfrak{D}^{AdS_5\times L^{1,7,3}}(z;\omega_1,\omega_2)
&=\prod_{I=1}^4 
\frac{\rho_I^{\frac12+\frac{z}{\rho_I}} \Gamma\left(1+\frac{z}{\rho_I}\right)}
{\sqrt{2\pi}}, 
\end{align}
where 
\begin{align}
\rho_I&=\frac13(b,m_I)(\omega_1+\omega_2). 
\end{align}
The vacuum exponent is 
\begin{align}
\label{vac_L173}
&
\mathfrak{D}^{AdS_5\times L^{1,7,3}}(0;1,1)
\nonumber\\
&=\frac{\sqrt{(2-x_1+4x_2)(7x_1-4x_3)(x_1+3x_2-x_3)(14-7(x_1+x_2)+5x_3)}}
{4\pi^2}. 
\end{align}

We would like to propose the generalization of these results for the non-singular Sasaki-Einstein manifold $L^{a,b,c}$. 
The total number of the $SU(N)$ gauge nodes is $a+b$. 
There are six types of the bifundamental chiral multiplets $U_1$, $U_2$, $V_1$, $V_2$, $Y$ and $Z$ 
whose multiplicities and the R-charges are given by \cite{Franco:2005sm}: 
\begin{align}
\begin{array}{c|cc}
&\textrm{number}&U(1)_R \\ \hline 
U_1&a+b-c&x_1-x_3 \\
U_2&c&2-x_1-x_2 \\
V_1&b-c&2-x_1-x_2+x_3 \\
V_2&c-a&x_1 \\
Y&b&x_2 \\
Z&a&x_3 \\
\end{array}
\end{align}
The theory contains six types of bifundamental chiral multiplets. 
The total number of the chiral multiplets is $a+3b$. 
Among them, the four types $U_2, Y, U_1,$ and $Z$, which form the primary baryonic operators. 
The theory is holographically dual to Type IIB string theory on $AdS_5\times L^{a,b,c}$. 
The toric diagram of the cone over $L^{a,b,c}$ is generated by the following four vectors
\begin{align}
v_1&=(1,0,0),& 
v_2&=(1,1,0),& 
v_3&=(1,ak,b),&
v_4&=(1,-al,c), 
\end{align}
where $k>0$ and $l\le0$ are integers obeying the B\'{e}zout's identity
\begin{align}
\label{Bezouts_id}
ck+bl&=1. 
\end{align}
The solutions exist if and only if $\mathrm{GCD}(b,c)$ $=$ $1$. 
The diagram is depicted as
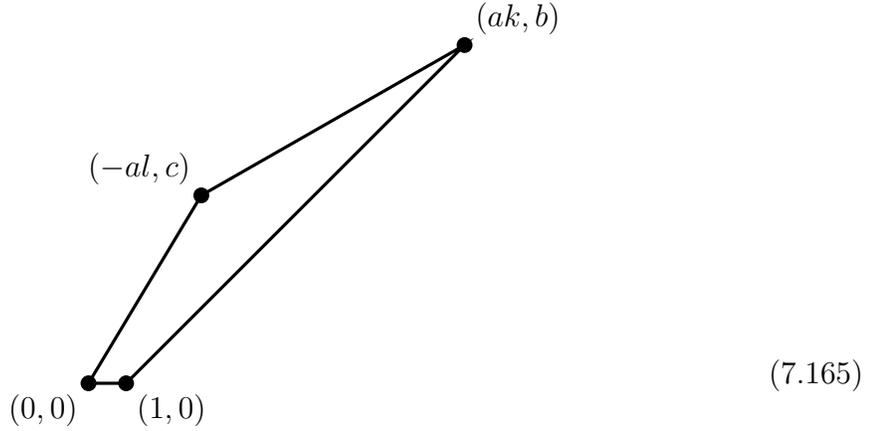
\begin{align}
\label{toric_Labc}
\begin{tikzpicture}[scale=0.5, baseline={(0,0)}]
  \fill (0,0) circle (6pt) node[below left] {$(0,0)$};
  \fill (1,0) circle (6pt) node[below right] {$(1,0)$};
  \fill (10,9) circle (6pt) node[above right] {$(ak,b)$};
  \fill (3,5) circle (6pt) node[above left] {$(-al,c)$};
  \draw[very thick] (0,0) -- (1,0) -- (10,9) -- (3,5) -- cycle;
\end{tikzpicture}
\end{align}
The multiplicities of four types of the chiral multiplets 
$U_2$, $Y$, $U_1$ and $Z$ are evaluated as
\begin{align}
\det (v_{4},v_{1},v_{2})&=c, \\
\det (v_{1},v_{2},v_{3})&=b, \\
\det (v_{2},v_{3},v_{4})&=a+b-c, \\
\det (v_{3},v_{4},v_{1})&=a.
\end{align}
The area of the polygon is
\begin{align}
\label{A_Labc}
A&=\frac{a+b}{2}. 
\end{align}
The dual cone is generated by
\begin{align}
\label{dualG_Labc}
m_1&=(0,0,1),& 
m_2&=(b,-b,ak-1),\nonumber\\
m_3&=\left(a,b-c,-a(k+l)\right),& 
m_4&=(0,c,al). 
\end{align}
The volume of $L^{a,b,c}$ takes the form\footnote{Also see \cite{Cvetic:2005ft}. }
\begin{align}
\label{vol_Labc}
\mathrm{Vol}(L^{a,b,c})
&=\frac{3abc+bc(b-c)b_1+a(a+b^2l-bcl-c)b_2}
{b_2\Bigl(3b-bb_1+(ak-1)b_2\Bigr) \Bigl(3a+(b-c)b_1-a(k+l)b_2\Bigr) \Bigl(cb_1+alb_2\Bigr)}
\pi^3, 
\end{align}
where $b=(3,b_1,b_2)$ is the Reeb vector. 
The components $b_1$ and $b_2$ are determined by the volume minimization \cite{Martelli:2005tp}. 
The R-charges of the chiral multiplet fields are given by
\begin{align}
R(U_2)&=\frac{\pi}{3}\frac{\mathrm{Vol}(\Sigma_1)}{\mathrm{Vol}(L^{a,b,c})}
\nonumber\\
&=\frac{2}{3}\frac{c(3b-bb_1+(ak-1)b_2)(3a+(b-c)b_1-a(k+l)b_2)}
{3abc+bc(b-c)b_1+a(a+b^2l-bcl-c)b_2}
=2-x_1-x_2, \\
R(Y)&=\frac{\pi}{3}\frac{\mathrm{Vol}(\Sigma_2)}{\mathrm{Vol}(L^{a,b,c})}
\nonumber\\
&=\frac{2}{3}\frac{b(3a+(b-c)b_1-a(k+l)b_2)(cb_1+alb_2)}
{3abc+bc(b-c)b_1+a(a+b^2l-bcl-c)b_2}
=x_2, \\
R(U_1)&=\frac{\pi}{3}\frac{\mathrm{Vol}(\Sigma_3)}{\mathrm{Vol}(L^{a,b,c})}
\nonumber\\
&=\frac{2}{3}\frac{(a+b-c)b_2(cb_1+alb_2)}
{3abc+bc(b-c)b_1+a(a+b^2l-bcl-c)b_2}
=x_1-x_3, \\
R(Z)&=\frac{\pi}{3}\frac{\mathrm{Vol}(\Sigma_4)}{\mathrm{Vol}(L^{a,b,c})}
\nonumber\\
&=\frac{2}{3}\frac{ab_2(3a+(b-c)b_1-a(k+l)b_2)}
{3abc+bc(b-c)b_1+a(a+b^2l-bcl-c)b_2}
=x_3, 
\end{align}
where 
\begin{align}
\mathrm{Vol}(\Sigma_1)&=2\pi^2 \frac{\det(v_{4},v_{1},v_{2})}{\det(b,v_{4},v_1)\det(b,v_{1},v_{2})}
=\frac{2c\pi^2}{b_2(cb_1+alb_2)}, \\
\mathrm{Vol}(\Sigma_2)&=2\pi^2 \frac{\det(v_{1},v_{2},v_{3})}{\det(b,v_{1},v_2)\det(b,v_{2},v_{3})}
=\frac{2b\pi^2}{b_2(3b-bb_1+(ak-1)b_2)}, \\
\mathrm{Vol}(\Sigma_3)&=2\pi^2 \frac{\det(v_{2},v_{3},v_{4})}{\det(b,v_{2},v_3)\det(b,v_{3},v_{4})}
\nonumber\\
&=\frac{2(a+b-c)\pi^2}{(3a+(b-c)b_1-a(k+l)b_2)(3b-bb_1+(ak-1)b_2)}, \\
\mathrm{Vol}(\Sigma_4)&=2\pi^2 \frac{\det(v_{3},v_{4},v_{1})}{\det(b,v_{3},v_4)\det(b,v_{4},v_{1})}
\nonumber\\
&=\frac{2a\pi^2}{(3a+(b-c)b_1-a(k+l)b_2)(cb_1+alb_2)}.
\end{align}

We propose that the gravity index for the non-singular Sasaki-Einstein manifold $L^{a,b,c}$ is given by 
\begin{align}
\label{ind_Labc}
i^{AdS_5\times L^{a,b,c}}(p;q)
&=\frac{(pq)^{\frac{bx_1+(c-b)x_3}{2}}}{1-(pq)^{\frac{bx_1+(c-b)x_3}{2}}}
+\frac{(pq)^{\frac{2b-bx_1-bx_2+(a+b-c)x_3}{2}}}{1-(pq)^{\frac{2b-bx_1-bx_2+(a+b-c)x_3}{2}}}
\nonumber\\
&+\frac{(pq)^{\frac{2a-ax_1+(b-c)x_2}{2}}}{1-(pq)^{\frac{2a-ax_1+(b-c)x_2}{2}}}
+\frac{(pq)^{\frac{ax_1+cx_2-ax_3}{2}}}{1-(pq)^{\frac{ax_1+cx_2-ax_3}{2}}}. 
\end{align}
This counts four extremal BPS mesonic operators corresponding to the primitive generators (\ref{dualG_Labc}) of the dual cone. 
The R-charges of the extremal BPS mesons are 
\begin{align}
R(\mathcal{M}_1)&=\frac23(b,m_1)=\frac23 b_2=bx_1+(c-b)x_3, \\
R(\mathcal{M}_2)&=\frac23(b,m_2)=\frac23 \Bigl(3b-bb_1+(ak-1)b_2\Bigr)=2b-bx_1-bx_2+(a+b-c)x_3, \\
R(\mathcal{M}_3)&=\frac23(b,m_3)=\frac23 \Bigl(3a+(b-c)b_1-a(k+l)b_2\Bigr)=2a-ax_1+(b-c)x_2, \\
R(\mathcal{M}_4)&=\frac23(b,m_4)=\frac23 (cb_1+alb_2)=ax_1+cx_2-ax_3,
\end{align}
where the components of the Reeb vector and the R-charge parameters are related by
\begin{align}
b_1&=\frac32\left(akx_1+x_2-\frac{a(1+k(b-c))}{b}x_3\right), \\
b_2&=\frac32\Bigl(bx_1+(c-b)x_3\Bigr). 
\end{align}

The supersymmetric zeta function is
\begin{align}
\label{zeta_Labc}
&
\mathfrak{Z}^{AdS_5\times L^{a,b,c}}(s,z;\omega_1,\omega_2)
\nonumber\\
&=
\left(\frac{3a+(b-c)b_1-a(k+l)b_2}{3}(\omega_1+\omega_2) \right)^{-s}\zeta\left(s,1+\frac{3z}{(3a+(b-c)b_1-a(k+l)b_2)(\omega_1+\omega_2)}\right)
\nonumber\\
&+\left(\frac{cb_1+alb_2}{3}(\omega_1+\omega_2) \right)^{-s}\zeta\left(s,1+\frac{3z}{(cb_1+alb_2)(\omega_1+\omega_2)}\right)
\nonumber\\
&+\left(\frac{b_2}{3}(\omega_1+\omega_2) \right)^{-s}\zeta\left(s,1+\frac{3z}{b_2(\omega_1+\omega_2)}\right)
\nonumber\\
&+\left(\frac{3b-bb_1+(ak-1)b_2}{3}(\omega_1+\omega_2) \right)^{-s}\zeta\left(s,1+\frac{3z}{(3b-bb_1+(ak-1)b_2)(\omega_1+\omega_2)}\right). 
\end{align}
The residue at a simple pole $s=1$ is 
\begin{align}
\label{Res_Labc}
&
\mathrm{Res}_{s=1}\mathfrak{Z}^{AdS_5\times L^{a,b,c}}(s,z;\omega_1,\omega_2)
\nonumber\\
&=3\left(
\frac{1}{3a+(b-c)b_1-a(k+l)b_2}
+\frac{1}{cb_1+alb_2}
+\frac{1}{b_2}
+\frac{1}{3b-bb_1+(ak-1)b_2}
\right). 
\end{align}
Substitution of the residue coefficient (\ref{Res_Labc}) and the volume (\ref{vol_Labc}) 
into the formula (\ref{Riem2_conj}) yields the following result for the integral of the squared Riemann curvature: 
\begin{align}
\label{Riem2_Labc}
&
\int_{L^{a,b,c}}\mathrm{Riem}^2
\nonumber\\
&=64\pi^3 \left(
\frac{1}{3a+(b-c)b_1-a(k+l)b_2}
+\frac{1}{cb_1+alb_2}
+\frac{1}{b_2}
+\frac{1}{3b-bb_1+(ak-1)b_2}
\right)
\nonumber\\
&-152\pi^3 \frac{3abc+bc(b-c)b_1+a(a+b^2l-bcl-c)b_2}
{b_2\Bigl(3b-bb_1+(ak-1)b_2\Bigr) \Bigl(3a+(b-c)b_1-a(k+l)b_2\Bigr) \Bigl(cb_1+alb_2\Bigr)}. 
\end{align}
The Zeta-index is
\begin{align}
\label{0_Labc}
\mathfrak{Z}^{AdS_5\times L^{a,b,c}}(0,0;\omega_1,\omega_2)
&=-2. 
\end{align}
The supersymmetric zeta value for $s=-1$ and $z=0$ is
\begin{align}
\label{-1_Labc}
\mathfrak{Z}^{AdS_5\times L^{a,b,c}}(-1,0;\omega_1,\omega_2)
&=-\frac{a+b}{12} (\omega_1+\omega_2). 
\end{align}
We find from the formulae (\ref{c0_formula}) and (\ref{a0_formula})
\begin{align}
c_{\mathcal{O}(N^0)}&=-\frac{a+b}{8}, \\
a_{\mathcal{O}(N^0)}&=-\frac{3(a+b)}{16}, 
\end{align}
and 
\begin{align}
c_{\mathcal{O}(N^0)}-a_{\mathcal{O}(N^0)}&=\frac{a+b}{16}. 
\end{align}

The supersymmetric determinant is given by
\begin{align}
\label{det_Labc}
\mathfrak{D}^{AdS_5\times L^{a,b,c}}(z;\omega_1,\omega_2)
&=\prod_{I=1}^4 
\frac{\rho_I^{\frac12+\frac{z}{\rho_I}} \Gamma\left(1+\frac{z}{\rho_I}\right)}
{\sqrt{2\pi}}, 
\end{align}
where 
\begin{align}
\rho_I&=\frac13(b,m_I)(\omega_1+\omega_2). 
\end{align}
The vacuum exponent is 
\begin{align}
\label{vac_Labc}
&
\mathfrak{D}^{AdS_5\times L^{a,b,c}}(0;1,1)
\nonumber\\
&=\frac{\sqrt{b_2 (3b-bb_1+(ak-1)b_2) (3a+(b-c)b_1-a(k+l)b_2) (cb_1+alb_2)}}{9\pi^2}. 
\end{align}

When $\mathrm{GCD}(b,c)$ $=$ $h$ $\neq 1$, there is no solution to the B\'{e}zout's identity (\ref{Bezouts_id}). 
The corresponding Calabi-Yau cone can be constructed as a finite Abelian orbifold of a primitive one. 
For example, when $a=2$, $b=6$ and $c=3$, 
the theory has 8 gauge nodes and 20 chiral multiplets. 
The field content is 
\begin{align}
\begin{array}{c|cc}
&\prod_{i=1}^8 SU(N)_i&U(1)_R\\ \hline 
X_{12}&(\bf{N},\overline{\bf{N}},\bf{1},\bf{1},\bf{1},\bf{1},\bf{1},\bf{1})&x_1-x_3 \\
X_{14}&(\bf{N},\bf{1},\bf{1},\overline{\bf{N}},\bf{1},\bf{1},\bf{1},\bf{1})&2-x_1-x_2 \\
X_{17}&(\bf{N},\bf{1},\bf{1},\bf{1},\bf{1},\bf{1},\overline{\bf{N}},\bf{1})&x_2 \\
X_{21}&(\overline{\bf{N}},\bf{N},\bf{1},\bf{1},\bf{1},\bf{1},\bf{1},\bf{1})&x_1 \\
X_{23}&(\bf{1},\bf{N},\overline{\bf{N}},\bf{1},\bf{1},\bf{1},\bf{1},\bf{1})&2-x_1-x_2+x_3 \\
X_{28}&(\bf{1},\bf{N},\bf{1},\bf{1},\bf{1},\bf{1},\bf{1},\overline{\bf{N}})&x_2 \\
X_{31}&(\overline{\bf{N}},\bf{1},\bf{N},\bf{1},\bf{1},\bf{1},\bf{1},\bf{1})&x_2 \\
X_{34}&(\bf{1},\bf{1},\bf{N},\overline{\bf{N}},\bf{1},\bf{1},\bf{1},\bf{1})&x_1-x_3 \\
X_{38}&(\bf{1},\bf{1},\bf{N},\bf{1},\bf{1},\bf{1},\bf{1},\overline{\bf{N}})&2-x_1-x_2+x_3 \\
X_{42}&(\bf{1},\overline{\bf{N}},\bf{1},\bf{N},\bf{1},\bf{1},\bf{1},\bf{1})&x_2 \\
X_{45}&(\bf{1},\bf{1},\bf{1},\bf{N},\overline{\bf{N}},\bf{1},\bf{1},\bf{1})&x_3 \\
X_{53}&(\bf{1},\bf{1},\overline{\bf{N}},\bf{1},\bf{N},\bf{1},\bf{1},\bf{1})&x_1-x_3 \\
X_{56}&(\bf{1},\bf{1},\bf{1},\bf{1},\bf{N},\overline{\bf{N}},\bf{1},\bf{1})&2-x_1-x_2 \\
X_{63}&(\bf{1},\bf{1},\overline{\bf{N}},\bf{1},\bf{1},\bf{N},\bf{1},\bf{1})&x_2 \\
X_{67}&(\bf{1},\bf{1},\bf{1},\bf{1},\bf{1},\bf{N},\overline{\bf{N}},\bf{1})&x_3 \\
X_{72}&(\bf{1},\overline{\bf{N}},\bf{1},\bf{1},\bf{1},\bf{1},\bf{N},\bf{1})&2-x_1-x_2 \\
X_{78}&(\bf{1},\bf{1},\bf{1},\bf{1},\bf{1},\bf{1},\bf{N},\overline{\bf{N}})&x_1-x_3 \\
X_{81}&(\overline{\bf{N}},\bf{1},\bf{1},\bf{1},\bf{1},\bf{1},\bf{1},\bf{N})&2-x_1-x_2+x_3 \\
X_{85}&(\bf{1},\bf{1},\bf{1},\bf{1},\overline{\bf{N}},\bf{1},\bf{1},\bf{N})&x_2 \\
X_{86}&(\bf{1},\bf{1},\bf{1},\bf{1},\bf{1},\overline{\bf{N}},\bf{1},\bf{N})&x_1-x_3 \\
\end{array}
\end{align}
The quiver diagram is depicted as
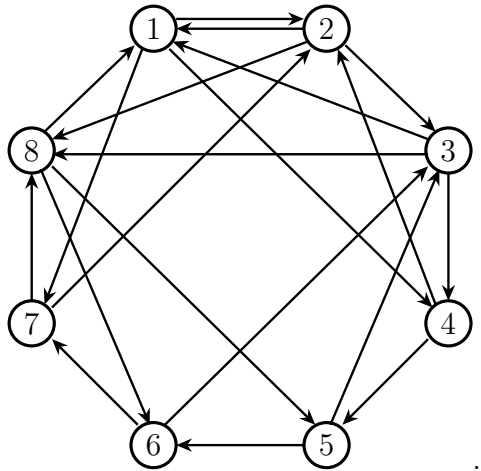
\begin{align}
\label{quiver_L263}
\begin{tikzpicture}[<->]
\draw[very thick] (-1.15,2.77) circle (3mm);
\node at (-1.15,2.77) {1};
\draw[very thick] (1.15,2.77) circle (3mm);
\node at (1.15,2.77) {2};
\draw[very thick] (2.77,1.15) circle (3mm);
\node at (2.77,1.15) {3};
\draw[very thick] (2.77,-1.15) circle (3mm);
\node at (2.77,-1.15) {4};
\draw[very thick] (1.15,-2.77) circle (3mm);
\node at (1.15,-2.77) {5};
\draw[very thick] (-1.15,-2.77) circle (3mm);
\node at (-1.15,-2.77) {6};
\draw[very thick] (-2.77,-1.15) circle (3mm);
\node at (-2.77,-1.15) {7};
\draw[very thick] (-2.77,1.15) circle (3mm);
\node at (-2.77,1.15) {8};
\draw[-{Stealth[length=6pt,width=5pt]}, line width=0.9pt] (-0.85,2.9) -- (0.85,2.9);
\draw[-{Stealth[length=6pt,width=5pt]}, line width=0.9pt] (-0.94,2.50) -- (2.56,-0.94);
\draw[-{Stealth[length=6pt,width=5pt]}, line width=0.9pt] (-1.3,2.5) -- (-2.6,-0.9);
\draw[-{Stealth[length=6pt,width=5pt]}, line width=0.9pt] (0.85,2.77) -- (-0.85,2.77);
\draw[-{Stealth[length=6pt,width=5pt]}, line width=0.9pt] (1.4,2.55) -- (2.6,1.4);
\draw[-{Stealth[length=6pt,width=5pt]}, line width=0.9pt] (0.9,2.6) -- (-2.5,1.3);
\draw[-{Stealth[length=6pt,width=5pt]}, line width=0.9pt] (2.5,1.3) -- (-0.9,2.6);
\draw[-{Stealth[length=6pt,width=5pt]}, line width=0.9pt] (2.77,0.85) -- (2.77,-0.85);
\draw[-{Stealth[length=6pt,width=5pt]}, line width=0.9pt] (2.50,1.1) -- (-2.50,1.1);
\draw[-{Stealth[length=6pt,width=5pt]}, line width=0.9pt] (2.6,-0.9) -- (1.3,2.5);
\draw[-{Stealth[length=6pt,width=5pt]}, line width=0.9pt] (2.50,-1.36) -- (1.36,-2.50);
\draw[-{Stealth[length=6pt,width=5pt]}, line width=0.9pt] (1.2,-2.50) -- (2.65,0.9);
\draw[-{Stealth[length=6pt,width=5pt]}, line width=0.9pt] (0.85,-2.77) -- (-0.85,-2.77);
\draw[-{Stealth[length=6pt,width=5pt]}, line width=0.9pt] (-1,-2.50) -- (2.50,0.94);
\draw[-{Stealth[length=6pt,width=5pt]}, line width=0.9pt] (-1.36,-2.50) -- (-2.50,-1.36);
\draw[-{Stealth[length=6pt,width=5pt]}, line width=0.9pt] (-2.50,-0.94) -- (0.94,2.50);
\draw[-{Stealth[length=6pt,width=5pt]}, line width=0.9pt] (-2.77,-0.85) -- (-2.77,0.85);
\draw[-{Stealth[length=6pt,width=5pt]}, line width=0.9pt] (-2.6,1.4) -- (-1.4,2.55);
\draw[-{Stealth[length=6pt,width=5pt]}, line width=0.9pt] (-2.50,0.94) -- (1,-2.50);
\draw[-{Stealth[length=6pt,width=5pt]}, line width=0.9pt] (-2.65,0.9) -- (-1.2,-2.50);
\end{tikzpicture}. 
\end{align}
The theory is holographically dual to Type IIB string theory on $AdS_5\times L^{2,6,3}$. 
The toric diagram for the toric Calabi-Yau cone over $L^{2,6,3}$ is generated by
\begin{align}
v_1&=(1,1,0),\qquad 
v_2=(1,2,0), \qquad 
v_3=(1,2,1), \\
v_4&=(1,2,2), \qquad 
v_5=(1,2,3), \qquad 
v_6=(1,0,2). 
\end{align}
It is shown as
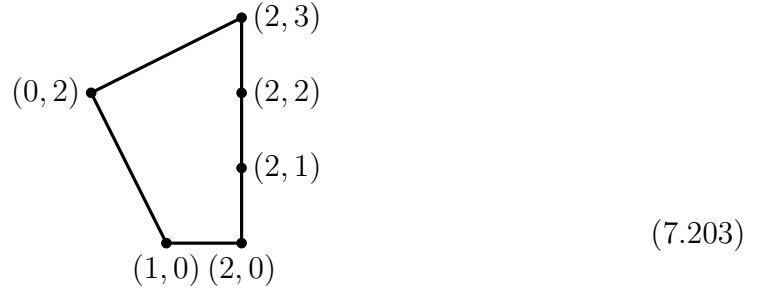
\begin{align}
\label{toric_L263}
\begin{tikzpicture}[scale=1, baseline={(0,0)}]
  \fill (1,0) circle (2pt) node[below] {$(1,0)$};
  \fill (2,0) circle (2pt) node[below] {$(2,0)$};
  \fill (2,1) circle (2pt) node[right] {$(2,1)$};
  \fill (2,2) circle (2pt) node[right] {$(2,2)$};
  \fill (2,3) circle (2pt) node[right] {$(2,3)$};
  \fill (0,2) circle (2pt) node[left] {$(0,2)$};
  \draw[very thick] (1,0) -- (2,0) -- (2,1) -- (2,2) -- (2,3) -- (0,2) -- cycle;
\end{tikzpicture}
\end{align}
The multiplicities of four types of the chiral multiplets 
$Z$ $=$ $\{X_{45}$,$X_{67}\}$, 
$U_2$ $=$ $\{X_{14}$,$X_{56}$,$X_{72}\}$, 
$Y$ $=$ $\{X_{17}$,$X_{28}$,$X_{31}$,$X_{42}$,$X_{63}$,$X_{85}\}$
and $U_1$ $=$ $\{X_{12},X_{34},X_{53},X_{78},X_{86}\}$ 
forming the baryons are evaluated as
\begin{align}
\det (v_{6},v_{1},v_{2})&=2, \\
\det (v_{1},v_{2},v_{5})&=3, \\
\det (v_{2},v_{5},v_{6})&=6, \\
\det (v_{5},v_{6},v_{1})&=5.
\end{align}
The area of the polygon is
\begin{align}
\label{A_L263}
A&=4. 
\end{align}
The dual cone is generated by
\begin{align}
\label{dualG_L263}
m_1&=(0,0,1),& 
m_2&=(2,-1,0),&
m_3&=(2,-1,0), \nonumber \\
m_4&=(2,-1,0),&
m_5&=(4,1,-2),& 
m_6&=(-2,2,1). 
\end{align}
The volume of $L^{2,6,3}$ is 
\begin{align}
\label{vol_L263}
\mathrm{Vol}(L^{2,6,3})
&=
\frac{2(b_1+3b_2+12)}
{b_2(6-b_1)(b_1-2b_2+12)(2b_1+b_2-6)}\pi^3, 
\end{align}
where $b=(3,b_1,b_2)$ is the Reeb vector. 
According to the volume minimization \cite{Martelli:2005tp} we numerically find the Reeb vector
\begin{align}
b_1&=3.76049\ldots, \qquad 
b_2=4.43439\ldots, 
\end{align}
and 
\begin{align}
\mathrm{Vol}(L^{2,6,3})
&=4.42188\ldots. 
\end{align}
The R-charges of the chiral multiplet fields are given by
\begin{align}
R(Z)&=\frac{\pi}{3}\frac{\mathrm{Vol}(\Sigma_1)}{\mathrm{Vol}(L^{2,6,3})}
=\frac{2(6-b_1)(12+b_1-2b_2)}{3(12+b_1+3b_2)}
=x_3=0.354028\ldots, \\
R(U_2)&=\frac{\pi}{3}\frac{\mathrm{Vol}(\Sigma_2)}{\mathrm{Vol}(L^{2,6,3})}
=\frac{(12+b_2-2b_2)(-6+2b_1+b_2)}{3(12+b_1+3b_2)}
=2-x_1-x_2=0.470721\ldots, \\
R(Y)&=\frac{\pi}{3}\frac{\mathrm{Vol}(\Sigma_5)}{\mathrm{Vol}(L^{2,6,3})}
=\frac{2b_2(-6+2b_1+b_2)}{3(12+b_1+3b_2)}
=x_2=0.605761\ldots, \\
R(U_1)&=\frac{\pi}{3}\frac{\mathrm{Vol}(\Sigma_4)}{\mathrm{Vol}(L^{2,6,3})}
=\frac{5b_2(6-b_1)}{3(12+b_1+3b_2)}
=x_1-x_3=0.569489\ldots, 
\end{align}
where 
\begin{align}
\mathrm{Vol}(\Sigma_1)&=2\pi^2 \frac{\det(v_{6},v_{1},v_{2})}{\det(b,v_{6},v_1)\det(b,v_{1},v_{2})}
=\frac{4\pi^2}{b_2(-6+2b_1+b_2)}, \\
\mathrm{Vol}(\Sigma_2)&=2\pi^2 \frac{\det(v_{1},v_{2},v_{5})}{\det(b,v_{1},v_2)\det(b,v_{2},v_{5})}
=\frac{2\pi^2}{6b_1-b_1b_2}, \\
\mathrm{Vol}(\Sigma_5)&=2\pi^2 \frac{\det(v_{2},v_{5},v_{6})}{\det(b,v_{2},v_5)\det(b,v_{5},v_{6})}
=\frac{4\pi^2}{(6-b_1)(12+b_1-2b_2)}, \\
\mathrm{Vol}(\Sigma_6)&=2\pi^2 \frac{\det(v_{5},v_{6},v_{1})}{\det(b,v_{5},v_6)\det(b,v_{6},v_{1})}
=\frac{10\pi^2}{(12+b_1-2b_2)(-6+2b_1+b_2)}.
\end{align}

We have the weighted adjacency matrix
\begin{align}
\label{M_L263}
&
M^{AdS_5\times L^{2,6,3}}
\nonumber\\
&=\left(
\begin{matrix}
i_v&{i}_{\textrm{cm}_{u_1}}+\tilde{i}_{\textrm{cm}_{v_2}}&\tilde{i}_{\textrm{cm}_{y}}&i_{\textrm{cm}_{u_2}}&0&0&{i}_{\textrm{cm}_{y}}&\tilde{i}_{\textrm{cm}_{v_1}} \\
\tilde{i}_{\textrm{cm}_{u_1}}+i_{\textrm{cm}_{v_2}}&i_v&i_{\textrm{cm}_{v_1}}&\tilde{i}_{\textrm{cm}_{y}}&0&0&\tilde{i}_{\textrm{cm}_{u_2}}&i_{\textrm{cm}_{y}} \\
i_{\textrm{cm}_{y}}&\tilde{i}_{\textrm{cm}_{v_1}}&i_v&i_{\textrm{cm}_{u_1}}&\tilde{i}_{\textrm{cm}_{u_1}}&\tilde{i}_{\textrm{cm}_{y}}&0&i_{\textrm{cm}_{v_1}} \\
\tilde{i}_{\textrm{cm}_{u_2}}&i_{\textrm{cm}_{y}}&\tilde{i}_{\textrm{cm}_{u_1}}&i_v&i_{\textrm{cm}_{z}}&0&0&0 \\
0&0&i_{\textrm{cm}_{u_1}}&\tilde{i}_{\textrm{cm}_{z}}&i_v&{i}_{\textrm{cm}_{u_2}}&0&\tilde{i}_{\textrm{cm}_{y}} \\
0&0&i_{\textrm{cm}_{y}}&0&\tilde{i}_{\textrm{cm}_{u_2}}&i_v&{i}_{\textrm{cm}_{z}}&i_{\textrm{cm}_{u_1}} \\
\tilde{i}_{\textrm{cm}_{y}}&{i}_{\textrm{cm}_{u_2}}&0&0&0&\tilde{i}_{\textrm{cm}_{z}}&i_v&i_{\textrm{cm}_{u_1}} \\
i_{\textrm{cm}_{v_1}}&\tilde{i}_{\textrm{cm}_{y}}&\tilde{i}_{\textrm{cm}_{v_1}} &0&i_{\textrm{cm}_{y}}&i_{\textrm{cm}_{u_1}}&\tilde{i}_{\textrm{cm}_{u_1}}&i_v \\
\end{matrix}
\right). 
\end{align}
From (\ref{M_L263}) we obtain the gravity index 
\begin{align}
\label{ind_L263}
i^{AdS_5\times L^{2,6,3}}(p;q)
&=\frac{(pq)^{\frac{4-2x_1+3x_2}{2}}}{1-(pq)^{\frac{4-2x_1+3x_2}{2}}}
+\frac{(pq)^{\frac{2x_1+3x_2-2x_3}{2}}}{1-(pq)^{\frac{2x_1+3x_2-2x_3}{2}}}
\nonumber\\
&+3\frac{(pq)^{\frac{2x_1-x_3}{2}}}{1-(pq)^{\frac{2x_1-x_3}{2}}}
+\frac{(pq)^{6+\frac{-6(x_1+x_2)+5x_3}{2}}}{1-(pq)^{6+\frac{-6(x_1+x_2)+5x_3}{2}}}. 
\end{align}
There exist six extremal BPS mesonic operators 
corresponding to the primitive generators (\ref{dualG_L263}) of the dual cone. 
The R-charges of the extremal BPS mesons are evaluated from the formula (\ref{Rch_meson}) as
\begin{align}
R(\mathcal{M}_1)&=\frac23(b,m_1)=\frac23 b_2=2x_1+3x_2-2x_3=2.95626\ldots, \\
R(\mathcal{M}_2)&=\frac23(b,m_2)=\frac23 (6-b_1)=2x_1-x_3=1.49301\ldots, \\
R(\mathcal{M}_3)&=\frac23(b,m_3)=\frac23 (6-b_1)=2x_1-x_3=1.49301\ldots, \\
R(\mathcal{M}_4)&=\frac23(b,m_4)=\frac23 (6-b_1)=2x_1-x_3=1.49301\ldots, \\
R(\mathcal{M}_5)&=\frac23(b,m_5)=\frac23 (b_1-2b_2+12)=(12-6(x_1+x_2)+5x_3)=4.59447\ldots, \\
R(\mathcal{M}_6)&=\frac23(b,m_6)=\frac23 (2b_1+b_2-6)=(4-2x_1+3x_2)=3.97025\ldots, 
\end{align}
where the Reeb vector and the R-charges are related by
\begin{align}
b_1&=6-3(x_1-\frac{x_3}{2}), \\
b_2&=3(x_1+\frac{3x_2}{2}-x_3). 
\end{align}

The supersymmetric zeta function reads
\begin{align}
\label{zeta_L263}
&
\mathfrak{Z}^{AdS_5\times L^{2,6,3}}(s,z;\omega_1,\omega_2)
\nonumber\\
&=
\left(\left(\frac{4-2x_1+3x_2}{2}\right)(\omega_1+\omega_2) \right)^{-s}\zeta\left(s,1+\frac{2z}{(4-2x_1+3x_2) (\omega_1+\omega_2))}\right)
\nonumber\\
&+\left(\left(\frac{2x_1+3x_2-2x_3}{2}\right)(\omega_1+\omega_2) \right)^{-s}\zeta\left(s,1+\frac{2z}{(2x_1+3x_2-2x_3) (\omega_1+\omega_2))}\right)
\nonumber\\
&+3\left(\left(\frac{2x_1-x_3}{2}\right)(\omega_1+\omega_2) \right)^{-s}\zeta\left(s,1+\frac{2z}{(2x_1-x_3) (\omega_1+\omega_2))}\right)
\nonumber\\
&+\left(\left(\frac{12-6(x_1+x_2)+5x_3}{2}\right)(\omega_1+\omega_2) \right)^{-s}\zeta\left(s,1+\frac{2z}{(12-6(x_1+x_2)+5x_3) (\omega_1+\omega_2))}\right). 
\end{align}
It has a simple pole at $s=1$ with the residue 
\begin{align}
\label{Res_L263}
&
\mathrm{Res}_{s=1}\mathfrak{Z}^{AdS_5\times L^{2,6,3}}(s,z;\omega_1,\omega_2)
\nonumber\\
&=
\left(
\frac{2}{4-2x_1+3x_2}+\frac{2}{2x_1+3x_2-2x_3}
+\frac{6}{2x_1-x_3}+\frac{2}{12-6(x_1+x_2)+5x_3}
\right)\frac{1}{\omega_1+\omega_2}
\nonumber\\
&=3\left(
\frac{3}{6-b_1}+\frac{1}{12+b_1-2b_2}+\frac{1}{b_2}+\frac{1}{-6+2b_1+b_2}
\right)\frac{1}{\omega_1+\omega_2}. 
\end{align}
Using the residue coefficient (\ref{Res_L263}) and the volume (\ref{vol_L263}), 
one arrives at the following result for the integrated curvature 
\begin{align}
\int_{L^{2,6,3}}\mathrm{Riem}^2
&=\frac{16\pi^3}{b_2(6-b_1)(12+b_1-2b_2)(-6+2b_1+b_2)}
\nonumber\\
&\times (-8b_1^3-24b_2^3+24b_1^2b_2-24b_1b_2^2
\nonumber\\
&-24b_1^2+216b_2^2+96b_1b_2+701b_1-201b_2-1956
)\nonumber\\
&=3054.79\ldots. 
\end{align}
The Zeta-index is given by
\begin{align}
\label{0_L263}
\mathfrak{Z}^{AdS_5\times L^{2,6,3}}(0,0;\omega_1,\omega_2)
&=-3. 
\end{align}
The supersymmetric zeta value with $s=-1$ and $z=0$ is
\begin{align}
\label{-1_L263}
\mathfrak{Z}^{AdS_5\times L^{2,6,3}}(-1,0;\omega_1,\omega_2)
&=-\frac23(\omega_1+\omega_2). 
\end{align}
We find from the formulae (\ref{c0_formula}) and (\ref{a0_formula})
\begin{align}
c_{\mathcal{O}(N^0)}&=-1, \\
a_{\mathcal{O}(N^0)}&=-\frac32, 
\end{align}
and 
\begin{align}
c_{\mathcal{O}(N^0)}-a_{\mathcal{O}(N^0)}&=\frac12. 
\end{align}

We have the supersymmetric determinant of the form
\begin{align}
\label{det_L263}
\mathfrak{D}^{AdS_5\times L^{2,6,3}}(z;\omega_1,\omega_2)
&=\prod_{I=1}^6 
\frac{\rho_I^{\frac12+\frac{z}{\rho_I}} \Gamma\left(1+\frac{z}{\rho_I}\right)}
{\sqrt{2\pi}}, 
\end{align}
where 
\begin{align}
\rho_I&=\frac13(b,m_I)(\omega_1+\omega_2). 
\end{align}
The vacuum exponent is 
\begin{align}
\label{vac_L263}
&
\mathfrak{D}^{AdS_5\times L^{2,6,3}}(0;1,1)
\nonumber\\
&=\frac{\sqrt{(4-2x_1+3x_2)(2x_1-x_3)^3(2x_1+3x_2-2x_3)(12-6(x_1+x_2)+5x_3)}}
{8\pi^{3}}. 
\end{align}

\section{Orbifolds of $L^{a,b,c}$}
\label{sec_orbLabc}

\subsection{$AdS_5\times L^{1,2,1}/\mathbb{Z}_2$}
Let us study the world-volume theory on a stack of D3-branes 
probing the geometry $\textrm{SPP}/\mathbb{Z}_2$ with orbifold action $(0,1,1,1)$, 
which is identified with the complex cone over the third Pseudo del Pezzo surface $PdP_{3c}$ 
\cite{Beasley:2001zp,Feng:2002fv,Feng:2004uq,Hanany:2012hi,Bianchi:2014qma,Antinucci:2020yki}. 
The theory is described by the 4d $\mathcal{N}=1$ quiver gauge theory 
with $\prod_{i=1}^6 SU(N)_i$ gauge group and 14 bifundamental chiral multiplets. 
The field content is
\begin{align}
\label{field_L121Z2}
\begin{array}{c|cc}
&\prod_{i=1}^6 SU(N)_i&U(1)_R\\ \hline 
X_{12}&(\bf{N},\overline{\bf{N}},\bf{1},\bf{1},\bf{1},\bf{1})&\frac{1}{\sqrt{3}} \\
X_{14}&(\bf{N},\bf{1},\bf{1},\overline{\bf{N}},\bf{1},\bf{1})&2\left(1-\frac{1}{\sqrt{3}}\right) \\
X_{16}&(\bf{N},\bf{1},\bf{1},\bf{1},\bf{1},\overline{\bf{N}})&\frac{1}{\sqrt{3}} \\
X_{24}&(\bf{1},\bf{N},\bf{1},\overline{\bf{N}},\bf{1},\bf{1})&\frac{1}{\sqrt{3}} \\
X_{26}&(\bf{1},\bf{N},\bf{1},\bf{1},\bf{1},\overline{\bf{N}})&1-\frac{1}{\sqrt{3}} \\
X_{31}&(\overline{\bf{N}},\bf{1},\bf{N},\bf{1},\bf{1},\bf{1})&\frac{1}{\sqrt{3}} \\
X_{32}&(\bf{1},\overline{\bf{N}},\bf{N},\bf{1},\bf{1},\bf{1})&1-\frac{1}{\sqrt{3}} \\
X_{41}&(\overline{\bf{N}},\bf{1},\bf{1},\bf{N},\bf{1},\bf{1})&2\left(1-\frac{1}{\sqrt{3}}\right) \\
X_{43}&(\bf{1},\bf{1},\overline{\bf{N}},\bf{N},\bf{1},\bf{1})&\frac{1}{\sqrt{3}} \\
X_{45}&(\bf{1},\bf{1},\bf{1},\bf{N},\overline{\bf{N}},\bf{1})&\frac{1}{\sqrt{3}} \\
X_{51}&(\overline{\bf{N}},\bf{1},\bf{1},\bf{1},\bf{N},\bf{1})&\frac{1}{\sqrt{3}} \\
X_{53}&(\bf{1},\bf{1},\overline{\bf{N}},\bf{1},\bf{N},\bf{1})&1-\frac{1}{\sqrt{3}} \\
X_{64}&(\bf{1},\bf{1},\bf{1},\overline{\bf{N}},\bf{1},\bf{N})&\frac{1}{\sqrt{3}} \\
X_{65}&(\bf{1},\bf{1},\bf{1},\bf{1},\overline{\bf{N}},\bf{N})&1-\frac{1}{\sqrt{3}} \\
\end{array}
\end{align}
The quiver diagram is drawn as
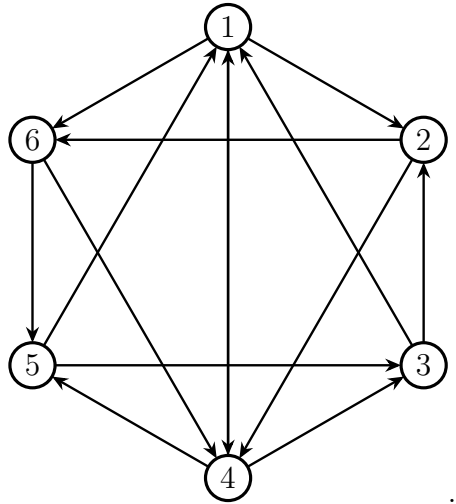
\begin{align}
\label{quiver_L121Z2}
\begin{tikzpicture}[<->]
\draw[very thick] ( 0, 3) circle (3mm); \node at (0, 3) {1};
\draw[very thick] ( 2.6, 1.5) circle (3mm); \node at ( 2.6, 1.5) {2};
\draw[very thick] ( 2.6,-1.5) circle (3mm); \node at ( 2.6,-1.5) {3};
\draw[very thick] ( 0,-3) circle (3mm); \node at ( 0,-3) {4};
\draw[very thick] (-2.6,-1.5) circle (3mm); \node at (-2.6,-1.5) {5};
\draw[very thick] (-2.6, 1.5) circle (3mm); \node at (-2.6, 1.5) {6};
\draw[-{Stealth[length=6pt,width=5pt]}, line width=0.9pt, shorten <=3mm, shorten >=3mm] ( 0, 3) -- ( 2.6, 1.5);
\draw[-{Stealth[length=6pt,width=5pt]}, line width=0.9pt, shorten <=3mm, shorten >=3mm] ( 0, 3) -- ( 0,-3);
\draw[-{Stealth[length=6pt,width=5pt]}, line width=0.9pt, shorten <=3mm, shorten >=3mm] ( 0, 3) -- (-2.6, 1.5);
\draw[-{Stealth[length=6pt,width=5pt]}, line width=0.9pt, shorten <=3mm, shorten >=3mm] ( 2.6, 1.5) -- ( 0,-3);
\draw[-{Stealth[length=6pt,width=5pt]}, line width=0.9pt, shorten <=3mm, shorten >=3mm] ( 2.6, 1.5) -- (-2.6, 1.5);
\draw[-{Stealth[length=6pt,width=5pt]}, line width=0.9pt, shorten <=3mm, shorten >=3mm] ( 2.6,-1.5) -- ( 0, 3);
\draw[-{Stealth[length=6pt,width=5pt]}, line width=0.9pt, shorten <=3mm, shorten >=3mm] ( 2.6,-1.5) -- ( 2.6, 1.5);
\draw[-{Stealth[length=6pt,width=5pt]}, line width=0.9pt, shorten <=3mm, shorten >=3mm] ( 0,-3) -- ( 0, 3);
\draw[-{Stealth[length=6pt,width=5pt]}, line width=0.9pt, shorten <=3mm, shorten >=3mm] ( 0,-3) -- ( 2.6,-1.5);
\draw[-{Stealth[length=6pt,width=5pt]}, line width=0.9pt, shorten <=3mm, shorten >=3mm] ( 0,-3) -- (-2.6,-1.5);
\draw[-{Stealth[length=6pt,width=5pt]}, line width=0.9pt, shorten <=3mm, shorten >=3mm] (-2.6,-1.5) -- ( 0, 3);
\draw[-{Stealth[length=6pt,width=5pt]}, line width=0.9pt, shorten <=3mm, shorten >=3mm] (-2.6,-1.5) -- ( 2.6,-1.5);
\draw[-{Stealth[length=6pt,width=5pt]}, line width=0.9pt, shorten <=3mm, shorten >=3mm] (-2.6, 1.5) -- ( 0,-3);
\draw[-{Stealth[length=6pt,width=5pt]}, line width=0.9pt, shorten <=3mm, shorten >=3mm] (-2.6, 1.5) -- (-2.6,-1.5);
\end{tikzpicture}. 
\end{align}
The superpotential is 
\begin{align}
\mathcal{W}
&=\Tr (X_{41}X_{16}X_{64}+X_{14}X_{43}X_{31}+X_{53}X_{32}X_{24}X_{45}+X_{65}X_{51}X_{12}X_{26}
\nonumber\\
&-X_{41}X_{12}X_{24}-X_{14}X_{45}X_{51}-X_{53}X_{31}X_{16}X_{65}-X_{64}X_{43}X_{32}X_{26}). 
\end{align}
The theory admits a holographic dual description in terms of Type IIB string theory on $AdS_5\times L^{1,2,1}/\mathbb{Z}_2$. 
The toric diagram for the $\textrm{SPP}/\mathbb{Z}_2$ is generated by
\begin{align}
v_1&=(1,0,0),\qquad 
v_2=(1,1,0), \qquad 
v_3=(1,2,0), \nonumber\\
v_4&=(1,2,1), \qquad
v_5=(1,2,2), \qquad 
v_6=(1,0,1). 
\end{align}
It is shown as
\begin{align}
\label{toric_L121Z2}
\begin{tikzpicture}[scale=1, baseline={(0,0)}]
  \fill (0,0) circle (2pt) node[below left] {$(0,0)$};
  \fill (1,0) circle (2pt) node[below] {$(1,0)$};
  \fill (2,0) circle (2pt) node[below right] {$(2,0)$};
  \fill (2,1) circle (2pt) node[right] {$(2,1)$};
  \fill (2,2) circle (2pt) node[above right] {$(2,2)$};
  \fill (0,1) circle (2pt) node[left] {$(0,1)$};
  \draw[very thick] (0,0) -- (2,0) -- (2,1) -- (2,2) -- (0,1) -- cycle;
\end{tikzpicture}
\end{align}
The multiplicities of four types of the chiral multiplets 
$U_2$ $=$ $\{X_{32},X_{65}\}$, 
$U_1$ $=$ $\{X_{12},X_{16},$ $X_{43},X_{45}\}$, 
$Y$ $=$ $\{X_{24},X_{31},$ $X_{51},X_{64}\}$, 
$Z$ $=$ $\{X_{26},X_{53}\}$ 
forming the baryonic operators are evaluated as
\begin{align}
\det (v_{6},v_{1},v_{3})&=2, \\
\det (v_{1},v_{3},v_{5})&=4, \\
\det (v_{3},v_{5},v_{6})&=4, \\
\det (v_{5},v_{6},v_{1})&=2.
\end{align}
The area of the polygon is
\begin{align}
\label{A_L121Z2}
A&=3. 
\end{align}
The dual cone is generated by
\begin{align}
\label{dualG_L121Z2}
m_1&=(0,0,1),& 
m_2&=(0,0,1),&
m_3&=(2,-1,0),\nonumber\\
m_4&=(2,-1,0),&
m_5&=(2,1,-2),&  
m_6&=(0,1,0). 
\end{align}
The Reeb vector is
\begin{align}
b&=\left(3,2\sqrt{3},\frac{3+\sqrt{3}}{2}\right). 
\end{align}
The volume of $L^{1,2,1}/\mathbb{Z}_2$ is 
\begin{align}
\label{vol_L121Z2}
\mathrm{Vol}(L^{1,2,1}/\mathbb{Z}_2)
&=\frac{\sqrt{3}\pi^3}{9}=5.96716\ldots. 
\end{align}
This is equal to one half of the volume (\ref{vol_L121}) of $L^{1,2,1}$. 
The R-charges of the chiral multiplet fields are 
\begin{align}
R(U_{2})&=\frac{\pi}{3}\frac{\mathrm{Vol}(\Sigma_1)}{\mathrm{Vol}(L^{1,2,1}/\mathbb{Z}_2)}=1-\frac{1}{\sqrt{3}}\\
R(U_1)&=\frac{\pi}{3}\frac{\mathrm{Vol}(\Sigma_3)}{\mathrm{Vol}(L^{1,2,1}/\mathbb{Z}_2)}=\frac{1}{\sqrt{3}}, \\
R(Y)&=\frac{\pi}{3}\frac{\mathrm{Vol}(\Sigma_5)}{\mathrm{Vol}(L^{1,2,1}/\mathbb{Z}_2)}=\frac{1}{\sqrt{3}}, \\
R(Z)&=\frac{\pi}{3}\frac{\mathrm{Vol}(\Sigma_6)}{\mathrm{Vol}(L^{1,2,1}/\mathbb{Z}_2)}=1-\frac{1}{\sqrt{3}}, 
\end{align}
where 
\begin{align}
\mathrm{Vol}(\Sigma_1)&=2\pi^2 \frac{\det(v_{6},v_{1},v_{3})}{\det(b,v_{6},v_1)\det(b,v_{1},v_{3})}
=\frac{2\pi^2}{3(1+\sqrt{3})}, \\
\mathrm{Vol}(\Sigma_3)&=2\pi^2 \frac{\det(v_{1},v_{3},v_{5})}{\det(b,v_{1},v_3)\det(b,v_{3},v_{5})}
=\frac{\pi^2}{3}, \\
\mathrm{Vol}(\Sigma_5)&=2\pi^2 \frac{\det(v_{3},v_{5},v_{6})}{\det(b,v_{3},v_5)\det(b,v_{5},v_{6})}
=\frac{\pi^2}{3}, \\
\mathrm{Vol}(\Sigma_6)&=2\pi^2 \frac{\det(v_{5},v_{6},v_{1})}{\det(b,v_{5},v_6)\det(b,v_{6},v_{1})}
=\frac{2\pi^2}{3(1+\sqrt{3})}.
\end{align}

From the field content (\ref{field_L121Z2}) we obtain the weighted adjacency matrix 
\begin{align}
\label{M_L121Z2}
&
M^{AdS_5\times L^{1,2,1}/\mathbb{Z}_2}
\nonumber\\
&=\left(
\begin{matrix}
i_v&i_{\frac{1}{\sqrt{3}}}&\tilde{i}_{\frac{1}{\sqrt{3}}}&{i}_{2(1-\frac{1}{\sqrt{3}})}+\tilde{i}_{2(1-\frac{1}{\sqrt{3}})}&\tilde{i}_{\frac{1}{\sqrt{3}}}&i_{\frac{1}{\sqrt{3}}} \\
\tilde{i}_{\frac{1}{\sqrt{3}}}&i_v&\tilde{i}_{1-\frac{1}{\sqrt{3}}}&i_{\frac{1}{\sqrt{3}}}&0&i_{1-\frac{1}{\sqrt{3}}} \\
i_{\frac{1}{\sqrt{3}}}&i_{1-\frac{1}{\sqrt{3}}}&i_v&\tilde{i}_{\frac{1}{\sqrt{3}}}&\tilde{i}_{1-\frac{1}{\sqrt{3}}}&0 \\
{i}_{2(1-\frac{1}{\sqrt{3}})}+\tilde{i}_{2(1-\frac{1}{\sqrt{3}})}&\tilde{i}_{\frac{1}{\sqrt{3}}}&i_{\frac{1}{\sqrt{3}}}&i_v&i_{\frac{1}{\sqrt{3}}}&\tilde{i}_{\frac{1}{\sqrt{3}}} \\
i_{\frac{1}{\sqrt{3}}}&0&i_{1-\frac{1}{\sqrt{3}}}&\tilde{i}_{\frac{1}{\sqrt{3}}}&i_v&\tilde{i}_{1-\frac{1}{\sqrt{3}}} \\
\tilde{i}_{\frac{1}{\sqrt{3}}}&\tilde{i}_{1-\frac{1}{\sqrt{3}}}&0&i_{\frac{1}{\sqrt{3}}}&i_{1-\frac{1}{\sqrt{3}}}&i_v \\
\end{matrix}
\right). 
\end{align}
We then find the gravity index 
\begin{align}
\label{ind_L121Z2}
&
i^{AdS_5\times L^{1,2,1}/\mathbb{Z}_2}(p;q)
\nonumber\\
&=\frac{(pq)^{1+\frac{1}{\sqrt{3}}}}{1-(pq)^{1+\frac{1}{\sqrt{3}}}}
+\frac{(pq)^{\frac{2}{\sqrt{3}}}}{1-(pq)^{\frac{2}{\sqrt{3}}}}
+2\frac{(pq)^{\frac{3+\sqrt{3}}{6}}}{1-(pq)^{\frac{3+\sqrt{3}}{6}}}
+2\frac{(pq)^{\frac{2(3-\sqrt{3})}{3}}}{1-(pq)^{\frac{2(3-\sqrt{3})}{3}}}. 
\end{align}
It enumerates six extremal BPS mesonic operators, 
corresponding to the primitive generators (\ref{dualG_L121Z2}) of the dual cone. 
The R-charges are computed from the formula (\ref{Rch_meson})
\begin{align}
R(\mathcal{M}_1)&=\frac23(b,m_1)=1+\frac{1}{\sqrt{3}}, \\
R(\mathcal{M}_2)&=\frac23(b,m_2)=1+\frac{1}{\sqrt{3}}, \\
R(\mathcal{M}_3)&=\frac23(b,m_3)=4\left(1-\frac{1}{\sqrt{3}}\right), \\
R(\mathcal{M}_4)&=\frac23(b,m_4)=4\left(1-\frac{1}{\sqrt{3}}\right), \\
R(\mathcal{M}_5)&=\frac23(b,m_5)=2\left(1+\frac{1}{\sqrt{3}}\right), \\ 
R(\mathcal{M}_6)&=\frac23(b,m_6)=\frac{4}{\sqrt{3}}.  
\end{align}

We find the supersymmetric zeta function 
\begin{align}
\label{zeta_L121Z2}
&
\mathfrak{Z}^{AdS_5\times L^{1,2,1}/\mathbb{Z}_2}(s,z;\omega_1,\omega_2)
\nonumber\\
&=
\left( \left(1+\frac{1}{\sqrt{3}}\right) (\omega_1+\omega_2) \right)^{-s}
\zeta\left(s,1+\frac{z}{\left(1+\frac{1}{\sqrt{3}}\right)(\omega_1+\omega_2)}\right)
\nonumber\\
&+\left( \frac{2}{\sqrt{3}} (\omega_1+\omega_2) \right)^{-s}
\zeta\left(s,1+\frac{\sqrt{3}z}{2(\omega_1+\omega_2)}\right)
\nonumber\\
&+2\left( \frac{3+\sqrt{3}}{6} (\omega_1+\omega_2) \right)^{-s}
\zeta\left(s,1+\frac{6z}{\left( 3+\sqrt{3} \right)(\omega_1+\omega_2)}\right)
\nonumber\\
&+2\left( \frac{2(3-\sqrt{3})}{3} (\omega_1+\omega_2) \right)^{-s}
\zeta\left(s,1+\frac{3z}{2\left( 3-\sqrt{3} \right)(\omega_1+\omega_2)}\right). 
\end{align}
The supersymmetric zeta function admits a simple pole at $s=1$ with the residue
\begin{align}
\label{Res_L121Z2}
\mathrm{Res}_{s=1}\mathfrak{Z}^{AdS_5\times L^{1,2,1}/\mathbb{Z}_2}(s,z;\omega_1,\omega_2)
&=\frac{3(6-\sqrt{3})}{2}\frac{1}{\omega_1+\omega_2}. 
\end{align}
Substituting the residue coefficient (\ref{Res_L121Z2}) and the volume (\ref{vol_L121Z2}) into the formula (\ref{Riem2_conj}), 
we obtain the integrated curvature invariant 
\begin{align}
\label{Riem2_L121Z2}
\int_{L^{1,2,1}/\mathbb{Z}_2}\mathrm{Riem}^2
&=\frac{8(216-55\sqrt{3})\pi^3}{9}
\nonumber\\
&=3327.65\ldots. 
\end{align}
We note that this value is not obtained by simply dividing the result (\ref{Riem2_L121}) for $L^{1,2,1}$ by two. 
Rather, it contains additional contributions arising from the singularities. 
We find that the value (\ref{Riem2_L121Z2}) fully agrees with the Hilbert series result (see Appendix \ref{app_HS}). 
The Zeta-index is
\begin{align}
\label{0_L121Z2}
\mathfrak{Z}^{AdS_5\times L^{1,2,1}/\mathbb{Z}_2}(0,0;\omega_1,\omega_2)
&=-3. 
\end{align}
Also we find the supersymmetric zeta value with $s=-1$ and $z=0$ 
\begin{align}
\label{-1_L121Z2}
\mathfrak{Z}^{AdS_5\times L^{1,2,1}/\mathbb{Z}_2}(-1,0;\omega_1,\omega_2)
&=-\frac12 (\omega_1+\omega_2). 
\end{align}
According to the formulae (\ref{c0_formula}) and (\ref{a0_formula}) we obtain
\begin{align}
c_{\mathcal{O}(N^0)}&=-\frac34, \\
a_{\mathcal{O}(N^0)}&=-\frac98, 
\end{align}
and 
\begin{align}
c_{\mathcal{O}(N^0)}-a_{\mathcal{O}(N^0)}&=\frac38. 
\end{align}

The supersymmetric determinant is
\begin{align}
\label{det_L121Z2}
&
\mathfrak{D}^{AdS_5\times L^{1,2,1}/\mathbb{Z}_2}(z;\omega_1,\omega_2)
\nonumber\\
&=\frac{
\left(1+\frac{1}{\sqrt{3}}\right)^{\frac{3z}{(3+\sqrt{3})(\omega_1+\omega_2)}}
\left(3-\sqrt{3}\right)^{\frac{3z}{(3-\sqrt{3})(\omega_1+\omega_2)}}
\left(3+\sqrt{3}\right)^{\frac12+\frac{12z}{(3+\sqrt{3})(\omega_1+\omega_2)}}
(\omega_1+\omega_2)^{3+\frac{3(6-\sqrt{3})z}{2(\omega_1+\omega_2)}}
}{\pi^3 2^{\frac32\left(1+\frac{(3-2\sqrt{3})z}{\omega_1+\omega_2}\right)} 3^{\frac74+\frac{5(6-\sqrt{3})z}{4(\omega_1+\omega_2)}}}
\nonumber\\
&\times 
\Gamma\left(1+\frac{\sqrt{3}z}{2(\omega_1+\omega_2)}\right)
\Gamma\left(1+\frac{3z}{(3+\sqrt{3})(\omega_1+\omega_2)}\right)
\nonumber\\
&\times 
\Gamma\left(1+\frac{3z}{2(3-\sqrt{3}) (\omega_1+\omega_2)}\right)^2
\Gamma\left(1+ \frac{6z}{(3+\sqrt{3})(\omega_1+\omega_2)}\right)^2. 
\end{align}
The vacuum exponent is
\begin{align}
\label{vac_L121Z2}
\mathfrak{D}^{AdS_5\times L^{1,2,1}/\mathbb{Z}_2}(0;1,1)
&=\frac{2\sqrt{2(1+\sqrt{3})}}{3\sqrt{3}\pi^3}
\nonumber\\
&=0.0290174\ldots. 
\end{align}

\subsection{$AdS_5\times L^{1,3,1}/\mathbb{Z}_2$}
Consider the 4d $\mathcal{N}=1$ quiver gauge theory on a stack of $N$ D3-branes 
probing the Calabi-Yau cone over the Sasaki-Einstein manifold $L^{1,3,1}/\mathbb{Z}_2$ with orbifold action $(0,1,1,1)$ \cite{Franco:2005sm,Butti:2005vn,Hanany:2012hi,Bao:2024nyu}. 
The theory has 8 gauge nodes and 20 chiral multiplets.\footnote{Another description related by the Seiberg duality contains 22 chiral multiplets \cite{Hanany:2012hi}. }
The field content is 
\begin{align}
\begin{array}{c|cc}
&\prod_{i=1}^8 SU(N)_i&U(1)_R\\ \hline 
X_{14}&(\bf{N},\bf{1},\bf{1},\overline{\bf{N}},\bf{1},\bf{1},\bf{1},\bf{1})&y \\
X_{17}&(\bf{N},\bf{1},\bf{1},\bf{1},\bf{1},\bf{1},\overline{\bf{N}},\bf{1})&y \\
X_{18}&(\bf{N},\bf{1},\bf{1},\bf{1},\bf{1},\bf{1},\bf{1},\overline{\bf{N}})&2x \\
X_{24}&(\bf{1},\bf{N},\bf{1},\overline{\bf{N}},\bf{1},\bf{1},\bf{1},\bf{1})&x \\
X_{27}&(\bf{1},\bf{N},\bf{1},\bf{1},\bf{1},\bf{1},\overline{\bf{N}},\bf{1})&y \\
X_{31}&(\overline{\bf{N}},\bf{1},\bf{N},\bf{1},\bf{1},\bf{1},\bf{1},\bf{1})&y \\
X_{32}&(\bf{1},\overline{\bf{N}},\bf{N},\bf{1},\bf{1},\bf{1},\bf{1},\bf{1})&y \\
X_{37}&(\bf{1},\bf{1},\bf{N},\bf{1},\bf{1},\bf{1},\overline{\bf{N}},\bf{1})&2x \\
X_{45}&(\bf{1},\bf{1},\bf{1},\bf{N},\overline{\bf{N}},\bf{1},\bf{1},\bf{1})&x \\
X_{48}&(\bf{1},\bf{1},\bf{1},\bf{N},\bf{1},\bf{1},\bf{1},\overline{\bf{N}})&y \\
X_{53}&(\bf{1},\bf{1},\overline{\bf{N}},\bf{1},\bf{N},\bf{1},\bf{1},\bf{1})&y \\
X_{56}&(\bf{1},\bf{1},\bf{1},\bf{1},\bf{N},\overline{\bf{N}},\bf{1},\bf{1})&x \\
X_{61}&(\overline{\bf{N}},\bf{1},\bf{1},\bf{1},\bf{1},\bf{N},\bf{1},\bf{1})&y \\
X_{62}&(\bf{1},\overline{\bf{N}},\bf{1},\bf{1},\bf{1},\bf{N},\bf{1},\bf{1})&x \\
X_{73}&(\bf{1},\bf{1},\overline{\bf{N}},\bf{1},\bf{1},\bf{1},\bf{N},\bf{1})&2x \\
X_{75}&(\bf{1},\bf{1},\bf{1},\bf{1},\overline{\bf{N}},\bf{1},\bf{N},\bf{1})&y \\
X_{78}&(\bf{1},\bf{1},\bf{1},\bf{1},\bf{1},\bf{1},\bf{N},\overline{\bf{N}})&y \\
X_{81}&(\overline{\bf{N}},\bf{1},\bf{1},\bf{1},\bf{1},\bf{1},\bf{1},\bf{N})&2x \\
X_{83}&(\bf{1},\bf{1},\overline{\bf{N}},\bf{1},\bf{1},\bf{1},\bf{1},\bf{N})&y \\
X_{86}&(\bf{1},\bf{1},\bf{1},\bf{1},\bf{1},\overline{\bf{N}},\bf{1},\bf{N})&y \\
\end{array}
\end{align}
where
\begin{align}
x&=\frac16(5-\sqrt{7}), \qquad 
y=\frac{1}{6}(1+\sqrt{7}). 
\end{align}
The quiver diagram is
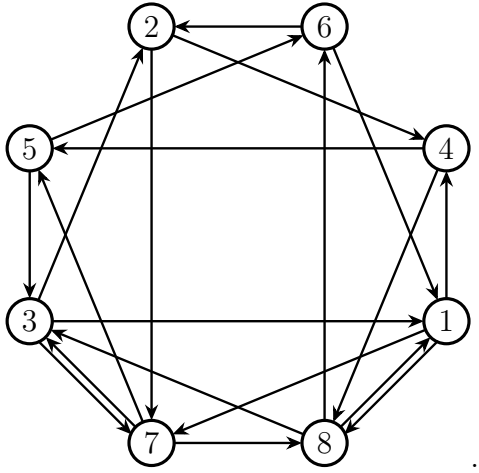
\begin{align}
\label{quiver_L131Z2}
\begin{tikzpicture}[<->]
\draw[very thick] (2.77,-1.15) circle (3mm);
\node at (2.77,-1.15) {1};
\draw[very thick] (-1.15,2.77) circle (3mm);
\node at (-1.15,2.77) {2};
\draw[very thick] (-2.77,-1.15) circle (3mm);
\node at (-2.77,-1.15) {3};
\draw[very thick] (2.77,1.15) circle (3mm);
\node at (2.77,1.15) {4};
\draw[very thick] (-2.77,1.15) circle (3mm);
\node at (-2.77,1.15) {5};
\draw[very thick] (1.15,2.77) circle (3mm);
\node at (1.15,2.77) {6};
\draw[very thick] (-1.15,-2.77) circle (3mm);
\node at (-1.15,-2.77) {7};
\draw[very thick] (1.15,-2.77) circle (3mm);
\node at (1.15,-2.77) {8};
\draw[-{Stealth[length=6pt,width=5pt]}, line width=0.9pt, shorten <=3mm, shorten >=3mm] (2.77,-1.15) -- (2.77,1.15);      
\draw[-{Stealth[length=6pt,width=5pt]}, line width=0.9pt, shorten <=3mm, shorten >=3mm] (2.77,-1.15) -- (-1.15,-2.77);   
\draw[-{Stealth[length=6pt,width=5pt]}, line width=0.9pt, shorten <=3mm, shorten >=3mm] (2.87,-1.2) -- (1.2,-2.87);    
\draw[-{Stealth[length=6pt,width=5pt]}, line width=0.9pt, shorten <=3mm, shorten >=3mm] (-1.15,2.77) -- (2.77,1.15);     
\draw[-{Stealth[length=6pt,width=5pt]}, line width=0.9pt, shorten <=3mm, shorten >=3mm] (-1.15,2.77) -- (-1.15,-2.77);   
\draw[-{Stealth[length=6pt,width=5pt]}, line width=0.9pt, shorten <=3mm, shorten >=3mm] (-2.77,-1.15) -- (2.77,-1.15);   
\draw[-{Stealth[length=6pt,width=5pt]}, line width=0.9pt, shorten <=3mm, shorten >=3mm] (-2.77,-1.15) -- (-1.15,2.77);   
\draw[-{Stealth[length=6pt,width=5pt]}, line width=0.9pt, shorten <=3mm, shorten >=3mm] (-2.87,-1.2) -- (-1.2,-2.87);  
\draw[-{Stealth[length=6pt,width=5pt]}, line width=0.9pt, shorten <=3mm, shorten >=3mm] (2.77,1.15) -- (-2.77,1.15);     
\draw[-{Stealth[length=6pt,width=5pt]}, line width=0.9pt, shorten <=3mm, shorten >=3mm] (2.77,1.15) -- (1.15,-2.77);     
\draw[-{Stealth[length=6pt,width=5pt]}, line width=0.9pt, shorten <=3mm, shorten >=3mm] (-2.77,1.15) -- (-2.77,-1.15);   
\draw[-{Stealth[length=6pt,width=5pt]}, line width=0.9pt, shorten <=3mm, shorten >=3mm] (-2.77,1.15) -- (1.15,2.77);     
\draw[-{Stealth[length=6pt,width=5pt]}, line width=0.9pt, shorten <=3mm, shorten >=3mm] (1.15,2.77) -- (2.77,-1.15);     
\draw[-{Stealth[length=6pt,width=5pt]}, line width=0.9pt, shorten <=3mm, shorten >=3mm] (1.15,2.77) -- (-1.15,2.77);      
\draw[-{Stealth[length=6pt,width=5pt]}, line width=0.9pt, shorten <=3mm, shorten >=3mm] (-1.15,-2.77) -- (-2.77,-1.15);  
\draw[-{Stealth[length=6pt,width=5pt]}, line width=0.9pt, shorten <=3mm, shorten >=3mm] (-1.15,-2.77) -- (-2.77,1.15);   
\draw[-{Stealth[length=6pt,width=5pt]}, line width=0.9pt, shorten <=3mm, shorten >=3mm] (-1.15,-2.77) -- (1.15,-2.77);   
\draw[-{Stealth[length=6pt,width=5pt]}, line width=0.9pt, shorten <=3mm, shorten >=3mm] (1.15,-2.77) -- (2.77,-1.15);    
\draw[-{Stealth[length=6pt,width=5pt]}, line width=0.9pt, shorten <=3mm, shorten >=3mm] (1.15,-2.77) -- (-2.77,-1.15);   
\draw[-{Stealth[length=6pt,width=5pt]}, line width=0.9pt, shorten <=3mm, shorten >=3mm] (1.15,-2.77) -- (1.15,2.77);     
\end{tikzpicture}. 
\end{align}
The superpotential takes the form \cite{Hanany:2012hi}
\begin{align}
\mathcal{W}&=\Tr
(
X_{31}X_{18}X_{83}+X_{32}X_{27}X_{73}+X_{53}X_{37}X_{75}+X_{78}X_{81}X_{17}
\nonumber\\
&-X_{14}X_{48}X_{81}-X_{31}X_{17}X_{73}-X_{78}X_{83}X_{37}-X_{86}X_{61}X_{18}
\nonumber\\
&+X_{14}X_{45}X_{56}X_{61}+X_{62}X_{24}X_{48}X_{86}
-X_{32}X_{24}X_{45}X_{53}-X_{62}X_{27}X_{75}X_{56}
). 
\end{align}
The theory is holographically dual to Type IIB string theory on $AdS_5\times L^{1,3,1}/\mathbb{Z}_2$. 
The toric diagram for the Calabi-Yau cone is generated by
\begin{align}
v_1&=(1,0,0),\qquad 
v_2=(1,1,0), \qquad 
v_3=(1,2,0), \qquad 
v_4=(1,2,1), \nonumber\\
v_5&=(1,1,2), \qquad 
v_6=(1,0,3), \qquad 
v_7=(1,0,2), \qquad 
v_8=(1,0,1). 
\end{align}
It is depicted as
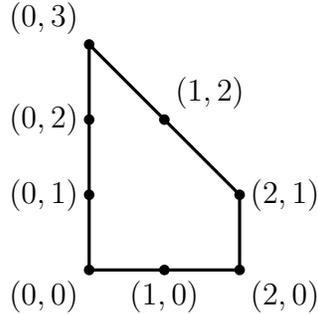
\begin{align}
\label{toric_L131Z2}
\begin{tikzpicture}[scale=1, baseline={(0,0)}]
  \fill (0,0) circle (2pt) node[below left] {$(0,0)$};
  \fill (1,0) circle (2pt) node[below] {$(1,0)$};
  \fill (2,0) circle (2pt) node[below right] {$(2,0)$};
  \fill (2,1) circle (2pt) node[right] {$(2,1)$};
  \fill (1,2) circle (2pt) node[above right] {$(1,2)$};
  \fill (0,3) circle (2pt) node[above left] {$(0,3)$};
  \fill (0,2) circle (2pt) node[left] {$(0,2)$};
  \fill (0,1) circle (2pt) node[left] {$(0,1)$};
  \draw[very thick] (0,0) -- (2,0) -- (2,1) -- (1,2) -- (0,3) -- (0,1) -- cycle;
\end{tikzpicture}
\end{align}
The multiplicities of four types of the chiral multiplets 
$U_1$ $=$ $\{X_{14},$ $X_{27},$ $X_{31},$ $X_{53},$ $X_{78},$ $X_{86}\}$, 
$U_2$ $=$ $\{X_{24},X_{45}\}$, 
$Z$ $=$ $\{X_{56},X_{62}\}$, 
$Y$ $=$ $\{X_{17},X_{32},X_{48},X_{61},X_{75},X_{83}\}$ 
forming the baryonic operators are evaluated as
\begin{align}
\det (v_{6},v_{1},v_{3})&=6, \\
\det (v_{1},v_{3},v_{4})&=2, \\
\det (v_{3},v_{4},v_{6})&=2, \\
\det (v_{4},v_{6},v_{1})&=6.
\end{align}
The area of the polygon is
\begin{align}
\label{A_L131Z2}
A&=4. 
\end{align}
The dual cone is generated by
\begin{align}
\label{dualG_L131Z2}
m_1&=(0,0,1),& 
m_2&=(0,0,1),&
m_3&=(2,-1,0),& 
m_4&=(3,-1,-1),\nonumber\\
m_5&=(3,-1,-1),&  
m_6&=(0,1,0), &
m_7&=(0,1,0), &
m_8&=(0,1,0). 
\end{align}
The Reeb vector is
\begin{align}
b&=\left(3,5-\sqrt{7},\frac{4+\sqrt{7}}{2}\right). 
\end{align}
The volume of $L^{1,3,1}/\mathbb{Z}_2$ is evaluated as
\begin{align}
\label{vol_L131Z2}
\mathrm{Vol}(L^{1,3,1}/\mathbb{Z}_2)
&=\frac{4 \pi^3}{10+7\sqrt{7}}. 
\end{align}
Note that this is exactly one half of the volume of $L^{1,3,1}$, which is evaluated from (\ref{vol_Laba}). 
The R-charges of the chiral multiplet fields are 
\begin{align}
R(U_{1})&=\frac{\pi}{3}\frac{\mathrm{Vol}(\Sigma_1)}{\mathrm{Vol}(L^{1,3,1}/\mathbb{Z}_2)}=\frac{1+\sqrt{7}}{6} \\
R(U_2)&=\frac{\pi}{3}\frac{\mathrm{Vol}(\Sigma_3)}{\mathrm{Vol}(L^{1,3,1}/\mathbb{Z}_2)}=\frac{5-\sqrt{7}}{6}, \\
R(Z)&=\frac{\pi}{3}\frac{\mathrm{Vol}(\Sigma_4)}{\mathrm{Vol}(L^{1,3,1}/\mathbb{Z}_2)}=\frac{5-\sqrt{7}}{6}, \\
R(Y)&=\frac{\pi}{3}\frac{\mathrm{Vol}(\Sigma_6)}{\mathrm{Vol}(L^{1,3,1}/\mathbb{Z}_2)}=\frac{1+\sqrt{7}}{6},
\end{align}
where 
\begin{align}
\mathrm{Vol}(\Sigma_1)&=2\pi^2 \frac{\det(v_{6},v_{1},v_{3})}{\det(b,v_{6},v_1)\det(b,v_{1},v_{3})}
=\frac{4\pi^2}{13+\sqrt{7}}, \\
\mathrm{Vol}(\Sigma_3)&=2\pi^2 \frac{\det(v_{1},v_{3},v_{4})}{\det(b,v_{1},v_3)\det(b,v_{3},v_{4})}
=\frac{4\pi^2}{11+5\sqrt{7}}, \\
\mathrm{Vol}(\Sigma_4)&=2\pi^2 \frac{\det(v_{3},v_{4},v_{6})}{\det(b,v_{3},v_4)\det(b,v_{4},v_{6})}
=\frac{4\pi^2}{11+5\sqrt{7}}, \\
\mathrm{Vol}(\Sigma_6)&=2\pi^2 \frac{\det(v_{4},v_{6},v_{1})}{\det(b,v_{4},v_6)\det(b,v_{6},v_{1})}
=\frac{4\pi^2}{13+\sqrt{7}}.
\end{align}

From the quiver diagram (\ref{quiver_L131Z2}) and the field content, the weighted adjacency matrix reads 
\begin{align}
\label{M_L131Z2}
&
M^{AdS_5\times L^{1,3,1}/\mathbb{Z}_2}
\nonumber\\
&=\left(
\begin{matrix}
i_v&0&\tilde{i}_{\textrm{cm}_y}&i_{\textrm{cm}_y}&0&\tilde{i}_{\textrm{cm}_y}&i_{\textrm{cm}_y}&i_{\textrm{cm}_{2x}}+\tilde{i}_{\textrm{cm}_{2x}} \\
0&i_v&\tilde{i}_{\textrm{cm}_y}&i_{\textrm{cm}_x}&0&\tilde{i}_{\textrm{cm}_x}&i_{\textrm{cm}_y}&0 \\
i_{\textrm{cm}_y}&i_{\textrm{cm}_y}&i_v&0&\tilde{i}_{\textrm{cm}_y}&0&i_{\textrm{cm}_{2x}}+\tilde{i}_{\textrm{cm}_{2x}}&\tilde{i}_{\textrm{cm}_y} \\
\tilde{i}_{\textrm{cm}_y}&\tilde{i}_{\textrm{cm}_x}&0&i_v&i_{\textrm{cm}_x}&0&0&i_{\textrm{cm}_y} \\
0&0&i_{\textrm{cm}_y}&\tilde{i}_{\textrm{cm}_x}&i_v&i_{\textrm{cm}_x}&\tilde{i}_{\textrm{cm}_y}&0 \\
i_{\textrm{cm}_y}&i_{\textrm{cm}_x}&0&0&\tilde{i}_{\textrm{cm}_x}&i_v&0&\tilde{i}_{\textrm{cm}_y} \\
\tilde{i}_{\textrm{cm}_y}&\tilde{i}_{\textrm{cm}_y}&i_{\textrm{cm}_{2x}}+\tilde{i}_{\textrm{cm}_{2x}}&0&i_{\textrm{cm}_y}&0&i_v&i_{\textrm{cm}_y} \\
i_{\textrm{cm}_{2x}}+\tilde{i}_{\textrm{cm}_{2x}}&0&i_{\textrm{cm}_y}&\tilde{i}_{\textrm{cm}_y}&0&i_{\textrm{cm}_y}&\tilde{i}_{\textrm{cm}_y}&i_v \\
\end{matrix}
\right). 
\end{align}
We obtain the gravity index 
\begin{align}
\label{ind_L131Z2}
&
i^{AdS_5\times L^{1,3,1}/\mathbb{Z}_2}(p;q)
\nonumber\\
&=\frac{(pq)^{\frac{1+\sqrt{7}}{3}}}{1-(pq)^{\frac{1+\sqrt{7}}{3}}}
+3\frac{(pq)^{\frac{5-\sqrt{7}}{3}}}{1-(pq)^{\frac{5-\sqrt{7}}{3}}}
+4\frac{(pq)^{\frac{4+\sqrt{7}}{6}}}{1-(pq)^{\frac{4+\sqrt{7}}{6}}}. 
\end{align}
It enumerates eight extremal BPS mesonic operators. 
They correspond to the primitive vectors (\ref{dualG_L131Z2}) of the dual cone. 
In fact, the R-charges are obtained from the formula (\ref{Rch_meson})
\begin{align}
R(\mathcal{M}_1)&=\frac23(b,m_1)=\frac{4+\sqrt{7}}{3}, \\
R(\mathcal{M}_2)&=\frac23(b,m_2)=\frac{4+\sqrt{7}}{3}, \\
R(\mathcal{M}_3)&=\frac23(b,m_3)=\frac{2(1+\sqrt{7})}{3}, \\
R(\mathcal{M}_4)&=\frac23(b,m_4)=\frac{4+\sqrt{7}}{3}, \\
R(\mathcal{M}_5)&=\frac23(b,m_5)=\frac{4+\sqrt{7}}{3}, \\
R(\mathcal{M}_6)&=\frac23(b,m_6)=\frac{2(5-\sqrt{7})}{3}, \\
R(\mathcal{M}_7)&=\frac23(b,m_7)=\frac{2(5-\sqrt{7})}{3}, \\
R(\mathcal{M}_8)&=\frac23(b,m_8)=\frac{2(5-\sqrt{7})}{3}. 
\end{align}

The supersymmetric zeta function is given by
\begin{align}
\label{zeta_L131Z2}
&
\mathfrak{Z}^{AdS_5\times L^{1,3,1}/\mathbb{Z}_2}(s,z;\omega_1,\omega_2)
\nonumber\\
&=
\left(\left(\frac{1+\sqrt{7}}{3}\right)(\omega_1+\omega_2) \right)^{-s}\zeta\left(s,1+\frac{3z}{(1+\sqrt{7}) (\omega_1+\omega_2)}\right)
\nonumber\\
&+3\left(\left(\frac{5-\sqrt{7}}{3}\right)(\omega_1+\omega_2) \right)^{-s}\zeta\left(s,1+\frac{3z}{(5-\sqrt{7}) (\omega_1+\omega_2)}\right)
\nonumber\\
&+4\left(\left(\frac{4+\sqrt{7}}{6}\right)(\omega_1+\omega_2) \right)^{-s}\zeta\left(s,1+\frac{6z}{(4+\sqrt{7}) (\omega_1+\omega_2)}\right). 
\end{align}
It has a simple pole at $s=1$ with the residue
\begin{align}
\label{Res_L131Z2}
\mathrm{Res}_{s=1}\mathfrak{Z}^{AdS_5\times L^{1,3,1}/\mathbb{Z}_2}(s,z;\omega_1,\omega_2)
&=\frac{38-5\sqrt{7}}{3(\omega_1+\omega_2)}. 
\end{align}
Plugging the residue coefficient (\ref{Res_L131Z2}) and the volume (\ref{vol_L131Z2}) into the formula (\ref{Riem2_conj}), 
we find that the integral of the squared Riemann tensor is given by
\begin{align}
\label{Riem2_L131Z2}
\int_{L^{1,3,1}/\mathbb{Z}_2}\mathrm{Riem}^2
&=\frac{32(2242-403\sqrt{7})}{243}\pi^3
\nonumber\\
&=4800.79\ldots. 
\end{align}
It is important to stress that this result is not merely one half of the integrated curvature invariant for $L^{1,3,1}$. 
There exist extra localized contributions associated with the singular loci. 
The computed value (\ref{Riem2_L131Z2}) is found to be in complete accordance with the Hilbert series calculation \cite{Bao:2024nyu} 
(also see Appendix \ref{app_HS}). 
The Zeta-index is
\begin{align}
\label{0_L131Z2}
\mathfrak{Z}^{AdS_5\times L^{1,3,1}/\mathbb{Z}_2}(0,0;\omega_1,\omega_2)
&=-4. 
\end{align}
The supersymmetric zeta value with $s=-1$ and $z=0$ is given by
\begin{align}
\label{-1_L131Z2}
\mathfrak{Z}^{AdS_5\times L^{1,3,1}/\mathbb{Z}_2}(-1,0;\omega_1,\omega_2)
&=-\frac23 (\omega_1+\omega_2). 
\end{align}
From the formulae (\ref{c0_formula}) and (\ref{a0_formula}) we obtain
\begin{align}
c_{\mathcal{O}(N^0)}&=-1, \\
a_{\mathcal{O}(N^0)}&=-\frac32
\end{align}
and 
\begin{align}
c_{\mathcal{O}(N^0)}-a_{\mathcal{O}(N^0)}&=\frac12. 
\end{align}

The supersymmetric determinant is
\begin{align}
\label{det_L131Z2}
&
\mathfrak{D}^{AdS_5\times L^{1,3,1}/\mathbb{Z}_2}(z;\omega_1,\omega_2)
\nonumber\\
&=\frac{
(5-\sqrt{7})^{\frac32+\frac{9z}{(5-\sqrt{7})(\omega_1+\omega_2)}}
(1+\sqrt{7})^{\frac12+\frac{3z}{(1+\sqrt{7})(\omega_1+\omega_2)}}
(4+\sqrt{7})^{2+\frac{24z}{(4+\sqrt{7})(\omega_1+\omega_2)}}
}
{2^{6+\frac{24z}{(4+\sqrt{7})(\omega_1+\omega_2)}} 3^{4+\frac{9(5+8\sqrt{7})z}{(10+7\sqrt{7})(\omega_1+\omega_2)}}\pi^4}
\nonumber\\
&\times 
(\omega_1+\omega_2)^{4+\frac{9(5+8\sqrt{7})z}{(10+7\sqrt{7})(\omega_1+\omega_2)}}
\Gamma\left(1+\frac{3z}{(1+\sqrt{7})(\omega_1+\omega_2)}\right)
\nonumber\\
&\times 
\Gamma\left(1+\frac{3z}{(5-\sqrt{7})(\omega_1+\omega_2)}\right)^3
\Gamma\left(1+\frac{6z}{(4+\sqrt{7})(\omega_1+\omega_2)}\right)^4. 
\end{align}
The vacuum exponent is
\begin{align}
\label{vac_L131Z2}
\mathfrak{D}^{AdS_5\times L^{1,3,1}/\mathbb{Z}_2}(0;1,1)
&=\frac{\sqrt{11290+4501\sqrt{7}}}{162\pi^4}
\nonumber\\
&=0.00965196\ldots. 
\end{align}

\section{Sasaki-Einstein manifolds from (Pseudo) del Pezzo surfaces}
\label{sec_dP}
We now turn to other Sasaki-Einstein manifolds 
that can be realized as circle bundles over (Pseudo) del Pezzo surfaces. 
The del Pezzo surface $dP_k$ is obtained by blowing up $\mathbb{P}^2$ at $k\le 8$ points in generic position. 
It is a complex surface whose anticanonical divisor $-K$ is \textit{ample}, 
meaning that it has strictly positive intersection with every curve on the surface \cite{MR463157}. 
In addition to these geometries one may consider Pseudo del Pezzo surfaces $PdP_k$, 
which arise when the blow-up points satisfy special relations in such a way that 
additional curves appear and the anticanonical divisor $-K$ is no longer ample but still has non-negative intersection with all curves, 
equivalently, $-K$ is \textit{nef} but not ample \cite{MR463157}.  
Such surfaces are known as weak Fano surfaces in algebraic geometry. 
We examine the geometries associated with 
$dP_2$, $dP_3$, $PdP_{3a}$, $PdP_{4a}$, $PdP_{4b}$ and $PdP_5$, 
which lie outside the toric infinite families discussed in other sections.\footnote{The cone over $dP_0$ $=$ $\mathbb{P}^2$ corresponds to $S^5/\mathbb{Z}_3$, that over $dP_1$ $=$ $F_1$ to $Y^{2,1}$ and that over $PdP_1$ to $S^5/\mathbb{Z}_4$. }

\subsection{$AdS_5\times Y_{dP_2}$}
\label{sec_dP2}
Let us consider the 4d $\mathcal{N}=1$ quiver gauge theory living on a stack of $N$ D3-branes at the tip of the complex cone over the second del Pezzo surface $dP_2$. 
The theory has five $SU(N)$ gauge nodes and bifundamental chiral multiplets \cite{Feng:2000mi} 
(also see e.g. \cite{Feng:2002zw,Franco:2005rj,Bertolini:2004xf,Forcella:2008bb,Forcella:2008ng,Hanany:2012hi}).  
There are two descriptions, Model I and Model II related by the Seiberg duality. 
For Model I the field content and R-charges are given by \cite{Bertolini:2004xf}
\begin{align}
\begin{array}{c|cc}
&\prod_{i=1}^5SU(N)_i&U(1)_R\\ \hline 
X_{15}^1&(\bf{N},\bf{1},\bf{1},\bf{1},\overline{\bf{N}})&\frac{1}{16}(17-\sqrt{33}) \\
X_{15}^2&(\bf{N},\bf{1},\bf{1},\bf{1},\overline{\bf{N}})&\frac{1}{16}(-21+5\sqrt{33}) \\
X_{25}^1&(\bf{1},\bf{N},\bf{1},\bf{1},\overline{\bf{N}})&\frac{1}{16}(17-\sqrt{33}) \\
X_{25}^2&(\bf{1},\bf{N},\bf{1},\bf{1},\overline{\bf{N}})&\frac{1}{16}(-21+5\sqrt{33}) \\
X_{31}&(\overline{\bf{N}},\bf{1},\bf{N},\bf{1},\bf{1})&\frac{1}{16}(-21+5\sqrt{33}) \\
X_{32}&(\bf{1},\overline{\bf{N}},\bf{N},\bf{1},\bf{1})&\frac{1}{16}(-21+5\sqrt{33}) \\
X_{34}&(\bf{1},\bf{1},\bf{N},\overline{\bf{N}},\bf{1})&\frac{1}{2}(-5+\sqrt{33}) \\
X_{41}&(\overline{\bf{N}},\bf{1},\bf{1},\bf{N},\bf{1})&\frac{3}{16}(19-3\sqrt{33}) \\
X_{42}&(\bf{1},\overline{\bf{N}},\bf{1},\bf{N},\bf{1})&\frac{3}{16}(19-3\sqrt{33}) \\
X_{53}^{\alpha}&(\bf{1},\bf{1},\overline{\bf{N}},\bf{1},\bf{N})&\frac{1}{4}(9-\sqrt{33}) \\
X_{53}^{3}&(\bf{1},\bf{1},\overline{\bf{N}},\bf{1},\bf{N})&\frac{1}{8}(37-5\sqrt{33}) \\
X_{54}&(\bf{1},\bf{1},\bf{1},\overline{\bf{N}},\bf{N})&\frac{1}{8}(-21+5\sqrt{33}) \\
\end{array}
\end{align}
and the quiver diagram is
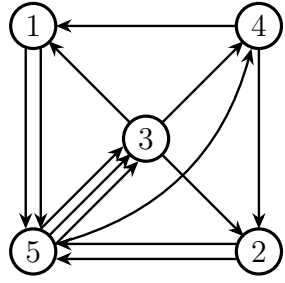
\begin{align}
\begin{tikzpicture}[<->]
    \draw[very thick] (-1.5,3) circle (3mm);
    \node at (-1.5,3) {1};
    \draw[very thick] (1.5,3) circle (3mm);
    \node at (1.5,3) {4};
    \draw[very thick] (0,1.5) circle (3mm);
    \node at (0,1.5) {3};
    \draw[very thick] (1.5,0) circle (3mm);
    \node at (1.5,0) {2};
    \draw[very thick] (-1.5,0) circle (3mm);
    \node at (-1.5,0) {5};
    \draw[-{Stealth[length=6pt,width=5pt]}, line width=0.9pt]
          (-1.6,2.7) to node[]{} (-1.6,0.3);
     \draw[-{Stealth[length=6pt,width=5pt]}, line width=0.9pt]
          (-1.4,2.7) to node[]{} (-1.4,0.3);
     \draw[-{Stealth[length=6pt,width=5pt]}, line width=0.9pt]
          (1.2,0.1) to node[]{} (-1.2,0.1);
     \draw[-{Stealth[length=6pt,width=5pt]}, line width=0.9pt]
          (1.2,-0.1) to node[]{} (-1.2,-0.1);
    \draw[-{Stealth[length=6pt,width=5pt]}, line width=0.9pt]
          (-0.2,1.7) to node[]{} (-1.3,2.8);
    \draw[-{Stealth[length=6pt,width=5pt]}, line width=0.9pt]
          (0.2,1.7) to node[]{} (1.3,2.8); 
    \draw[-{Stealth[length=6pt,width=5pt]}, line width=0.9pt]
          (0.2,1.3) to node[]{} (1.3,0.2);
    \draw[-{Stealth[length=6pt,width=5pt]}, line width=0.9pt]
          (1.2,3) to node[]{} (-1.2,3);
    \draw[-{Stealth[length=6pt,width=5pt]}, line width=0.9pt]
          (1.5,2.7) to node[]{} (1.5,0.3);
    \draw[-{Stealth[length=6pt,width=5pt]}, line width=0.9pt]
          (-1.3,0.2) to node[]{} (-0.2,1.3);
    \draw[-{Stealth[length=6pt,width=5pt]}, line width=0.9pt]
          (-1.4,0.3) to node[]{} (-0.3,1.4);
    \draw[-{Stealth[length=6pt,width=5pt]}, line width=0.9pt]
          (-1.2,0.1) to node[]{} (-0.1,1.2);
    \draw[-{Stealth[length=6pt,width=5pt]}, line width=0.9pt, bend right=30]
         (-1.2,0.1) to node[]{} (1.4,2.7);
\end{tikzpicture}. 
\end{align}
The field content and the R-charges for Model II are \cite{Bertolini:2004xf}
\begin{align}
\begin{array}{c|cc}
&\prod_{i=1}^5SU(N)_i&U(1)_R\\ \hline 
X_{14}&(\bf{N},\bf{1},\bf{1},\overline{\bf{N}},\bf{1})&\frac{3}{16}(19-3\sqrt{33}) \\
X_{15}&(\bf{N},\bf{1},\bf{1},\bf{1},\overline{\bf{N}})&\frac{1}{16}(-21+5\sqrt{33}) \\
X_{23}^1&(\bf{1},\bf{N},\overline{\bf{N}},\bf{1},\bf{1})&\frac{1}{16}(-21+5\sqrt{33}) \\
X_{23}^2&(\bf{1},\bf{N},\overline{\bf{N}},\bf{1},\bf{1})&\frac{1}{16}(17-\sqrt{33}) \\
X_{31}^1&(\overline{\bf{N}},\bf{1},\bf{N},\bf{1},\bf{1})&\frac{1}{16}(-21+5\sqrt{33}) \\
X_{31}^2&(\overline{\bf{N}},\bf{1},\bf{N},\bf{1},\bf{1})&\frac{1}{16}(17-\sqrt{33}) \\
X_{34}&(\bf{1},\bf{1},\bf{N},\overline{\bf{N}},\bf{1})&\frac{1}{4}(9-\sqrt{33}) \\
X_{42}&(\bf{1},\overline{\bf{N}},\bf{1},\bf{N},\bf{1})&\frac{1}{16}(-21+5\sqrt{33}) \\
X_{45}&(\bf{1},\bf{1},\bf{1},\bf{N},\overline{\bf{N}})&\frac{1}{2}(-5+\sqrt{33}) \\
X_{52}&(\bf{1},\overline{\bf{N}},\bf{1},\bf{1},\bf{N})&\frac{3}{16}(19-3\sqrt{33}) \\
X_{53}&(\bf{1},\bf{1},\overline{\bf{N}},\bf{1},\bf{N})&\frac{1}{4}(9-\sqrt{33}) \\
\end{array}
\end{align}
The quiver diagram for Model II is 
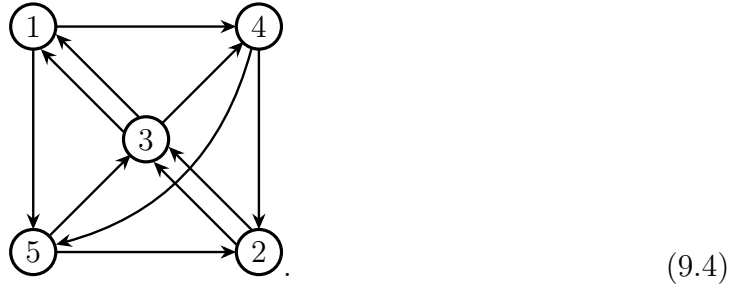
\begin{align}
\begin{tikzpicture}[<->]
    \draw[very thick] (-1.5,3) circle (3mm);
    \node at (-1.5,3) {1};
    \draw[very thick] (1.5,3) circle (3mm);
    \node at (1.5,3) {4};
    \draw[very thick] (0,1.5) circle (3mm);
    \node at (0,1.5) {3};
    \draw[very thick] (1.5,0) circle (3mm);
    \node at (1.5,0) {2};
    \draw[very thick] (-1.5,0) circle (3mm);
    \node at (-1.5,0) {5};
    \draw[-{Stealth[length=6pt,width=5pt]}, line width=0.9pt]
          (-1.5,2.7) to node[]{} (-1.5,0.3);
    \draw[-{Stealth[length=6pt,width=5pt]}, line width=0.9pt]
          (-1.2,3) to node[]{} (1.2,3);
    \draw[-{Stealth[length=6pt,width=5pt]}, line width=0.9pt]
          (1.2,0.1) to node[]{} (0.1,1.2);
    \draw[-{Stealth[length=6pt,width=5pt]}, line width=0.9pt]
          (1.4,0.3) to node[]{} (0.3,1.4);
    \draw[-{Stealth[length=6pt,width=5pt]}, line width=0.9pt]
          (-0.1,1.8) to node[]{} (-1.2,2.9);
    \draw[-{Stealth[length=6pt,width=5pt]}, line width=0.9pt]
          (-0.3,1.6) to node[]{} (-1.4,2.7);
    \draw[-{Stealth[length=6pt,width=5pt]}, line width=0.9pt]
          (0.2,1.7) to node[]{} (1.3,2.8); 
    \draw[-{Stealth[length=6pt,width=5pt]}, line width=0.9pt]
          (1.5,2.7) to node[]{} (1.5,0.3);
     \draw[-{Stealth[length=6pt,width=5pt]}, line width=0.9pt]
          (-1.2,0) to node[]{} (1.2,0);
    \draw[-{Stealth[length=6pt,width=5pt]}, line width=0.9pt]
          (-1.3,0.2) to node[]{} (-0.2,1.3);
    \draw[-{Stealth[length=6pt,width=5pt]}, line width=0.9pt, bend left=30]
         (1.4,2.7) to node[]{} (-1.2,0.1);
\end{tikzpicture}. 
\end{align}
The superpotential takes the form
\begin{align}
\mathcal{W}&=\Tr
(X_{34}X_{45}X_{53}-X_{53}X_{31}^2X_{15}+X_{34}X_{42}X_{23}^2
\nonumber\\
&+X_{23}^2X_{31}^1X_{15}X_{52}+X_{42}X_{23}^1X_{31}^2X_{14}-X_{23}^1X_{31}^1X_{14}X_{45}X_{52}
). 
\end{align}
The theory is holographically dual to Type IIB string theory on $AdS_5\times Y_{dP_2}$, 
where $Y_{dP_2}$ is the Sasaki-Einstein base of the complex cone over $dP_2$. 
The toric diagram of the cone over $dP_2$ is generated by 
\begin{align}
v_1&=(1,0,-1),& v_2&=(1,1,-1),& v_3&=(1,1,0),\nonumber\\
v_4&=(1,0,1),& v_5&=(1,-1,0). 
\end{align}
It is illustrated as
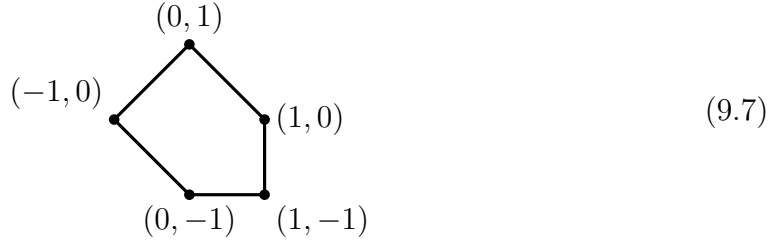
\begin{align}
\label{toric_dP2}
\begin{tikzpicture}[scale=1, baseline={(0,0)}]
  \fill (0,-1) circle (2pt) node[below] {$(0,-1)$};
  \fill (1,-1) circle (2pt) node[below right] {$(1,-1)$};
  \fill (1,0) circle (2pt) node[right] {$(1,0)$};
  \fill (0,1) circle (2pt) node[above] {$(0,1)$};
  \fill (-1,0) circle (2pt) node[above left] {$(-1,0)$};
  \draw[very thick] (0,-1) -- (1,-1) -- (1,0) -- (0,1) -- (-1,0) -- cycle;
\end{tikzpicture}
\end{align}
The multiplicities of five types of the chiral multiplets 
\begin{align}
\textrm{Model I :}\qquad 
X_1&=\{X_{41}\}, \quad X_2=\{X_{34}\}, \quad X_3=\{X_{42}\}, \nonumber\\
X_4&=\{X_{15}^2,X_{25}^2\}, \quad X_5=\{X_{31},X_{32}\}, \nonumber\\
\textrm{Model II :}\qquad 
X_1&=\{X_{14}\}, \quad X_2=\{X_{45}\}, \quad X_3=\{X_{52}\}, \nonumber\\
X_{4}&=\{X_{15},X_{42}\}, \quad X_5=\{X_{23}^1,X_{31}^1\}, 
\end{align}
forming the baryonic operators are evaluated as
\begin{align}
\det (v_{5},v_{1},v_{2})&=1, \\
\det (v_{1},v_{2},v_{3})&=1, \\
\det (v_{2},v_{3},v_{4})&=1, \\
\det (v_{3},v_{4},v_{5})&=2, \\
\det (v_{4},v_{5},v_{1})&=2. 
\end{align}
The area of the polygon is
\begin{align}
\label{A_dP2}
A&=\frac52. 
\end{align}
The dual cone is generated by
\begin{align}
\label{dualG_dP2}
m_1&=(1,0,1),& 
m_2&=(1,-1,0),&
m_3&=(1,-1,-1),\nonumber\\ 
m_4&=(1,1,-1),&
m_5&=(1,1,1). 
\end{align}
The Reeb vector is
\begin{align}
b&=\left(3,\frac{3(19-3\sqrt{33})}{16},\frac{3(-19+3\sqrt{33})}{16}\right). 
\end{align}
The volume of the Sasaki-Einstein base $Y_{dP_2}$ is evaluated as
\begin{align}
\label{vol_dP2}
\mathrm{Vol}(Y_{dP_2})
&=\frac{59+11\sqrt{33}}{486}\pi^3. 
\end{align}
As $dP_2$ does not admit a K\"{a}hler Einstein metric \cite{MR904143}, $Y_{dP_2}$ is non-regular and its volume is not rational \cite{Bergman:2001qi}. 
The R-charges of the chiral multiplet fields are 
\begin{align}
R(X_{1})&=\frac{\pi}{3}\frac{\mathrm{Vol}(\Sigma_1)}{\mathrm{Vol}(Y_{dP_2})}=\frac{3}{16}(19-3\sqrt{33}), \\
R(X_2)&=\frac{\pi}{3}\frac{\mathrm{Vol}(\Sigma_2)}{\mathrm{Vol}(Y_{dP_2})}=\frac12(-5+\sqrt{33}), \\
R(X_{3})&=\frac{\pi}{3}\frac{\mathrm{Vol}(\Sigma_3)}{\mathrm{Vol}(Y_{dP_2})}=\frac{3}{16}(19-3\sqrt{33}), \\
R(X_4)&=\frac{\pi}{3}\frac{\mathrm{Vol}(\Sigma_4)}{\mathrm{Vol}(Y_{dP_2})}=\frac{1}{16}(-21+5\sqrt{33}), \\
R(X_5)&=\frac{\pi}{3}\frac{\mathrm{Vol}(\Sigma_5)}{\mathrm{Vol}(Y_{dP_2})}=\frac{1}{16}(-21+5\sqrt{33}), 
\end{align}
where the volumes of the 3-cycles are 
\begin{align}
\mathrm{Vol}(\Sigma_1)&=2\pi^2 \frac{\det(v_{5},v_{1},v_{2})}{\det(b,v_{5},v_1)\det(b,v_{1},v_{2})}
=\frac{1+\sqrt{33}}{27}\pi^2, \\
\mathrm{Vol}(\Sigma_2)&=2\pi^2 \frac{\det(v_{1},v_{2},v_{3})}{\det(b,v_{1},v_2)\det(b,v_{2},v_{3})}
=\frac{17+\sqrt{33}}{81}\pi^2, \\
\mathrm{Vol}(\Sigma_3)&=2\pi^2 \frac{\det(v_{2},v_{3},v_{4})}{\det(b,v_{2},v_3)\det(b,v_{3},v_{4})}
=\frac{1+\sqrt{33}}{27}\pi^2, \\
\mathrm{Vol}(\Sigma_4)&=2\pi^2 \frac{\det(v_{3},v_{4},v_{5})}{\det(b,v_{3},v_4)\det(b,v_{4},v_{5})}
=\frac{2(9+\sqrt{33})}{81}\pi^2, \\
\mathrm{Vol}(\Sigma_5)&=2\pi^2 \frac{\det(v_{4},v_{5},v_{1})}{\det(b,v_{4},v_5)\det(b,v_{5},v_{1})}
=\frac{2(9+\sqrt{33})}{81}\pi^2. 
\end{align}

For Model I we have the following weighted adjacency matrix: 
\begin{align}
\label{M_dP2_I}
&
M^{AdS_5\times Y_{dP_2^{(I)}}}
\nonumber\\
&=\left(
\begin{smallmatrix}
i_v&0&\tilde{i}_{\textrm{cm}_{\frac{1}{16}(-21+5\sqrt{33})}}
&\tilde{i}_{\textrm{cm}_{\frac{3}{16}(19-3\sqrt{33})}}&i_A\\
0&i_v&\tilde{i}_{\textrm{cm}_{\frac{1}{16}(-21+5\sqrt{33})}}
&\tilde{i}_{\textrm{cm}_{\frac{3}{16}(19-3\sqrt{33})}}&i_A \\
i_{\textrm{cm}_{\frac{1}{16}(-21+5\sqrt{33})}}&i_{\textrm{cm}_{\frac{1}{16}(-21+5\sqrt{33})}}&i_v
&i_{\textrm{cm}_{\frac12(-5+\sqrt{33})}}&\tilde{i}_B \\
i_{\textrm{cm}_{\frac{3}{16}(19-3\sqrt{33})}}&i_{\textrm{cm}_{\frac{3}{16}(19-3\sqrt{33})}}&\tilde{i}_{\textrm{cm}_{\frac12(-5+\sqrt{33})}}
&i_v&\tilde{i}_{\textrm{cm}_{\frac18(-21+5\sqrt{33})}} \\
\tilde{i}_A&\tilde{i}_A&i_B&i_{\textrm{cm}_{\frac18(-21+5\sqrt{33})}}&i_v \\
\end{smallmatrix}
\right),
\end{align}
where
\begin{align}
i_A&=i_{\textrm{cm}_{\frac{1}{16}(-21+5\sqrt{33})}}+i_{\textrm{cm}_{\frac{1}{16}(17-\sqrt{33})}}, \\
\tilde{i}_A&=\tilde{i}_{\textrm{cm}_{\frac{1}{16}(-21+5\sqrt{33})}}+\tilde{i}_{\textrm{cm}_{\frac{1}{16}(17-\sqrt{33})}}, \\
i_B&=2i_{\textrm{cm}_{\frac14(9-\sqrt{33})}}+i_{\textrm{cm}_{\frac18(37-5\sqrt{33})}}. 
\end{align}
Similarly, for Model II the weighted adjacency matrix reads
\begin{align}
\label{M_dP2_II}
&
M^{AdS_5\times Y_{dP_2^{(II)}}}
\nonumber\\
&=\left(
\begin{smallmatrix}
i_v&0&\tilde{i}_A
&i_{\textrm{cm}_{\frac{3}{16}(19-3\sqrt{33})}}&i_{\textrm{cm}_{\frac{1}{16}(-21+5\sqrt{33})}}\\
0&i_v&i_A
&\tilde{i}_{\textrm{cm}_{\frac{1}{16}(-21+5\sqrt{33})}}&\tilde{i}_{\textrm{cm}_{\frac{3}{16}(19-3\sqrt{33})}} \\
i_A&\tilde{i}_A&i_v&i_{\textrm{cm}_{\frac14(9-\sqrt{33})}}&\tilde{i}_{\textrm{cm}_{\frac14(9-\sqrt{33})}} \\
\tilde{i}_{\textrm{cm}_{\frac{3}{16}(19-3\sqrt{33})}}&i_{\textrm{cm}_{\frac{1}{16}(-21+5\sqrt{33})}}
&\tilde{i}_{\textrm{cm}_{\frac14(9-\sqrt{33})}}&i_v&i_{\textrm{cm}_{\frac12(-5+\sqrt{33})}}\\
\tilde{i}_{\textrm{cm}_{\frac{1}{16}(-21+5\sqrt{33})}}&i_{\textrm{cm}_{\frac{3}{16}(19-3\sqrt{33})}}
&i_{\textrm{cm}_{\frac14(9-\sqrt{33})}}&\tilde{i}_{\textrm{cm}_{\frac12(-5+\sqrt{33})}}&i_v\\
\end{smallmatrix}
\right). 
\end{align}
The gravity indices obtained by the two matrices (\ref{M_dP2_I}) and (\ref{M_dP2_II}) coincide. 
We find 
\begin{align}
\label{ind_dP2}
i^{AdS_5\times Y_{dP_2}}(p;q)
&=2\frac{pq}{1-pq}+2\frac{(pq)^{\frac{3\sqrt{33}-3}{16}}}{1-(pq)^{\frac{3\sqrt{33}-3}{16}}}
+\frac{(pq)^{\frac{27-3\sqrt{33}}{8}}}{1-(pq)^{\frac{27-3\sqrt{33}}{8}}}. 
\end{align}
There are five extremal BPS mesonic operators 
corresponding to the five generators (\ref{dualG_dP2}) of the dual cone. 
The R-charges are obtained from the formula (\ref{Rch_meson})
\begin{align}
R(\mathcal{M}_1)&=\frac23(b,m_1)=\frac{3\sqrt{33}-3}{8}, \\
R(\mathcal{M}_2)&=\frac23(b,m_2)=\frac{3\sqrt{33}-3}{8}, \\
R(\mathcal{M}_3)&=\frac23(b,m_3)=2, \\
R(\mathcal{M}_4)&=\frac23(b,m_4)=\frac{27-3\sqrt{33}}{4}, \\
R(\mathcal{M}_5)&=\frac23(b,m_5)=2. 
\end{align}

We have the supersymmetric zeta function
\begin{align}
\label{zeta_dP2}
&
\mathfrak{Z}^{AdS_5\times Y_{dP_2}}(s,z;\omega_1,\omega_2)
\nonumber\\
&=2(\omega_1+\omega_2)^{-s}
\zeta\left(s,1+\frac{z}{\omega_1+\omega_2}\right)
+2\left(\frac{3\sqrt{33}-3}{16}(\omega_1+\omega_2)\right)^{-s}
\zeta\left(s,1+\frac{\frac{16z}{3\sqrt{33}-3}}{\omega_1+\omega_2}\right)
\nonumber\\
&+\left(\frac{27-3\sqrt{33}}{8}(\omega_1+\omega_2)\right)^{-s}
\zeta\left(s, 1+\frac{\frac{8z}{27-3\sqrt{33}}}{\omega_1+\omega_2}\right). 
\end{align}
The residue at a simple pole $s=1$ is given by
\begin{align}
\label{Res_dP2}
\mathrm{Res}_{s=1}\mathfrak{Z}^{AdS_5\times Y_{dP_2}}(s,z;\omega_1,\omega_2)
&=\frac{51+7\sqrt{33}}{18}\frac{1}{\omega_1+\omega_2}. 
\end{align}
Substituting the residue coefficient (\ref{Res_dP2}) and the volume (\ref{vol_dP2}) into formula (\ref{Riem2_conj}), 
we obtain the integral of the squared Riemann tensor
\begin{align}
\int_{Y_{dP_2}} \mathrm{Riem}^2
&=\frac{4(2551+295\sqrt{33})}{243}\pi^3
\nonumber\\
&=2166.94\ldots. 
\end{align}
This value coincides precisely with the Hilbert series computation (see Appendix \ref{app_HS}).
The Zeta-index is
\begin{align}
\mathfrak{Z}^{AdS_5\times Y_{dP_2}}(0,0;\omega_1,\omega_2)
&=-\frac52. 
\end{align}
The special supersymmetric zeta value with $s=-1$ and $z=0$ is
\begin{align}
\mathfrak{Z}^{AdS_5\times Y_{dP_2}}(-1,0;\omega_1,\omega_2)
&=-\frac{5}{12}(\omega_1+\omega_2). 
\end{align}
From the formulae (\ref{c0_formula}) and (\ref{a0_formula}) we obtain
\begin{align}
c_{\mathcal{O}(N^0)}&=-\frac{5}{8}, \\
a_{\mathcal{O}(N^0)}&=-\frac{15}{16}, 
\end{align}
and 
\begin{align}
c_{\mathcal{O}(N^0)}-a_{\mathcal{O}(N^0)}&=\frac{5}{16}. 
\end{align}

The supersymmetric determinant is
\begin{align}
&
\mathfrak{D}^{AdS_5\times Y_{dP_2}}(z;\omega_1,\omega_2)
\nonumber\\
&=\frac{3^{\frac16\left(9+\frac{16z}{(9-\sqrt{33})(\omega_1+\omega_2)}+\frac{64z}{(-1+\sqrt{33})(\omega_1+\omega_2)} \right)}
(-1+\sqrt{33})^{1+\frac{32z}{3(-1+\sqrt{33})(\omega_1+\omega_2)}}
(9-\sqrt{33})^{\frac12+\frac{8z}{3(9-\sqrt{33})(\omega_1+\omega_2)}}
}
{2^{8+\frac{17+9\sqrt{33}}{6(\omega_1+\omega_2)}z} \pi^{\frac52}}
\nonumber\\
&\times 
(\omega_1+\omega_2)^{\frac52+\frac{4(7+9\sqrt{33})z}{3(9-\sqrt{33})(-1+\sqrt{33})(\omega_1+\omega_2)}}
\Gamma\left(1+\frac{z}{\omega_1+\omega_2}\right)^2
\Gamma\left(1+\frac{16z}{3(-1+\sqrt{33})(\omega_1+\omega_2)}\right)^2
\nonumber\\
&\times 
\Gamma\left(1+\frac{8z}{3(9-\sqrt{33})(\omega_1+\omega_2)}\right). 
\end{align}
The vacuum exponent is
 \begin{align}
\mathfrak{D}^{AdS_5\times Y_{dP_2}}(0;1,1)
&=\frac{3\sqrt{558-78\sqrt{33}}}{32\pi^{\frac52}}
=0.0561879\ldots. 
\end{align}

\subsection{$AdS_5\times Y_{dP_3}$}
\label{sec_dP3}
For a stack of $N$ D3-branes probing the toric Calabi-Yau 3-fold, which is a complex cone over the third del Pezzo surface $dP_3$, 
there are four Seiberg dual descriptions of the 4d $\mathcal{N}=1$ quiver gauge theory with six $SU(N)$ gauge nodes, Model I, II, III and IV 
\cite{Feng:2000mi,Feng:2001xr,Feng:2001bn,Beasley:2001zp,Feng:2002fv,Feng:2002zw,Franco:2005rj,Forcella:2008bb,Forcella:2008ng,Hanany:2012hi}. 
For Model I\footnote{This is called Model II in \cite{Beasley:2001zp}. }, we have $14$ bifundamental chiral multiplets with charges
\begin{align}
\begin{array}{c|cc}
&\prod_{i=1}^6 SU(N)_i&U(1)_R\\ \hline 
X_{12}&(\bf{N},\overline{\bf{N}},\bf{1},\bf{1},\bf{1},\bf{1})&\frac13\\
X_{13}&(\bf{N},\bf{1},\overline{\bf{N}},\bf{1},\bf{1},\bf{1})&\frac23\\
X_{16}&(\bf{N},\bf{1},\bf{1},\bf{1},\bf{1},\overline{\bf{N}})&\frac23\\
X_{23}&(\bf{1},\bf{N},\overline{\bf{N}},\bf{1},\bf{1},\bf{1})&\frac13\\
X_{26}&(\bf{1},\bf{N},\bf{1},\bf{1},\bf{1},\overline{\bf{N}})&\frac13\\
X_{34}&(\bf{1},\bf{1},\bf{N},\overline{\bf{N}},\bf{1},\bf{1})&\frac13\\
X_{35}&(\bf{1},\bf{1},\bf{N},\bf{1},\overline{\bf{N}},\bf{1})&\frac23\\
X_{41}&(\overline{\bf{N}},\bf{1},\bf{1},\bf{N},\bf{1},\bf{1})&1\\
X_{45}&(\bf{1},\bf{1},\bf{1},\bf{N},\overline{\bf{N}},\bf{1})&\frac13\\
X_{51}^{\alpha}&(\overline{\bf{N}},\bf{1},\bf{1},\bf{1},\bf{N},\bf{1})&\frac23\\
X_{52}&(\bf{1},\overline{\bf{N}},\bf{1},\bf{1},\bf{N},\bf{1})&1\\
X_{64}&(\bf{1},\bf{1},\bf{1},\overline{\bf{N}},\bf{1},\bf{N})&\frac13\\
X_{65}&(\bf{1},\bf{1},\bf{1},\bf{1},\overline{\bf{N}},\bf{N})&\frac23\\
\end{array}
\end{align}
with $\alpha=1,2$. 
The quiver diagram is 
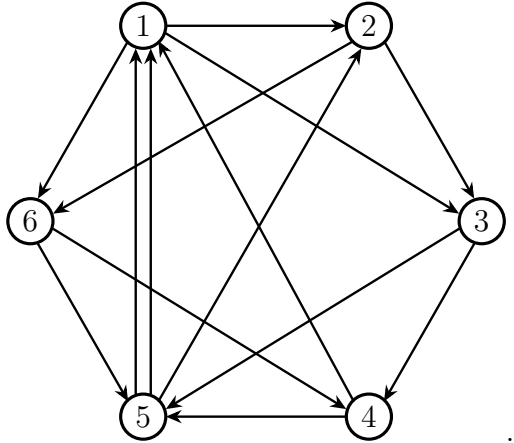
\begin{align}
\begin{tikzpicture}[<->]
    \draw[very thick] (-1.5,2.6) circle (3mm);
    \node at (-1.5,2.6) {1};    
    \draw[very thick] (1.5,2.6) circle (3mm);
    \node at (1.5,2.6) {2};
    \draw[very thick] (3,0) circle (3mm);
    \node at (3,0) {3};
    \draw[very thick] (1.5,-2.6) circle (3mm);
    \node at (1.5,-2.6) {4};
    \draw[very thick] (-1.5,-2.6) circle (3mm);
    \node at (-1.5,-2.6) {5};
    \draw[very thick] (-3,0) circle (3mm);
    \node at (-3,0) {6};
    \draw[-{Stealth[length=6pt,width=5pt]}, line width=0.9pt]
          (-1.2,2.6) to node[]{} (1.2,2.6);
    \draw[-{Stealth[length=6pt,width=5pt]}, line width=0.9pt]
          (-1.2,2.5) to node[]{} (2.7,0.1);
    \draw[-{Stealth[length=6pt,width=5pt]}, line width=0.9pt]
          (-1.7,2.4) to node[]{} (-2.9,0.3);
    \draw[-{Stealth[length=6pt,width=5pt]}, line width=0.9pt]
          (1.7,2.4) to node[]{} (2.9,0.3);
    \draw[-{Stealth[length=6pt,width=5pt]}, line width=0.9pt]
          (1.3,2.4) to node[]{} (-2.7,0.1);
    \draw[-{Stealth[length=6pt,width=5pt]}, line width=0.9pt]
          (2.9,-0.3) to node[]{} (1.7,-2.4);
    \draw[-{Stealth[length=6pt,width=5pt]}, line width=0.9pt]
          (2.7,-0.1) to node[]{} (-1.2,-2.5);
    \draw[-{Stealth[length=6pt,width=5pt]}, line width=0.9pt]
          (1.3,-2.4) to node[]{} (-1.3,2.4);
    \draw[-{Stealth[length=6pt,width=5pt]}, line width=0.9pt]
          (1.2,-2.6) to node[]{} (-1.2,-2.6);
    \draw[-{Stealth[length=6pt,width=5pt]}, line width=0.9pt]
          (-1.6,-2.3) to node[]{} (-1.6,2.3);
    \draw[-{Stealth[length=6pt,width=5pt]}, line width=0.9pt]
          (-1.4,-2.3) to node[]{} (-1.4,2.3);
    \draw[-{Stealth[length=6pt,width=5pt]}, line width=0.9pt]
          (-1.3,-2.4) to node[]{} (1.4,2.3);
    \draw[-{Stealth[length=6pt,width=5pt]}, line width=0.9pt]
          (-2.7,-0.1) to node[]{} (1.2,-2.5);
    \draw[-{Stealth[length=6pt,width=5pt]}, line width=0.9pt]
          (-2.9,-0.3) to node[]{} (-1.7,-2.4);
\end{tikzpicture}. 
\end{align}
The superpotential is \cite{Feng:2000mi}
\begin{align}
\mathcal{W}
&=\Tr (
X_{34}X_{41}X_{13}-X_{41}X_{16}X_{64}-X_{51}^1X_{13}X_{35}-X_{12}X_{23}X_{34}X_{45}X_{51}^1
\nonumber\\
&+X_{65}X_{51}^2X_{16}+X_{23}X_{35}X_{52}+X_{12}X_{26}X_{64}X_{45}X_{51}^1-X_{26}X_{65}X_{52}
). 
\end{align}
The toric diagram of the cone over $dP_3$ is generated by
\begin{align}
v_1&=(1,0,-1),& 
v_2&=(1,1,-1),& 
v_3&=(1,1,0), \nonumber\\
v_4&=(1,0,1),& 
v_5&=(1,-1,1),& 
v_6&=(1,-1,0). 
\end{align}
It is illustrated as
\begin{align}
\label{toric_dP3}
\begin{tikzpicture}[scale=1, baseline={(0,0)}]
  \fill (0,-1) circle (2pt) node[below] {$(0,-1)$};
  \fill (1,-1) circle (2pt) node[below right] {$(1,-1)$};
  \fill (1,0) circle (2pt) node[right] {$(1,0)$};
  \fill (0,1) circle (2pt) node[above] {$(0,1)$};
  \fill (-1,1) circle (2pt) node[above left] {$(-1,1)$};
  \fill (-1,0) circle (2pt) node[above left] {$(-1,0)$};
  \draw[very thick] (0,-1) -- (1,-1) -- (1,0) -- (0,1) -- (-1,1) -- (-1,0) -- cycle;
\end{tikzpicture}
\end{align}
There are six types of the chiral multiplets 
$X_1$ $=$ $X_{12}$, 
$X_2$ $=$ $X_{23}$, 
$X_3$ $=$ $X_{26}$, 
$X_4$ $=$ $X_{34}$, 
$X_5$ $=$ $X_{45}$ and 
$X_6$ $=$ $X_{64}$, 
which form the baryonic operators. 
We have 
\begin{align}
\det (v_{6},v_{1},v_{2})&=1, \\
\det (v_{1},v_{2},v_{3})&=1, \\
\det (v_{2},v_{3},v_{4})&=1, \\
\det (v_{3},v_{4},v_{5})&=1, \\
\det (v_{4},v_{5},v_{6})&=1, \\ 
\det (v_{5},v_{6},v_{1})&=1. 
\end{align}
The area of the polygon is
\begin{align}
\label{A_dP3}
A&=3. 
\end{align}
The dual cone is generated by
\begin{align}
\label{dualG_dP3}
m_1&=(1,0,1),& 
m_2&=(1,-1,0),&
m_3&=(1,-1,-1),\nonumber\\ 
m_4&=(1,0,-1),&
m_5&=(1,1,0),&
m_6&=(1,1,1). 
\end{align}
The Reeb vector is
\begin{align}
b&=\left(3,0,0\right). 
\end{align}
The volume of the Sasaki-Einstein base $Y_{dP_3}$ is 
\begin{align}
\label{vol_dP3}
\mathrm{Vol}(Y_{dP_3})
&=\frac{2\pi^3}{9}. 
\end{align}
The R-charges of the chiral multiplet fields are 
\begin{align}
R(X_{I})&=\frac{\pi}{3}\frac{\mathrm{Vol}(\Sigma_I)}{\mathrm{Vol}(Y_{dP_3})}=\frac13. 
\end{align}
where 
\begin{align}
\mathrm{Vol}(\Sigma_I)&=2\pi^2 \frac{\det(v_{I-1},v_{I},v_{I+1})}{\det(b,v_{I-1},v_I)\det(b,v_{I},v_{I+1})}
=\frac{2\pi^2}{9}, 
\end{align}
for $I=1,\cdots, 6$.

For Model I we have the weighted adjacency matrix 
\begin{align}
\label{M_dP3}
M^{AdS_5\times Y_{dP_3^{(I)}}}
&=\left(
\begin{matrix}
i_v&i_{\textrm{cm}_{\frac13}}&i_{\textrm{cm}_{\frac23}}&\tilde{i}_{\textrm{cm}_{1}}&2\tilde{i}_{\textrm{cm}_{\frac23}}&\tilde{i}_{\textrm{cm}_{\frac23}} \\
\tilde{i}_{\textrm{cm}_{\frac13}}&i_v&i_{\textrm{cm}_{\frac13}}&0&\tilde{i}_{\textrm{cm}_{1}}&i_{\textrm{cm}_{\frac13}} \\
\tilde{i}_{\textrm{cm}_{\frac23}}&\tilde{i}_{\textrm{cm}_{\frac13}}&i_v&i_{\textrm{cm}_{\frac13}}&i_{\textrm{cm}_{\frac23}}&0 \\
i_{\textrm{cm}_{1}}&0&\tilde{i}_{\textrm{cm}_{\frac13}}&i_v&i_{\textrm{cm}_{\frac13}}&i_{\textrm{cm}_{\frac13}} \\
2i_{\textrm{cm}_{\frac23}}&i_{\textrm{cm}_{1}}&\tilde{i}_{\textrm{cm}_{\frac23}}&\tilde{i}_{\textrm{cm}_{\frac13}}&i_v&\tilde{i}_{\textrm{cm}_{\frac23}} \\
\tilde{i}_{\textrm{cm}_{\frac23}}&\tilde{i}_{\textrm{cm}_{\frac13}}&0&\tilde{i}_{\textrm{cm}_{\frac13}}&i_{\textrm{cm}_{\frac23}}&i_v \\
\end{matrix}
\right). 
\end{align}
The supersymmetric index is evaluated as
\begin{align}
\label{ind_dP3}
i^{AdS_5\times Y_{dP_3}}(p;q)
&=6\frac{pq}{1-pq}. 
\end{align}
They enumerate six extremal BPS mesonic operators. 
They correspond to the primitive generators (\ref{dualG_dP3}) of the dual cone. 
The R-charges are given by
\begin{align}
R(\mathcal{M}_I)&=\frac23 (b,m_I)=2, \qquad I=1,\cdots, 6. 
\end{align}

The supersymmetric zeta function is given by
\begin{align}
\label{zeta_dP3}
\mathfrak{Z}^{AdS_5\times Y_{dP_3}}(s,z;\omega_1,\omega_2)
&=6 (\omega_1+\omega_2)^{-s}\zeta\left(s,1+\frac{z}{\omega_1+\omega_2}\right). 
\end{align}
It has the residue 
\begin{align}
\label{Res_dP3}
\mathrm{Res}_{s=1}\mathfrak{Z}^{AdS_5\times Y_{dP_3}}(s,z;\omega_1,\omega_2)
&=\frac{6}{\omega_1+\omega_2} 
\end{align}
at a simple pole $s=1$. 
Upon inserting the residue coefficient (\ref{Res_dP3}) together with the volume (\ref{vol_dP3}) into the formula (\ref{Riem2_conj}), 
the integral of the squared Riemann tensor is evaluated as
\begin{align}
\label{Riem2_dP3}
\int_{Y_{dP_3}}\mathrm{Riem}^2
&=\frac{848}{9}\pi^3=2921.48\ldots. 
\end{align}
We confirm that the value (\ref{Riem2_dP3}) agrees with the Hilbert series calculation (see Appendix \ref{app_HS}).
The Zeta-index is
\begin{align}
\mathfrak{Z}^{AdS_5\times Y_{dP_3}}(0,0;\omega_1,\omega_2)&=-3. 
\end{align}
The supersymmetric zeta value with $s=-1$ and $z=0$ is
\begin{align}
\mathfrak{Z}^{AdS_5\times Y_{dP_3}}(-1,0;\omega_1,\omega_2)
&=-\frac12 (\omega_1+\omega_2). 
\end{align}
According to the formulae (\ref{c0_formula}) and (\ref{a0_formula}), we find
\begin{align}
c_{\mathcal{O}(N^0)}&=-\frac34, \\
a_{\mathcal{O}(N^0)}&=-\frac98,  
\end{align}
and 
\begin{align}
c_{\mathcal{O}(N^0)}-a_{\mathcal{O}(N^0)}&=\frac38. 
\end{align}

We obtain the supersymmetric determinant of the form
\begin{align}
\mathfrak{D}^{AdS_5\times Y_{dP_3}}(z;\omega_1,\omega_2)
&=\frac{(\omega_1+\omega_2)^{3+\frac{6z}{\omega_1+\omega_2}} \Gamma\left(1+\frac{z}{\omega_1+\omega_2}\right)^6}
{8\pi^3}. 
\end{align}
The vacuum exponent is
 \begin{align}
\mathfrak{D}^{AdS_5\times Y_{dP_3}}(0;1,1)
&=\frac{1}{\pi^3}=0.0322515\ldots. 
\end{align}

\subsection{$AdS_5\times Y_{PdP_{3a}}$}
\label{sec_PdP3a}
The toric Calabi-Yau 3-fold given by the complex cone over the third Pseudo del Pezzo surface, 
when probed by a stack of $N$ D3-branes, leads to a 4d $\mathcal{N}=1$ quiver gauge theory with 
six $SU(N)$ gauge nodes and 18 bifundamental chiral multiplets \cite{Feng:2002fv,Feng:2004uq,Hanany:2012hi}. 
The field content is
\begin{align}
\begin{array}{c|cc}
&\prod_{i=1}^6 SU(N)_i&U(1)_R\\ \hline 
X_{12}&(\bf{N},\overline{\bf{N}},\bf{1},\bf{1},\bf{1},\bf{1})&\frac23 \\
X_{13}&(\bf{N},\bf{1},\overline{\bf{N}},\bf{1},\bf{1},\bf{1})&\frac23 \\
X_{15}&(\bf{N},\bf{1},\bf{1},\bf{1},\overline{\bf{N}},\bf{1})&\frac23 \\
X_{24}&(\bf{1},\bf{N},\bf{1},\overline{\bf{N}},\bf{1},\bf{1})&\frac23 \\
X_{25}&(\bf{1},\bf{N},\bf{1},\bf{1},\overline{\bf{N}},\bf{1})&\frac23 \\
X_{26}&(\bf{1},\bf{N},\bf{1},\bf{1},\bf{1},\overline{\bf{N}})&\frac23 \\
X_{32}&(\bf{1},\overline{\bf{N}},\bf{N},\bf{1},\bf{1},\bf{1})&\frac23 \\
X_{34}&(\bf{1},\bf{1},\bf{N},\overline{\bf{N}},\bf{1},\bf{1})&\frac23 \\
X_{35}&(\bf{1},\bf{1},\bf{N},\bf{1},\overline{\bf{N}},\bf{1})&\frac23 \\
X_{41}&(\overline{\bf{N}},\bf{1},\bf{1},\bf{N},\bf{1},\bf{1})&\frac23 \\
X_{43}&(\bf{1},\bf{1},\overline{\bf{N}},\bf{N},\bf{1},\bf{1})&\frac23 \\
X_{46}&(\bf{1},\bf{1},\bf{1},\bf{N},\bf{1},\overline{\bf{N}})&\frac23 \\
X_{51}&(\overline{\bf{N}},\bf{1},\bf{1},\bf{1},\bf{N},\bf{1})&\frac23 \\
X_{54}&(\bf{1},\bf{1},\bf{1},\overline{\bf{N}},\bf{N},\bf{1})&\frac23 \\
X_{56}&(\bf{1},\bf{1},\bf{1},\bf{1},\bf{N},\overline{\bf{N}})&\frac23 \\
X_{61}&(\overline{\bf{N}},\bf{1},\bf{1},\bf{1},\bf{1},\bf{N})&\frac23 \\
X_{62}&(\bf{1},\overline{\bf{N}},\bf{1},\bf{1},\bf{1},\bf{N})&\frac23 \\
X_{63}&(\bf{1},\bf{1},\overline{\bf{N}},\bf{1},\bf{1},\bf{N})&\frac23 \\
\end{array}
\end{align}
The quiver diagram is 
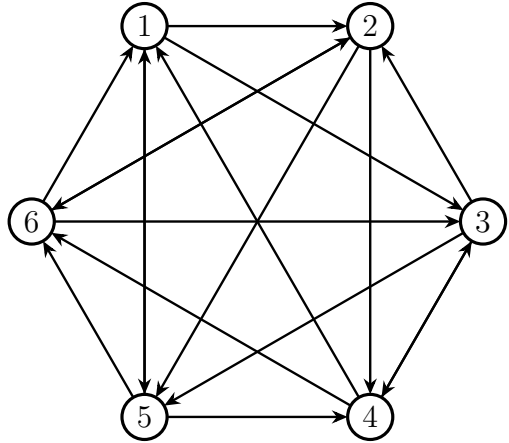
\begin{align}
\begin{tikzpicture}
    \draw[very thick] (-1.5,2.6) circle (3mm);
    \node at (-1.5,2.6) {1};    
    \draw[very thick] (1.5,2.6) circle (3mm);
    \node at (1.5,2.6) {2};
    \draw[very thick] (3,0) circle (3mm);
    \node at (3,0) {3};
    \draw[very thick] (1.5,-2.6) circle (3mm);
    \node at (1.5,-2.6) {4};
    \draw[very thick] (-1.5,-2.6) circle (3mm);
    \node at (-1.5,-2.6) {5};
    \draw[very thick] (-3,0) circle (3mm);
    \node at (-3,0) {6};
\draw[-{Stealth[length=6pt,width=5pt]}, line width=0.9pt, shorten <=3mm, shorten >=3mm] (-1.5,2.6) -- (1.5,2.6);   
\draw[-{Stealth[length=6pt,width=5pt]}, line width=0.9pt, shorten <=3mm, shorten >=3mm] (-1.5,2.6) -- (3,0);       
\draw[-{Stealth[length=6pt,width=5pt]}, line width=0.9pt, shorten <=3mm, shorten >=3mm] (-1.5,2.6) -- (-1.5,-2.6); 
\draw[-{Stealth[length=6pt,width=5pt]}, line width=0.9pt, shorten <=3mm, shorten >=3mm] (1.5,2.6) -- (1.5,-2.6);   
\draw[-{Stealth[length=6pt,width=5pt]}, line width=0.9pt, shorten <=3mm, shorten >=3mm] (1.5,2.6) -- (-1.5,-2.6);  
\draw[-{Stealth[length=6pt,width=5pt]}, line width=0.9pt, shorten <=3mm, shorten >=3mm] (1.5,2.6) -- (-3,0);       
\draw[-{Stealth[length=6pt,width=5pt]}, line width=0.9pt, shorten <=3mm, shorten >=3mm] (3,0) -- (1.5,2.6);        
\draw[-{Stealth[length=6pt,width=5pt]}, line width=0.9pt, shorten <=3mm, shorten >=3mm] (3,0) -- (1.5,-2.6);       
\draw[-{Stealth[length=6pt,width=5pt]}, line width=0.9pt, shorten <=3mm, shorten >=3mm] (3,0) -- (-1.5,-2.6);      
\draw[-{Stealth[length=6pt,width=5pt]}, line width=0.9pt, shorten <=3mm, shorten >=3mm] (1.5,-2.6) -- (-1.5,2.6);  
\draw[-{Stealth[length=6pt,width=5pt]}, line width=0.9pt, shorten <=3mm, shorten >=3mm] (1.5,-2.6) -- (3,0);       
\draw[-{Stealth[length=6pt,width=5pt]}, line width=0.9pt, shorten <=3mm, shorten >=3mm] (1.5,-2.6) -- (-3,0);      
\draw[-{Stealth[length=6pt,width=5pt]}, line width=0.9pt, shorten <=3mm, shorten >=3mm] (-1.5,-2.6) -- (-1.5,2.6); 
\draw[-{Stealth[length=6pt,width=5pt]}, line width=0.9pt, shorten <=3mm, shorten >=3mm] (-1.5,-2.6) -- (1.5,-2.6); 
\draw[-{Stealth[length=6pt,width=5pt]}, line width=0.9pt, shorten <=3mm, shorten >=3mm] (-1.5,-2.6) -- (-3,0);     
\draw[-{Stealth[length=6pt,width=5pt]}, line width=0.9pt, shorten <=3mm, shorten >=3mm] (-3,0) -- (-1.5,2.6);      
\draw[-{Stealth[length=6pt,width=5pt]}, line width=0.9pt, shorten <=3mm, shorten >=3mm] (-3,0) -- (1.5,2.6);       
\draw[-{Stealth[length=6pt,width=5pt]}, line width=0.9pt, shorten <=3mm, shorten >=3mm] (-3,0) -- (3,0);           
\end{tikzpicture}
\end{align}
The superpotential takes the form
\begin{align}
\mathcal{W}
&=\Tr(X_{12}X_{26}X_{61}+X_{63}X_{34}X_{46}+X_{24}X_{43}X_{32}+X_{35}X_{51}X_{13}+X_{41}X_{15}X_{54}+X_{56}X_{62}X_{25}
\nonumber\\
&-X_{12}X_{25}X_{51}-X_{63}X_{32}X_{26}-X_{24}X_{46}X_{62}-X_{35}X_{54}X_{43}-X_{41}X_{13}X_{34}-X_{56}X_{61}X_{15}
). 
\end{align}
The toric diagram of the cone over $PdP_{3a}$ is generated by
\begin{align}
v_1&=(1,0,0),& 
v_2&=(1,2,1),& 
v_3&=(1,1,2), \nonumber\\
v_4&=(1,0,3),&
v_5&=(1,0,2),& 
v_6&=(1,0,1). 
\end{align}
It is shown as
\begin{align}
\label{toric_PdP3a}
\begin{tikzpicture}[scale=1, baseline={(0,0)}]
  \fill (0,0) circle (2pt) node[below] {$(0,0)$};
  \fill (2,1) circle (2pt) node[right] {$(2,1)$};
  \fill (1,2) circle (2pt) node[right] {$(1,2)$};
  \fill (0,3) circle (2pt) node[above] {$(0,3)$};
  \fill (0,2) circle (2pt) node[left] {$(0,2)$};
  \fill (0,1) circle (2pt) node[left] {$(0,1)$};
  \draw[very thick] (0,0) -- (2,1) -- (1,2) -- (0,3) -- (0,2) -- (0,1) -- cycle;
\end{tikzpicture}
\end{align}
There are three types of the chiral multiplets $X_1$, $X_2$ and $X_4$ 
corresponding to the vertices $v_1$, $v_2$ and $v_4$, 
which form the baryons. 
The multiplicities are 
\begin{align}
\det (v_{4},v_{1},v_{2})&=6, \\
\det (v_{1},v_{2},v_{4})&=6, \\
\det (v_{2},v_{4},v_{1})&=6. 
\end{align}
The area of the lattice polygon is
\begin{align}
\label{A_PdP3a}
A&=3. 
\end{align}
The dual cone is generated by
\begin{align}
\label{dualG_PdP3a}
m_1&=(0,-1,2),& 
m_2&=(3,-1,-1),&
m_3&=(3,-1,-1),\nonumber\\ 
m_4&=(0,1,0),&
m_5&=(0,1,0),&
m_6&=(0,1,0). 
\end{align}
Applying the volume minimization \cite{Martelli:2005tp}, 
we find the Reeb vector 
\begin{align}
b&=(3,2,4).
\end{align} 
and the volume of the Sasaki-Einstein base 
\begin{align}
\label{vol_PdP3a}
\mathrm{Vol}(Y_{PdP_{3a}})
&=\frac{\pi^3}{6}. 
\end{align}
The volume is the same volume as $S^5/\mathbb{Z}_6$ 
as the complex cone over $PdP_{3a}$ can be described as the orbifold $\mathbb{C}^3/\mathbb{Z}_6 (1,2,-3)$. 
The R-charges of the chiral multiplet fields are evaluated as
\begin{align}
R(X_{1})&=\frac{\pi}{3}\frac{\mathrm{Vol}(\Sigma_1)}{\mathrm{Vol}(Y_{PdP_{3a}})}
=\frac23, \\
R(X_2)&=\frac{\pi}{3}\frac{\mathrm{Vol}(\Sigma_2)}{\mathrm{Vol}(Y_{PdP_{3a}})}
=\frac23, \\
R(X_{4})&=\frac{\pi}{3}\frac{\mathrm{Vol}(\Sigma_4)}{\mathrm{Vol}(Y_{PdP_{4a}})}
=\frac23, 
\end{align}
where 
\begin{align}
\mathrm{Vol}(\Sigma_1)&=2\pi^2 \frac{\det(v_{4},v_{1},v_{2})}{\det(b,v_{4},v_1)\det(b,v_{1},v_{2})}
=\frac{\pi^2}{3}, \\
\mathrm{Vol}(\Sigma_2)&=2\pi^2 \frac{\det(v_{1},v_{2},v_{4})}{\det(b,v_{1},v_2)\det(b,v_{2},v_{4})}
=\frac{\pi^2}{3}, \\
\mathrm{Vol}(\Sigma_4)&=2\pi^2 \frac{\det(v_{2},v_{4},v_{1})}{\det(b,v_{2},v_4)\det(b,v_{4},v_{1})}
=\frac{\pi^2}{3}. 
\end{align}

We have the weighted adjacency matrix 
\begin{align}
\label{M_PdP3a}
&
M^{AdS_5\times Y_{PdP_{3a}}}
\nonumber\\
&=\left(
\begin{matrix}
i_v&i_{\textrm{cm}_{\frac23}}&i_{\textrm{cm}_{\frac23}}&\tilde{i}_{\textrm{cm}_{\frac23}}&i_{\textrm{cm}_{\frac23}}+\tilde{i}_{\textrm{cm}_{\frac23}}&\tilde{i}_{\textrm{cm}_{\frac23}} \\
\tilde{i}_{\textrm{cm}_{\frac23}}&i_v&\tilde{i}_{\textrm{cm}_{\frac23}}&i_{\textrm{cm}_{\frac23}}&i_{\textrm{cm}_{\frac23}}&i_{\textrm{cm}_{\frac23}}+\tilde{i}_{\textrm{cm}_{\frac23}} \\
\tilde{i}_{\textrm{cm}_{\frac23}}&i_{\textrm{cm}_{\frac23}}&i_v&i_{\textrm{cm}_{\frac23}}+\tilde{i}_{\textrm{cm}_{\frac23}}&i_{\textrm{cm}_{\frac23}}&\tilde{i}_{\textrm{cm}_{\frac23}} \\
i_{\textrm{cm}_{\frac23}}&\tilde{i}_{\textrm{cm}_{\frac23}}&i_{\textrm{cm}_{\frac23}}+\tilde{i}_{\textrm{cm}_{\frac23}}&i_v&\tilde{i}_{\textrm{cm}_{\frac23}}&i_{\textrm{cm}_{\frac23}} \\
i_{\textrm{cm}_{\frac23}}+\tilde{i}_{\textrm{cm}_{\frac23}}&\tilde{i}_{\textrm{cm}_{\frac23}}&\tilde{i}_{\textrm{cm}_{\frac23}}&i_{\textrm{cm}_{\frac23}}&i_v&i_{\textrm{cm}_{\frac23}} \\
i_{\textrm{cm}_{\frac23}}&i_{\textrm{cm}_{\frac23}}+\tilde{i}_{\textrm{cm}_{\frac23}}&i_{\textrm{cm}_{\frac23}}&\tilde{i}_{\textrm{cm}_{\frac23}}&\tilde{i}_{\textrm{cm}_{\frac23}}&i_v \\
\end{matrix}
\right)
\end{align}
The gravity index is evaluated from (\ref{M_PdP3a}) as
\begin{align}
\label{ind_PdP3a}
i^{AdS_5\times Y_{PdP_{3a}}}(p;q)
&=\frac{(pq)^2}{1-(pq)^2}+2\frac{pq}{1-pq}+3\frac{(pq)^{\frac23}}{1-(pq)^{\frac23}}. 
\end{align}
The R-charges of the six extremal BPS mesons, 
which correspond to the six vectors (\ref{dualG_PdP3a}) of the dual cone, are given by
\begin{align}
R(\mathcal{M}_1)&=\frac23(b,m_1)=4, \\
R(\mathcal{M}_2)&=\frac23(b,m_2)=2, \\
R(\mathcal{M}_3)&=\frac23(b,m_3)=2, \\
R(\mathcal{M}_4)&=\frac23(b,m_4)=\frac43, \\
R(\mathcal{M}_5)&=\frac23(b,m_5)=\frac43, \\
R(\mathcal{M}_6)&=\frac23(b,m_6)=\frac43.
\end{align}

We obtain the supersymmetric zeta function 
\begin{align}
\label{zeta_PdP3a}
&
\mathfrak{Z}^{AdS_5\times Y_{PdP_{3a}}}(s,z;\omega_1,\omega_2)
\nonumber\\
&=\left(2(\omega_1+\omega_2)\right)^{-s}
\zeta\left(s,1+\frac{z}{2(\omega_1+\omega_2)}\right)
+2(\omega_1+\omega_2)^{-s}
\zeta\left(s,1+\frac{z}{\omega_1+\omega_2}\right)
\nonumber\\
&+3\left(\frac23(\omega_1+\omega_2)\right)^{-s}
\zeta\left(s,1+\frac{3z}{2(\omega_1+\omega_2)}\right). 
\end{align}
It has the residue 
\begin{align}
\label{Res_PdP3a}
\mathrm{Res}_{s=1}\mathfrak{Z}^{AdS_5\times Y_{PdP_{3a}}}(s,z;\omega_1,\omega_2)
&=\frac{7}{\omega_1+\omega_2} 
\end{align}
at a simple pole $s=1$. 
Substituting the residue coefficient (\ref{Res_PdP3a}) and the volume (\ref{vol_PdP3a}) into the formula (\ref{Riem2_conj}), 
the integral of the squared Riemann tensor is evaluated as
\begin{align}
\label{Riem2_PdP3a}
\int_{Y_{PdP_{3a}}}\mathrm{Riem}^2
&=124\pi^3. 
\end{align}
There is exact agreement between the value (\ref{Riem2_PdP3a}) and the result from the Hilbert series (see Appendix \ref{app_HS}).
The Zeta-index is
\begin{align}
\label{0_PdP3a}
\mathfrak{Z}^{AdS_5\times Y_{PdP_{3a}}}(0,0;\omega_1,\omega_2)&=-3. 
\end{align}
The supersymmetric zeta value with $s=-1$ and $z=0$ is
\begin{align}
\mathfrak{Z}^{AdS_5\times Y_{PdP_{3a}}}(-1,0;\omega_1,\omega_2)
&=-\frac12 (\omega_1+\omega_2). 
\end{align}
According to the formulae (\ref{c0_formula}) and (\ref{a0_formula}), we find
\begin{align}
c_{\mathcal{O}(N^0)}&=-\frac34, \\
a_{\mathcal{O}(N^0)}&=-\frac98,  
\end{align}
and 
\begin{align}
c_{\mathcal{O}(N^0)}-a_{\mathcal{O}(N^0)}&=\frac38. 
\end{align}

The supersymmetric determinant takes the form 
\begin{align}
\label{det_PdP3a}
\mathfrak{D}^{AdS_5\times Y_{PdP_{3a}}}(z;\omega_1,\omega_2)
&=\prod_{I=1}^6 \frac{\rho_I^{\frac12+\frac{z}{\rho_I}} \Gamma\left(1+\frac{z}{\rho_I}\right)}
{\sqrt{2\pi}}, 
\end{align}
where 
\begin{align}
\rho_I&=\frac13(b,m_I)(\omega_1+\omega_2). 
\end{align}
The vacuum exponent is
\begin{align}
\label{vac_PdP3a}
\mathfrak{D}^{AdS_5\times Y_{PdP_{3a}}}(0;1,1)
&=\frac{4}{3\sqrt{3} \pi^3}. 
\end{align}

\subsection{$AdS_5\times Y_{PdP_{4a}}$}
\label{sec_PdP4a}
While the del Pezzo surface $dP_4$ defined as the blow-up of $\mathbb{P}^2$ at generic four points is not toric, 
when the four points are in specific positions one finds the toric variety, the fourth Pseudo del Pezzo surface. 
The gauge theory on a stack of $N$ D3-branes probing the Calabi-Yau cone over 
$PdP_{4a}$ has 7 gauge nodes and 15 chiral multiplets 
\cite{Feng:2002fv,Butti:2006hc,Butti:2006nk,Benvenuti:2006qr,Forcella:2008bb,Hanany:2012hi,Bianchi:2014qma}. 
The field content is
\begin{align}
\begin{array}{c|cc}
&\prod_{i=1}^7 SU(N)_i&U(1)_R\\ \hline 
X_{13}&(\bf{N},\bf{1},\overline{\bf{N}},\bf{1},\bf{1},\bf{1},\bf{1})&x+y \\
X_{14}&(\bf{N},\bf{1},\bf{1},\overline{\bf{N}},\bf{1},\bf{1},\bf{1})&x+y \\
X_{17}&(\bf{N},\bf{1},\bf{1},\bf{1},\bf{1},\bf{1},\overline{\bf{N}})&2(1-x-y) \\
X_{21}&(\overline{\bf{N}},\bf{N},\bf{1},\bf{1},\bf{1},\bf{1},\bf{1})&2(1-x-y) \\
X_{27}&(\bf{1},\bf{N},\bf{1},\bf{1},\bf{1},\bf{1},\overline{\bf{N}})&2y \\
X_{35}&(\bf{1},\bf{1},\bf{N},\bf{1},\overline{\bf{N}},\bf{1},\bf{1})&y \\
X_{36}&(\bf{1},\bf{1},\bf{N},\bf{1},\bf{1},\overline{\bf{N}},\bf{1})&2(1-x-y) \\
X_{45}&(\bf{1},\bf{1},\bf{1},\bf{N},\overline{\bf{N}},\bf{1},\bf{1})&2(1-x-y) \\
X_{46}&(\bf{1},\bf{1},\bf{1},\bf{N},\bf{1},\overline{\bf{N}},\bf{1})&y \\
X_{51}&(\overline{\bf{N}},\bf{1},\bf{1},\bf{1},\bf{N},\bf{1},\bf{1})&x+y \\
X_{52}&(\bf{1},\overline{\bf{N}},\bf{1},\bf{1},\bf{N},\bf{1},\bf{1})&x \\
X_{61}&(\overline{\bf{N}},\bf{1},\bf{1},\bf{1},\bf{1},\bf{N},\bf{1})&x+y \\
X_{62}&(\bf{1},\overline{\bf{N}},\bf{1},\bf{1},\bf{1},\bf{N},\bf{1})&x \\
X_{73}&(\bf{1},\bf{1},\overline{\bf{N}},\bf{1},\bf{1},\bf{1},\bf{N})&x \\
X_{74}&(\bf{1},\bf{1},\bf{1},\overline{\bf{N}},\bf{1},\bf{1},\bf{N})&x \\
\end{array}
\end{align}
The quiver diagram is 
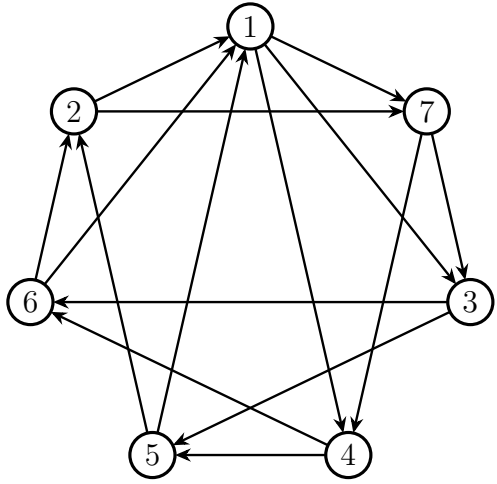
\begin{align}
\begin{tikzpicture}[<->]
\draw[very thick] (0.000,3.000) circle (3mm);
\node at (0.000,3.000) {1};
\draw[very thick] (-2.345,1.870) circle (3mm);
\node at (-2.345,1.870) {2};
\draw[very thick] (2.924,-0.667) circle (3mm);
\node at (2.924,-0.667) {3};
\draw[very thick] (1.302,-2.702) circle (3mm);
\node at (1.302,-2.702) {4};
\draw[very thick] (-1.302,-2.702) circle (3mm);
\node at (-1.302,-2.702) {5};
\draw[very thick] (-2.924,-0.667) circle (3mm);
\node at (-2.924,-0.667) {6};
\draw[very thick] (2.345,1.870) circle (3mm);
\node at (2.345,1.870) {7};
\draw[-{Stealth[length=6pt,width=5pt]}, line width=0.9pt, shorten <=3mm, shorten >=3mm] (0.000,3.000) -- (2.924,-0.667);   
\draw[-{Stealth[length=6pt,width=5pt]}, line width=0.9pt, shorten <=3mm, shorten >=3mm] (0.000,3.000) -- (1.302,-2.702);   
\draw[-{Stealth[length=6pt,width=5pt]}, line width=0.9pt, shorten <=3mm, shorten >=3mm] (0.000,3.000) -- (2.345,1.870);    
\draw[-{Stealth[length=6pt,width=5pt]}, line width=0.9pt, shorten <=3mm, shorten >=3mm] (-2.345,1.870) -- (0.000,3.000);  
\draw[-{Stealth[length=6pt,width=5pt]}, line width=0.9pt, shorten <=3mm, shorten >=3mm] (-2.345,1.870) -- (2.345,1.870);  
\draw[-{Stealth[length=6pt,width=5pt]}, line width=0.9pt, shorten <=3mm, shorten >=3mm] (2.924,-0.667) -- (-1.302,-2.702); 
\draw[-{Stealth[length=6pt,width=5pt]}, line width=0.9pt, shorten <=3mm, shorten >=3mm] (2.924,-0.667) -- (-2.924,-0.667); 
\draw[-{Stealth[length=6pt,width=5pt]}, line width=0.9pt, shorten <=3mm, shorten >=3mm] (1.302,-2.702) -- (-1.302,-2.702); 
\draw[-{Stealth[length=6pt,width=5pt]}, line width=0.9pt, shorten <=3mm, shorten >=3mm] (1.302,-2.702) -- (-2.924,-0.667); 
\draw[-{Stealth[length=6pt,width=5pt]}, line width=0.9pt, shorten <=3mm, shorten >=3mm] (-1.302,-2.702) -- (0.000,3.000);  
\draw[-{Stealth[length=6pt,width=5pt]}, line width=0.9pt, shorten <=3mm, shorten >=3mm] (-1.302,-2.702) -- (-2.345,1.870); 
\draw[-{Stealth[length=6pt,width=5pt]}, line width=0.9pt, shorten <=3mm, shorten >=3mm] (-2.924,-0.667) -- (0.000,3.000);  
\draw[-{Stealth[length=6pt,width=5pt]}, line width=0.9pt, shorten <=3mm, shorten >=3mm] (-2.924,-0.667) -- (-2.345,1.870); 
\draw[-{Stealth[length=6pt,width=5pt]}, line width=0.9pt, shorten <=3mm, shorten >=3mm] (2.345,1.870) -- (2.924,-0.667);   
\draw[-{Stealth[length=6pt,width=5pt]}, line width=0.9pt, shorten <=3mm, shorten >=3mm] (2.345,1.870) -- (1.302,-2.702);   
\end{tikzpicture}
\end{align}
The superpotential takes the form \cite{Forcella:2008bb}
\begin{align}
\mathcal{W}
&=\Tr(X_{61}X_{17}X_{74}X_{46}+X_{21}X_{13}X_{35}X_{52}+X_{27}X_{73}X_{36}X_{62}+X_{14}X_{45}X_{51}
\nonumber\\
&-X_{51}X_{17}X_{73}X_{35}-X_{21}X_{14}X_{46}X_{62}-X_{27}X_{74}X_{45}X_{52}-X_{13}X_{36}X_{61}
). 
\end{align}
The toric diagram of the cone over $PdP_{4a}$ is generated by
\begin{align}
v_1&=(1,0,-1),& 
v_2&=(1,1,-1),& 
v_3&=(1,1,0), &
v_4&=(1,0,1), \nonumber\\
v_5&=(1,-1,1),& 
v_6&=(1,-1,0),&
v_7&=(1,-1,-1). 
\end{align}
It is shown as
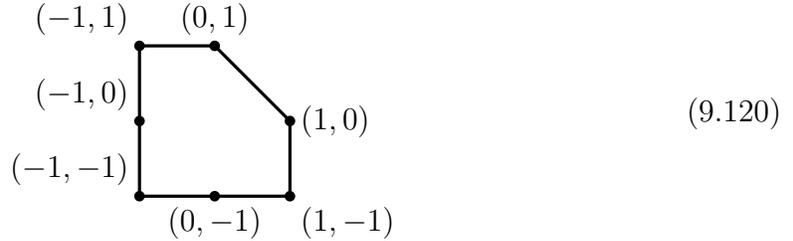
\begin{align}
\label{toric_PdP4a}
\begin{tikzpicture}[scale=1, baseline={(0,0)}]
  \fill (0,-1) circle (2pt) node[below] {$(0,-1)$};
  \fill (1,-1) circle (2pt) node[below right] {$(1,-1)$};
  \fill (1,0) circle (2pt) node[right] {$(1,0)$};
  \fill (0,1) circle (2pt) node[above] {$(0,1)$};
  \fill (-1,1) circle (2pt) node[above left] {$(-1,1)$};
  \fill (-1,0) circle (2pt) node[above left] {$(-1,0)$};
    \fill (-1,-1) circle (2pt) node[above left] {$(-1,-1)$};
  \draw[very thick] (0,-1) -- (1,-1) -- (1,0) -- (0,1) -- (-1,1) -- (-1,0) -- (-1,-1) -- cycle;
\end{tikzpicture}
\end{align}
The multiplicities of five types of the chiral multiplets 
$X_2$ $=$ $\{X_{52},X_{74}\}$, 
$X_3$ $=$ $\{X_{35}\}$, 
$X_4$ $=$ $\{X_{46}\}$, 
$X_5$ $=$ $\{X_{62},X_{73}\}$ 
and $X_7$ $=$ $\{X_{17},X_{21},X_{36},X_{45}\}$  
forming the baryons are given by
\begin{align}
\det (v_{7},v_{2},v_{3})&=2, \\
\det (v_{2},v_{3},v_{4})&=1, \\
\det (v_{3},v_{4},v_{5})&=1, \\
\det (v_{3},v_{5},v_{7})&=2, \\
\det (v_{5},v_{7},v_{2})&=4. 
\end{align}
The area of the polygon is
\begin{align}
\label{A_PdP4a}
A&=\frac72. 
\end{align}
The dual cone is generated by
\begin{align}
\label{dualG_PdP4a}
m_1&=(1,0,1),& 
m_2&=(1,-1,0),&
m_3&=(1,-1,-1),&
m_4&=(1,0,-1),\nonumber\\ 
m_5&=(1,1,0),&
m_6&=(1,1,0),&
m_7&=(1,0,1).  
\end{align}
Since the toric diagram admits a reflection symmetry exchanging two lattice directions, 
we can write the Reeb vector as $b=(3,b_1,b_1)$.  
The volume of the Sasaki-Einstein base of the cone over $PdP_{4a}$ is given by
\begin{align}
\label{vol_PdP4a}
\mathrm{Vol}(Y_{PdP_{4a}})
&=
\frac{15-b_1}
{(3-b_1)(3+b_1)^2(3-2b_1)}\pi^3. 
\end{align}
Applying the volume minimization \cite{Martelli:2005tp}, 
the Reeb vector is numerically determined as
\begin{align}
\label{Reeb_PdP4a}
b_1&=-0.379079\ldots, 
\end{align}
and 
\begin{align}
\mathrm{Vol}(Y_{PdP_{4a}})
&=5.46637\ldots. 
\end{align}
The R-charges of the chiral multiplet fields are evaluated as
\begin{align}
R(X_{2})&=\frac{\pi}{3}\frac{\mathrm{Vol}(\Sigma_2)}{\mathrm{Vol}(Y_{PdP_{4a}})}
=\frac{2(3+b_1)(3-2b_1)}{3(15-b_1)}=x=0.42698\ldots, \\
R(X_3)&=\frac{\pi}{3}\frac{\mathrm{Vol}(\Sigma_3)}{\mathrm{Vol}(Y_{PdP_{4a}})}
=\frac{2(3+b_1)^2}{3(15-b_1)}=y=0.297774\ldots, \\
R(X_{4})&=\frac{\pi}{3}\frac{\mathrm{Vol}(\Sigma_4)}{\mathrm{Vol}(Y_{PdP_{4a}})}
=\frac{2(3+b_1)^2}{3(15-b_1)}=y=0.297774\ldots, \\
R(X_5)&=\frac{\pi}{3}\frac{\mathrm{Vol}(\Sigma_5)}{\mathrm{Vol}(Y_{PdP_{4a}})}
=\frac{2(3+b_1)(3-2b_1)}{3(15-b_1)}=x=0.42698\ldots, \\
R(X_7)&=\frac{\pi}{3}\frac{\mathrm{Vol}(\Sigma_7)}{\mathrm{Vol}(Y_{PdP_{4a}})}
=\frac{2(3-b_1)(3-2b_1)}{3(15-b_1)}=2(1-x-y)=0.550493\ldots, 
\end{align}
where 
\begin{align}
\mathrm{Vol}(\Sigma_2)&=2\pi^2 \frac{\det(v_{7},v_{2},v_{3})}{\det(b,v_{7},v_2)\det(b,v_{2},v_{3})}
=\frac{2\pi^2}{9-b_1^2}=2.22883\ldots, \\
\mathrm{Vol}(\Sigma_3)&=2\pi^2 \frac{\det(v_{2},v_{3},v_{4})}{\det(b,v_{2},v_3)\det(b,v_{3},v_{4})}
=\frac{2\pi^2}{9-9b_1+2b_1^2}=1.55438\ldots, \\
\mathrm{Vol}(\Sigma_4)&=2\pi^2 \frac{\det(v_{3},v_{4},v_{5})}{\det(b,v_{3},v_4)\det(b,v_{4},v_{5})}
=\frac{2\pi^2}{9-9b_1+2b_1^2}=1.55438\ldots, \\
\mathrm{Vol}(\Sigma_5)&=2\pi^2 \frac{\det(v_{4},v_{5},v_{7})}{\det(b,v_{4},v_5)\det(b,v_{5},v_{7})}
=\frac{2\pi^2}{9-b_1^2}=2.22883\ldots, \\
\mathrm{Vol}(\Sigma_7)&=2\pi^2 \frac{\det(v_{5},v_{7},v_{2})}{\det(b,v_{5},v_7)\det(b,v_{7},v_{2})}
=\frac{2\pi^2}{(3+b_1)^2}=2.87357\ldots. 
\end{align}

The weighted adjacency matrix is 
\begin{align}
\label{M_PdP4a}
&
M^{AdS_5\times Y_{PdP_{4a}}}
\nonumber\\
&=\left(
\begin{matrix}
i_v&\tilde{i}_{\textrm{cm}_{2(1-x-y)}}&i_{\textrm{cm}_{x+y}}&i_{\textrm{cm}_{x+y}}&\tilde{i}_{\textrm{cm}_{x+y}}&\tilde{i}_{\textrm{cm}_{x+y}}&i_{\textrm{cm}_{2(1-x-y)}} \\
i_{\textrm{cm}_{2(1-x-y)}}&i_v&0&0&\tilde{i}_{\textrm{cm}_{x}}&\tilde{i}_{\textrm{cm}_{x}}&i_{\textrm{cm}_{2y}} \\
\tilde{i}_{\textrm{cm}_{x+y}}&0&i_v&0&i_{\textrm{cm}_{y}}&i_{\textrm{cm}_{2(1-x-y)}}&\tilde{i}_{\textrm{cm}_{x}} \\
\tilde{i}_{\textrm{cm}_{x+y}}&0&0&i_v&i_{\textrm{cm}_{2(1-x-y)}}&i_{\textrm{cm}_{y}}&\tilde{i}_{\textrm{cm}_{x}} \\
i_{\textrm{cm}_{x+y}}&i_{\textrm{cm}_{x}}&\tilde{i}_{\textrm{cm}_{y}}&\tilde{i}_{\textrm{cm}_{2(1-x-y)}}&i_v&0&0 \\
i_{\textrm{cm}_{x+y}}&i_{\textrm{cm}_{x}}&\tilde{i}_{\textrm{cm}_{2(1-x-y)}}&\tilde{i}_{\textrm{cm}_{y}}&0&i_v&0 \\
\tilde{i}_{\textrm{cm}_{2(1-x-y)}}&\tilde{i}_{\textrm{cm}_{2y}}&i_{\textrm{cm}_{x}}&i_{\textrm{cm}_{x}}&0&0&i_v \\
\end{matrix}
\right). 
\end{align}
The gravity index is evaluated as
\begin{align}
\label{ind_PdP4a}
&
i^{AdS_5\times Y_{PdP_{4a}}}(p;q)
\nonumber\\
&=\frac{(pq)^{3-2x-3y}}{1-(pq)^{3-2x-3y}}
+4\frac{(pq)^{x+\frac{3y}{2}}}{1-(pq)^{x+\frac{3y}{2}}}
+2\frac{(pq)^{2-x-\frac{3y}{2}}}{1-(pq)^{2-x-\frac{3y}{2}}}. 
\end{align}
There exist seven extremal BPS mesons, 
which correspond to the primitive generators (\ref{dualG_PdP4a}) of the dual cone. 
The R-charges are given by the pairings
\begin{align}
R(\mathcal{M}_1)&=\frac23(b,m_1)=\frac23(3+b_1)=2x+3y=1.74728\ldots, \\
R(\mathcal{M}_2)&=\frac23(b,m_2)=\frac23(3-b_1)=4-2x-3y=2.25272\ldots, \\
R(\mathcal{M}_3)&=\frac23(b,m_3)=\frac23(3-2b_1)=6-4x-6y=2.50544\ldots, \\
R(\mathcal{M}_4)&=\frac23(b,m_4)=\frac23(3-b_1)=4-2x-3y=2.25272\ldots, \\
R(\mathcal{M}_5)&=\frac23(b,m_5)=\frac23(3+b_1)=2x+3y=1.74728\ldots, \\
R(\mathcal{M}_6)&=\frac23(b,m_6)=\frac23(3+b_1)=2x+3y=1.74728\ldots, \\
R(\mathcal{M}_7)&=\frac23(b,m_7)=\frac23(3+b_1)=2x+3y=1.74728\ldots, 
\end{align}
where
\begin{align}
b_1&=\frac32(2x+3y)-3. 
\end{align}

The supersymmetric zeta function is given by
\begin{align}
\label{zeta_PdP4a}
&
\mathfrak{Z}^{AdS_5\times Y_{PdP_{4a}}}(s,z;\omega_1,\omega_2)
\nonumber\\
&=\left(\left( 3-2x-3y \right)(\omega_1+\omega_2) \right)^{-s}\zeta\left(s,1+\frac{z}{(3-2x-3y)(\omega_1+\omega_2)}\right)
\nonumber\\
&+4\left(\left( x+\frac{3y}{2} \right)(\omega_1+\omega_2) \right)^{-s}\zeta\left(s,1+\frac{z}{(x+\frac{3y}{2})(\omega_1+\omega_2)}\right)
\nonumber\\
&+2\left(\left(2- x-\frac{3y}{2} \right)(\omega_1+\omega_2) \right)^{-s}\zeta\left(s,1+\frac{z}{(2-x-\frac{3y}{2})(\omega_1+\omega_2)}\right). 
\end{align}
The residue at a simple pole $s=1$ is
\begin{align}
&
\label{Res_PdP4a}
\mathrm{Res}_{s=1}\mathfrak{Z}^{AdS_5\times Y_{PdP_{4a}}}(s,z;\omega_1,\omega_2)
\nonumber\\
&=\frac{12x^2+27y^2+36xy-80x-120y+96}
{(2x+3y-4)(2x+3y-3)(2x+3y)}\frac{1}{\omega_1+\omega_2}
\nonumber\\
&=\frac{9(21+b_1^2-14b_1)}{(3-b_1)(3+b_1)(3-2b_1)}\frac{1}{\omega_1+\omega_2}. 
\end{align}
Substituting the residue coefficient (\ref{Res_PdP4a}) and the volume (\ref{vol_PdP4a}) into the formula (\ref{Riem2_conj}), 
we obtain the integrated Riemann-tensor squared
\begin{align}
\label{Riem2_PdP4a}
\int_{Y_{PdP_{4a}}} \mathrm{Riem}^2
&=\frac{8\pi^3(24b_1^3-264b_1^2-485b_1+1227)}{(3-b_1)(3+b_1)^2(3-2b_1)}
\nonumber\\
&=3900.22\ldots.
\end{align}
Perfect agreement is found between the value (\ref{Riem2_PdP4a}) and the Hilbert series computation (see Appendix \ref{app_HS}).
The Zeta-index is
\begin{align}
\label{0_PdP4a}
\mathfrak{Z}^{AdS_5\times Y_{PdP_{4a}}}(0,0;\omega_1,\omega_2)&=-\frac72. 
\end{align}
The supersymmetric zeta value with $s=-1$ and $z=0$ is evaluated as
\begin{align}
\label{-1_PdP4a}
\mathfrak{Z}^{AdS_5\times Y_{PdP_{4a}}}(-1,0;\omega_1,\omega_2)
&=-\frac{7}{12} (\omega_1+\omega_2). 
\end{align}
From (\ref{c0_formula}) and (\ref{a0_formula}) we find
\begin{align}
c_{\mathcal{O}(N^0)}&=-\frac78, \\
a_{\mathcal{O}(N^0)}&=-\frac{21}{16}, 
\end{align}
and 
\begin{align}
c_{\mathcal{O}(N^0)}-a_{\mathcal{O}(N^0)}&=\frac{7}{16}. 
\end{align}

The supersymmetric determinant takes the form 
\begin{align}
\label{det_PdP4a}
\mathfrak{D}^{AdS_5\times Y_{PdP_{4a}}}(z;\omega_1,\omega_2)
&=\prod_{I=1}^7 \frac{\rho_I^{\frac12+\frac{z}{\rho_I}} \Gamma\left(1+\frac{z}{\rho_I}\right)}
{\sqrt{2\pi}}, 
\end{align}
where 
\begin{align}
\rho_I&=\frac13(b,m_I)(\omega_1+\omega_2). 
\end{align}
The vacuum exponent is
\begin{align}
\label{vac_PdP4a}
\mathfrak{D}^{AdS_5\times Y_{PdP_{4a}}}(0;1,1)
&=\frac{\sqrt{3-2x-3y}(2x+3y)^2(4-2x-3y)}{8\pi^{\frac72}}
\nonumber\\
&=0.0175083\ldots. 
\end{align}

\subsection{$AdS_5\times Y_{PdP_{4b}}$}
\label{sec_PdP4b}
The 4d $\mathcal{N}=1$ quiver gauge theory describing $N$ D3-branes probing the toric Calabi-Yau 3-fold as the complex cone over $PdP_{4b}$ 
has seven $SU(N)$ gauge nodes and 19 bifundamental chiral multiplets \cite{Hanany:2012hi,Bianchi:2014qma}. 
The field content is
\begin{align}
\begin{array}{c|cc}
&\prod_{i=1}^7 SU(N)_i&U(1)_R\\ \hline 
X_{13}&(\bf{N},\bf{1},\overline{\bf{N}},\bf{1},\bf{1},\bf{1},\bf{1})&z \\
X_{16}&(\bf{N},\bf{1},\bf{1},\bf{1},\bf{1},\overline{\bf{N}},\bf{1})&y \\
X_{17}&(\bf{N},\bf{1},\bf{1},\bf{1},\bf{1},\bf{1},\overline{\bf{N}})&x \\
X_{21}&(\overline{\bf{N}},\bf{N},\bf{1},\bf{1},\bf{1},\bf{1},\bf{1})&2-x-y \\
X_{25}&(\bf{1},\bf{N},\bf{1},\bf{1},\overline{\bf{N}},\bf{1},\bf{1})&y \\
X_{26}&(\bf{1},\bf{N},\bf{1},\bf{1},\bf{1},\overline{\bf{N}},\bf{1})&2-y-z \\
X_{34}&(\bf{1},\bf{1},\bf{N},\overline{\bf{N}},\bf{1},\bf{1},\bf{1})&2-x-y-z \\
X_{37}&(\bf{1},\bf{1},\bf{N},\bf{1},\bf{1},\bf{1},\overline{\bf{N}})&y \\
X_{42}&(\bf{1},\overline{\bf{N}},\bf{1},\bf{N},\bf{1},\bf{1},\bf{1})&z \\
X_{45}&(\bf{1},\bf{1},\bf{1},\bf{N},\overline{\bf{N}},\bf{1},\bf{1})&x \\
X_{51}&(\overline{\bf{N}},\bf{1},\bf{1},\bf{1},\bf{N},\bf{1},\bf{1})&y \\
X_{53}&(\bf{1},\bf{1},\overline{\bf{N}},\bf{1},\bf{N},\bf{1},\bf{1})&x \\
X_{56}&(\bf{1},\bf{1},\bf{1},\bf{1},\bf{N},\overline{\bf{N}},\bf{1})&2-x-y \\
X_{62}&(\bf{1},\overline{\bf{N}},\bf{1},\bf{1},\bf{1},\bf{N},\bf{1})&x \\
X_{64}&(\bf{1},\bf{1},\bf{1},\overline{\bf{N}},\bf{1},\bf{N},\bf{1})&y \\
X_{67}&(\bf{1},\bf{1},\bf{1},\bf{1},\bf{1},\bf{N},\overline{\bf{N}})&z \\
X_{71}&(\overline{\bf{N}},\bf{1},\bf{1},\bf{1},\bf{1},\bf{1},\bf{N})&2-y-z \\
X_{72}&(\bf{1},\overline{\bf{N}},\bf{1},\bf{1},\bf{1},\bf{1},\bf{N})&y \\
X_{75}&(\bf{1},\bf{1},\bf{1},\bf{1},\overline{\bf{N}},\bf{1},\bf{N})&2-x-y \\
\end{array}
\end{align}
The quiver diagram is 
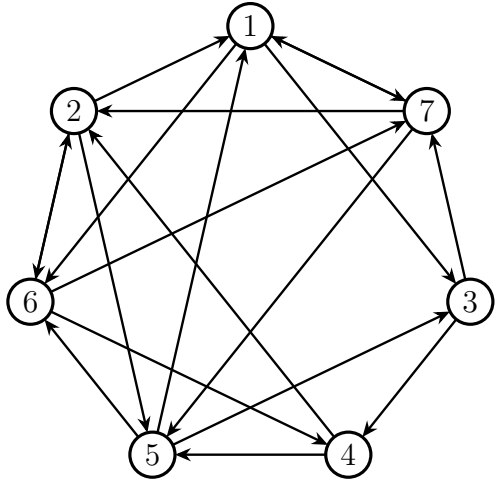
\begin{align}
\begin{tikzpicture}[<->]
\draw[very thick] (0.000,3.000) circle (3mm);
\node at (0.000,3.000) {1};
\draw[very thick] (-2.345,1.870) circle (3mm);
\node at (-2.345,1.870) {2};
\draw[very thick] (2.924,-0.667) circle (3mm);
\node at (2.924,-0.667) {3};
\draw[very thick] (1.302,-2.702) circle (3mm);
\node at (1.302,-2.702) {4};
\draw[very thick] (-1.302,-2.702) circle (3mm);
\node at (-1.302,-2.702) {5};
\draw[very thick] (-2.924,-0.667) circle (3mm);
\node at (-2.924,-0.667) {6};
\draw[very thick] (2.345,1.870) circle (3mm);
\node at (2.345,1.870) {7};
\draw[-{Stealth[length=6pt,width=5pt]}, line width=0.9pt, shorten <=3mm, shorten >=3mm] (0.000,3.000) -- (2.924,-0.667);   
\draw[-{Stealth[length=6pt,width=5pt]}, line width=0.9pt, shorten <=3mm, shorten >=3mm] (0.000,3.000) -- (-2.924,-0.667);  
\draw[-{Stealth[length=6pt,width=5pt]}, line width=0.9pt, shorten <=3mm, shorten >=3mm] (0.000,3.000) -- (2.345,1.870);    
\draw[-{Stealth[length=6pt,width=5pt]}, line width=0.9pt, shorten <=3mm, shorten >=3mm] (-2.345,1.870) -- (0.000,3.000);   
\draw[-{Stealth[length=6pt,width=5pt]}, line width=0.9pt, shorten <=3mm, shorten >=3mm] (-2.345,1.870) -- (-1.302,-2.702); 
\draw[-{Stealth[length=6pt,width=5pt]}, line width=0.9pt, shorten <=3mm, shorten >=3mm] (-2.345,1.870) -- (-2.924,-0.667); 
\draw[-{Stealth[length=6pt,width=5pt]}, line width=0.9pt, shorten <=3mm, shorten >=3mm] (2.924,-0.667) -- (1.302,-2.702);  
\draw[-{Stealth[length=6pt,width=5pt]}, line width=0.9pt, shorten <=3mm, shorten >=3mm] (2.924,-0.667) -- (2.345,1.870);   
\draw[-{Stealth[length=6pt,width=5pt]}, line width=0.9pt, shorten <=3mm, shorten >=3mm] (1.302,-2.702) -- (-2.345,1.870);  
\draw[-{Stealth[length=6pt,width=5pt]}, line width=0.9pt, shorten <=3mm, shorten >=3mm] (1.302,-2.702) -- (-1.302,-2.702); 
\draw[-{Stealth[length=6pt,width=5pt]}, line width=0.9pt, shorten <=3mm, shorten >=3mm] (-1.302,-2.702) -- (0.000,3.000);  
\draw[-{Stealth[length=6pt,width=5pt]}, line width=0.9pt, shorten <=3mm, shorten >=3mm] (-1.302,-2.702) -- (2.924,-0.667); 
\draw[-{Stealth[length=6pt,width=5pt]}, line width=0.9pt, shorten <=3mm, shorten >=3mm] (-1.302,-2.702) -- (-2.924,-0.667);
\draw[-{Stealth[length=6pt,width=5pt]}, line width=0.9pt, shorten <=3mm, shorten >=3mm] (-2.924,-0.667) -- (-2.345,1.870); 
\draw[-{Stealth[length=6pt,width=5pt]}, line width=0.9pt, shorten <=3mm, shorten >=3mm] (-2.924,-0.667) -- (1.302,-2.702); 
\draw[-{Stealth[length=6pt,width=5pt]}, line width=0.9pt, shorten <=3mm, shorten >=3mm] (-2.924,-0.667) -- (2.345,1.870);  
\draw[-{Stealth[length=6pt,width=5pt]}, line width=0.9pt, shorten <=3mm, shorten >=3mm] (2.345,1.870) -- (0.000,3.000);    
\draw[-{Stealth[length=6pt,width=5pt]}, line width=0.9pt, shorten <=3mm, shorten >=3mm] (2.345,1.870) -- (-2.345,1.870);  
\draw[-{Stealth[length=6pt,width=5pt]}, line width=0.9pt, shorten <=3mm, shorten >=3mm] (2.345,1.870) -- (-1.302,-2.702); 
\end{tikzpicture}
\end{align}
The superpotential takes the form
\begin{align}
\mathcal{W}
&=\Tr(X_{21}X_{17}X_{72}+X_{42}X_{26}X_{64}+X_{56}X_{62}X_{25}+X_{67}X_{71}X_{16}
\nonumber\\
&+X_{75}X_{53}X_{37}+X_{13}X_{34}X_{45}X_{51}-X_{13}X_{37}X_{71}-X_{16}X_{62}X_{21}
\nonumber\\
&-X_{56}X_{64}X_{45}-X_{67}X_{72}X_{26}-X_{75}X_{51}X_{17}-X_{25}X_{53}X_{34}X_{42}
). 
\end{align}
The toric diagram of the cone over $PdP_{4b}$ is generated by
\begin{align}
v_1&=(1,0,0),& 
v_2&=(1,1,0),& 
v_3&=(1,2,1), &
v_4&=(1,1,2), \nonumber\\
v_5&=(1,0,3),& 
v_6&=(1,0,2),&
v_7&=(1,0,1). 
\end{align}
It is shown as
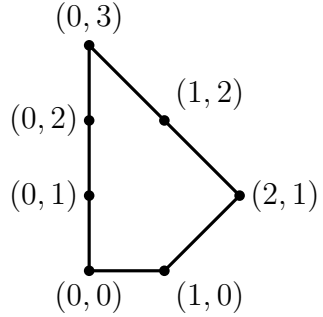
\begin{align}
\label{toric_PdP4b}
\begin{tikzpicture}[scale=1, baseline={(0,0)}]
  \fill (0,0) circle (2pt) node[below] {$(0,0)$};
  \fill (1,0) circle (2pt) node[below right] {$(1,0)$};
  \fill (2,1) circle (2pt) node[right] {$(2,1)$};
  \fill (1,2) circle (2pt) node[above right] {$(1,2)$};
  \fill (0,3) circle (2pt) node[above] {$(0,3)$};
  \fill (0,2) circle (2pt) node[left] {$(0,2)$};
  \fill (0,1) circle (2pt) node[left] {$(0,1)$};
  \draw[very thick] (0,0) -- (1,0) -- (2,1) -- (1,2) -- (0,3) -- (0,2) -- (0,1) -- cycle;
\end{tikzpicture}
\end{align}
The multiplicities of four types of the chiral multiplets 
$X_1$ $=$ $\{X_{13}$, $X_{42}$, $X_{67}\}$, 
$X_2$ $=$ $\{X_{34}\}$, 
$X_3$ $=$ $\{X_{17}$, $X_{45}$, $X_{53}$, $X_{62}\}$, 
$X_5$ $=$ $\{X_{16}$, $X_{25}$, $X_{37}$, $X_{51}$, $X_{64}$, $X_{72}\}$, 
forming the baryons are given by
\begin{align}
\det (v_{5},v_{1},v_{2})&=3, \\
\det (v_{1},v_{2},v_{3})&=1, \\
\det (v_{2},v_{3},v_{5})&=4, \\
\det (v_{3},v_{5},v_{1})&=6. 
\end{align}
The area of the polygon is
\begin{align}
\label{A_PdP4b}
A&=\frac72. 
\end{align}
The dual cone is generated by
\begin{align}
\label{dualG_PdP4b}
m_1&=(0,0,1),& 
m_2&=(1,-1,1),&
m_3&=(3,-1,-1),&
m_4&=(3,-1,-1),\nonumber\\ 
m_5&=(0,1,0),&
m_6&=(0,1,0),&
m_7&=(0,1,0).  
\end{align}
The volume of the Sasaki-Einstein base of the cone over $PdP_{4b}$ is given by
\begin{align}
\label{vol_PdP4b}
\mathrm{Vol}(Y_{PdP_{4b}})
&=
\frac{9-b_1+3b_2}{b_1b_2(3-b_1+b_2)(9-b_1-b_2)}\pi^3, 
\end{align}
where $b=(3,b_1,b_2)$ is the Reeb vector. 
Applying the volume minimization \cite{Martelli:2005tp}, 
the Reeb vector is numerically determined as
\begin{align}
\label{Reeb_PdP4b}
b_1&=2.09747\ldots, \\
b_2&=3.74544\ldots, 
\end{align}
and 
\begin{align}
\mathrm{Vol}(Y_{PdP_{4b}})
&=4.87877\ldots. 
\end{align}
The R-charges of the chiral multiplet fields are evaluated as
\begin{align}
R(X_{1})&=\frac{\pi}{3}\frac{\mathrm{Vol}(\Sigma_1)}{\mathrm{Vol}(Y_{PdP_{4b}})}
=\frac{2(3-b_1+b_2)(9-b_1-b_2)}{3(9-b_1+3b_2)}=z=0.539323\ldots, \\
R(X_2)&=\frac{\pi}{3}\frac{\mathrm{Vol}(\Sigma_2)}{\mathrm{Vol}(Y_{PdP_{4b}})}
=\frac{2b_1(9-b_1-b_2)}{3(9-b_1+3b_2)}=2-x-y-z=0.243378\ldots, \\
R(X_{3})&=\frac{\pi}{3}\frac{\mathrm{Vol}(\Sigma_3)}{\mathrm{Vol}(Y_{PdP_{4b}})}
=\frac{4b_1b_2}{3(9-b_1+3b_2)}=x=0.577468\ldots, \\
R(X_5)&=\frac{\pi}{3}\frac{\mathrm{Vol}(\Sigma_5)}{\mathrm{Vol}(Y_{PdP_{4b}})}
=\frac{2b_2(3-b_1+b_2)}{3(9-b_1+3b_2)}=y=0.639832\ldots,
\end{align}
where 
\begin{align}
\mathrm{Vol}(\Sigma_1)&=2\pi^2 \frac{\det(v_{5},v_{1},v_{2})}{\det(b,v_{5},v_1)\det(b,v_{1},v_{2})}
=\frac{2\pi^2}{b_1b_2}=2.51264\ldots, \\
\mathrm{Vol}(\Sigma_2)&=2\pi^2 \frac{\det(v_{1},v_{2},v_{3})}{\det(b,v_{1},v_2)\det(b,v_{2},v_{3})}
=\frac{2\pi^2}{b_2(3-b_1+b_2)}=1.13387\ldots, \\
\mathrm{Vol}(\Sigma_3)&=2\pi^2 \frac{\det(v_{2},v_{3},v_{5})}{\det(b,v_{2},v_3)\det(b,v_{3},v_{5})}
=\frac{4\pi^2}{(9-b_1-b_2)(3-b_1+b_2)}=2.69035\ldots, \\
\mathrm{Vol}(\Sigma_5)&=2\pi^2 \frac{\det(v_{3},v_{5},v_{1})}{\det(b,v_{3},v_5)\det(b,v_{5},v_{1})}
=\frac{2\pi^2}{b_1(9-b_1-b_2)}=2.9809\ldots. 
\end{align}

One has the weighted adjacency matrix
\begin{align}
\label{M_PdP4b}
&
M^{AdS_5\times Y_{PdP_{4b}}}
\nonumber\\
&=\left(
\begin{matrix}
i_v&\tilde{i}_{2-x-y}&i_z&0&\tilde{i}_{y}&i_y&i_x+\tilde{i}_{2-y-z}\\
i_{2-x-y}&i_v&0&\tilde{i}_z&i_y&i_{2-y-z}+\tilde{i}_x&\tilde{i}_y\\
\tilde{i}_z&0&i_v&i_{2-x-y-z}&\tilde{i}_x&0&i_y\\
0&i_z&\tilde{i}_{2-x-y-z}&i_v&i_x&\tilde{i}_y&0 \\
i_y&\tilde{i}_y&i_x&\tilde{i}_x&i_v&i_{2-x-y}&\tilde{i}_{2-x-y} \\
\tilde{i}_y&\tilde{i}_{2-y-z}+i_{x}&0&i_y&\tilde{i}_{2-x-y}&i_v&i_z \\
\tilde{i}_x+i_{2-y-z}&i_y&\tilde{i}_y&0&i_{2-x-y}&\tilde{i}_z&i_v \\
\end{matrix}
\right). 
\end{align}
The gravity index reads
\begin{align}
\label{ind_PdP4b}
&
i^{AdS_5\times Y_{PdP_{4b}}}(p;q)
\nonumber\\
&=\frac{(pq)^{\frac{b_2}{3}}}{1-(pq)^{\frac{b_2}{3}}}
+\frac{(pq)^{\frac{3-b_1+b_2}{3}}}{1-(pq)^{\frac{3-b_1+b_2}{3}}}
+2\frac{(pq)^{\frac{9-b_1-b_2}{3}}}{1-(pq)^{\frac{9-b_1-b_2}{3}}}
+3\frac{(pq)^{\frac{b_1}{3}}}{1-(pq)^{\frac{b_1}{3}}}. 
\end{align}
The primitive generators (\ref{dualG_PdP4b}) of the dual cone
correspond to seven extremal BPS mesons, whose R-charges are given as follows.
\begin{align}
R(\mathcal{M}_1)&=\frac23(b,m_1)=\frac{2b_2}{3}=x+3y=2.49696\ldots, \\
R(\mathcal{M}_2)&=\frac23(b,m_2)=\frac{2(3-b_1+b_2)}{3}=z+4y=3.09865\ldots, \\
R(\mathcal{M}_3)&=\frac23(b,m_3)=\frac{2(9-b_1-b_2)}{3}=4-2x-2y+z=2.10472\ldots, \\
R(\mathcal{M}_4)&=\frac23(b,m_4)=\frac{2(9-b_1-b_2)}{3}=4-2x-2y+z=2.10472\ldots, \\
R(\mathcal{M}_5)&=\frac23(b,m_5)=\frac{2b_1}{3}=2+x-y-z=1.39831\ldots, \\
R(\mathcal{M}_6)&=\frac23(b,m_6)=\frac{2b_1}{3}=2+x-y-z=1.39831\ldots, \\
R(\mathcal{M}_7)&=\frac23(b,m_7)=\frac{2b_1}{3}=2+x-y-z=1.39831\ldots, 
\end{align}
where
\begin{align}
b_1&=\frac32(2+x-y-z), \qquad 
b_2=\frac32(x+3y). 
\end{align}

The supersymmetric zeta function is 
\begin{align}
\label{zeta_PdP4b}
&
\mathfrak{Z}^{AdS_5\times Y_{PdP_{4b}}}(s,z;\omega_1,\omega_2)
\nonumber\\
&=\left(\frac{b_2}{3}(\omega_1+\omega_2)\right)^{-s}
\zeta\left(s,1+\frac{3z}{b_2(\omega_1+\omega_2)}\right)
\nonumber\\
&+\left(\frac{3-b_1+b_2}{3}(\omega_1+\omega_2)\right)^{-s}
\zeta\left(s,1+\frac{3z}{(3-b_1+b_2)(\omega_1+\omega_2)}\right)
\nonumber\\
&+2\left(\frac{9-b_1-b_2}{3}(\omega_1+\omega_2)\right)^{-s}
\zeta\left(s,1+\frac{3z}{(9-b_1-b_2)(\omega_1+\omega_2)}\right)
\nonumber\\
&+3\left(\frac{b_1}{3}(\omega_1+\omega_2)\right)^{-s}
\zeta\left(s,1+\frac{3z}{b_1(\omega_1+\omega_2)}\right). 
\end{align}
The residue at a simple pole $s=1$ is
\begin{align}
&
\label{Res_PdP4b}
\mathrm{Res}_{s=1}\mathfrak{Z}^{AdS_5\times Y_{PdP_{4b}}}(s,z;\omega_1,\omega_2)
\nonumber\\
&=\frac{3b_1^3-9b_2^3-36b_1^2+54b_2^2-45b_1b_2+81b_1+243b_2}
{b_1b_2(3-b_1+b_2)(9-b_1-b_2)}\frac{1}{\omega_1+\omega_2}. 
\end{align}
Inserting the residue coefficient (\ref{Res_PdP4b}) and the volume (\ref{vol_PdP4b}) into the formula (\ref{Riem2_conj}) gives
the integrated Riemann-tensor squared
\begin{align}
\label{Riem2_PdP4b}
\int_{Y_{PdP_{4b}}} \mathrm{Riem}^2
&=\frac{64b_1^3-192b_2^3-768b_1^2+1152b_2^2-960b_1b_2+1880b_1+4728b_2-1368}
{b_1b_2(3-b_1+b_2)(9-b_1-b_2)}
\nonumber\\
&=4310.57\ldots.
\end{align}
We have verified that the value (\ref{Riem2_PdP4b}) exactly agrees with the Hilbert series result (see Appendix \ref{app_HS}).
The Zeta-index is
\begin{align}
\label{0_PdP4b}
\mathfrak{Z}^{AdS_5\times Y_{PdP_{4b}}}(0,0;\omega_1,\omega_2)&=-\frac72. 
\end{align}
The supersymmetric zeta value with $s=-1$ and $z=0$ is 
\begin{align}
\label{-1_PdP4b}
\mathfrak{Z}^{AdS_5\times Y_{PdP_{4b}}}(-1,0;\omega_1,\omega_2)
&=-\frac{7}{12} (\omega_1+\omega_2). 
\end{align}
From (\ref{c0_formula}) and (\ref{a0_formula}) we find
\begin{align}
c_{\mathcal{O}(N^0)}&=-\frac78, \\
a_{\mathcal{O}(N^0)}&=-\frac{21}{16}, 
\end{align}
and 
\begin{align}
c_{\mathcal{O}(N^0)}-a_{\mathcal{O}(N^0)}&=\frac{7}{16}. 
\end{align}

The supersymmetric determinant takes the form 
\begin{align}
\label{det_PdP4b}
\mathfrak{D}^{AdS_5\times Y_{PdP_{4b}}}(z;\omega_1,\omega_2)
&=\prod_{I=1}^7 \frac{\rho_I^{\frac12+\frac{z}{\rho_I}} \Gamma\left(1+\frac{z}{\rho_I}\right)}
{\sqrt{2\pi}}, 
\end{align}
where 
\begin{align}
\rho_I&=\frac13(b,m_I)(\omega_1+\omega_2). 
\end{align}
The vacuum exponent is
\begin{align}
\label{vac_PdP4b}
\mathfrak{D}^{AdS_5\times Y_{PdP_{4b}}}(0;1,1)
&=\frac{\sqrt{x+3y}(2+x-y-z)^{\frac32}(4-2x-2y+z)\sqrt{4y+z}}
{8\sqrt{2}\pi^{\frac72}}
\nonumber\\
&=0.0155691\ldots. 
\end{align}

\subsection{$AdS_5\times Y_{PdP_5}$}
Again the del Pezzo surface $dP_5$ defined as the blow-up of $\mathbb{P}^2$ at generic four points is not toric. 
Nevertheless, there exists a toric blow-up of $\mathbb{P}^2$ at five points in specific positions, 
$PdP_5$, the fifth Pseudo del Pezzo surface \cite{Feng:2002fv,Franco:2005rj,Forcella:2008bb,Hanany:2012hi}. 
It can be viewed as the $\mathbb{Z}_2\times \mathbb{Z}_2$ orbifold of the conifold \cite{Hanany:2005ve}. 
When we consider a stack of $N$ D3-branes probing the toric Calabi-Yau cone over $PdP_5$, 
the low-energy effective gauge theory has 8 gauge nodes and 16 chiral multiplets of the same R-charge $r=1/2$. 
The quiver diagram is shown as
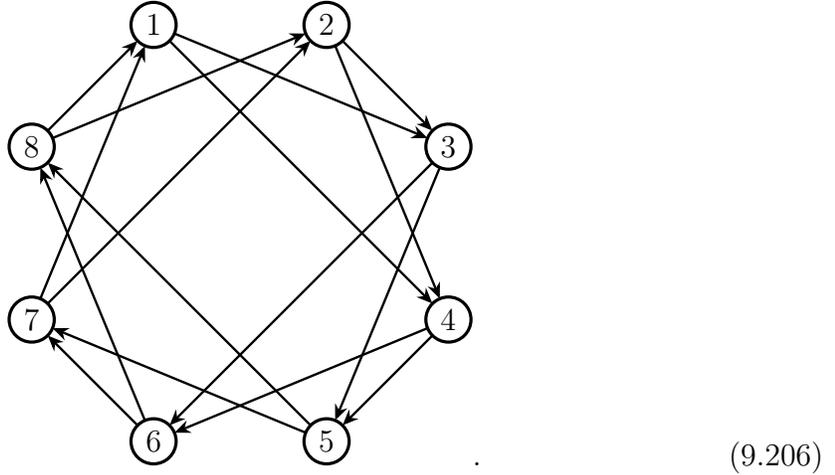
\begin{align}
\label{quiver_PdP5}
\begin{tikzpicture}[<->]
\draw[very thick] (-1.15,2.77) circle (3mm);
\node at (-1.15,2.77) {1};
\draw[very thick] (1.15,2.77) circle (3mm);
\node at (1.15,2.77) {2};
\draw[very thick] (2.77,1.15) circle (3mm);
\node at (2.77,1.15) {3};
\draw[very thick] (2.77,-1.15) circle (3mm);
\node at (2.77,-1.15) {4};
\draw[very thick] (1.15,-2.77) circle (3mm);
\node at (1.15,-2.77) {5};
\draw[very thick] (-1.15,-2.77) circle (3mm);
\node at (-1.15,-2.77) {6};
\draw[very thick] (-2.77,-1.15) circle (3mm);
\node at (-2.77,-1.15) {7};
\draw[very thick] (-2.77,1.15) circle (3mm);
\node at (-2.77,1.15) {8};
\draw[-{Stealth[length=6pt,width=5pt]}, line width=0.9pt, shorten <=3mm, shorten >=3mm] (-1.15, 2.77) -- ( 2.77, 1.15);
\draw[-{Stealth[length=6pt,width=5pt]}, line width=0.9pt, shorten <=3mm, shorten >=3mm] (-1.15, 2.77) -- ( 2.77,-1.15);
\draw[-{Stealth[length=6pt,width=5pt]}, line width=0.9pt, shorten <=3mm, shorten >=3mm] ( 1.15, 2.77) -- ( 2.77, 1.15);
\draw[-{Stealth[length=6pt,width=5pt]}, line width=0.9pt, shorten <=3mm, shorten >=3mm] ( 1.15, 2.77) -- ( 2.77,-1.15);
\draw[-{Stealth[length=6pt,width=5pt]}, line width=0.9pt, shorten <=3mm, shorten >=3mm] ( 2.77, 1.15) -- ( 1.15,-2.77);
\draw[-{Stealth[length=6pt,width=5pt]}, line width=0.9pt, shorten <=3mm, shorten >=3mm] ( 2.77, 1.15) -- (-1.15,-2.77);
\draw[-{Stealth[length=6pt,width=5pt]}, line width=0.9pt, shorten <=3mm, shorten >=3mm] ( 2.77,-1.15) -- ( 1.15,-2.77);
\draw[-{Stealth[length=6pt,width=5pt]}, line width=0.9pt, shorten <=3mm, shorten >=3mm] ( 2.77,-1.15) -- (-1.15,-2.77);
\draw[-{Stealth[length=6pt,width=5pt]}, line width=0.9pt, shorten <=3mm, shorten >=3mm] ( 1.15,-2.77) -- (-2.77,-1.15);
\draw[-{Stealth[length=6pt,width=5pt]}, line width=0.9pt, shorten <=3mm, shorten >=3mm] ( 1.15,-2.77) -- (-2.77, 1.15);
\draw[-{Stealth[length=6pt,width=5pt]}, line width=0.9pt, shorten <=3mm, shorten >=3mm] (-1.15,-2.77) -- (-2.77,-1.15);
\draw[-{Stealth[length=6pt,width=5pt]}, line width=0.9pt, shorten <=3mm, shorten >=3mm] (-1.15,-2.77) -- (-2.77, 1.15);
\draw[-{Stealth[length=6pt,width=5pt]}, line width=0.9pt, shorten <=3mm, shorten >=3mm] (-2.77,-1.15) -- (-1.15, 2.77);
\draw[-{Stealth[length=6pt,width=5pt]}, line width=0.9pt, shorten <=3mm, shorten >=3mm] (-2.77,-1.15) -- ( 1.15, 2.77);
\draw[-{Stealth[length=6pt,width=5pt]}, line width=0.9pt, shorten <=3mm, shorten >=3mm] (-2.77, 1.15) -- (-1.15, 2.77);
\draw[-{Stealth[length=6pt,width=5pt]}, line width=0.9pt, shorten <=3mm, shorten >=3mm] (-2.77, 1.15) -- ( 1.15, 2.77);
\end{tikzpicture}. 
\end{align}
The superpotential is
\begin{align}
\mathcal{W}&=\Tr
(
-X_{13}X_{35}X_{58}X_{81}+X_{14}X_{46}X_{68}X_{81}
+X_{35}X_{57}X_{72}X_{23}-X_{46}X_{67}X_{72}X_{24}
\nonumber\\
&+X_{67}X_{71}X_{13}X_{36}-X_{57}X_{71}X_{14}X_{45}
+X_{58}X_{82}X_{24}X_{45}-X_{68}X_{82}X_{23}X_{36}
). 
\end{align}
The theory is holographically dual to Type IIB string theory on $AdS_5\times Y_{PdP_5}$, 
where $Y_{PdP_5}$ is the Sasaki-Einstein base of the Calabi-Yau cone over $PdP_5$. 
The toric diagram for the toric Calabi-Yau cone is generated by
\begin{align}
v_1&=(1,0,-1),&
v_2&=(1,1,-1), & 
v_3&=(1,1,0), & 
v_4&=(1,1,1), \nonumber\\
v_5&=(1,0,1), & 
v_6&=(1,-1,1), & 
v_7&=(1,-1,0), & 
v_8&=(1,-1,-1). 
\end{align}
It is depicted as
\begin{align}
\label{toric_PdP5}
\begin{tikzpicture}[scale=1, baseline={(0,0)}]
  \fill (0,-1) circle (2pt) node[below] {$(0,-1)$};
  \fill (1,-1) circle (2pt) node[below right] {$(1,-1)$};
  \fill (1,0) circle (2pt) node[right] {$(1,0)$};
  \fill (1,1) circle (2pt) node[above right] {$(1,1)$};
  \fill (0,1) circle (2pt) node[above] {$(0,1)$};
  \fill (-1,1) circle (2pt) node[above left] {$(-1,1)$};
  \fill (-1,0) circle (2pt) node[left] {$(-1,0)$};
  \fill (-1,-1) circle (2pt) node[below left] {$(-1,-1)$};
  \draw[very thick] (0,-1) -- (1,-1) -- (1,0) -- (1,1) -- (0,1) -- (-1,1) -- (-1,0) -- (-1,-1)-- cycle;
\end{tikzpicture}
\end{align}
The multiplicities of four types of the chiral multiplets 
$X_2$ $=$ $\{X_{13},X_{35},X_{57},X_{71}\}$, 
$X_4$ $=$ $\{X_{23},X_{45},X_{67},X_{81}\}$, 
$X_6$ $=$ $\{X_{24},X_{46},X_{68},X_{82}\}$ 
and  
$X_8$ $=$ $\{X_{14},X_{36},X_{58},X_{72}\}$ 
forming the baryons are given by
\begin{align}
\det (v_{8},v_{2},v_{4})&=4, \\
\det (v_{2},v_{4},v_{6})&=4, \\
\det (v_{4},v_{6},v_{8})&=4, \\
\det (v_{6},v_{8},v_{2})&=4. 
\end{align}
The area of the polygon is
\begin{align}
\label{A_PdP5}
A&=4. 
\end{align}
The dual cone is generated by
\begin{align}
\label{dualG_PdP5}
m_1&=(1,0,1),& 
m_2&=(1,-1,0),&
m_3&=(1,-1,0),& 
m_4&=(1,0,-1),\nonumber\\
m_5&=(1,0,-1),&  
m_6&=(1,1,0), &
m_7&=(1,1,0), &
m_8&=(1,0,1). 
\end{align}
The Reeb vector is
\begin{align}
b&=\left(3,0,0\right). 
\end{align}
The volume of $Y_{PdP_5}$ is evaluated as
\begin{align}
\label{vol_PdP5}
\mathrm{Vol}(Y_{PdP_5})
&=\frac{4\pi^3}{27}. 
\end{align}
The resulting volume is one quarter of the volume (\ref{vol_T11}) of the conifold, 
in agreement with the expectation for the $\mathbb{Z}_2\times \mathbb{Z}_2$ orbifold of the conifold. 
The R-charges of the chiral multiplet fields are evaluated as
\begin{align}
R(X_{I})&=\frac{\pi}{3}\frac{\mathrm{Vol}(\Sigma_I)}{\mathrm{Vol}(Y_{PdP_5})}
=\frac{1}{2}, \qquad I=2,4,6,8, 
\end{align}
where 
\begin{align}
\mathrm{Vol}(\Sigma_I)&=2\pi^2 \frac{\det(v_{I-2},v_{I},v_{I+2})}{\det(b,v_{I-2},v_I)\det(b,v_{I},v_{I+2})}
=\frac{2\pi^2}{9}, \qquad I=2,4,6,8. 
\end{align}
Here the indices $I$ are understood modulo $8$. 

We have the weighted adjacency matrix 
\begin{align}
\label{M_PdP5}
&
M^{AdS_5\times Y_{PdP_5}}
\nonumber\\
&=\left(
\begin{matrix}
i_v&0&i_{\textrm{cm}_{\frac12}}&i_{\textrm{cm}_{\frac12}}&0&0&\tilde{i}_{\textrm{cm}_{\frac12}}&\tilde{i}_{\textrm{cm}_{\frac12}} \\
0&i_v&i_{\textrm{cm}_{\frac12}} &i_{\textrm{cm}_{\frac12}} &0&0&\tilde{i}_{\textrm{cm}_{\frac12}} &\tilde{i}_{\textrm{cm}_{\frac12}}  \\
\tilde{i}_{\textrm{cm}_{\frac12}} &\tilde{i}_{\textrm{cm}_{\frac12}} &i_v&0&i_{\textrm{cm}_{\frac12}} &i_{\textrm{cm}_{\frac12}} &0&0 \\
\tilde{i}_{\textrm{cm}_{\frac12}} &\tilde{i}_{\textrm{cm}_{\frac12}} &0&i_v&i_{\textrm{cm}_{\frac12}} &i_{\textrm{cm}_{\frac12}} &0&0 \\
0&0&\tilde{i}_{\textrm{cm}_{\frac12}} &\tilde{i}_{\textrm{cm}_{\frac12}}&i_v&0&i_{\textrm{cm}_{\frac12}}&i_{\textrm{cm}_{\frac12}} \\
0&0&\tilde{i}_{\textrm{cm}_{\frac12}} &\tilde{i}_{\textrm{cm}_{\frac12}}&0&i_v&i_{\textrm{cm}_{\frac12}}&i_{\textrm{cm}_{\frac12}} \\
i_{\textrm{cm}_{\frac12}}&i_{\textrm{cm}_{\frac12}}&0&0&\tilde{i}_{\textrm{cm}_{\frac12}} &\tilde{i}_{\textrm{cm}_{\frac12}}&i_v&0 \\
i_{\textrm{cm}_{\frac12}}&i_{\textrm{cm}_{\frac12}}&0&0&\tilde{i}_{\textrm{cm}_{\frac12}} &\tilde{i}_{\textrm{cm}_{\frac12}}&0&i_v \\
\end{matrix}
\right). 
\end{align}
The gravity index is 
\begin{align}
\label{ind_PdP5}
i^{AdS_5\times Y_{PdP_5}}(p;q)&=8\frac{pq}{1-pq}. 
\end{align}
This enumerates eight extremal BPS mesonic operators with R-charge
\begin{align}
R(\mathcal{M}_I)&=\frac23(b,m_I)=2, \qquad I=1,\cdots, 8, 
\end{align}
corresponding to the generators (\ref{dualG_PdP5}) of the dual cone. 

The supersymmetric zeta function is 
\begin{align}
\label{zeta_PdP5}
&
\mathfrak{Z}^{AdS_5\times Y_{PdP_5}}(s,z;\omega_1,\omega_2)
\nonumber\\
&=8(\omega_1+\omega_2)^{-s}\zeta\left(s,1+\frac{z}{\omega_1+\omega_2}\right). 
\end{align}
The residue at a simple pole $s=1$ is
\begin{align}
\label{Res_PdP5}
\mathrm{Res}_{s=1}\mathfrak{Z}^{AdS_5\times Y_{PdP_5}}(s,z;\omega_1,\omega_2)
&=\frac{8}{\omega_1+\omega_2}. 
\end{align}
Plugging the residue coefficient (\ref{Res_PdP5}) and the volume (\ref{vol_PdP5}) into the formula (\ref{Riem2_conj}), 
the integrated Riemann square is given by
\begin{align}
\label{Riem2_PdP5}
\int_{Y_{PdP_5}}\mathrm{Riem}^2 &=\frac{4000\pi^3}{27}
=4593.52\ldots.
\end{align}
The computed value (\ref{Riem2_PdP5}) is in perfect concordance with the Hilbert series result (see Appendix \ref{app_HS}).
The Zeta-index is
\begin{align}
\label{0_PdP5}
\mathfrak{Z}^{AdS_5\times Y_{PdP_5}}(0,0;\omega_1,\omega_2)&=-4. 
\end{align}
The supersymmetric zeta value with $s=-1$ and $z=0$ is 
\begin{align}
\label{-1_PdP5}
\mathfrak{Z}^{AdS_5\times Y_{PdP_5}}(-1,0;\omega_1,\omega_2)
&=-\frac{2}{3} (\omega_1+\omega_2). 
\end{align}
From the formulae (\ref{c0_formula}) and (\ref{a0_formula}) we get
\begin{align}
c_{\mathcal{O}(N^0)}&=-1, \\
a_{\mathcal{O}(N^0)}&=-\frac{3}{2}, 
\end{align}
and 
\begin{align}
c_{\mathcal{O}(N^0)}-a_{\mathcal{O}(N^0)}&=\frac{1}{2}. 
\end{align}

The supersymmetric determinant is
\begin{align}
\label{det_PdP5}
\mathfrak{D}^{AdS_5\times Y_{PdP_5}}(z;\omega_1,\omega_2)
&=\frac{(\omega_1+\omega_2)^{4+\frac{8z}{\omega_1+\omega_2}}}{16\pi^4}
\Gamma\left(1+\frac{z}{\omega_1+\omega_2}\right)^8. 
\end{align}
The vacuum exponent is
\begin{align}
\label{vac_PdP5}
\mathfrak{D}^{AdS_5\times Y_{PdP_5}}(0;1,1)
&=\frac{1}{\pi^4}=0.010266\ldots. 
\end{align}

\section{Other toric Sasaki-Einstein families}
\label{sec_XZ}
So far, we have computed the supersymmetric indices, supersymmetric zeta functions, 
supersymmetric determinants for the toric Sasaki-Einstein geometries 
by starting from the data of the quiver gauge theory and verified that 
the analytic structure of the supersymmetric zeta functions encodes the curvature-squared integral and the subleading holographic anomaly 
and that they are consistent with the combinatorial formulae discussed in section \ref{sec_spectral}. 
In this section, we present examples for which they are evaluated directly from the formulae using only the data of the toric diagram, 
without relying on the gauge theory description. 

\subsection{$X^{\mathsf{p},\mathsf{q}}$}
\label{sec_Xpq}
Hanany, Kazakopoulos and Wecht \cite{Hanany:2005hq} introduced 
an infinite class of 4d $\mathcal{N}=1$ quiver gauge theories, 
the $X^{\mathsf{p},\mathsf{q}}$ quiver gauge theories, 
which undergo Higgsing to the theories dual to the Sasaki-Einstein manifolds $Y^{\mathsf{p},\mathsf{q}}$ and $Y^{\mathsf{p},\mathsf{q-1}}$
upon turning on appropriate vevs for bifundamental chirals (see \cite{Feng:2002fv} for the prescription). 
The $X^{\mathsf{p},\mathsf{q}}$ quiver gauge theory has $2\mathsf{p}+1$ $SU(N)$ gauge nodes and $4\mathsf{p}+2\mathsf{q}+1$ bifundamental chirals. 
It is conjectured that 
the gravity dual is Type IIB string theory on the $AdS_5$ $\times$ $X^{\mathsf{p},\mathsf{q}}$, 
where $X^{\mathsf{p},\mathsf{q}}$ are the Sasaki-Einstein manifolds diffeomorphic to connected sums $(S^2\times S^3)\# (S^2\times S^3)$. 
A salient feature of the geometries $X^{\mathsf{p},\mathsf{q}}$ is that 
they have not yet been written with known explicit Einstein metrics. 
The geometry $X^{1,1}$ coincides with $L^{1,2,1}$ in section \ref{sec_L121}. 
The $X^{2,1}$ provides the cone over $dP_2$ in section \ref{sec_dP2}. 
For $X^{2,2}$ one finds the second Pseudo del Pezzo surface $PdP_2$ \cite{Feng:2002fv,Feng:2004uq,Hanany:2012hi} (also see \cite{Eager:2011ns}). 
The toric diagram is generated by 
\begin{align}
v_1&=(1,1,0),& v_2&=(1,2,0),& v_3&=(1,1,\mathsf{p}), \nonumber\\
v_4&=(1,0,\mathsf{p}-\mathsf{q}+1),& v_5&=(1,0,\mathsf{p}-\mathsf{q}). 
\end{align}
It is illustrated as
\begin{align}
\label{toric_Xpq}
\begin{tikzpicture}[scale=1.0, baseline={(0,0)}]
  \fill (1,0) circle (3pt) node[below left] {$(1,0)$};
  \fill (2,0) circle (3pt) node[below right] {$(2,0)$};
  \fill (1,4) circle (3pt) node[above] {$(1,\mathsf{p})$};
  \fill (0,3) circle (3pt) node[above left] {$(0,\mathsf{p}-\mathsf{q}+1)$};
  \fill (0,2) circle (3pt) node[below left] {$(0,\mathsf{p}-\mathsf{q})$};
  \draw[very thick] (1,0) -- (2,0) -- (1,4) -- (0,3) -- (0,2) -- cycle;
\end{tikzpicture}
\end{align}
The multiplicities of five types of the chiral multiplets $X^{(I)}$, $I=1,\cdots,5$ 
forming the baryonic operators are evaluated as
\begin{align}
\det (v_{5},v_{1},v_{2})&=\mathsf{p}-\mathsf{q}, \\
\det (v_{1},v_{2},v_{3})&=\mathsf{p}, \\
\det (v_{2},v_{3},v_{4})&=\mathsf{p}+\mathsf{q}-1, \\
\det (v_{3},v_{4},v_{5})&=1, \\
\det (v_{4},v_{5},v_{1})&=1.
\end{align}
The area of the lattice polygon is
\begin{align}
\label{A_Xpq}
A&=\frac{2\mathsf{p}+1}{2}. 
\end{align}
The dual cone is generated by
\begin{align}
\label{dualG_Xpq}
m_1&=(0,0,1),& 
m_2&=(2\mathsf{p},-\mathsf{p},-1),&
m_3&=(\mathsf{p}-\mathsf{q}+1,\mathsf{q}-1,-1),\nonumber\\
m_4&=(0,1,0),&
m_5&=(-\mathsf{p}+\mathsf{q},\mathsf{p}-\mathsf{q},1). 
\end{align}
Note that in the special case $\mathsf{p}=\mathsf{q}$ the generator $m_5$ $=$ $(-\mathsf{p}+\mathsf{q},\mathsf{p}-\mathsf{q},1)$ reduces to $(0,0,1)$, coinciding with $m_1$.
Geometrically this implies that the vertices $v_5$, $v_1$, $v_2$ become collinear, and the toric polygon degenerates from a pentagon to a quadrilateral. 
The volume of $X^{\mathsf{p},\mathsf{q}}$ takes the form
\begin{align}
\label{vol_Xpq}
&
\mathrm{Vol}(X^{\mathsf{p},\mathsf{q}})
\nonumber\\
&=\frac{\alpha(b_1,b_2) \pi^3}{b_1b_2
(6\mathsf{p}-\mathsf{p}b_1-b_2) 
\Bigl((\mathsf{p}-\mathsf{q})b_1+b_2-3(\mathsf{p}-\mathsf{q})\Bigr) 
\Bigl((\mathsf{q}-1)b_1-b_2+3(\mathsf{p}-\mathsf{q}+1)\Bigr) }, 
\end{align}
where 
\begin{align}
\label{XpqConst1}
\alpha(b_1,b_2)&=
\mathsf{p}(\mathsf{p}-\mathsf{q})(\mathsf{q}-1)b_1^2-b_2^2
+2\mathsf{p}(\mathsf{q}-1)b_1b_2
+3\mathsf{p}(\mathsf{p}-\mathsf{q})(\mathsf{p}-\mathsf{q}+1)b_1
+6\mathsf{p}b_2
\end{align}
and 
$b=(3,b_1,b_2)$ is the Reeb vector. 
The components $b_1$ and $b_2$ are fixed by the volume minimization \cite{Martelli:2005tp}. 
For example, when $(\mathsf{p},\mathsf{q})$ $=$ $(2,2)$, 
the associated Calabi-Yau 3-fold coincides with the toric cone over $PdP_2$. 
The Reeb vector is numerically obtained as 
\begin{align}
\label{Reeb_X22}
b_1&=2.7217\ldots, \qquad 
b_2=2.16876\ldots. 
\end{align}
Substituting the above values into the volume formula (\ref{vol_Xpq}), 
we find 
\begin{align}
\label{vol_X22}
\mathrm{Vol}(X^{2,2})
&=\frac{(12+4b_1-b_2)\pi^3}{b_1b_2(3+b_1-b_2)(12-2b_1-b_2)}
\nonumber\\
&=6.98084\ldots. 
\end{align}
The R-charges of five types of the chiral multiplet fields $X^{(I)}$, $I=1,\cdots,5$ are given by
\begin{align}
R(X^{(1)})&=\frac{\pi}{3}\frac{\mathrm{Vol}(\Sigma_1)}{\mathrm{Vol}(X^{\mathsf{p},\mathsf{q}})}
=\frac{2(\mathsf{p}-\mathsf{q}) b_1 (6\mathsf{p}-\mathsf{p}b_1-b_2) \Bigl((\mathsf{q}-1)b_1-b_2+3(\mathsf{p}-\mathsf{q}+1)\Bigr)}
{3\alpha(b_1,b_2)}, \\
R(X^{(2)})&=\frac{\pi}{3}\frac{\mathrm{Vol}(\Sigma_2)}{\mathrm{Vol}(X^{\mathsf{p},\mathsf{q}})}
=\frac{2\mathsf{p} b_1 \Bigl((\mathsf{p}-\mathsf{q})b_1+b_2-3(\mathsf{p}-\mathsf{q})\Bigr)  \Bigl((\mathsf{q}-1)b_1-b_2+3(\mathsf{p}-\mathsf{q}+1)\Bigr)}
{3\alpha(b_1,b_2)}, \\
R(X^{(3)})&=\frac{\pi}{3}\frac{\mathrm{Vol}(\Sigma_3)}{\mathrm{Vol}(X^{\mathsf{p},\mathsf{q}})}
=\frac{2(\mathsf{p}+\mathsf{q}-1)b_1b_2\Bigl((\mathsf{p}-\mathsf{q})b_1+b_2-3(\mathsf{p}-\mathsf{q})\Bigr)}
{3\alpha(b_1,b_2)}, \\
R(X^{(4)})&=\frac{\pi}{3}\frac{\mathrm{Vol}(\Sigma_4)}{\mathrm{Vol}(X^{\mathsf{p},\mathsf{q}})}
=\frac{2b_2 (6\mathsf{p}-\mathsf{p}b_1-b_2) \Bigl((\mathsf{p}-\mathsf{q})b_1+b_2-3(\mathsf{p}-\mathsf{q})\Bigr)}
{3\alpha(b_1,b_2)}, \\
R(X^{(5)})&=\frac{\pi}{3}\frac{\mathrm{Vol}(\Sigma_5)}{\mathrm{Vol}(X^{\mathsf{p},\mathsf{q}})}
=\frac{2b_2 (6\mathsf{p}-\mathsf{p}b_1-b_2) \Bigl((\mathsf{q}-1)b_1-b_2+3(\mathsf{p}-\mathsf{q}+1)\Bigr)}
{3\alpha(b_1,b_2)},
\end{align}
where the volumes of the supersymmetric 3-cycles are 
\begin{align}
\mathrm{Vol}(\Sigma_1)&=2\pi^2 \frac{\det(v_{5},v_{1},v_{2})}{\det(b,v_{5},v_1)\det(b,v_{1},v_{2})}
=\frac{2(\mathsf{p}-\mathsf{q})\pi^2}{b_2\Bigl((\mathsf{p}-\mathsf{q})b_1+b_2-3(\mathsf{p}-\mathsf{q})\Bigr)}, \\
\mathrm{Vol}(\Sigma_2)&=2\pi^2 \frac{\det(v_{1},v_{2},v_{3})}{\det(b,v_{1},v_2)\det(b,v_{2},v_{3})}
=\frac{2\mathsf{p}\pi^2}{b_2(6\mathsf{p}-\mathsf{p}b_1-b_2) }, \\
\mathrm{Vol}(\Sigma_3)&=2\pi^2 \frac{\det(v_{2},v_{3},v_{4})}{\det(b,v_{2},v_3)\det(b,v_{3},v_{4})}
=\frac{2(\mathsf{p}+\mathsf{q}-1)\pi^2}
{(6\mathsf{p}-\mathsf{p}b_1-b_2) \Bigl((\mathsf{q}-1)b_1-b_2+3(\mathsf{p}-\mathsf{q}+1)\Bigr)}, \\
\mathrm{Vol}(\Sigma_4)&=2\pi^2 \frac{\det(v_{3},v_{4},v_{5})}{\det(b,v_{3},v_4)\det(b,v_{4},v_{5})}
=\frac{2\pi^2}{b_1  \Bigl((\mathsf{q}-1)b_1-b_2+3(\mathsf{p}-\mathsf{q}+1)\Bigr)}, \\
\mathrm{Vol}(\Sigma_5)&=2\pi^2 \frac{\det(v_{4},v_{5},v_{1})}{\det(b,v_{4},v_5)\det(b,v_{5},v_{1})}
=\frac{2\pi^2}{b_1 \Bigl((\mathsf{p}-\mathsf{q})b_1+b_2-3(\mathsf{p}-\mathsf{q})\Bigr)}.
\end{align}

According to the formula (\ref{ind_gravity_conj}), 
the gravity index for the Sasaki-Einstein manifold $X^{\mathsf{p},\mathsf{q}}$ is given by 
\begin{align}
\label{ind_Xpq}
i^{AdS_5\times X^{\mathsf{p},\mathsf{q}}}(p;q)
&=\frac{(pq)^{\frac{b_1}{3}} }{1-(pq)^{\frac{b_1}{3}}}
+\frac{(pq)^{\frac{b_2}{3}} }{1-(pq)^{\frac{b_2}{3}}}
+\frac{(pq)^{\frac{6\mathsf{p}-\mathsf{p}b_1-b_2}{3}} }{1-(pq)^{\frac{6\mathsf{p}-\mathsf{p}b_1-b_2}{3}}}
\nonumber\\
&+\frac{(pq)^{\frac{(\mathsf{q}-1)b_1-b_2+3(\mathsf{p}-\mathsf{q}+1)}{3}} }{1-(pq)^{\frac{(\mathsf{q}-1)b_1-b_2+3(\mathsf{p}-\mathsf{q}+1)}{3}} }
+\frac{(pq)^{\frac{(\mathsf{p}-\mathsf{q})b_1+b_2-3(\mathsf{p}-\mathsf{q})}{3}}}{1-(pq)^{\frac{(\mathsf{p}-\mathsf{q})b_1+b_2-3(\mathsf{p}-\mathsf{q})}{3}}}. 
\end{align}
When $(\mathsf{p},\mathsf{q})$ $=$ $(1,1)$, 
it reproduces the $L^{1,2,1}$ gravity index (\ref{ind_L121}) with $(b_1,b_2)$ $=$ $(\frac12(3+\sqrt{3}), 3-\sqrt{3})$. 
For $(\mathsf{p},\mathsf{q})$ $=$ $(2,1)$ and $(b_1,b_2)$ $=$ $(9(\sqrt{33}-1)/16,3)$, 
it coincides with the gravity index (\ref{ind_dP2}) for the cone over $dP_2$, as expected. 
There are five extremal BPS mesonic operators $\mathcal{M}_I$, $I=1,\cdots, 5$ with R-charges 
\begin{align}
R(\mathcal{M}_1)&=\frac23(b,m_1)=\frac23 b_2, \\
R(\mathcal{M}_2)&=\frac23(b,m_2)=\frac23 (6\mathsf{p}-\mathsf{p}b_1-b_2), \\
R(\mathcal{M}_3)&=\frac23(b,m_3)=\frac23 \Bigl( (\mathsf{q}-1)b_1-b_2+3(\mathsf{p}-\mathsf{q}+1)\Bigr) , \\
R(\mathcal{M}_4)&=\frac23(b,m_4)=\frac23 b_1, \\
R(\mathcal{M}_5)&=\frac23(b,m_5)=\frac23 \Bigl( (\mathsf{p}-\mathsf{q})b_1+b_2-3(\mathsf{p}-\mathsf{q})\Bigr). 
\end{align}

The supersymmetric zeta function reads
\begin{align}
\label{zeta_Xpq}
&
\mathfrak{Z}^{AdS_5\times X^{\mathsf{p},\mathsf{q}}}(s,z;\omega_1,\omega_2)
\nonumber\\
&=\left(\frac{b_1}{3}(\omega_1+\omega_2)\right)^{-s}
\zeta\left(s,1+\frac{3z}{b_1(\omega_1+\omega_2)}\right)
\nonumber\\
&
+\left(\frac{b_2}{3}(\omega_1+\omega_2)\right)^{-s}
\zeta\left(s,1+\frac{3z}{b_2(\omega_1+\omega_2)}\right)
\nonumber\\
&+\left(\frac{6\mathsf{p}-\mathsf{p}b_1-b_2}{3}(\omega_1+\omega_2)\right)^{-s}
\zeta\left(s,1+\frac{3z}{(6\mathsf{p}-\mathsf{p}b_1-b_2)(\omega_1+\omega_2)}\right)
\nonumber\\
&+\left(\frac{(\mathsf{q}-1)b_1-b_2+3(\mathsf{p}-\mathsf{q}+1)}{3}(\omega_1+\omega_2)\right)^{-s}
\nonumber\\
&\times 
\zeta\left(s,1+\frac{3z}{((\mathsf{q}-1)b_1-b_2+3(\mathsf{p}-\mathsf{q}+1))(\omega_1+\omega_2)}\right)
\nonumber\\
&+\left(\frac{(\mathsf{p}-\mathsf{q})b_1+b_2-3(\mathsf{p}-\mathsf{q})}{3}(\omega_1+\omega_2)\right)^{-s}
\nonumber\\
&\times 
\zeta\left(s,1+\frac{3z}{((\mathsf{p}-\mathsf{q})b_1+b_2-3(\mathsf{p}-\mathsf{q}))(\omega_1+\omega_2)}\right). 
\end{align}
The residue at a simple pole $s=1$ is
\begin{align}
&
\label{Res_Xpq}
\mathrm{Res}_{s=1}\mathfrak{Z}^{AdS_5\times X^{\mathsf{p},\mathsf{q}}}(s,z;\omega_1,\omega_2)
=3\Biggl(
\frac{1}{b_1}+\frac{1}{b_2}+\frac{1}{6\mathsf{p}-\mathsf{p}b_1-b_2}
\nonumber\\
&
+\frac{1}{(\mathsf{q}-1)b_1-b_2+3(\mathsf{p}-\mathsf{q}+1)}
+\frac{1}{(\mathsf{p}-\mathsf{q})b_1+b_2-3(\mathsf{p}-\mathsf{q})}
\Biggr)
\frac{1}{\omega_1+\omega_2}. 
\end{align}
Making use of the formula (\ref{Riem2_conj}), 
the integrated Riemann square is given by
\begin{align}
\label{Riem2_Xpq}
&
\int_{X^{\mathsf{p},\mathsf{q}}}\mathrm{Riem}^2
=64\pi^3 \Biggl(
\frac{1}{b_1}+\frac{1}{b_2}+\frac{1}{6\mathsf{p}-\mathsf{p}b_1-b_2}
\nonumber\\
&+\frac{1}{(\mathsf{q}-1)b_1-b_2+3(\mathsf{p}-\mathsf{q}+1)}
+\frac{1}{(\mathsf{p}-\mathsf{q})b_1+b_2-3(\mathsf{p}-\mathsf{q})}
\Biggr)
\nonumber\\
&-152\pi^3
\frac{\alpha(b_1,b_2)}{b_1b_2
(6\mathsf{p}-\mathsf{p}b_1-b_2) 
\Bigl((\mathsf{p}-\mathsf{q})b_1+b_2-3(\mathsf{p}-\mathsf{q})\Bigr) 
\Bigl((\mathsf{q}-1)b_1-b_2+3(\mathsf{p}-\mathsf{q}+1)\Bigr) }, 
\end{align}
where $\alpha(b_1,b_2)$ is given by (\ref{XpqConst1}). 
For example, for $X^{2,2}$ we obtain
\begin{align}
\label{Riem2_X22}
&
\int_{X^{2,2}}\mathrm{Riem}^2
=\frac{\pi^3}{b_1b_2(3+b_1-b_2)(12-2b_1-b_2)}
(-256b_1^3+64b_2^3
\nonumber\\
&-64b_1^2b_2+64b_1b_2^2+768b_1^2-960b_2^2-576b_1b_2+4000b_1+2456b_2-1824
)
\nonumber\\
&=2580.78\ldots. 
\end{align}
This value matches the Hilbert series result (see Appendix \ref{app_HS}). 
The Zeta-index is
\begin{align}
\mathfrak{Z}^{AdS_5\times X^{\mathsf{p},\mathsf{q}}}(0,0;1,1)
&=-\frac52. 
\end{align}
The supersymmetric zeta value with $s=-1$ and $z=0$ is
\begin{align}
\mathfrak{Z}^{AdS_5\times X^{\mathsf{p},\mathsf{q}}}(-1,0;\omega_1,\omega_2)
&=-\frac{2\mathsf{p}+1}{12}(\omega_1+\omega_2). 
\end{align}
According to the formulae (\ref{c0_formula}) and (\ref{a0_formula}), we find 
\begin{align}
c_{\mathcal{O}(N^0)}&=-\frac{2\mathsf{p}+1}{8}, \\
a_{\mathcal{O}(N^0)}&=-\frac{3(2\mathsf{p}+1)}{16}, 
\end{align}
and 
\begin{align}
c_{\mathcal{O}(N^0)}-a_{\mathcal{O}(N^0)}&=\frac{2\mathsf{p}+1}{16}. 
\end{align}

The supersymmetric determinant is written as
\begin{align}
\label{det_Xpq}
\mathfrak{D}^{AdS_5\times X^{\mathsf{p},\mathsf{q}}}(z;\omega_1,\omega_2)
&=\prod_{I=1}^5 \frac{\rho_I^{\frac12+\frac{z}{\rho_I}} \Gamma\left(1+\frac{z}{\rho_I}\right)}
{\sqrt{2\pi}}, 
\end{align}
where
\begin{align}
\rho_I&=\frac13 (b,m_I)(\omega_1+\omega_2). 
\end{align}
The vacuum exponent is given by
\begin{align}
\label{vac_Xpq}
&
\mathfrak{D}^{AdS_5\times X^{\mathsf{p},\mathsf{q}}}(0;1,1)
\nonumber\\
&=\frac{\sqrt{b_1b_2(6\mathsf{p}-\mathsf{p}b_1-b_2) ((\mathsf{q}-1)b_1-b_2+3(\mathsf{p}-\mathsf{q}+1)) ((\mathsf{p}-\mathsf{q})b_1+b_2-3(\mathsf{p}-\mathsf{q}))}}
{9\sqrt{3}\pi^{\frac52}}. 
\end{align}

\subsection{$Z^{\mathsf{p},\mathsf{q}}$}
\label{sec_Zpq}
Further class of 4d $\mathcal{N}=1$ quiver gauge theories, 
which yields the $X^{\mathsf{p},\mathsf{q}}$ quiver gauge theories upon the Higgsing, was introduced by Oota and Yasui \cite{Oota:2006eg}. 
The theory is referred to as the $Z^{\mathsf{p},\mathsf{q}}$ quiver gauge theory. 
The theory has $2(\mathsf{p}+1)$ $SU(N)$ gauge nodes and $4\mathsf{p}+2\mathsf{q}+2$ bifundamental chiral multiplets. 
The gravity duals of the quiver gauge theories are conjectured to involve the Sasaki-Einstein manifolds $Z^{\mathsf{p},\mathsf{q}}$ 
which are diffeomorphic to $\#3(S^2\times S^3)$, the connected sum of $3$ copies of $S^2\times S^3$.\footnote{It is shown \cite{Boyer:2000pg} that $\#3(S^2\times S^3)$ admits a countably infinite family of inequivalent non-regular Sasaki-Einstein structures. }  
In particular, the geometry $Z^{2,1}$ coincides with the cone over $dP_3$ in section \ref{sec_dP3}. 
For the geometry $Z^{2,2}$, the toric data agree with those characterizing $PdP_{3b}$ 
\cite{Beasley:2001zp,Feng:2002fv,Feng:2004uq,Hanany:2012hi,Antinucci:2020yki}, 
establishing the equivalence at the level of the complex cone.
The toric diagram is generated by 
\begin{align}
v_1&=(1,1,0),& v_2&=(1,2,0),& v_3&=(1,2,1), \nonumber\\
v_4&=(1,1,\mathsf{p}),& v_5&=(1,0,\mathsf{p}-\mathsf{q}+1),& v_6&=(1,0,\mathsf{p}-\mathsf{q}). 
\end{align}
It is depicted as
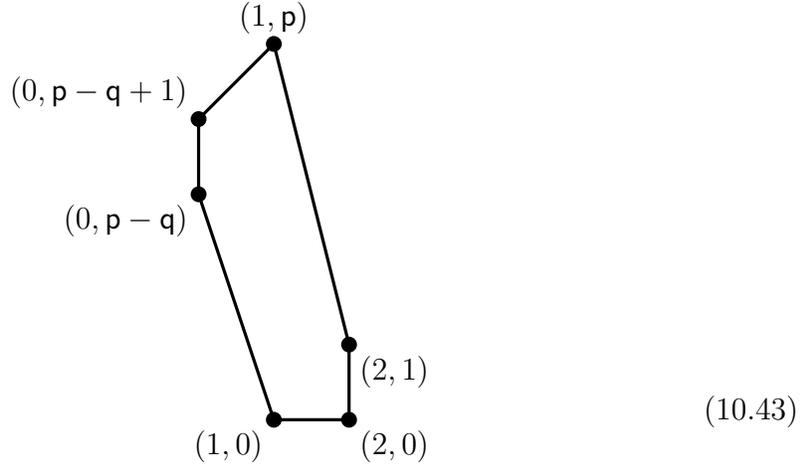
\begin{align}
\label{toric_Zpq}
\begin{tikzpicture}[scale=1.0, baseline={(0,0)}]
  \fill (1,0) circle (3pt) node[below left] {$(1,0)$};
  \fill (2,0) circle (3pt) node[below right] {$(2,0)$};
  \fill (2,1) circle (3pt) node[below right] {$(2,1)$};
  \fill (1,5) circle (3pt) node[above] {$(1,\mathsf{p})$};
  \fill (0,4) circle (3pt) node[above left] {$(0,\mathsf{p}-\mathsf{q}+1)$};
  \fill (0,3) circle (3pt) node[below left] {$(0,\mathsf{p}-\mathsf{q})$};
  \draw[very thick] (1,0) -- (2,0) -- (2,1) -- (1,5) -- (0,4) -- (0,3) -- cycle;
\end{tikzpicture}
\end{align}
The multiplicities of six types of the chiral multiplets $X^{(I)}$, $I=1,\cdots,6$ forming the baryonic operators are
\begin{align}
\det (v_{6},v_{1},v_{2})&=\mathsf{p}-\mathsf{q}, \\
\det (v_{1},v_{2},v_{3})&=1, \\
\det (v_{2},v_{3},v_{4})&=1, \\
\det (v_{3},v_{4},v_{5})&=\mathsf{p}+\mathsf{q}-2, \\
\det (v_{4},v_{5},v_{6})&=1, \\
\det (v_{5},v_{6},v_{1})&=1. 
\end{align}
The area of the lattice polygon is
\begin{align}
\label{A_Zpq}
A&=\mathsf{p}+1. 
\end{align}
The dual cone is generated by
\begin{align}
\label{dualG_Zpq}
m_1&=(0,0,1),& 
m_2&=(2,-1,0),&
m_3&=(2\mathsf{p}-1,-\mathsf{p}+1,-1),\nonumber\\
m_4&=(\mathsf{p}-\mathsf{q}+1,\mathsf{q}-1,-1),&
m_5&=(0,1,0),& 
m_6&=(-\mathsf{p}+\mathsf{q},\mathsf{p}-\mathsf{q},1). 
\end{align}
The volume of $Z^{\mathsf{p},\mathsf{q}}$ is given by
\begin{align}
\label{vol_Zpq}
&
\mathrm{Vol}(Z^{\mathsf{p},\mathsf{q}})
=\frac{\pi^3}{3}
\Biggl(
\frac{\mathsf{p}-\mathsf{q}}{b_2\Bigl((\mathsf{p}-\mathsf{q})b_1+b_2-3(\mathsf{p}-\mathsf{q})\Bigr)}
+\frac{1}{b_2(6-b_1)}
\nonumber\\
&+\frac{1}{(6-b_1)\Bigl((1-\mathsf{p})b_1-b_2+3(2\mathsf{p}-1)\Bigr)}
\nonumber\\
&
+\frac{\mathsf{p}+\mathsf{q}-2}{\Bigl((1-\mathsf{p})b_1-b_2+3(2\mathsf{p}-1)\Bigr) \Bigl((\mathsf{q}-1)b_1-b_2+3(\mathsf{p}-\mathsf{q}+1)\Bigr)}
\nonumber\\
&
+\frac{1}{b_1 \Bigl((\mathsf{q}-1)b_1-b_2+3(\mathsf{p}-\mathsf{q}+1)\Bigr)}
+\frac{1}{b_1 \Bigl((\mathsf{p}-\mathsf{q})b_1+b_2-3(\mathsf{p}-\mathsf{q})\Bigr)}
\Biggr), 
\end{align}
where $b=(3,b_1,b_2)$ is the Reeb vector 
with $b_1$ and $b_2$ being determined by the volume minimization \cite{Martelli:2005tp}. 
For example, for the $Z^{2,2}$ Sasaki-Einstein manifold, 
the corresponding Calabi-Yau cone exhibits isolated singularities that coincide with those of the third Pseudo del Pezzo geometry $PdP_{3b}$. 
One finds the Reeb vector 
\begin{align}
\label{Reeb_Z22}
b&=(3,3,3(3-\sqrt{5}))
\end{align}
and the resulting volume
\begin{align}
\label{vol_Z22}
\mathrm{Vol}(Z^{2,2})
&=\frac{11+5\sqrt{5}}{108}\pi^3=6.36787\ldots. 
\end{align}
The R-charges of the six chiral multiplets $X^{(I)}$, $I=1,\cdots, 6$ are given by
\begin{align}
R(X^{(I)})&=\frac{\pi}{3}\frac{\mathrm{Vol}(\Sigma_I)}{\mathrm{Vol}(Z^{\mathsf{p},\mathsf{q}})}, 
\end{align}
where the volumes of the supersymmetric 3-cycles are given by
\begin{align}
\mathrm{Vol}(\Sigma_1)&=2\pi^2 \frac{\det(v_{6},v_{1},v_{2})}{\det(b,v_{6},v_1)\det(b,v_{1},v_{2})}
=\frac{2\pi^2 (\mathsf{p}-\mathsf{q})}{b_2 \Bigl((\mathsf{p}-\mathsf{q})b_1+b_2-3(\mathsf{p}-\mathsf{q})\Bigr)}, \\
\mathrm{Vol}(\Sigma_2)&=2\pi^2 \frac{\det(v_{1},v_{2},v_{3})}{\det(b,v_{1},v_2)\det(b,v_{2},v_{3})}
=\frac{2\pi^2}{b_2(6-b_1)}, \\
\mathrm{Vol}(\Sigma_3)&=2\pi^2 \frac{\det(v_{2},v_{3},v_{4})}{\det(b,v_{2},v_3)\det(b,v_{3},v_{4})}
=\frac{2\pi^2}{(6-b_1) \Bigl((1-\mathsf{p})b_1-b_2+3(2\mathsf{p}-1)\Bigr)}, \\
\mathrm{Vol}(\Sigma_4)&=2\pi^2 \frac{\det(v_{3},v_{4},v_{5})}{\det(b,v_{3},v_4)\det(b,v_{4},v_{5})}
\nonumber\\
&=\frac{2\pi^2 (\mathsf{p}+\mathsf{q}-2)}{\Bigl((1-\mathsf{p})b_1-b_2+3(2\mathsf{p}-1)\Bigr) \Bigl((\mathsf{q}-1)b_1-b_2+3(\mathsf{p}-\mathsf{q}+1)\Bigr)}, \\
\mathrm{Vol}(\Sigma_5)&=2\pi^2 \frac{\det(v_{4},v_{5},v_{6})}{\det(b,v_{4},v_5)\det(b,v_{5},v_{6})}
=\frac{2\pi^2}{b_1 \Bigl((\mathsf{q}-1)b_1-b_2+3(\mathsf{p}-\mathsf{q}+1)\Bigr)}, \\
\mathrm{Vol}(\Sigma_6)&=2\pi^2 \frac{\det(v_{5},v_{6},v_{1})}{\det(b,v_{5},v_6)\det(b,v_{6},v_{1})}
=\frac{2\pi^2}{b_1 \Bigl((\mathsf{p}-\mathsf{q})b_1+b_2-3(\mathsf{p}-\mathsf{q})\Bigr)}. 
\end{align}

From the formula (\ref{ind_gravity_conj}), we obtain the gravity index 
\begin{align}
\label{ind_Zpq}
&
i^{AdS_5\times Z^{\mathsf{p},\mathsf{q}}}(p;q)
\nonumber\\
&=\frac{(pq)^{\frac{b_1}{3}} }{1-(pq)^{\frac{b_1}{3}}}
+\frac{(pq)^{\frac{b_2}{3}} }{1-(pq)^{\frac{b_2}{3}}}
+\frac{(pq)^{\frac{6-b_1}{3}}}{1-(pq)^{\frac{6-b_1}{3}}}
+\frac{(pq)^{\frac{(1-\mathsf{p})b_1-b_2+3(2\mathsf{p}-1)}{3}}}{1-(pq)^{\frac{(1-\mathsf{p})b_1-b_2+3(2\mathsf{p}-1)}{3}}}
\nonumber\\
&
+\frac{(pq)^{\frac{(\mathsf{q}-1)b_1-b_2+3(\mathsf{p}-\mathsf{q}+1)}{3}} }{1-(pq)^{\frac{(\mathsf{q}-1)b_1-b_2+3(\mathsf{p}-\mathsf{q}+1)}{3}} }
+\frac{(pq)^{\frac{(\mathsf{p}-\mathsf{q})b_1+b_2-3(\mathsf{p}-\mathsf{q})}{3}}}{1-(pq)^{\frac{(\mathsf{p}-\mathsf{q})b_1+b_2-3(\mathsf{p}-\mathsf{q})}{3}}}. 
\end{align}
For $(\mathsf{p},\mathsf{q})$ $=$ $(2,1)$, 
it reproduces the gravity index (\ref{ind_dP3}) with $(b_1,b_2)$ $=$ $(3,3)$, as expected. 
For $Z^{2,2}$ associated with $PdP_{3b}$, it reduces to 
\begin{align}
i^{AdS_5\times Z^{2,2}}(p;q)
&=2\frac{pq}{1-pq}
+2\frac{(pq)^{\sqrt{5}-1}}{1-(pq)^{\sqrt{5}-1}}
+2\frac{(pq)^{3-\sqrt{5}}}{1-(pq)^{3-\sqrt{5}}}. 
\end{align}
There are six extremal BPS mesonic operators $\mathcal{M}_I$, $I=1,\cdots, 6$ with R-charges 
\begin{align}
R(\mathcal{M}_1)&=\frac23(b,m_1)=\frac23 b_2, \\
R(\mathcal{M}_2)&=\frac23(b,m_2)=\frac23(6-b_1), \\
R(\mathcal{M}_3)&=\frac23(b,m_3)=\frac23\Bigl( (1-\mathsf{p})b_1-b_2+3(2\mathsf{p}-1)\Bigr), \\
R(\mathcal{M}_4)&=\frac23(b,m_4)=\frac23 \Bigl( (\mathsf{q}-1)b_1-b_2+3(\mathsf{p}-\mathsf{q}+1)\Bigr) , \\
R(\mathcal{M}_5)&=\frac23(b,m_5)=\frac23 b_1, \\
R(\mathcal{M}_6)&=\frac23(b,m_6)=\frac23 \Bigl( (\mathsf{p}-\mathsf{q})b_1+b_2-3(\mathsf{p}-\mathsf{q})\Bigr). 
\end{align}

We find the supersymmetric zeta function 
\begin{align}
\label{zeta_Zpq}
&
\mathfrak{Z}^{AdS_5\times Z^{\mathsf{p},\mathsf{q}}}(s,z;\omega_1,\omega_2)
\nonumber\\
&=\left(\frac{b_1}{3}(\omega_1+\omega_2)\right)^{-s}
\zeta\left(s,1+\frac{3z}{b_1(\omega_1+\omega_2)}\right)
\nonumber\\
&+\left(\frac{b_2}{3}(\omega_1+\omega_2)\right)^{-s}
\zeta\left(s,1+\frac{3z}{b_2(\omega_1+\omega_2)}\right)
\nonumber\\
&+\left(\frac{6-b_1}{3}(\omega_1+\omega_2)\right)^{-s}
\zeta\left(s,1+\frac{3z}{(6-b_1)(\omega_1+\omega_2)}\right)
\nonumber\\
&+\left(\frac{(1-\mathsf{p})b_1-b_2+3(2\mathsf{p}-1)}{3}(\omega_1+\omega_2)\right)^{-s}
\nonumber\\
&\times 
\zeta\left(s,1+\frac{3z}{((1-\mathsf{p})b_1-b_2+3(2\mathsf{p}-1))(\omega_1+\omega_2)}\right)
\nonumber\\
&+\left(\frac{(\mathsf{q}-1)b_1-b_2+3(\mathsf{p}-\mathsf{q}+1)}{3}(\omega_1+\omega_2)\right)^{-s}
\nonumber\\
&\times 
\zeta\left(s,1+\frac{3z}{((\mathsf{q}-1)b_1-b_2+3(\mathsf{p}-\mathsf{q}+1))(\omega_1+\omega_2)}\right)
\nonumber\\
&+\left(\frac{(\mathsf{p}-\mathsf{q})b_1+b_2-3(\mathsf{p}-\mathsf{q})}{3}(\omega_1+\omega_2)\right)^{-s}
\nonumber\\
&\times 
\zeta\left(s,1+\frac{3z}{((\mathsf{p}-\mathsf{q})b_1+b_2-3(\mathsf{p}-\mathsf{q}))(\omega_1+\omega_2)}\right). 
\end{align}
It has the residue at a simple pole $s=1$
\begin{align}
&
\label{Res_Zpq}
\mathrm{Res}_{s=1}\mathfrak{Z}^{AdS_5\times Z^{\mathsf{p},\mathsf{q}}}(s,z;\omega_1,\omega_2)
=3\Biggl(
\frac{1}{b_1}+\frac{1}{b_2}+\frac{1}{6-b_1}
+\frac{1}{(1-\mathsf{p})b_1-b_2+3(2\mathsf{p}-1)}
\nonumber\\
&
+\frac{1}{(\mathsf{q}-1)b_1-b_2+3(\mathsf{p}-\mathsf{q}+1)}
+\frac{1}{(\mathsf{p}-\mathsf{q})b_1+b_2-3(\mathsf{p}-\mathsf{q})}
\Biggr)
\frac{1}{\omega_1+\omega_2}. 
\end{align}
Applying the formula (\ref{Riem2_conj}), 
we obtain the integrated Riemann square of the form 
\begin{align}
\label{Riem2_Zpq}
&
\int_{Z^{\mathsf{p},\mathsf{q}}}\mathrm{Riem}^2
=64\pi^3 \Biggl(
\frac{1}{b_1}+\frac{1}{b_2}+\frac{1}{6-b_1}+\frac{1}{(1-\mathsf{p})b_1-b_2+3(2\mathsf{p}-1)}
\nonumber\\
&+\frac{1}{(\mathsf{q}-1)b_1-b_2+3(\mathsf{p}-\mathsf{q}+1)}
+\frac{1}{(\mathsf{p}-\mathsf{q})b_1+b_2-3(\mathsf{p}-\mathsf{q})}
\Biggr)
\nonumber\\
&-\frac{152\pi^3}{3}
 \Biggl(
\frac{\mathsf{p}-\mathsf{q}}{b_2\Bigl((\mathsf{p}-\mathsf{q})b_1+b_2-3(\mathsf{p}-\mathsf{q})\Bigr)}
+\frac{1}{b_2(6-b_1)}
\nonumber\\
&+\frac{1}{(6-b_1)\Bigl((1-\mathsf{p})b_1-b_2+3(2\mathsf{p}-1)\Bigr)}
\nonumber\\
&
+\frac{\mathsf{p}+\mathsf{q}-2}{\Bigl((1-\mathsf{p})b_1-b_2+3(2\mathsf{p}-1)\Bigr) \Bigl((\mathsf{q}-1)b_1-b_2+3(\mathsf{p}-\mathsf{q}+1)\Bigr)}
\nonumber\\
&
+\frac{1}{b_1 \Bigl((\mathsf{q}-1)b_1-b_2+3(\mathsf{p}-\mathsf{q}+1)\Bigr)}
+\frac{1}{b_1 \Bigl((\mathsf{p}-\mathsf{q})b_1+b_2-3(\mathsf{p}-\mathsf{q})\Bigr)}
\Biggr). 
\end{align}
For example, in the case of $Z^{2,2}$ associated with $PdP_{3b}$, 
it is evaluated as
\begin{align}
\label{Riem2_Z22}
\int_{Z^{2,2}}\mathrm{Riem}^2
&=\frac{2(943+193\sqrt{5})}{27}\pi^3
\nonumber\\
&=3157.04\ldots. 
\end{align}
We find that the value (\ref{Riem2_Z22}) fully agrees with the Hilbert series result (see Appendix \ref{app_HS}).
We find the Zeta-index
\begin{align}
\mathfrak{Z}^{AdS_5\times Z^{\mathsf{p},\mathsf{q}}}(0,0;1,1)
&=-3. 
\end{align}
The supersymmetric zeta value with $s=-1$ and $z=0$ is
\begin{align}
\mathfrak{Z}^{AdS_5\times Z^{\mathsf{p},\mathsf{q}}}(-1,0;\omega_1,\omega_2)
&=-\frac{\mathsf{p}+1}{6}(\omega_1+\omega_2). 
\end{align}
According to the formulae (\ref{c0_formula}) and (\ref{a0_formula}), we get
\begin{align}
c_{\mathcal{O}(N^0)}&=-\frac{\mathsf{p}+1}{4}, \\
a_{\mathcal{O}(N^0)}&=-\frac{3(\mathsf{p}+1)}{8}, 
\end{align}
and 
\begin{align}
c_{\mathcal{O}(N^0)}-a_{\mathcal{O}(N^0)}&=\frac{\mathsf{p}+1}{8}. 
\end{align}

The supersymmetric determinant is written as
\begin{align}
\label{det_Zpq}
\mathfrak{D}^{AdS_5\times Z^{\mathsf{p},\mathsf{q}}}(z;\omega_1,\omega_2)
&=\prod_{I=1}^6 \frac{\rho_I^{\frac12+\frac{z}{\rho_I}} \Gamma\left(1+\frac{z}{\rho_I}\right)}
{\sqrt{2\pi}}, 
\end{align}
where
\begin{align}
\rho_I&=\frac13 (b,m_I)(\omega_1+\omega_2). 
\end{align}
The vacuum exponent is given by
\begin{align}
\label{vac_Zpq}
&
\mathfrak{D}^{AdS_5\times Z^{\mathsf{p},\mathsf{q}}}(0;1,1)
=\frac{1}
{27\pi^{3}}
\sqrt{b_1b_2(6-b_1) ((1-\mathsf{p})b_1-b_2+3(2\mathsf{p}-1))}
\nonumber\\
&\times \sqrt{ ((\mathsf{q}-1)b_1-b_2+3(\mathsf{p}-\mathsf{q}+1)) ((\mathsf{p}-\mathsf{q})b_1+b_2-3(\mathsf{p}-\mathsf{q})) }. 
\end{align}

\subsection*{Acknowledgements}
We would like to thank Katsushi Ito, Hai Lin and Jaewon Song for useful discussions and comments. 
YN is in part supported by JSPS KAKENHI Grant Number 21K03581. 
TO was supported by the Startup Funding no.\ 4007012317 of Southeast University.  

\appendix

\section{Hilbert series}
\label{app_HS}
In this appendix, we present an alternative derivation of the curvature-squared integral based on the Hilbert series for the Calabi-Yau cone $X$. 
While the main text derives the result using the formula (\ref{Riem2_conj2}) and its toric specialization (\ref{Riem2_conj}), 
the same invariant can be reconstructed from the asymptotic structure of the Hilbert series as proposed in \cite{Eager:2010dk}. 
This agreement provides a non-trivial consistency check and strongly supports these formulae. 

Let $X$ be the toric Calabi-Yau 3-fold realized as a complex cone over a projective variety. 
As an affine variety, $X$ can be embedded in $\mathbb{C}^k$ and described by polynomial equations $f_i(x_1,\cdots, x_k)$ $=$ $0$. 
The Hilbert series $g[X](t)$ for $X$ is defined as a generating function for the dimension of the graded pieces of 
the coordinate ring given by $\mathbb{C}[x_1,\cdots, x_k]/\langle f_i\rangle$
\begin{align}
g[X](t)&=\sum_{i\ge 0}\dim_{\mathbb{C}}(X_i)t^i, 
\end{align}
where the spaces $X_i$ are the graded components of the coordinate ring 
and consist of homogeneous functions of degree $i$. 
For a 4d SCFT arising from a stack of D3-branes probing a Calabi-Yau cone $X$ $=$ $C(Y)$ of the Sasaki-Einstein manifold $Y$, 
the Hilbert series $g[X](t)$ for $X$ encodes the spectrum of mesonic chiral operators 
\cite{Benvenuti:2006qr,Feng:2007ur,Butti:2006au,Butti:2007jv,Hanany:2007zz,Gray:2008yu,Hanany:2012hi,Bianchi:2014qma,Bao:2024nyu}. 
The Hilbert series can be obtained from Molien-type integrals  
and many exact expressions can be found in \cite{Hanany:2012hi,Bianchi:2014qma}. 
Eager, Gary, and Roberts \cite{Eager:2010dk} propose that 
the Hilbert series has the following Laurent series expansion 
\begin{align}
g[X](t=e^{-\beta})
&=\frac{c_3}{\beta^3}+\frac{c_2}{\beta^2}+\frac{c_1}{\beta}+\cdots, 
\end{align}
where 
\begin{align}
c_3&=c_2=\frac{27}{8\pi^3}\mathrm{Vol}(Y), \\
c_1&=\frac{91}{64\pi^3}\mathrm{Vol}(Y)+\frac{1}{512\pi^3}\int_Y \mathrm{Riem}^2. 
\end{align}
Consequently, the geometric quantities of $Y$ are expressible as linear functionals of the expansion coefficients
\begin{align}
\label{HS_conj1}
\mathrm{Vol}(Y)&=\frac{8c_3}{27}\pi^3, \\
\label{HS_conj2}
\int_{Y}\mathrm{Riem}^2&=\left(512c_1-\frac{5824}{27}c_3\right)\pi^3. 
\end{align}
In the following, we show the Hilbert series and 
extract the volume and curvature-squared invariant by applying the formula (\ref{HS_conj2}). 
The results agree with the values obtained from the formulae (\ref{Riem2_conj}) and (\ref{Riem2_conj2}). 

\subsection{$S^5$}
The Hilbert series is 
\begin{align}
\label{HS_S5}
g[\mathbb{C}^3](t)
&=\frac{1}{(1-t^{\frac23})^3}
=\frac{27}{8\beta^3}+\frac{27}{8\beta^2}+\frac{3}{2\beta}+\cdots. 
\end{align}
We have
\begin{align}
\mathrm{Vol}(S^5)&=\pi^3,& 
\int_{S^5} \mathrm{Riem}^2&=40\pi^3. 
\end{align}

\subsection{$(S^5/\mathbb{Z}_{k})_{\textrm{ALE}}$}
The Calabi-Yau cone is $\mathbb{C}\times \mathbb{C}^2/\mathbb{Z}_{k}$. 
The Hilbert series is
\begin{align}
g[\mathbb{C}\times \mathbb{C}^2/\mathbb{Z}_{k}](t)
&=\frac{1-t^{\frac{4k}{3}}}{(1-t^{\frac23}) (1-t^{\frac43}) (1-t^{\frac{2k}{3}})^2}
=\frac{27}{8k\beta^3}+\frac{27}{8k\beta^2}+\frac{k^2+11}{8k\beta}+\cdots. 
\end{align}
We get
\begin{align}
\mathrm{Vol}((S^5/\mathbb{Z}_k)_{\textrm{ALE}})&=\frac{\pi^3}{k},& 
\int_{(S^5/\mathbb{Z}_k)_{\textrm{ALE}}} \mathrm{Riem}^2&=\frac{8(8k^2-3)\pi^3}{k}. 
\end{align}

\subsection{$S^5/\mathbb{Z}_3$}
The Calabi-Yau cone is $\mathbb{C}^3/\mathbb{Z}_3$. 
Equivalently it is viewed as the Calabi-Yau cone over the zeroth del Pezzo surface $dP_0$, i.e. $\mathbb{P}^2$. 
The Hilbert series is 
\begin{align}
\label{HS_S5/Z3}
g[\mathbb{C}^3/\mathbb{Z}_3](t)
&=\frac{1+7t^2+t^4}{(1-t^2)^3}
=\frac{9}{8\beta^3}+\frac{9}{8\beta^2}+\frac{1}{2\beta}+\cdots. 
\end{align}
One finds
\begin{align}
\mathrm{Vol}(S^5/\mathbb{Z}_3)&=\frac{\pi^3}{3},& 
\int_{S^5/\mathbb{Z}_3} \mathrm{Riem}^2&=\frac{40\pi^3}{3}. 
\end{align}

\subsection{$S^5/\mathbb{Z}_4$}
The Calabi-Yau cone is $\mathbb{C}^3/\mathbb{Z}_4$. 
The Hilbert series is
\begin{align}
g[\mathbb{C}^3/\mathbb{Z}_4](t)
&=\frac{1+3t^2+3t^{\frac83}+t^{\frac{14}{3}}}
{(1-t^{\frac43})(1-t^{\frac83})^2}
=\frac{27}{32\beta^3}+\frac{27}{32\beta^2}+\frac{15}{32\beta}+\cdots. 
\end{align}
We get
\begin{align}
\mathrm{Vol}(S^5/\mathbb{Z}_4)&=\frac{\pi^3}{4},& 
\int_{S^5/\mathbb{Z}_4} \mathrm{Riem}^2&=58\pi^3. 
\end{align}

\subsection{$T^{1,1}$}
The Calabi-Yau cone $C(T^{1,1})$ is a conifold $\mathcal{C}$. 
The Hilbert series is
\begin{align}
g[\mathcal{C}](t)
&=\frac{1-t^2}{(1-t)^4}
=\frac{2}{\beta^3}+\frac{2}{\beta^2}+\frac{1}{\beta}+\cdots. 
\end{align}
One finds
\begin{align}
\mathrm{Vol}(T^{1,1})&=\frac{16\pi^3}{27},& 
\int_{T^{1,1}} \mathrm{Riem}^2&=\frac{2176\pi^3}{27}. 
\end{align}

\subsection{$T^{2,2}$}
The Calabi-Yau cone is a cone over the zeroth Hirzebruch surface 
$F_0=\mathbb{P}^1\times \mathbb{P}^1$. 
It can be realized as the total space $\mathrm{Tot}(K_{F_0})$ of the canonical bundle over $F_0$. 
The Hilbert series is
\begin{align}
g[\mathrm{Tot}(K_{F_0})](t)
&=\frac{1+5t^2-5t^4-t^6}{(1-t^2)^4}
=\frac{1}{\beta^3}+\frac{1}{\beta^2}+\frac{1}{2\beta}+\cdots. 
\end{align}
We get
\begin{align}
\mathrm{Vol}(T^{2,2})&=\frac{8\pi^3}{27},& 
\int_{T^{2,2}} \mathrm{Riem}^2&=\frac{1088\pi^3}{27}. 
\end{align}

\subsection{$Y^{2,1}$}
The Calabi-Yau cone is a cone over the first del Pezzo surface $dP_1$, 
realized as the total space of its canonical bundle, $X=\mathrm{Tot}(K_{dP_1})$. 
The Hilbert series is
\begin{align}
g[\mathrm{Tot}(K_{dP_1})](t)
&=\frac{1+3t^2-3t^4-t^6+2t^{\frac23(7-\sqrt{13})}-2t^{\frac23(2+\sqrt{13})}}
{\left(1-t^{\frac23 (-1+\sqrt{13})}\right)^2 \left(1-t^{\frac23(7-\sqrt{13})}\right)^2}
\nonumber\\
&=\frac{46+13\sqrt{13}}{96\beta^3}+\frac{46+13\sqrt{13}}{96\beta^2}
+\frac{22+7\sqrt{13}}{96\beta}+\cdots. 
\end{align}
One finds
\begin{align}
\mathrm{Vol}(Y^{2,1})&=\frac{46+13\sqrt{13}}{324}\pi^3,& 
\int_{Y^{2,1}} \mathrm{Riem}^2&=\frac{2(566+329\sqrt{13})}{81}\pi^3.  
\end{align}

\subsection{$L^{1,2,1}/\mathbb{Z}_2$}
The Calabi-Yau cone is $\mathrm{SPP}/\mathbb{Z}_2$, 
which corresponds to a cone over the third Pseudo del Pezzo surface $PdP_{3c}$. 
The Hilbert series is
\begin{align}
g[\mathrm{SPP}/\mathbb{Z}_2](t)
&=\frac{1+t^2-t^4+t^{3-\frac{1}{\sqrt{3}}}-t^{2+\frac{4}{\sqrt{3}}}+t^{1+\sqrt{3}}-t^{3+\sqrt{3}}-t^{5+\sqrt{3}}}
{\left(1-t^{\frac{4}{\sqrt{3}}}\right) \left(1-t^{4-\frac{4}{\sqrt{3}}}\right) 
\left(1-t^{1+\frac{1}{\sqrt{3}}}\right) \left(1-t^{2(1+\frac{1}{\sqrt{3}})}\right)}
\nonumber\\
&=\frac{3\sqrt{3}}{8\beta^3}+\frac{3\sqrt{3}}{8\beta^2}+\frac{6+\sqrt{3}}{16\beta}+\cdots. 
\end{align}
We find
\begin{align}
\mathrm{Vol}(L^{1,2,1}/\mathbb{Z}_2)&=\frac{\pi^3}{3\sqrt{3}},& 
\int_{L^{1,2,1}/\mathbb{Z}_2} \mathrm{Riem}^2&=\frac{8(216-55\sqrt{3})}{9}\pi^3.  
\end{align}

\subsection{$L^{1,3,1}/\mathbb{Z}_2$}
Let $C(L^{1,3,1}/\mathbb{Z}_2)$ denote the Calabi-Yau cone over the Sasaki-Einstein orbifold $L^{1,3,1}/\mathbb{Z}_2$. 
The Hilbert series is
\begin{align}
g[C(L^{1,3,1}/\mathbb{Z}_2)](t)
&=\frac{1+t^2+t^{\frac{4+\sqrt{7}}{3}}+t^{\frac{10+\sqrt{7}}{3}}}
{\left(1-t^{\frac{10-2\sqrt{7}}{3}}\right) \left(1-t^{\frac{2(1+\sqrt{7})}{3}}\right) \left(1-t^{\frac{4+\sqrt{7}}{3}}\right)}
\nonumber\\
&=\frac{7\sqrt{7}-10}{18\beta^3}+\frac{7\sqrt{7}-10}{18\beta^2}+\frac{74+13\sqrt{7}}{216\beta}+\cdots. 
\end{align}
One obtains
\begin{align}
\mathrm{Vol}(L^{1,3,1}/\mathbb{Z}_2)&=\frac{4(7\sqrt{7}-10)}{243}\pi^3,& 
\int_{L^{1,3,1}/\mathbb{Z}_2} \mathrm{Riem}^2&=\frac{32(2242-403\sqrt{7})}{243}\pi^3.  
\end{align}

\subsection{$Y_{dP_2}$}
The Calabi-Yau cone is $X=\mathrm{Tot}(K_{dP_2})$, 
the total space of the canonical bundle over the second del Pezzo surface $dP_2$. 
The Hilbert series is
\begin{align}
g[\mathrm{Tot}(K_{dP_2})](t)
&=\frac{1+2t^2-2t^4-t^6+2t^{\frac{35-3\sqrt{33}}{8}}-2t^{\frac{13+3\sqrt{33}}{8}} }
{(1-t^2) (1-t^{\frac{27-3\sqrt{33}}{4}}) (1-t^{\frac{3\sqrt{33}-3}{8}})^2}
\nonumber\\
&=\frac{59+11\sqrt{33}}{144\beta^3}+\frac{59+11\sqrt{33}}{144\beta^2}+\frac{55+9\sqrt{33}}{216\beta}+\cdots. 
\end{align}
We have
\begin{align}
\mathrm{Vol}(Y_{dP_2})&=\frac{59+11\sqrt{33}}{486}\pi^3,& 
\int_{Y_{dP_2}} \mathrm{Riem}^2&=\frac{4(2551+295\sqrt{33})}{243}\pi^3.  
\end{align}

\subsection{$Y_{dP_3}$}
The Calabi-Yau cone is $X=\mathrm{Tot}(K_{dP_3})$, 
the total space of the canonical bundle over the third del Pezzo surface $dP_3$. 
The Hilbert series is
\begin{align}
g[\mathrm{Tot}(K_{dP_3})](t)
&=\frac{1+t^2-8t^4+8t^6-t^8-t^{10}}{(1-t^2)^6}
\nonumber\\
&=\frac{3}{4\beta^3}+\frac{3}{4\beta^2}+\frac{1}{2\beta}+\cdots. 
\end{align}
One finds
\begin{align}
\mathrm{Vol}(Y_{dP_3})&=\frac{2\pi^3}{9},& 
\int_{Y_{dP_3}} \mathrm{Riem}^2&=\frac{848\pi^3}{9}.  
\end{align}

\subsection{$Y_{PdP_{3a}}$}
The Calabi-Yau cone is $\mathbb{C}^3/\mathbb{Z}_6 (1,2,-3)$. 
The Hilbert series is
\begin{align}
g[\mathbb{C}^3/\mathbb{Z}_6 (1,2,-3)](t)
&=\frac{1+t^2+2t^{\frac83}+t^{\frac{10}{3}}+t^{\frac{16}{3}}}
{(1-t^{\frac43}) (1-t^2) (1-t^4)}
\nonumber\\
&=\frac{9}{16\beta^3}+\frac{9}{16\beta^2}+\frac{23}{48\beta}+\cdots. 
\end{align}
We have
\begin{align}
\mathrm{Vol}(Y_{PdP_{3a}})&=\frac{\pi^3}{6},& 
\int_{Y_{PdP_{3a}}} \mathrm{Riem}^2&=124\pi^3.  
\end{align}

\subsection{$Y_{PdP_{4a}}$}
The Calabi-Yau cone is $X=\mathrm{Tot}(K_{PdP_{4a}})$, 
the total space of the canonical bundle over the fourth Pseudo del Pezzo surface $PdP_{4a}$. 
The Hilbert series takes the form 
\begin{align}
&
g[\mathrm{Tot}(K_{PdP_{4a}})](t)
\nonumber\\
&=\frac{1}{(1-t^{2r_1+r_2+2r_3}) (1-t^{2r_1+2r_2+r_4}) (1-t^{r_1+3r_3+r_5}) (1-t^{2r_3+r_4+2r_5}) (1-t^{r_2+2r_4+2r_5})}
\nonumber\\
&\times (1+t^{r_1+r_2+r_3+r_4+r_5}-t^{2(r_1+r_2+r_3+r_4+r_5)}-t^{3r_1+2r_2+3r_3+r_4+r_5}
\nonumber\\
&-t^{2r_1+r_2+4r_3+r_4+2r_5}-t^{r_1+r_2+3r_3+2r_4+3r_5}+t^{3r_1+2r_2+5r_3+2r_4+3r_5}
+t^{4r_1+3r_2+6r_3+3r_4+4r_5}), 
\end{align}
where 
\begin{align}
r_1&=r_5=\frac{2(3+b_1)(3-2b_1)}{3(15-b_1)}, \\
r_2&=r_4=\frac{2(3+b_1)^2}{3(15-b_1)}, \\
r_3&=\frac{2(3-b_1)(3-2b_1)}{3(15-b_1)}, 
\end{align}
and $b_1$ is numerically computed as (\ref{Reeb_PdP4a}). 
It can be expanded as
\begin{align}
&
g[\mathrm{Tot}(K_{PdP_{4a}})](t)
\nonumber\\
&=\frac{27(15-b_1)}{8(3-b_1)(3+b_1)^2(3-2b_1)\beta^3}
+\frac{27(15-b_1)}{8(3-b_1)(3+b_1)^2(3-2b_1)\beta^2}
\nonumber\\
&+\frac{3(b_1^3-11b_1^2-24b_1+108)}{8(3-b_1)(3+b_1)^2(3-2b_1)\beta}+\cdots. 
\end{align}
We obtain
\begin{align}
\mathrm{Vol}(Y_{PdP_{4a}})
&=\frac{15-b_1}{(3-b_1)(3+b_1)^2(3-2b_1)}\pi^3=5.46637\ldots, \\
\int_{Y_{PdP_{4a}}} \mathrm{Riem}^2
&=\frac{8\pi^3(24b_1^3-264b_1^2-485b_1+1227)}{(3-b_1)(3+b_1)^2(3-2b_1)}\pi^3=3900.22\ldots.
\end{align}

\subsection{$Y_{PdP_{4b}}$}
We consider the Calabi-Yau cone $X=\mathrm{Tot}(K_{PdP_{4b}})$, 
the total space of the canonical bundle over the fourth Pseudo del Pezzo surface $PdP_{4b}$. 
The Hilbert series takes the form 
\begin{align}
&
g[\mathrm{Tot}(K_{PdP_{4b}})](t)
\nonumber\\
&=\frac{1+t^{r_1+r_2+r_3+r_4}-t^{2r_1+4r_2+r_3+r_4}+t^{2r_2+2r_3+r_4}-t^{r_1+5r_2+2r_3+r_4}-t^{2r_1+6r_2+3r_3+2r_4}}
{(1-t^{r_1+3r_2}) (1-t^{4r_2+r_3}) (1-t^{2r_1+r_4}) (1-t^{3r_3+2r_4})}, 
\end{align}
where 
\begin{align}
r_1&=\frac{4b_1b_2}{3(9-b_1+3b_2)}, \\
r_2&=\frac{2b_2(3-b_1+b_2)}{3(9-b_1+3b_2)}, \\
r_3&=\frac{2(3-b_1+b_2)(9-b_1-b_2)}{3(9-b_1+3b_2)}, \\
r_4&=\frac{2b_1(9-b_1-b_2)}{3(9-b_1+3b_2)}, 
\end{align}
and $b_1$ and $b_2$ are numerically computed as (\ref{Reeb_PdP4b}). 
We have 
\begin{align}
&
g[\mathrm{Tot}(K_{PdP_{4b}})](t)
\nonumber\\
&=\frac{27(9-b_1+3b_2)}{8b_1b_2(3-b_1+b_2)(9-b_1-b_2)\beta^3}
+\frac{27(9-b_1+3b_2)}{8b_1b_2(3-b_1+b_2)(9-b_1-b_2)\beta^2}
\nonumber\\
&+\frac{b_1^3-3b_2^3-12b_1^2+18b_2^2-15b_1b_2+18b_1+108b_2+81}
{8b_1b_2(3-b_1+b_2)(9-b_1-b_2)\beta}+\cdots. 
\end{align}
One finds
\begin{align}
\mathrm{Vol}(Y_{PdP_{4b}})
&=\frac{9-b_1+3b_2}{b_1b_2(3-b_1+b_2)(9-b_1-b_2)}\pi^3
=4.87877\ldots, \\
\int_{Y_{PdP_{4b}}} \mathrm{Riem}^2
&=\frac{64b_1^3-192b_2^3-768b_1^2+1152b_2^2-960b_1b_2+1880b_1+4728b_2-1368}
{b_1b_2(3-b_1+b_2)(9-b_1-b_2)}\pi^3
\nonumber\\
&=4310.57\ldots.
\end{align}

\subsection{$Y_{PdP_5}$}
The Calabi-Yau cone is the $\mathbb{Z}_2\times \mathbb{Z}_2$ orbifold of the conifold $\mathcal{C}/\mathbb{Z}_2\times \mathbb{Z}_2$ $(1,0,0,1)(0,1,1,0)$. 
The Hilbert series is
\begin{align}
g[\mathcal{C}/\mathbb{Z}_2\times \mathbb{Z}_2](t)
&=\frac{(1-t^4)^2}{(1-t^2)^5}
=\frac{1}{2\beta^3}+\frac{1}{2\beta^2}+\frac{1}{2\beta}+\cdots. 
\end{align}
We have
\begin{align}
\mathrm{Vol}(Y_{PdP_5})&=\frac{4\pi^3}{27},& 
\int_{Y_{PdP_5}} \mathrm{Riem}^2&=\frac{4000\pi^3}{27}.  
\end{align}

\subsection{$X^{2,2}$}
The Calabi-Yau cone is $X=\mathrm{Tot}(K_{PdP_2})$, 
the total space of the canonical bundle over the second Pseudo del Pezzo surface $PdP_2$. 
The Hilbert series takes the form 
\begin{align}
&
g[\mathrm{Tot}(K_{PdP_2})](t)
\nonumber\\
&=\frac{1}{(1-t^{3r_1+r_2}) (1-t^{4r_2+3r_3}) (1-t^{2r_1+r_4}) (1-t^{r_3+2r_4})}
\nonumber\\
&\times 
(1+t^{2r_1+2r_2+r_3}+t^{r_1+3r_2+2r_3}+t^{r_1+r_2+r_3+r_4}-t^{2(r_1+r_2+r_3+r_4)}
\nonumber\\
&-t^{4r_1+2r_2+r_3+r_4}+t^{2r_2+2r_3+r_4}-t^{3r_1+3r_2+2r_3+r_4}
-t^{3r_1+r_2+r_3+2r_4}-t^{4r_1+4r_2+3r_3+2r_4}
), 
\end{align}
where 
\begin{align}
r_1&=\frac{4b_1(3+b_1+b_2)}{36+12b_1-3b_2}, \\
r_2&=\frac{2(3+b_1-b_2)(12-2b_1-b_2)}{36+12b_1-3b_2}, \\
r_3&=\frac{2b_2(12-2b_1-b_2)}{36+12b_1-3b_2}, \\
r_4&=\frac{2b_1b_2}{12+4b_1-b_2}, 
\end{align}
and the parameters $b_1$ and $b_2$ are numerically evaluated as (\ref{Reeb_X22}). 
We find
\begin{align}
&
g[\mathrm{Tot}(K_{PdP_2})](t)
\nonumber\\
&=\frac{27(12+4b_1-b_2)}{8b_1b_2(3+b_1-b_2)(12-2b_1-b_2)\beta^3}
+\frac{27(12+4b_1-b_2)}{8b_1b_2(3+b_1-b_2)(12-2b_1-b_2)\beta^2}
\nonumber\\
&+\frac{4b_1^3-b_2^3+b_1^2b_2-b_1b_2^2-12b_1^2+15b_2^2+9b_1b_2-108b_1-27b_2-108}
{8b_1b_2(3+b_1-b_2)(12-2b_1-b_2)\beta}+\cdots, 
\end{align}
\begin{align}
\mathrm{Vol}(X^{2,2})
&=\frac{(12+4b_1-b_2)\pi^3}{b_1b_2(3+b_1-b_2)(12-2b_1-b_2)}=6.98084\ldots, \\
\int_{X^{2,2}}\mathrm{Riem}^2
&=\frac{\pi^3}{b_1b_2(3+b_1-b_2)(12-2b_1-b_2)}
(-256b_1^3+64b_2^3
\nonumber\\
&-64b_1^2b_2+64b_1b_2^2+768b_1^2-960b_2^2-576b_1b_2+4000b_1+2456b_2-1824
)
\nonumber\\
&=2580.78\ldots. 
\end{align}

\subsection{$Z^{2,2}$}
The Calabi-Yau cone is $X=\mathrm{Tot}(K_{PdP_{3b}})$, 
the total space of the canonical bundle over the third Pseudo del Pezzo surface $PdP_{3b}$. 
The Hilbert series is evaluated as
\begin{align}
g[\mathrm{Tot}(K_{PdP_{3b}})](t)
&=\frac{1+2t^2-t^4-2t^{2\sqrt{5}}+t^{2\sqrt{5}-2}-t^{2+2\sqrt{5}}}
{(1-t^2)(1-t^{2\sqrt{5}-2})^2(1-t^{6-2\sqrt{5}})}
\nonumber\\
&=\frac{11+5\sqrt{5}}{32\beta^3}
+\frac{11+5\sqrt{5}}{32\beta^2}
+\frac{3(3+\sqrt{5})}{32\beta}+\cdots. 
\end{align}
One obtains
\begin{align}
\mathrm{Vol}(Z^{2,2})&=\frac{11+5\sqrt{5}}{108}\pi^3,& 
\int_{Z^{2,2}} \mathrm{Riem}^2&=\frac{2(943+193\sqrt{5})}{27}\pi^3.  
\end{align}

\section{Curvature-squared integrals}

\subsection{Examples}
\label{app_Riem2}
In order to facilitate a quantitative comparison among Sasaki-Einstein geometries, 
we collect below the total volume and curvature-squared invariants, 
which are evaluated from the combinatorial formulae (\ref{vol_Y1}), (\ref{vol_Y2}), (\ref{Riem2_conj}) as well as (\ref{Riem2_conj2}). 
\begin{table}[H]
\centering
\begin{tabular}{|c|c|c|}
\hline
$Y$ & $\mathrm{Vol}(Y)$ & $\int_Y \mathrm{Riem}^2$ \\ \hline \hline 
$S^5$ & $\pi^3$ & $40\pi^3$ \\ \hline
$\mathbb{RP}^5$ & $\frac{\pi^3}{2}$ & $20\pi^3$ \\ \hline
$S^5/\mathbb{Z}_2$ & $\frac{\pi^3}{2}$ & $116\pi^3$ \\ \hline
$(S^5/\mathbb{Z}_k)_{\textrm{ALE}}$ & $\frac{\pi^3}{k}$ & $\frac{8(8k^2-3)\pi^3}{k}$ \\ \hline
$S^5/\mathbb{Z}_{2n}$ & $\frac{\pi^3}{2n}$ & $\frac{116\pi^3}{n}$ \\ \hline
$S^5/\mathbb{Z}_{2n-1}$ & $\frac{\pi^3}{2n-1}$ & $\frac{40\pi^3}{2n-1}$ \\ \hline
$T^{1,1}$ & $\frac{16\pi^3}{27}$ &$\frac{2176\pi^3}{27}$ \\ \hline
$Y^{\mathsf{p},\mathsf{q}}$ 
& $\frac{\mathsf{q}^2(2\mathsf{p}+\sqrt{4\mathsf{p}^2-3\mathsf{q}^2})\pi^3}{3\mathsf{p}^2(3\mathsf{q}^2-2\mathsf{p}^2+\mathsf{p}\sqrt{4\mathsf{p}^2-3\mathsf{q}^2})}$ 
& $\frac{8\mathsf{q}^2(106\mathsf{p}-19\sqrt{4\mathsf{p}^2-3\mathsf{q}^2})\pi^3}{3\mathsf{p}^2(3\mathsf{q}^2-2\mathsf{p}^2+\mathsf{p}\sqrt{4\mathsf{p}^2-3\mathsf{q}^2})}$ \\ \hline 
$L^{1,2,1}$ & $\frac{2\sqrt{3}\pi^3}{9}$ & $\frac{16(108-19\sqrt{3})\pi^3}{9}$ \\ \hline
$L^{a,b,a}$ & (\ref{vol_Laba})
& (\ref{Riem2_Laba})
\\ \hline 
$L^{a,b,c}$ &(\ref{vol_Labc})& (\ref{Riem2_Labc}) \\ \hline
$L^{1,3,1}/\mathbb{Z}_2$ & $\frac{4\pi^3}{10+7\sqrt{7}}$ & $\frac{32(2242-403\sqrt{7})\pi^3}{243}$ \\ \hline 
$Y_{dP_1}=Y^{2,1}$& $\frac{(46+13\sqrt{13})\pi^3}{324}$ & $\frac{2(566+329\sqrt{13})\pi^3}{81}$ \\ \hline 
$Y_{PdP_1}=S^5/\mathbb{Z}_4$& $\frac{\pi^3}{4}$ & $58\pi^3$ \\ \hline 
$Y_{dP_2}$ & $\frac{59+11\sqrt{33}}{486}\pi^3$ & $\frac{4(2551+295\sqrt{33})\pi^3}{243}$ \\ \hline 
$Y_{PdP_2}=X^{2,2}$ & (\ref{vol_X22}) & (\ref{Riem2_X22}) \\ \hline 
$Y_{dP_3}$ & $\frac{2\pi^3}{9}$ & $\frac{848\pi^3}{9}$ \\ \hline 
$Y_{PdP_{3a}}$ & $\frac{\pi^3}{6}$ & $124\pi^3$ \\ \hline 
$Y_{PdP_{3b}}=Z^{2,2}$ & $\frac{(11+5\sqrt{5})\pi^3}{108}$ & $\frac{2(943+193\sqrt{5})\pi^3}{27}$ \\ \hline 
$Y_{PdP_{3c}}=L^{1,2,1}/\mathbb{Z}_2$ & $\frac{\sqrt{3}\pi^3}{9}$ & $\frac{8(216-55\sqrt{3})\pi^3}{9}$ \\ \hline 
$Y_{PdP_{4a}}$ & (\ref{vol_PdP4a}) & (\ref{Riem2_PdP4a}) \\ \hline 
$Y_{PdP_{4b}}$ & (\ref{vol_PdP4b}) & (\ref{Riem2_PdP4b}) \\ \hline 
$Y_{PdP_{5}}$ & $\frac{4\pi^3}{27}$ & $\frac{4000\pi^3}{27}$ \\ \hline 
$X^{\mathsf{p},\mathsf{q}}$ &(\ref{vol_Xpq})& (\ref{Riem2_Xpq}) \\ \hline
$Z^{\mathsf{p},\mathsf{q}}$ &(\ref{vol_Zpq})& (\ref{Riem2_Zpq}) \\ \hline
\end{tabular}
\caption{Volumes and curvature squared integrals for Sasaki-Einstein manifolds $Y$.}
\end{table}

\subsection{Towards a proof}
\label{app_proof}
The combinatorial formula (\ref{Riem2_conj}) for the integrated Riemann tensor squared was obtained above 
by combining the residue formula (\ref{Res_formula}) with the conjecture (\ref{ALS_conj}). 
In this appendix, we indicate how it can also be established directly, for a compact
\textit{smooth} toric Sasaki-Einstein 5-manifold $Y$, without invoking (\ref{ALS_conj}). 
In terms of the toric data of section \ref{sec_SE}, smoothness means that no lattice point lies in the relative interior of an edge of the toric diagram, 
equivalently that the $d$ generators $m_I$ of the dual cone are pairwise distinct. 
The restrictions this leaves are discussed at the end, and they are the reason we continue to regard (\ref{Riem2_conj2}) as a proposal.
The skeleton of this partial proof proceeds through the following chain of relations:
\begin{align}
\begin{aligned}
\vert\mathrm{Riem}_{g}\vert^2&= \vert\mathrm{Riem}^T\vert^2 - 152
&& \text{(Step 1: 5d-4d reduction along the Reeb direction)} \\
\vert\mathrm{Riem}^T\vert^2 d\mu_g&= 32\pi^2\, \eta \wedge P
&& \text{(Step 2: transverse Gauss-Bonnet)} \\
I(b) &:= \int_Y \eta \wedge P
&& \text{(Step 3: foliation invariance)} \\
I(b) &= 2\pi \sum_{I=1}^d \frac{1}{(b,m_I)}
&& \text{(Step 4: Satake-Gauss-Bonnet theorem)}
\end{aligned}
\end{align}
Below, we detail each step. 

\begin{enumerate}

\item \textbf{5d-4d reduction along the Reeb direction}

Pointwise on $Y$, the 5d curvature reduces to a 4d transverse curvature up to a constant. 
The 5d Sasaki-Einstein manifold $Y$ is decomposed into a 4d transverse base space and a 1d Reeb vector field $\xi$. 
Writing $\mathcal{D}=\ker\eta$ and $d\eta=2\omega$, 
the O'Neill formula \cite{MR200865} relates $\mathrm{Riem}_g$ restricted to $\mathcal{D}$ to the transverse curvature $\mathrm{Riem}^T$. 
The components along $\xi$ are fixed by $\mathrm{Riem}(X,Y)\xi=0$ and $\mathrm{Riem}(e_a,\xi,\xi,e_b)=g^T_{ab}$, 
where we have a local orthonormal frame $\{e_0 = \xi, e_a\}$, $a=1,...,4$ with $e_a$ horizontal. 
Contracting, using $\mathrm{Ric}^T=6g^T$, and adding the four components with two indices along $\xi$ gives the Sasakian identity 
\begin{align}
\vert{}\mathrm{Riem}_{g}\vert{}^2 = \vert{}\mathrm{Riem}^T\vert{}^2 - 152
\end{align}
Integrating this over $Y$ yields
\begin{align}
\int_Y \mathrm{Riem}^2 = \int_Y \vert{}\mathrm{Riem}^T\vert{}^2 \, d\mu_g - 152 \, \mathrm{Vol}(Y), 
\end{align}
which already accounts for the constant coefficient in the formula (\ref{Riem2_conj}). 
A closely related computation on the metric cone appears as eq.(A.16) of \cite{Eager:2010dk}.

\item \textbf{Transverse Gauss-Bonnet} 

The transverse curvature-squared density is exactly the transverse Gauss-Bonnet density. 
Generally, the Euler density $E^T$ is 
\begin{align}
E^T = \vert{}\mathrm{Riem}^T\vert{}^2 - 4\vert{}\mathrm{Ric}^T\vert{}^2 + (R^T)^2.
\end{align}
In four transverse dimensions, the privileged K\"{a}hler-Einstein condition forces $\mathrm{Ric}^T=6g^T$, 
which fixes $\vert{}\mathrm{Ric}^T\vert{}^2=144$ and $R^T=24$. 
This leads to a precise cancellation: $-4(144)+(24)^2=0$, 
meaning $\vert{}\mathrm{Riem}^T\vert{}^2 = E^T$. 
This explains why a curvature-squared integral can be combinatorial at all. 
The dynamical curvature integrand collapses into the Euler density of the transverse geometry. 

\item \textbf{Foliation invariance} 

The integral 
\begin{align}
I(b) := \int_Y \eta\wedge P, \qquad P=\frac{1}{32\pi^2}E^T d\mu^T
\end{align}
depends only on the Reeb foliation, not on the transverse metric. 
Here $d\mu_g=\eta\wedge d\mu^T$, so that 
$\int_Y \vert\mathrm{Riem}^T\vert^2 d\mu_g = 32\pi^2 I(b)$.
By transverse Chern-Weil theory (see e.g. \cite{MR2382957}), 
$P$ is a closed basic $4$-form whose basic cohomology class is independent of the compatible transverse metric. 
A change in the transverse metric merely shifts $P$ by an exact basic form $d\beta$. 
Then the induced change in $I(b)$ is $\int_Y \eta\wedge d\beta = 2\int_Y \omega\wedge\beta$,
which vanishes because $\omega\wedge\beta$ is a basic form of transverse degree five on a foliation of transverse dimension four.
Thus $I(b)$ is defined for every $b$ in the Reeb cone, independent of the choice of compatible transverse metric, 
hence computable without knowing the Sasaki-Einstein metric explicitly. 

\item  \textbf{Satake-Gauss-Bonnet theorem}

Equivariant localization reduces the integral to a sum over closed Reeb orbits. 
For a quasi-regular $b$, $Y$ is a Seifert fibration over a compact transverse K\"{a}hler orbifold $B$. 
The Satake-Gauss-Bonnet theorem \cite{MR95520} gives $\int_B P = \chi_{\rm orb}(B)$. 
On the toric surface $B$, only the $d$ fixed points of the transverse torus action contribute to the Euler characteristic, yielding
\begin{align}
32\pi^2 I(b)= 64\pi^3 \sum_I \frac{1}{(b,m_I)}, \qquad \ell_I = \frac{2\pi}{(b,m_I)},
\end{align}
where $\ell_I$ is the period of the closed Reeb orbit $O_I$ lying over the $I$-th one-dimensional face of the moment cone. 
Physically, these orbits are the point-like BPS geodesics dual to the extremal BPS mesons \cite{Benvenuti:2005ja,Benvenuti:2005cz}. 
Because both sides of the equation are homogeneous of degree $-1$ under $b \to \lambda b$ and continuous on the Reeb cone, 
and quasi-regular directions are dense there, 
this formula holds for all $b$ by continuity, including the generically irregular critical Reeb vector, 
for which the Reeb orbits do not close and the Satake-Gauss-Bonnet theorem does not apply directly. 
Combining all four steps with the volume formula (\ref{vol_Y1}) reproduces (\ref{Riem2_conj}). 

\end{enumerate}

A restriction should be emphasized to clarify what is established and what remains open. 
Smoothness enters step 4 through the identification of the $d$ generators $m_I$ with distinct closed Reeb orbits. 
For an orbifold several $m_I$ coincide, the multiplicity of a given ray being the number of primitive segments on the corresponding edge of the toric diagram, 
and these multiplicities carry precisely the curvature localized on the singular strata. 
A derivation in the orbifold case requires a Kawasaki-type extension of steps 3 and 4, which we do not attempt here. 
This applies to every example in which the $m_I$ fail to be pairwise distinct, 
equivalently in which the toric diagram has a lattice point in the relative interior of an edge. 
Among the geometries treated here these are $S^5/\mathbb{Z}_2$ and $(S^5/\mathbb{Z}_k)_{\textrm{ALE}}$ of section \ref{sec_symSE}, 
$S^5/\mathbb{Z}_{2n}$ of section \ref{sec_orb}, $L^{1,2,1}$, $L^{a,b,a}$ and $L^{2,6,3}$ of section \ref{sec_Labc}, 
the quotients of section \ref{sec_orbLabc}, the geometries associated with the Pseudo del Pezzo surfaces in section \ref{sec_dP}, 
and the members $X^{\mathsf{p},\mathsf{p}}$ and $Z^{\mathsf{p},\mathsf{p}}$ of the families of section \ref{sec_XZ}. 
These provide independent verifications instead.

\bibliographystyle{utphys}
\bibliography{ref}

\end{document}